\documentclass[12pt,twoside,doublespace]{mitthesis}
\usepackage{graphicx}
\usepackage{avant}
\usepackage{lgrind}
\usepackage{graphics}
\usepackage{psfrag}
\begin{document}

\newcommand{\gc}{\ensuremath{\Gamma_{\mathrm c}}}
\newcommand{\tc}{\ensuremath{T_{\mathrm c}}}
\newcommand{\ga}{\ensuremath{\Gamma_{\mathrm a}}}
\newcommand{\tint}{\ensuremath{t_{\rm int}}}
\newcommand{\us}{\ensuremath{\mu{\rm sec}}}
\newcommand{\um}{\ensuremath{\mu{\rm m}}}
\newcommand{\uw}{\ensuremath{\mu{\rm W}}}
\newcommand{\ba}{\mbox{\ensuremath{^{138}{\rm Ba}}}}
\newcommand{\transitiontriplet}{\mbox{\ensuremath{^1{\rm S}_0\leftrightarrow\,\!^3{\rm P}_1}}}
\newcommand{\transitionsinglet}{\mbox{\ensuremath{^1{\rm S}_0\leftrightarrow\,\!^1{\rm P}_1}}}
\newcommand{\be}{\begin{equation}}
\newcommand{\ee}{\end{equation}}
\newcommand{\bfig}{\begin{figure}}
\newcommand{\efig}{\end{figure}}
\newcommand{\f}{\ensuremath{\mathcal{F}}}
\newcommand{\fsr}{\ensuremath{\mathrm FSR}}
\newcommand{\gtwo}{\ensuremath{g^{(2)}}}
\newcommand{\sno}{791 nm}
\newcommand{\fft}{553 nm}
\newcommand{\Nth}{\ensuremath{N_{\rm th}}}
\newcommand{\nth}{\ensuremath{n_{\rm th}}}
\newcommand{\nex}{\ensuremath{n_{\rm ex}}}
\newcommand{\Nex}{\ensuremath{N_{\rm ex}}}
\newcommand{\Neff}{\ensuremath{N_{\rm eff}}}
\newcommand{\sngt}{\ensuremath{\sqrt{n+1}\ g \tint}}
\newcommand{\mhz}{\ensuremath{{\rm MHz}}}
\newcommand{\khz}{\ensuremath{{\rm kHz}}}
\newcommand{\tem}{\ensuremath{{\rm TEM}_{00}}}
\newcommand{\ocav}{\ensuremath{\omega_{\rm cav}}}
\setlength{\unitlength}{1in}

\title{Multiple Thresholds and Many-Atom Dynamics in the Cavity QED Microlaser}
\author{Christopher Minwah Fang-Yen}
\prevdegrees{B.S. Physics and Mathematics\\ Stanford University, 1995}
\department{Department of Physics}
\degree{Doctor of Philosophy}
\degreemonth{February}
\degreeyear{2002}
\thesisdate{February 6, 2002}


\supervisor{Michael S. Feld}{Professor of Physics}


\chairman{Thomas J.~Greytak}
{Professor of Physics\\Associate Department Head for Education}

\maketitle



\cleardoublepage
\setcounter{savepage}{\thepage}
\begin{abstractpage}

%

This thesis describes a study of a cavity QED microlaser in which many
atoms are present simultaneously and atom-cavity interaction is
well-defined.
%
%
%
The microlaser is found to display multiple thresholds analogous to
first-order phase transitions of the cavity field.  Hysteresis is
observed as a function of atom-cavity detuning and number of atoms.
Data is compared with a rate equation model and fully quantized
treatment based on micromaser theory.  Good agreement between theory
and experiment is found when the cavity is resonant with atoms of the
most probable velocity, but long lifetimes of metastable states
preclude the observation of true steady-state transition points.  For
nonzero atom-cavity detuning the microlaser displays broadenings and
shifts which are not yet well-understood.

%

Quantum trajectory simulations are performed to investigate many-atom
and finite transit time effects in the microlaser.  We show that over
a wide range of parameters the many-atom microlaser scales with the
single-atom theory, with a perturbation in the photon statistics due
to cavity decay during the atom transit time.  
%

\end{abstractpage}

\cleardoublepage

\section*{Acknowledgments}



It is a pleasure to thank the many people who contributed directly or
indirectly to this work.  First I thank Michael Feld for his guidance
and support during my time as a graduate student.  His intuitive
approach to physics and optimistic attitude will be an inspiration to
me for years to come.  I also appreciate the freedom he has given me
as a graduate student to set the direction for my research.



This thesis would not have been possible without Kyungwon An, who
initiated the microlaser experiment as a graduate student.  
Kyungwon taught me a great deal about experimental physics during my
early years at MIT.

An equally large amount of credit for this thesis should go to
Chung-Chieh Yu, who worked on the project for three years.
His expertise in quantum optics and atomic beams helped move the
project along faster than ever before.

Ramachandra Dasari deserves thanks for many useful discussions, and
for managing lab equipment and funding issues.  From the beginning
Ramachandra expressed confidence in me and my ability to eventually
take on a leading role in the project.  This is something I have
appreciated very much.

I enjoyed working with several other students in the laboratory.
Abdulaziz Aljalal developed the optical velocity selection scheme and
second-order correlation experiment.  Bryndol Sones performed the
absorption-induced bistability experiments.  Alan Heins helped
construct the dye laser locking system and the supersonic oven.
Sangkeun Ha designed several of the electronic circuits and assisted
with data collection.

Professor John Thomas of Duke University deserves special thanks for
helping us develop the supersonic beam oven that has proved critical
to the success of these experiments.

Thanks to Steve Smith of Coherent who often went out of his way to
keep the laboratory equipped with working lasers.

I thank the members of my thesis committee, Marlan Scully, Daniel
Kleppner, and Erich Ippen, for their helpful questions and
suggestions.

This research was funded by a grant from the National Science
Foundation.  My research assistantship has been supported by a Lester
Wolfe Fellowship.

Thanks to all my previous teachers and mentors who have inspired me to
pursue a career in science, especially David Witt, Michael Russelle,
Doug Osheroff, and Mark Dykman.



I dedicate this thesis to my parents, John and Rosemay.

\begin{flushright} C.\,F.\,Y. \end{flushright}

\pagestyle{plain}
\tableofcontents
\listoffigures
\mbox{}
\newpage
\mbox{}

\chapter{Introduction}
\label{chap-intro}


\section{Cavity QED and the microlaser/micromaser}

One of the simplest models for light-matter interaction consists of a
single two-level atom interacting with a single electromagnetic field
mode of a cavity.  If the atom-cavity coupling is stronger than the
atomic or photon decay rates, the irreversible decay of an excited
atom is replaced by an oscillatory exchange of energy between the atom
and cavity, in a manner characteristic of two coupled oscillators.


In the 1950's this model was the subject of considerable theoretical
study,
primarily in connection with the then-recently invented maser
\cite{J-C-IEEE}.  Experimental realization of such a strongly
coupled system would not occur until several decades later, however,
with advances in Rydberg states and superconducting microwave
cavities.


In the early 1980's the micromaser, or single-atom maser, was invented
by Walther and collaborators \cite{Meschede-PRL85}.  In this
experiment, individual Rydberg atoms pass through and emit photons
into a superconducting microwave cavity with extremely high quality
factor.  The micromaser formed a close approximation to the
single-atom, single-mode ideal.  Subsequent experiments showed that
the micromaser exhibited a variety of interesting and unusual
phenomena, including trap states, near-number states, bistability, and
nonclassical statistics \cite{trap-state-OL88}-\cite{varcoe-nature00}.

The recent development of multilayer dielectric mirrors with very low
loss has made it possible to conduct similar experiments at optical
wavelengths.  In 1994 Kyungwon An and Michael Feld developed the
microlaser, the optical analogue of the micromaser~\cite{An-PRL94,
an-thesis}.  This device is also known as the cavity QED microlaser in
order to distinguish it from other microlasers, for example those using
semiconductor microresonators.

This thesis describes the first experiments exploring the microlaser's
truly unusual properties.  In particular, we show that in contrast to
conventional lasers which display a single laser threshold, the
microlaser exhibits multiple thresholds.

\section{Microlaser concept}

The basic scheme of the microlaser is illustrated in
Fig.~\ref{fig-orig-microlaser}.  A beam of two-level atoms pass
through a resonant high-finesse optical cavity.  Before entering the
cavity mode each atom is excited to its upper state via a laser
$\pi$-pulse.  The atom interacts with the mode for a time \tint,
during which it may emit a photon into the cavity.

The key characteristic of the microlaser is not only the strong
atom-cavity coupling but that different atoms experience almost the
same interaction with the cavity: both the atom-cavity coupling
strength $g$ and the atom-cavity interaction time
\tint\ are well-defined.
It will be shown in chapter \ref{chap-theory} that the resulting
emission probability can be written as \be P_{\rm emit} =
\sin^2(\sqrt{n+1}\ g\,\tint) \ee
\noindent where $n$ is the number of photons present in the cavity prior
to the entrance of the atom.
The unusual properties of the microlaser are primarily due to
this sinusoidal dependence of the atomic emission probability on the
cavity photon number.




\bfig
\centerline{\resizebox{\textwidth}{!}{\includegraphics{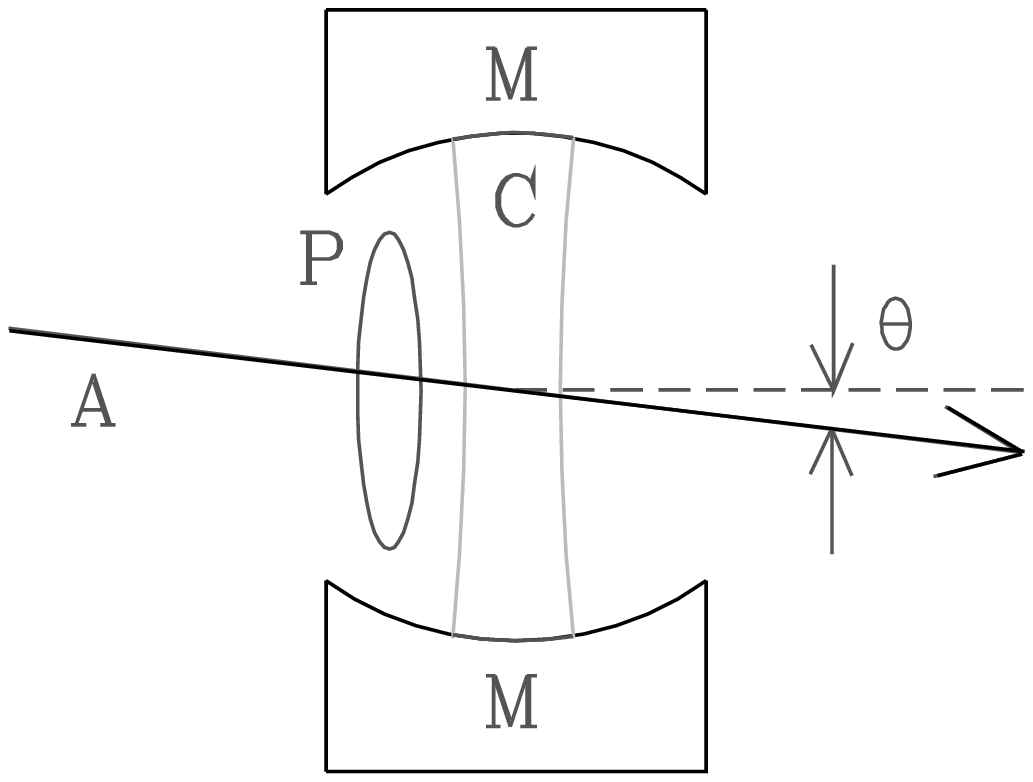}}}
\caption{Schematic of cavity QED microlaser.  A: atomic beam, M:
mirrors, C: cavity mode, P: pump field, $\theta$: cavity tilt angle
(cf. \cite{An-OL97b}).  }
\label{fig-orig-microlaser}
\efig

\section{Previous work}

\subsection{Micromaser}
Reviews of the micromaser can be found in a number of references
in the bibliography.
Most relevant for this experiment is the study by Benson et al
\cite{benson-prl94} of dynamic behavior near the micromaser field
phase transitions.
Spontaneous jumps and slow transitions between metastable states are
observed; hysteresis is seen as interaction times are varied.  The
results were explained by quantum Monte Carlo simulations and a
Fokker-Planck description of the micromaser dynamics.  The presence of
small stray fields prevented a full quantitative agreement between
theory and experiment from being reached; the calculated transition
rates were generally much slower than what was observed
experimentally.

%

\subsection{Microlaser vs. micromaser}

We comment here on how the microlaser differs from its microwave
counterpart.

\subsubsection{Many atoms required}

The threshold atom number as derived in Chapter \ref{chap-theory} is
\be \Nth = {{\gc} \over {g^2 \tint}} \ee 

For our current microlaser with supersonic beam we have $\gc \approx
180$ \khz, $\tint \approx 0.1 \us$, and $g \approx 193 \khz$, giving
$\Nth \approx 10$.

This is to be compared with the micromaser for which typical figures
are $\gc \approx 30$~Hz, $\tint \approx 47 \us$, and $g \approx 40$
kHz (values from \cite{benson-prl94}), so threshold atom number $\Nth
\approx 3 \times 10^{-4}$, almost five orders of magnitude lower.  

One consequence of the many-atom nature is that the trap state
resonances, corresponding to situations in which atoms perform an
integer number of entire Rabi oscillations, cannot observed; nor can
the associated near-number states of the micromaser \cite{varcoe-nature00}.
However, other interesting effects appear from the many-atom case.
The many atom nature of the current microlaser is treated in detail in
Chapter \ref{chap-simulations}.

\subsubsection{No thermal photons}

At finite temperature every field mode has a nonzero photon occupancy
\nth\ given by the Planck formula.  For the micromaser operating at
$\sim 20 $GHz, blackbody photons \nth\ is as high as 0.15 even at 0.5
K and have a significant effect on the micromaser, as discussed in
many micromaser articles.  In the microlaser at 300K, $\nth \sim
10^{-25}$ and can be ignored.

\subsubsection{Constant transit time}

Micromaser experiments are typically performed with a fixed atomic
injection rate and variable interaction times.  This roughly
corresponds to varying the pump parameter $\theta$ as described in
micromaser theory (see Chapter \ref{chap-theory}).  

Use of a supersonic beam of the microlaser prevents any significant
variation in the transit times.  Instead, the parameter most easily
varied is the number of intracavity atoms $N$.  As $N$ is varied, both
the expectation photon number \nex\ and the pump parameter $\theta$
change.

\subsubsection{Direct detection of photons}

Optical frequency photons can be easily outcoupled out of a cavity,
and have sufficiently high energy to be directly individually via a
photomultiplier tube or avalanche photodiode.  This is in contrast to
the micromaser experiments, in which maser oscillation must be
inferred via observation of the atom states.
Direct detection of photons opens up new possibilities for photon
statistics or laser linewidth experiments.

\subsubsection{Atom-cavity coupling}

The centimeter-scale wavelengths used in the micromaser allow the
atomic beam to largely avoid coupling nonuniformities due to mode
structure.  In the micromaser, standing-wave coupling variation
requires us to employ a traveling-wave interaction scheme (see Chapter
\ref{chap-methods}).  In addition, the small dimensions of the cavity mode 
and requirement of uniform coupling place strict requirements on the
position and geometry of the atomic beam coupled to it.  

\subsection{Optical cavity QED}

At optical wavelengths the experiments most similar to this one have
been spectroscopy and probe transmission experiments of high-finesse
optical cavities containing a small number of atoms.

Heinzen and Feld \cite{heinzen-prl1987a,heinzen-prl1987b} performed
the first observation of enhanced and supressed spontaneous emission
and level shifts in an optical cavity.

The vacuum normal-mode splitting of atoms in a cavity was observed 
in \cite{raizen-prl1989}.

Oscillatory transmission and nonclassical photon statistics of a
probed cavity was reported by the Kimble group \cite{Rempe-PRL91}.
Optical bistability was also observed.


To our knowledge, all these optical experiments employed atomic
excited states with lifetimes $\sim$ns much shorter than atom-cavity
transmit times.  They were performed in passive (non-laser) cavities
in which a probe beam was incident.  In addition, the atom-cavity
couplings were not well-determined due to standing-wave mode profiles
(cf.\ chapter \ref{chap-methods}).  

\subsection{Lasers and optical bistability}
Bistability in the microlaser bears some similarity to ``classical''
optical bistability (see, for example,
\cite{gibbs-optbistab}) in active and passive cavities.  
Multimode laser bistabilities due to mode competition in gas lasers
have been studied since the earliest days of the laser
\cite{lamb-pr64}.  Current-feedback bistability has been observed in
some semiconductor laser systems, e.g.\ \cite{lim-vcsel}.


\subsection{Phase transition analogies}

A detailed description of a laser is necessarily statistical in
nature, due to the presence of randomness in many forms and large
numbers of degrees of freedom.  
It is therefore natural to draw analogies between laser behavior and
other problems in statistical physics.
%
In particular, for some time physicists have drawn analogies between a
laser near threshold and matter near a second-order phase transition.
This comparison was first described in detail by DeGiorgio and Scully
in 1970
\cite{degiorgio-pra70, Scully-Lamb-LP}, although mention of this 
analogy date back considerably earlier (cf. references in ibid.).

In the present work we investigate jumps in the microlaser field which
can be shown to be analogous to {\it first-order} (discontinuous)
phase transitions \cite{Filipowicz-PRA86}.  To our knowledge no
similar behavior has been observed in any other laser.

\subsection{Many-atom dynamics in cavity QED}

The field of collective effects in radiative properties goes back at
least to Dicke \cite{Dicke54}, who showed that a collection of $N$
two-level radiators, localized within a distance smaller than their
radiative wavelength, may emit photons at a collective rate that
scales with $N^2$ instead of the usual $N$.  

Many-atom effects in the micromaser have been studied by a number of
authors (\cite{kolobov-pra97}-\cite{DAriano-PRL95}).  It is well-known
that the micromaser's trap state resonances are easily destroyed by
two-atom events (\cite{guevara-pra97} and \cite{orszag-pra94}).


A study by Kolobov and Haake
\cite{kolobov-pra97} of two-atom events predicts a change in average
photon number as well as location of the first-order thresholds,
relative to the single-atom theory.  On the other hand, Elk
\cite{Elk-PRA96}, considers a micromaser injected with multiple atoms,
equally spaced in time, and finds a scaling behavior very similar to
the single-atom theory.  D'Ariano \cite{DAriano-PRL95} considered a
micromaser pumped with clusters of up to several hundred atoms at
once, an interesting model but one that cannot be considered realistic
in the present context.

The issue of many-atom effects is central to this thesis and is
considered in detail in Chapter~\ref{chap-simulations}.  We find a
result similar to that of \cite{Elk-PRA96}, but we have a differing
interpretation of an observed extra variance.

\chapter{Theory of the microlaser}
\label{chap-theory}

We begin by considering the ideal model of a single atom in a cavity.
The microlaser is introduced by an semiclassical rate equation model
which describes the basic features of the microlaser in an intuitive
way.  We then give a fully quantized treatment following the density
matrix approach \cite{scully-zubairy-quantum-optics}.  A Fokker-Planck
analysis from \cite{Filipowicz-PRA86} is reviewed which reproduces the
rate equation solutions and also describes the transitions between
different solutions in terms of an effective potential.



\section{Atom in a cavity}

The Hamiltonian describing the interaction of a radiation field with a
single-electron atom is
\be \mathcal{H} = \mathcal{H}_A + \mathcal{H}_F - e {\bf r} \cdot {\bf E} 
\ee
where $\mathcal{H}_A$ and $\mathcal{H}_F$ are the atom and radiation
field energies, $\bf r$ is the position of the electron, and $\bf E$
is the electric field operator:

\be
\mathcal{H}_F = \sum_{\bf k} \hbar \omega_{\bf k} a_{\bf k}^\dag
a_{\bf k}
\ee

\be
\mathcal{H}_A = {1 \over 2} \hbar \omega_a \sigma_z 
\ee
\be
{\bf E} = \sum_{\bf k} \hat{\bf \epsilon_k} \mathcal{E}_{\bf k} (a_{\bf k}
+ a_{\bf k}^\dag)
\ee
Here the sum is over all allowed wave vectors $\bf k$ and
\be
\mathcal{E}_{\bf k} = \sqrt{{\hbar \omega_k} \over { 2 \epsilon_0 V_{\rm k}}}
\ee
is a function of mode volume $V_{\rm k}$ which for a near-planar cavity is
given by \be V = \frac{\pi}{4} w_m^2 L \ee \noindent where $L$ is the cavity
mirror separation and $w_m$ is the Gaussian mode waist at the center of the cavity.

The atom-field interaction is given by
\be
\mathcal{H}_{AF}= \sum_{\bf k} \hbar g^{ij}_{\bf k}(\sigma_+ a_{\bf k} +
a_{\bf k}^\dag \sigma_-) \ee
where
\be
g^{ij}_{\bf k} = {{e \langle i | {\bf r} | j \rangle   \cdot 
\hat{\epsilon_{\bf k}} \mathcal{E}_{\bf k}} \over \hbar}
\ee

For a two-level atom, we have in the rotating-wave approximation
(i.e. neglecting non-energy-conserving processes)

\be
\mathcal{H} = \sum_{\bf k} \hbar \omega_{\bf k} a_{\bf k}^\dag
a_{\bf k} + {1 \over 2} \hbar \omega_a \sigma_z + \sum_{\bf k}
\hbar g_{\bf k}(\sigma_+ a_{\bf k} + a_{\bf k}^\dag \sigma_-)
\ee

where $a_{\bf k}^\dag$, $a_{\bf k}$ are photon creation and
destruction operators, $\sigma_+$, $\sigma_+$ are atom creation and
destruction operators, $\hbar \omega_a$ is the atom level spacing.
 
For a single mode cavity of frequency $\omega_c$ we have 

\be
\mathcal{H} = \hbar \omega_c a^\dag
a + {1 \over 2} \hbar \omega_a \sigma_z + 
\hbar g(\sigma_+ a + a^\dag \sigma_-)
\ee

To go to the interaction picture we write
\be 
\begin{array}{ll}
H & = H_0 + H_1  \\
H_0 & = \hbar \omega_c a^\dag
a + {1 \over 2} \hbar \omega_a \sigma_z \\
H_1 & = \hbar g(\sigma_+ a + a^\dag \sigma_-)
\end{array}
\ee\
The Hamiltonian in the interaction picture is
\be V = e^{i H_0 t / \hbar} H_1 e^{- i H_0 t / \hbar} = \hbar
g(\sigma_+ a e^{i \Delta t} + a^\dag \sigma_- e^{-i \Delta t})
\ee
where $\Delta = \omega_c - \omega_a$ is the atom-cavity detuning.

\be i \hbar {\partial \over {\partial t}} |\psi \rangle = V
|\psi \rangle.
\ee

The atom-cavity wavefunction can be written
\be
|\psi(t) \rangle = \sum_{n=0}^\infty [ c_{e,n} (t) | e,n\rangle
+c_{g,n}(t)|g,n\rangle ]
\ee

where $|e,n\rangle$ corresponds to a state with an excited atom with
$n$ photons in the cavity, etc.  Clearly the Hamiltonian mixes only
the states $|e,n\rangle$ and $|g,n+1\rangle$.  Therefore the Schrodinger
equation can be written as the coupled equations
\be
\dot{c}_{e,n} = - i g \sqrt{n+1} e^{i \Delta t} c_{g,n+1},
\label{eq-bloch1}
\ee
\be
\dot{c}_{g,n+1} = - i g \sqrt{n+1} e^{-i \Delta t} c_{e,n}
\label{eq-bloch2}
\ee

A general solution is given by

\be
c_{e,n}(t) = \left\{c_{e,n}(0)\left[\cos (\Omega_n t/2) - 
{{i \Delta}\over{\Omega_n}} \sin (\Omega_n t/2)\right] - \right.
\ee 
\[
\left. {{2 i g \sqrt{n+1}}\over{\Omega_n}} c_{g,n+1}(0) \sin (\Omega_n t/2)
\right\} e^{i \Delta t / 2}
\]

\be
c_{g,n+1}(t) = \left\{c_{g,n+1}(0)\left[\cos (\Omega_n t/2) + 
{{i \Delta}\over{\Omega_n}} \sin (\Omega_n t/2)\right] - \right.
\ee
\[
\left. {{2 i g \sqrt{n+1}}\over{\Omega_n}} c_{e,n}(0) \sin (\Omega_n t/2)
\right\} e^{-i \Delta t / 2}
\]

\noindent where
\be
\Omega_n = \sqrt{\Delta^2 + 4 g^2(n+1)}
\ee

If the atom is initially in the excited state, then $c_{e,n}(0) = c_n(0)$
and $c_{g,n+1}(0)=0$, and

\be
c_{e,n}(t) = c_n(0)\left[\cos (\Omega_n t/2) - {{i
\Delta}\over{\Omega_n}} \sin (\Omega_n t/2)\right] e^{i
\Delta t / 2}
\ee

\be
c_{g,n+1}(t) =
- c_n(0) \left[{{ 2 i g \sqrt{n+1}}\over{\Omega_n}} \sin (\Omega_n t/2)
\right] e^{-i \Delta t / 2}
\ee

Consider the case of perfect resonance $\Delta=0$ and initial condition of
excited atom with exactly $n_0$ cavity photons.  The solution is then
given by

\be
\label{soln-no-detuning1}
c_{e,n_0}(t) = \cos (\sqrt{n_0+1}\ g t)
\ee
\be
\label{soln-no-detuning2}
c_{g,n_0+1}(t) = \sin (\sqrt{n_0+1}\ g t)
\ee
\noindent with all other $c_{e,n}(t)$, $c_{g,n}(t)$ equal to zero.

\section{Rate equation for microlaser}

Equations \ref{soln-no-detuning1} and \ref{soln-no-detuning2} describe
resonant quantized Rabi oscillation.  Note that oscillation proceeds
even for an initially empty cavity ($n_0=0$) (vacuum Rabi oscillation).

In the microlaser, atoms interact with the cavity for an interaction
time \tint.  The emission probability for an atom is given by the
squared ground state amplitude after its interaction with the cavity:
\be
P_{\rm emit} = | c_{g,n_0+1}(\tint)|^2 = \sin^2 (\sqrt{n_0+1}\ g t)
\ee
\noindent where for simplicity we suppose that the cavity initially
contained exactly $n_0$ photons.

Let us now write a rate equation for the cavity photon number $n$: the
time rate of change of $n$ is equal to to the difference between gain
and loss terms:

\be \frac{dn}{dt} = G - L = { N \over \tint}
 \sin^2(\sqrt{ n+1}\ g \tint) - \gc n = 0
\label{eq-rate-equation}
\ee

\noindent which is set equal to zero at steady-state.  Here $N$ is the
number of atoms in the cavity, which can be defined by $N/\tint = r$,
the injection rate.  Loss is equal to the photon number multiplied by
the cavity loss rate \gc.


The rate equation can be solved graphically as in
Fig.~\ref{fig-gain-loss-rate-equation}.  In general there exists more
than one solution.

A solution $n'$ is called stable if the system responds to a small
perturbation in $n$ in a manner that tends to restore the original
solution.  

\be \mbox{\rm Stability condition:}\ {\partial \over {\partial
n}} (G - L)|_{n'} < 0 \ee 

In terms of Fig.~\ref{fig-gain-loss-rate-equation} a solution is
stable if and only if the slope of the loss line exceeds that of the
gain line at their intersection.

\bfig
\centerline{$G,L$ \hspace*{4.5in}}
\centerline{\resizebox{5in}{!}{\includegraphics{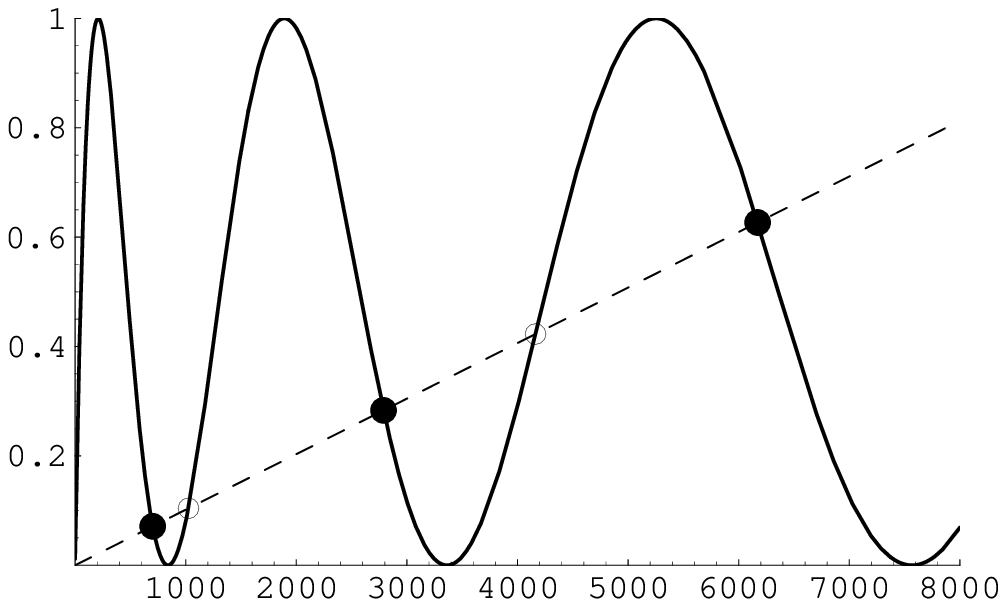}}}
\centerline{photon number $n$}
\caption{Dimensionless gain (solid line) and loss (dashed line)
in the semiclassical rate equation analysis for experimental $g$,
$\tint$, and $N=1000$.  Closed circles represent stable solutions,
open circles unstable solutions.}
\label{fig-gain-loss-rate-equation}
\efig

The threshold atom number $\Nth$ is given by the point at which the
gain and loss terms have equal slope, assuming small argument of the
sine function:
\be \Nth = {{\gc} \over {g^2 \tint}} \ee 

Solutions to \ref{eq-rate-equation} as the injection rate $r$ is
varied are plotted in Fig.\ref{fig-rate-equation-solns} , with $g
\tint = 0.1$.  The number of solutions increases indefinitely with
increased $r$.  Solutions are asymptotic to constant-$n$ lines
corresponding to integer number of Rabi oscillations $\sqrt{n+1}\ g
\tint = m \pi$, $m$ integer.

\bfig
\centerline{$n$ \hspace*{4.5in}}
\centerline{\resizebox{5in}{!}{\includegraphics{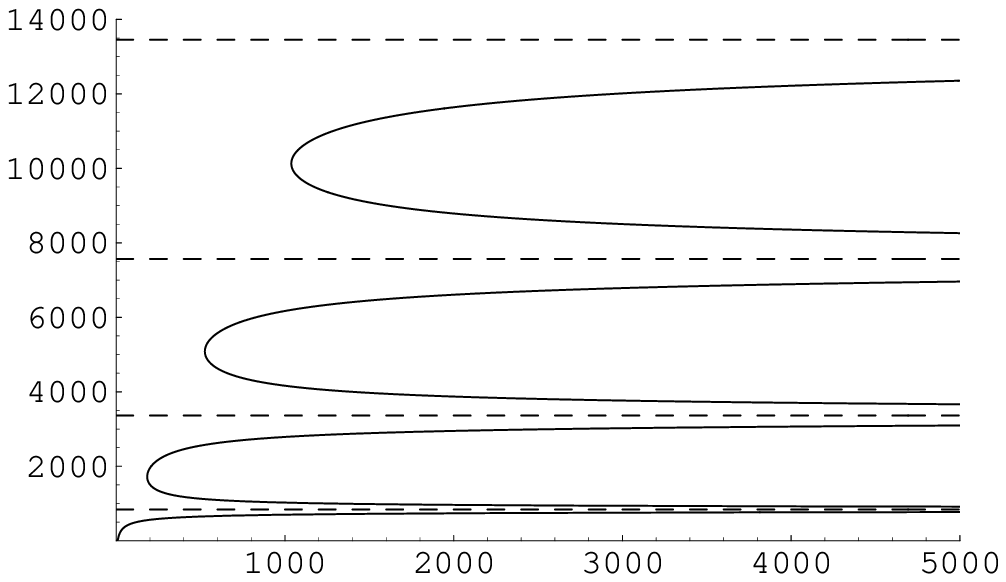}}}
\centerline{$N_{\rm atoms}$}
\caption{Multiple solutions of microlaser rate equation.  Dotted lines 
represent photon numbers corresponding to an integer number of complete
Rabi oscillations}
\label{fig-rate-equation-solns}
\efig

The rate equation analysis shows how the atoms' sinusoidal emission
probability leads to bistability and multistability.  However, its
derivation was less than rigorous and assumes that the cavity photon
number is well-determined.  In addition, it gives no information about
the relative stability of different solutions, or the cavity photon
statistics.

For a more complete theory we turn to the quantum theory developed for
the micromaser by Filipowicz, Meystre, Scully and others
(\cite{Filipowicz-PRA86})

\section{Quantum theory of microlaser}

We follow the density matrix approach
\cite{scully-zubairy-quantum-optics}.
This theory in based on the following assumptions:

\begin{enumerate}
\item Excited atoms are injected at a rate low enough so that at most
one atom at a time is present in the cavity.
\item Excited and ``ground'' state lifetimes are much longer than \tint.
\item Cavity damping during interaction time is small: $\gc \tint \ll 1$
\end{enumerate}
\subsection{Density matrix equation of motion}
\subsubsection{Gain term}
Consider the change in the cavity field density matrix due to the
injection at time $t_0$ of an excited atom, which then interacts with
the cavity for time \tint:
\be
\delta \rho_{nn'} = \rho_{nn'}(t_0 + \tint) - \rho_{nn'}(t_0)
\ee
The state $\rho_{nn'}(t_0 + \tint)$ may be found by determining the
atom-field evolution from time $t_0$ to $t_0 + \tint$ and then tracing
over atomic variables:
\be \rho_{nn'}(t_0 + \tint) = \sum_{\alpha=e,g}
\rho_{\alpha,n;\alpha,n'}(t_0+\tint) 
\ee
Note that density operators with four indices represent combined
atom-cavity systems; those with two represent the field only.  
A coarse-grained derivative for the gain contribution to the field
density matrix can now be written
\be
\left(\frac{d\rho_{nn'}}{dt}\right)_{\rm gain} = r[\rho_{en;en'}(t_0 + \tint)+
\rho_{gn;gn'}(t_0+ \tint) - \rho_{en;en'}(t_0)]  
\ee
We now replace $t_0$ with the arbitrary continuous time $t$.  The
time-evolved density matrices $\rho_{en;en'}(t + \tint)$ and
$\rho_{gn;gn'}(t + \tint)$ are determined from the solutions for the
excited state amplitudes as in section 1 of this chapter.  We have
\be
\begin{array}{lll}
c_{e,n}(t+\tint) &=& c_{en}(t) \cos(\sngt) \\
c_{g,n+1}(t+\tint)& =& c_{g,n+1}(t) \sin(\sngt)
\end{array}
\ee
\noindent which gives for the density matrices
\be
\begin{array}{ll}
\rho_{en;en'}(t + \tint)& = c_{e,n}(t+\tint) c_{e,n'}^*(t+\tint) = \\
& \rho_{nn'}(t) \cos(\sngt) \cos(\sqrt{n'+1}\ g \tint) \\
\rho_{gn;gn'}(t + \tint)& = c_{g,n}(t+\tint) c_{g,n'}^*(t+\tint) = \\
& \rho_{n-1,n-1'}(t) \cos(\sqrt{n}\ g \tint) \cos(\sqrt{n'}\ g \tint) \\
\end{array}
\ee
We then have the reduced field density matrix equation of motion:
\be
\label{eq-density-matrix-gain}
\begin{array}{ll}
\left(\frac{d\rho_{nn'}}{dt}\right)_{\rm gain} = & - r[1 - \cos(\sngt)
\cos(\sqrt{n'+1} g \tint)] \rho_{nn'} + \\ & r \sin(\sqrt{n} g \tint)
 \sin(\sqrt{n'} g \tint) \rho_{n-1,n'-1}.
\end{array}
\ee

Note that we have made no assumptions about the photon number distribution

\subsubsection{Loss term}

The contribution due to cavity loss is
\be
\label{eq-density-matrix-loss}
\left(\frac{d\rho_{nn'}}{dt}\right)_{\rm loss} = -
\frac{\gc}{2}\left(a^\dag a \rho - 2 a \rho a^\dag + \rho a^\dag
a\right)
\ee
\noindent for zero thermal photons.  This is a standard result from
the theory of oscillators coupled to a reservoir (c.f.
\cite{scully-zubairy-quantum-optics} chapter 8.)

Combining \ref{eq-density-matrix-gain} and
\ref{eq-density-matrix-loss}, the total density matrix equation of
motion is
\be
\dot{\rho}_{nn'} = a_{nn'}\rho_{nn'} + b_{n-1,n'-1} \rho_{n-1,n'-1}
+ c_{n+1,n'+1}\rho_{n+1,n'+1}
\ee
where
\be
\begin{array}{ll}
a_{nn'} = & - r[1 - \cos(\sngt) \cos(\sqrt{n'+1} g \tint)]
           - \frac{\gc}{2}(n+n') \\
b_{nn'} = & r \sin(\sqrt{n} g \tint) \sin(\sqrt{n'} g \tint) \\
c_{nn'} = & \gc \sqrt{nn'} \\
\end{array}
\ee
If we restrict ourselves to diagonal matrix elements this master equation
reduces to
\be
\dot{\rho}_{nn} = a_{nn}\rho_{nn} + b_{n-1,n-1} \rho_{n-1,n-1}
+ c_{n+1,n+1}\rho_{n+1,n+1}
\ee
where
\be
\begin{array}{ll}
a_{nn} = & - r\sin^2(\sngt) - \gc n \\
b_{nn} = & r \sin^2(\sqrt{n} g \tint) \\
c_{nn} = & \gc n \\
\end{array}
\ee
\subsection{Steady-state photon statistics}
Setting $\dot{p}_n(t)= \dot{\rho}_{nn}(t)=0$ gives the equation
\be 
\begin{array}{c}
\left\{ r\sin^2(\sngt) + \gc n \right\} p(n) + 
 r  \sin^2(\sqrt{n} g \tint)  p(n-1) + \\
\gc (n+1) p(n+1) = 0 \\
\end{array}
\ee
which leads to the following recurrence relations:
\be
\{ r\sin^2(\sngt) + \gc n \} p(n) = \gc (n+1) p(n+1) 
\ee
\be
r  \sin^2(\sqrt{n} g \tint) p(n-1) = \gc n p(n) 
\ee
We obtain the steady-state photon distribution of the microlaser:
\be
p(n) = p(0) \prod_{k=1}^{n}\frac{ r \sin^2(\sqrt{k}\,g \tint)}{\gc k}
\label{eq-microlaser-pn}
\ee
where $p(0)$ is determined from normalization.
\be
\sum_n p(n) = 1
\ee
It is convenient to define two quantities, the expectation photon number
\be
\Nex = r / \gc
\ee
equal to the number of atom injections per cavity decay time, and the 
normalized interaction time or pump parameter
\be
\theta = \sqrt{\Nex} g \tint
\ee
which is the Rabi phase if \Nex\ photons are present.  Linearization
of the rate equation for small $n$ and setting gain equal to loss
gives $\theta = 1$ as the laser threshold condition.

\subsubsection{Mean photon number}

In Fig.~\ref{fig-mt-n} the photon number 
%
%
has been plotted as function of pump parameter $\theta$ for $\Nex =
10$.  Note the dips due to integer number of Rabi oscillations (trap
states) and a ``jump'' near $\theta = 2 \pi$ which becomes sharper as
$\Nex$ becomes large.

\subsubsection{Photon statistics}
The width of the photon number distribution can be characterized by
the Mandel Q parameter, defined as
\be Q \equiv \frac{\langle n^2 \rangle - \langle n \rangle^2}{\langle
n \rangle} - 1 \ee
The Q parameter is equal to zero for Poisson statistics (for which the
variance $\langle n^2 \rangle - \langle n \rangle^2 = \langle n
\rangle$, greater than zero for super-Poisson statistics (e.g. thermal
light) and less than zero for sub-Poisson statistics.

\bfig
\centerline{\resizebox{5in}{!}{\includegraphics{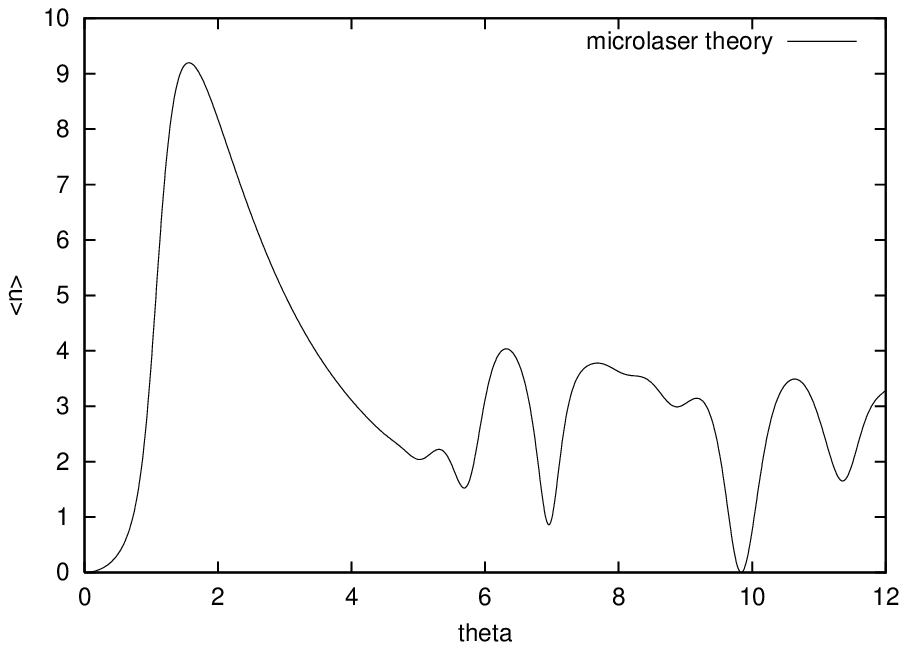}}}
\caption{Micromaser theory result for $\langle n \rangle$, $N_{ex}=10$}
\label{fig-mt-n}
\efig

\bfig
\centerline{\resizebox{5in}{!}{\includegraphics{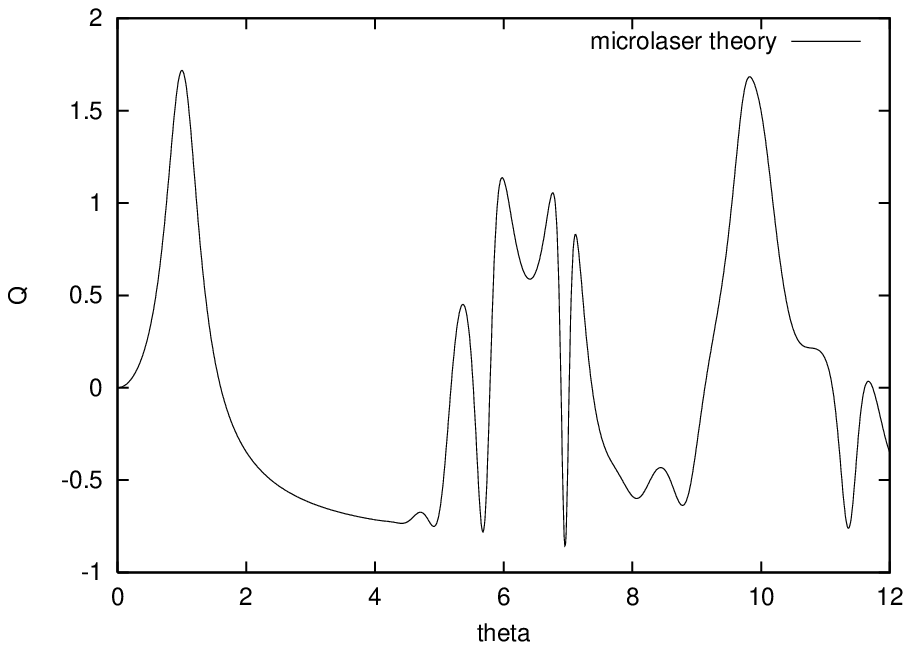}}}
\caption{Micromaser theory result for $Q$, with $N_{ex}=10$.  $Q<0$ indicates
sub-Poisson statistics.  }
\label{fig-mt-q}
\efig

In Fig.~\ref{fig-mt-q} the Mandel Q parameter is plotted as function of pump
 parameter $\theta$ for $\Nex = 10$.  The cavity photon
 statistics are sub-Poisson over wide ranges of $\theta$.

To compare with the rate equation result we now calculate the results
from Eq.~\ref{eq-microlaser-pn} as the number of cavity atoms is
varied, with fixed $g \tint$.

\bfig
\centerline{$n$ \hspace*{4.5in}}
\centerline{\resizebox{5in}{!}{\includegraphics{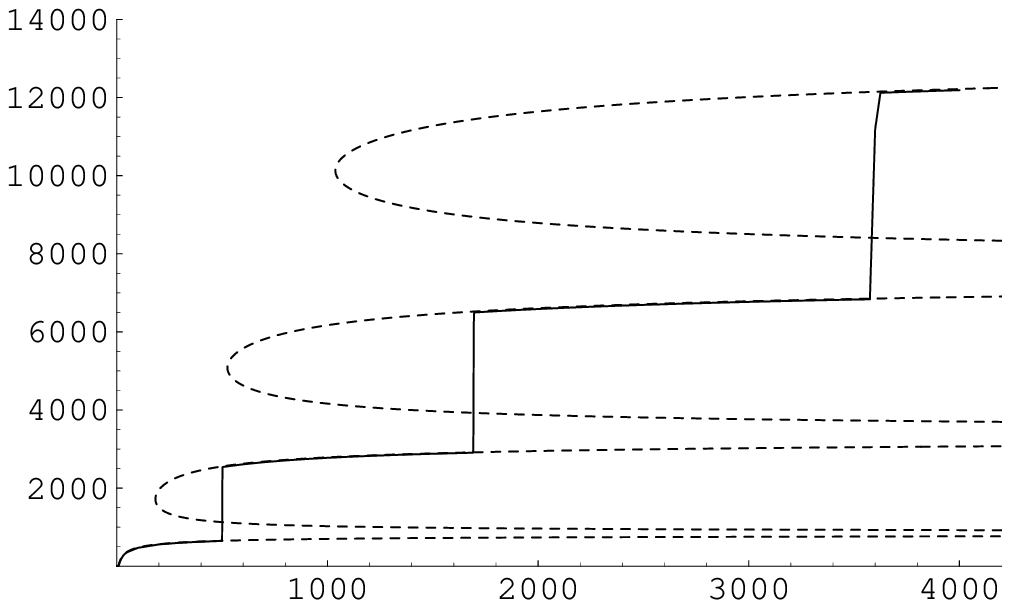}}}
\caption{Rate equation solutions (dotted lines) compared with quantum 
microlaser theory (solid line).  }
\label{fig-re-vs-mt}
\efig

\bfig
\centerline{$Q$ \hspace*{4.5in}}
\centerline{\resizebox{5in}{!}{\includegraphics{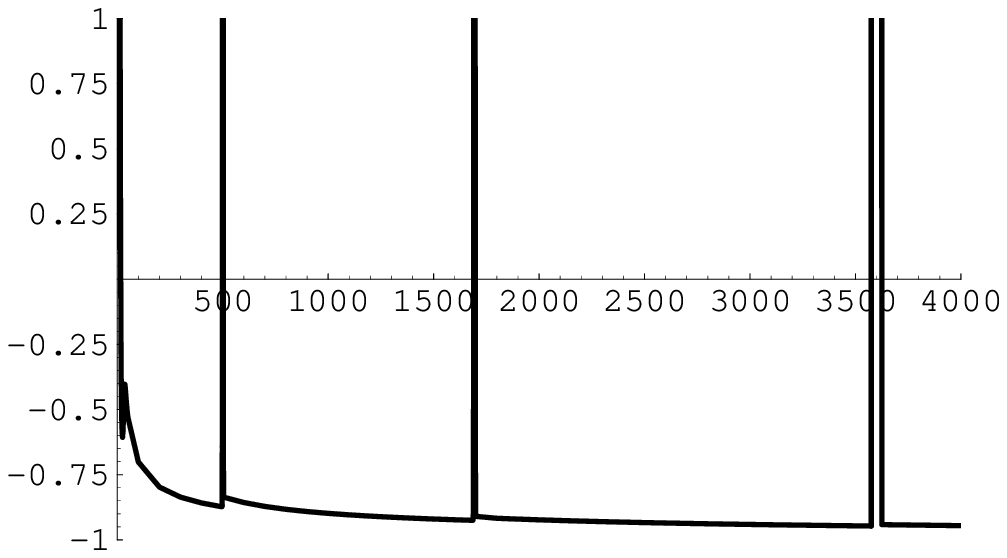}}}
\centerline{$N_{\rm atoms}$}
\caption{Q parameter according to  microlaser quantum theory.  Statistics
are strongly sub-Poissonian for almost every point over threshold.  Spikes
are due to bimodal photon distributions during phase transitions.}
\label{fig-qplotplain}
\efig

Fig.~\ref{fig-re-vs-mt} shows the average photon number from
Eq.~\ref{eq-microlaser-pn} together with the rate equation result.  We
see that the fully quantized theory agrees with one of the rate
equation stable solutions except during ``jumps'' corresponding to
transitions between the solutions.  The quantum theory essentially
``chooses'' one of the rate equation solutions.

During the transitions, which become increasingly narrow
as $N_ex$ is increased, the photon number distribution has two peaks
corresponding to the two stable solutions.

It can be shown that the rate equation solutions correspond to within
$\pm 1$ of a local maxima of the $p(n)$ distributions from the quantum
theory.  Suppose a local maximum in $p(n)$ exists at $n=n_0$; this
means that $p(n_0) > p(n_0 \pm 1)$ and the quantity inside the product
of Eq.~\ref{eq-microlaser-pn} passes through 1 between $k=n_0$ and
$k=n_0+1$:
\be
\frac{ r \sin^2(\sqrt{n_0}\,g \tint)}{\gc n_0} > 1 > 
\frac{ r \sin^2(\sqrt{n_0+1}\,g \tint)}{\gc (n_0+1)}
\ee
Setting 
\be
\frac{ r \sin^2(\sqrt{n_0+1}\,g \tint)}{\gc (n_0+1)} = 1
\ee
gives the original rate equation Eq.~\ref{eq-rate-equation}.

%



%
%






%
%
%
%

\section{Modification for realistic parameters}
\label{theory-realistic-effects}

We describe modifications to the rate equation and quantum theory in
order to accomodate the atomic velocity distribution, nonuniform
atom-cavity coupling, atom-cavity detuning, and imperfect pumping.

\subsection{Averaging of gain function}

The framework for all these modifications is the same: we calculate
the effect of the relevant broadening on the ``gain function''
$\beta(n)$ which for an ideal (monovelocity etc.) microlaser
is given by
\be
\beta_0(n) = \sin^2(\sngt).
\ee
For example, the effect of a velocity distribution $f_v(v)$ is the average
\be
\bar{\beta}(n) = \int_0^\infty \sin^2(\sqrt{n+1} g \tint(v)) f_v(v) dv
\ee
The steady-state photon statistics are then given by
\be
p(n) = p(0) \prod_{k=1}^{n}\frac{ r \bar{\beta}(k) }{\gc k}
\label{pn-dist-avgd}
\ee
Note that we essentially replace a process in which different atoms
experience different interactions, with a process in which all atoms
experience the same {\it averaged} interaction.  This may seem
implausible until we realize that the end result \ref{pn-dist-avgd}
represents an average itself.  Our averaging procedure follows from
the assumption that the random variables associated with the velocity,
etc. are independent (cf. \cite{Filipowicz-PRA86}).

The following expression is used to give the average of the gain
function over velocity distribution, coupling variation, and detuning:
\be
\bar{\beta}(n) = \int_0^\infty B(n, v, \Delta(v))\;f_v(v)\;f_g(g)\;dv\;dg 
\label{eq-betabar}
\ee
where $\Delta(v) = k v \theta$ and $B(n,v,\Delta)$ is the gain
function including detuning.  It might be expected that
$B(n,v,\Delta)$ is given by the well-known solution involving the
generalized Rabi frequency
\be
B(n,v,\Delta) = \frac{4 (n+1) g^2}{4 (n+1) g^2 + \Delta^2}
\sin^2 \left[4 (n+1) g^2 + \Delta^2 \tint \right] \ \  {\it (wrong)}
\ee
However, this expression is only valid for constant fields.  In
general $B(n,v,\Delta)$ must be calculated numerically and its form
will depend on the shape of the coupling profile $g(x=vt)$, not just
its area.  The function $B(n,v,\Delta)$ is found by integrating the
Bloch equations ( Eqns.~\ref{eq-bloch1} and \ref{eq-bloch2}) for a
initially excited atom in a field with detuning $\Delta$ and Rabi
frequency $\sqrt{n+1} g(t)$ during a simulated transit through the
cavity.
%
%
Integration of $B(n,v,\Delta)$ for every calculation would be very
time-consuming; therefore we tabulate this function on a grid and
construct an interpolating function to be called on by our theory
calculations.

\subsection{Imperfect inversion}
The adiabatic pumping process described in Chapter \ref{chap-methods}
is not perfect: for the data shown in the Chapter \ref{chap-results},
the excited state probability was about 80\%, measured by a
fluorescence experiment.  This includes a small contribution from
atomic decay during the pump process and the transit time between the
pump field and cavity mode.

Let $\rho_{ee}$ and $\rho_{gg}$ be the upper and lower state
probabilities, respectively.  We claim that for \Nex\ very large the
effect of nonzero $\rho_{gg}$ is simply to reduce the effective number
of atoms to
\be
\Neff \equiv N (\rho_{ee}-\rho_{gg}).
\ee
\noindent assuming $\rho_{ee} > \rho_{gg}$.  That is, instead of 
the number of atoms we may speak of the {\it inversion}.  This idea is
a familiar one in laser physics.

To support this claim we note that atoms initially in the ground state
have a probability of {\rm removing} (absorbing) a photon from the
field of
\be
P_{\rm absorb} = \sin^2(\sqrt{n}\;g\tint)
\ee
Intuitively it is clear that the case $\rho_{ee} =
\rho_{gg} = 1/2$ would produce essentially zero photons, since the
initially lower state atoms absorb photons in the same manner as the
upper state atoms emit them.

In the rate equation picture, the net emission rate per atom is then
modified to
\be
P_{\rm emit}' = \rho_{ee} \sin^2(\sqrt{n+1}\;g\tint) - \rho_{gg} \sin^2(\sqrt{n}\;g\tint) \approx (\rho_{ee} - \rho_{gg} ) \sin^2(\sqrt{n+1}\;g\tint)
\ee
\noindent with corrections on the order of $1/2\sqrt{n}$.  Corrections may 
affect the initial threshold region somewhat but for most of the range
of interest in our experiments, $n$ is very large.

\subsubsection{Realistic rate equation and quantum theory}

Fig.~\ref{fig-re-mt-realistic} gives the equivalent of
Fig.~\ref{fig-re-vs-mt} after realistic effects, appropriate to the
data presented in Chapter~\ref{chap-results} have been included.

\bfig
\centerline{$G,L$ \hspace*{4.5in}}
\centerline{\resizebox{5in}{!}{\includegraphics{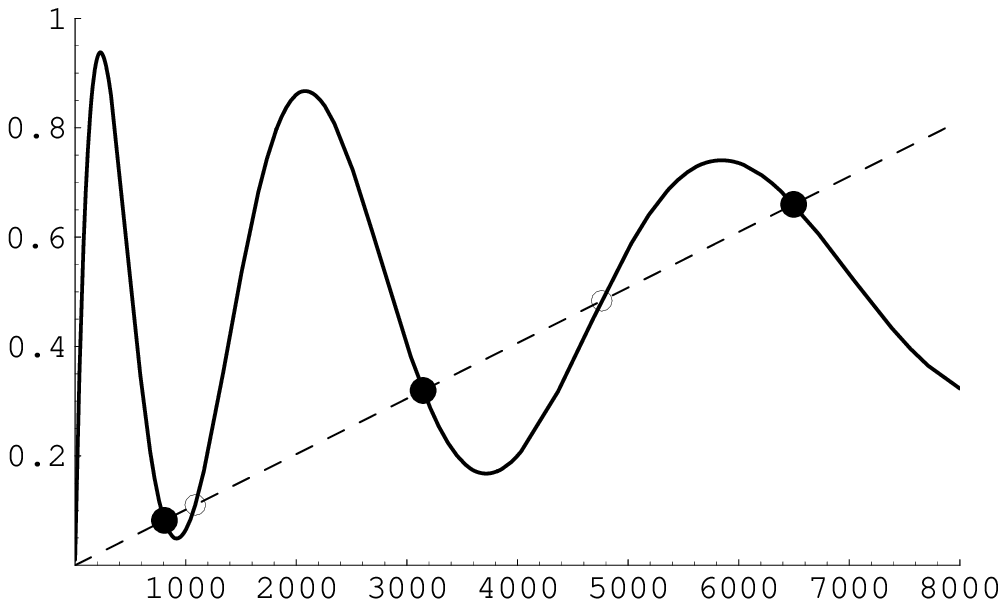}}}
\centerline{$n$}
\caption{Dimensionless gain (solid line) and loss (dashed line)
in the semiclassical rate equation analysis, including effects of
velocity distribution, nonuniform coupling, and detuning, with
$N=1000$.  Closed circles represent stable solutions, open circles
unstable solutions.}
\label{fig-gain-loss-rate-equation02}
\efig

\bfig
\centerline{$n$ \hspace*{4.5in}}
\centerline{\resizebox{5in}{!}{\includegraphics{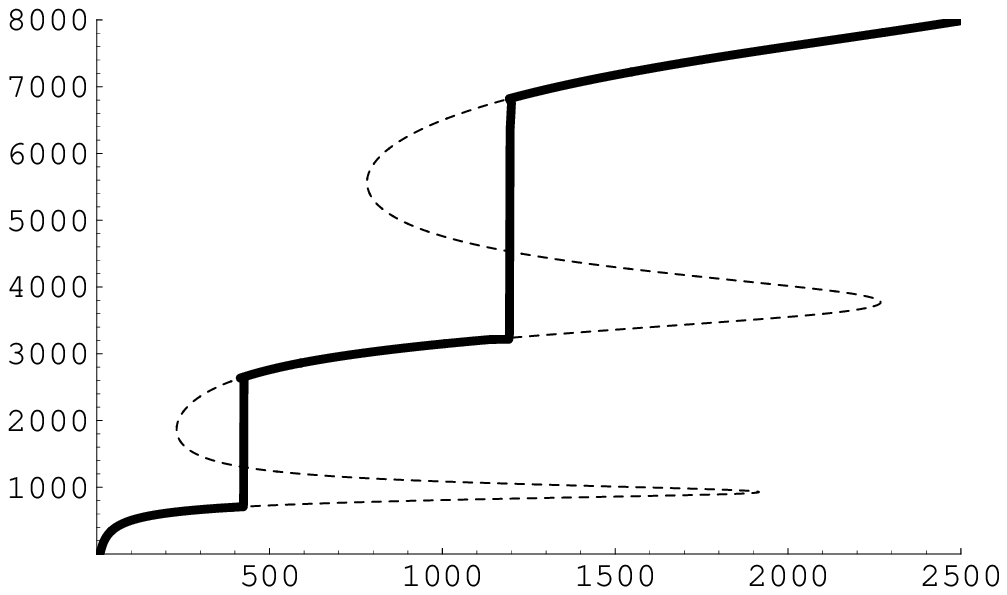}}}
\centerline{$N$}
\caption{Photon number vs.\ atom number, rate equation (dashed line) 
and microlaser theory (solid line) including effects of velocity 
distribution, nonuniform coupling, and detuning.}
\label{fig-re-mt-realistic}
\efig

\bfig
\centerline{$Q$ \hspace*{4.5in}}
\centerline{\resizebox{5in}{!}{\includegraphics{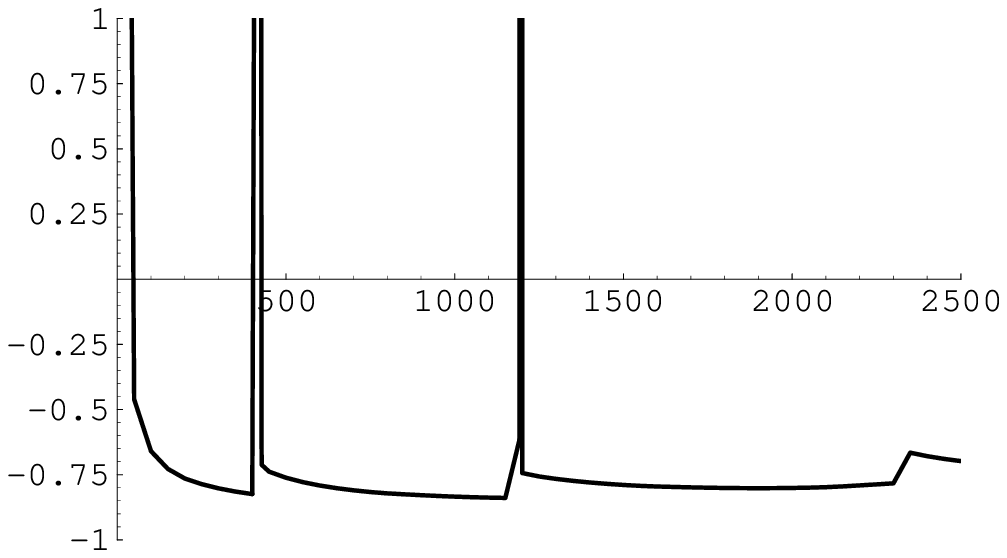}}}
\centerline{$N$}
\caption{Q parameter according to  microlaser quantum theory, including
realistic effects as in Fig.~\ref{fig-re-mt-realistic}.  Spikes
are due to bimodal photon distributions during transitions.}
\label{fig-qplot02}
\efig

The average photon number can also be plotted as a function of
detuning.  (Fig.~\ref{fig-re-detuning1}).

\bfig[t]
\centerline{$n$ \hspace*{4.5in}}
\centerline{\resizebox{5in}{!}{\includegraphics{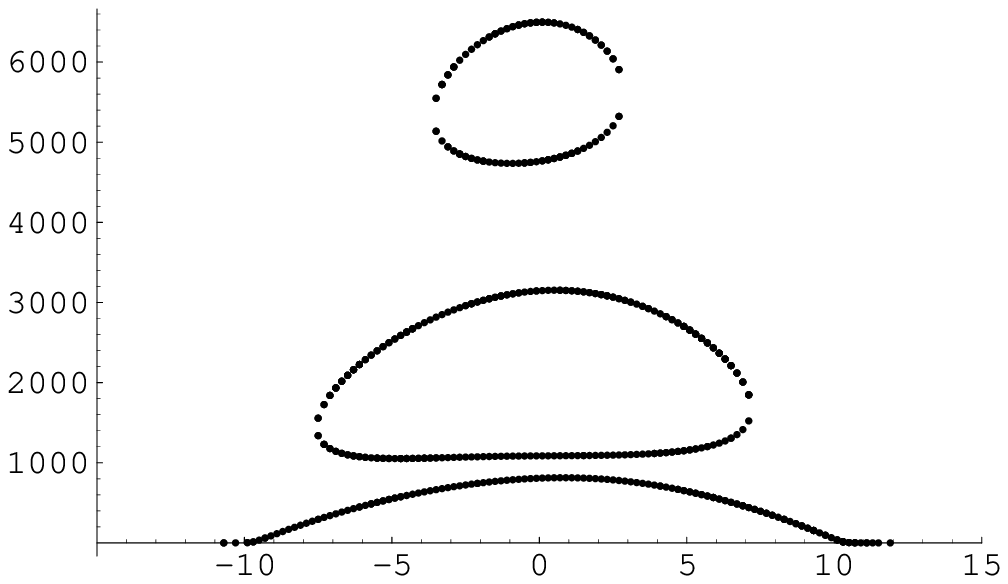}}}
\centerline{Detuning, \mhz}
\caption{Rate-equation solutions for $\Neff=1000$.  Second and third
branches appear as closed curves.  Upper halves of closed curves
represent stable solutions.  Small gaps are artifacts from algorithm
used to find solutions.}
\label{fig-ecleanplot}
\efig

\bfig[t]
\centerline{\resizebox{5in}{!}{\includegraphics{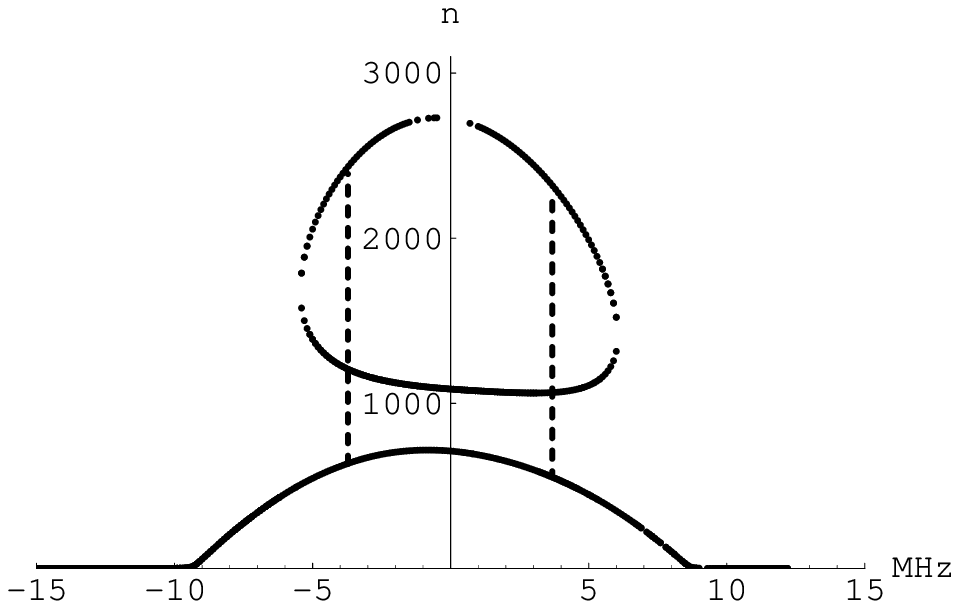}}}
\caption{Solutions to rate equation with realistic parameters, 
as a function of detuning, $\Neff=600$.  Dotted line, quantum theory shows
jumps in detuning curve. }
\label{fig-re-detuning1}
\efig

\section{Fokker-Planck analysis}

In \cite{Filipowicz-PRA86} a Fokker-Planck equation for the cavity
photon number distribution $p(n,t)$ is derived:
\be
\frac{\partial}{\partial t} p(n,t) = 
- \frac{\partial}{\partial n}{Q(n)p(n,t)} + 
\frac{1}{2}\frac{\partial^2}{\partial n^2}[G(n)p(n,t)]
\label{eq-fokker-planck}
\ee
where for zero thermal photon number
\be
Q(n) = R \sin^2(\sqrt{n} g \tint) - \gc n
\ee
\be
G(n) = R \sin^2(\sqrt{n} g \tint) + \gc n
\ee
and $R$ is the injection rate.  To generalize to our averaged case we
replace the $\sin^2(\sqrt{n} g \tint)$ terms with $\bar{\beta}(n)$
from \ref{eq-betabar}.  (The replacement of $n$ by $n+1$ is
insignificant on the photon number scale we will consider.)  Equation
\ref{eq-fokker-planck} was shown to reproduce the average photon number and
photon statistics of micromaser theory as long as (i) the pump
parameter $\theta \ll (n \Nex)^{1/2}$ and (ii) $n \gg 1$.  Both of
these conditions are easily met in our case.

We now rescale time to the cavity lifetime and photon number to
expectation photon number: $\tau \equiv t \gc$, $\nu \equiv n/\Nex$.
Then Eq.~\ref{eq-fokker-planck} becomes
\be
\frac{\partial}{\partial t} p(\nu,t) = 
- \frac{\partial}{\partial \nu}{q(\nu)p(\nu,\tau)} + 
\frac{1}{2}\frac{\partial^2}{\partial \nu^2}[g(\nu)p(\nu,\tau)]
\label{eq-fokker-planck2}
\ee
\noindent with
\be
q(\nu) = \bar{\beta}(\nu Nex) - \nu
\ee
\be
g(\nu) = \bar{\beta}(\nu Nex) + \nu
\ee
\noindent noting that $\Nex = R/\gc$.  

The stationary solution to the Fokker-Planck equation is then
\be
p(\nu) = \frac{C}{g(\nu)} \exp\left(2 \Nex
\int_0^\infty\;d\nu\;\frac{q(\nu)}{g(\nu)}\right)
\label{eq-sol-fp}
\ee
\noindent where $C$ is a normalization constant.  
For large $\Nex$ the photon number distribution will accumulate in the
global maximum of the exponent of Eq.~\ref{eq-sol-fp}, or equivalently
the global minimum of an effective potential defined by
\be
V(\nu) = - \int\;d\nu\;\frac{q(\nu)}{g(\nu)}
\ee
The local minima of $V(\nu)$ are the zeros of $q(\nu)$, which are the
solutions of 
\be
\bar{\beta}(\nu \Nex) = \nu
\ee
which is equivalent to the rate equation derived in \ref{eq-rate-equation}.

We now have an intuitive picture for the transitions between stable
points: they occur as the global minimum of the effective potential
$V(\nu)$ changes from one point to another.  This is a variation on
the Landau theory of phase transitions, with $\sqrt{\nu}$ as the order
parameter.  Note that as $\Nex \rightarrow \infty$ the transitions
become infinitely sharp in a similar manner to the sharpness
of phase transitions for a large number of particles.

We note that the initial threshold is of second-order: as the
injection rate is increased there is a continuous increase of the
global minimum in the effective potential from $\nu=0$.

\subsection{Metastability and hysteresis}


In general the microlaser does not automatically find the global
minimum but must fluctuate to it from out of a local minimum.
Therefore hysteresis is expected if parameters are varied faster than
the transition rate between metastable and stable states.

To give an example, let us consider the rate equation solutions as
detuning is varied.  As shown in Fig.~\ref{fig-re-detuning1} the first
branch appears as an inverted parabola-type shape centered around zero
detuning (relative to resonance of most probable velocity atoms);
higher branches first appear as closed curves, and with increased atom
number, join with the first branch solution. (See
Fig.~\ref{fig-re-detuning2}).  Suppose that metastable lifetimes are
very long and spontaneous transitions do not occur.  Then as detuning
is modulated we would expect a consistent hysteresis pattern following
the arrows in Fig.~\ref{fig-re-detuning2}).  These transitions occur
when parameters change so that a local minimum in the potential
disappears entirely, and the system is forced to the next potential
minimum it encounters (not necessarily the global minimum).

Similar features would be encountered for photon number as a function
of atom number.

\bfig
\centerline{$n$ \hspace*{3.5in}}
\centerline{\resizebox{4in}{!}{\includegraphics{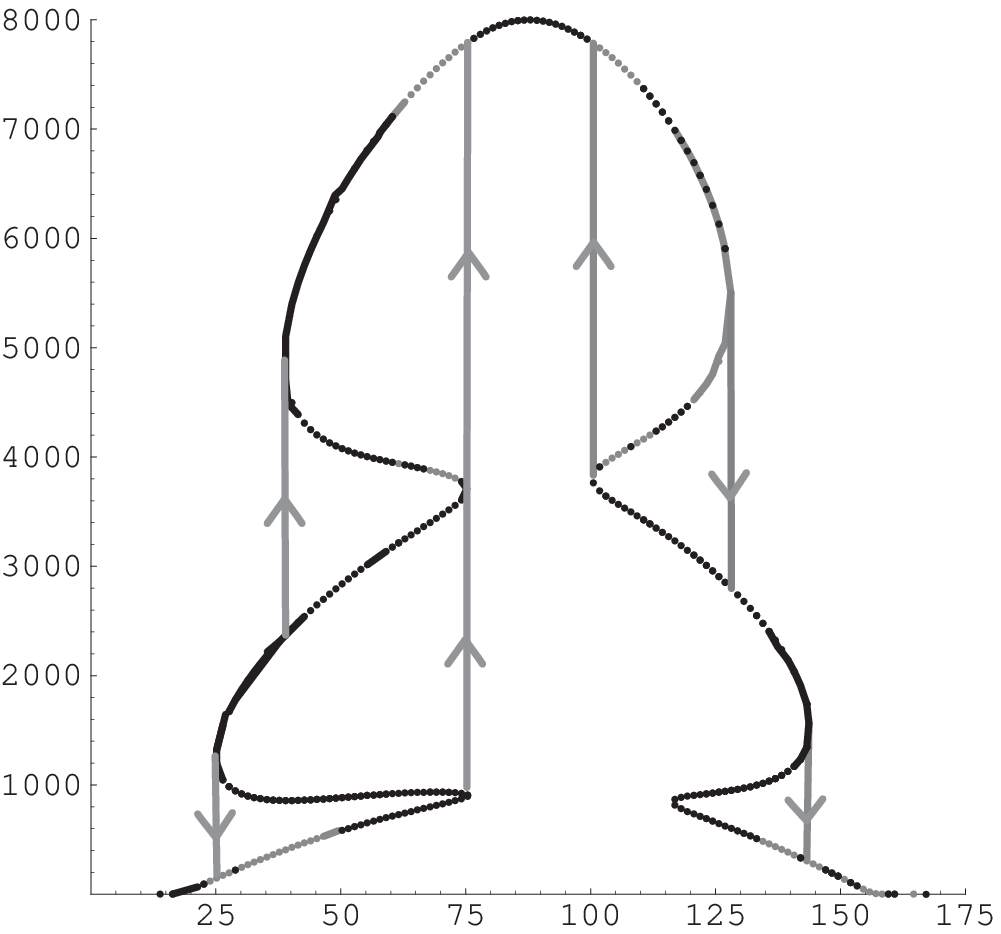}}}
\centerline{Detuning, $10^6\ s^{-1}$, for experimental parameters.}
\caption{Rate equation solutions (realistic parameters) 
as a function of detuning, $N=2500$.  Arrows show transitions that
could occur due to disappearance of solutions as detuning is changed.
}
\label{fig-re-detuning2}
\efig


\subsection{Transition rates}

The tunneling rate from a metastable solution to a global minimum is
estimated via the Kramers analysis \cite{Filipowicz-PRA86} to be, in
units of cavity linewidth,
\be
W = \frac{1}{2\pi}\left[|q'(\nu_m)| q'(\nu_M) 
\frac{g(\nu_M)}{g(\nu_m)}\right]^{1/2} \exp\{-2\Nex[V(\nu_M)-V(\nu_m)]\}
\label{eq-transition-rate}
\ee
\noindent where $\nu_m$ is the metastable local minima point and  $\nu_M$ 
is the local maximum over which the system must tunnel to reach a
global minimum (or possibly another, lower energy, metastable local
minimum).  (Fig.~\ref{fig-tunneling})

\bfig
\centerline{$V(\nu)$ \hspace*{3.5in}}
\centerline{\resizebox{4in}{!}{\includegraphics{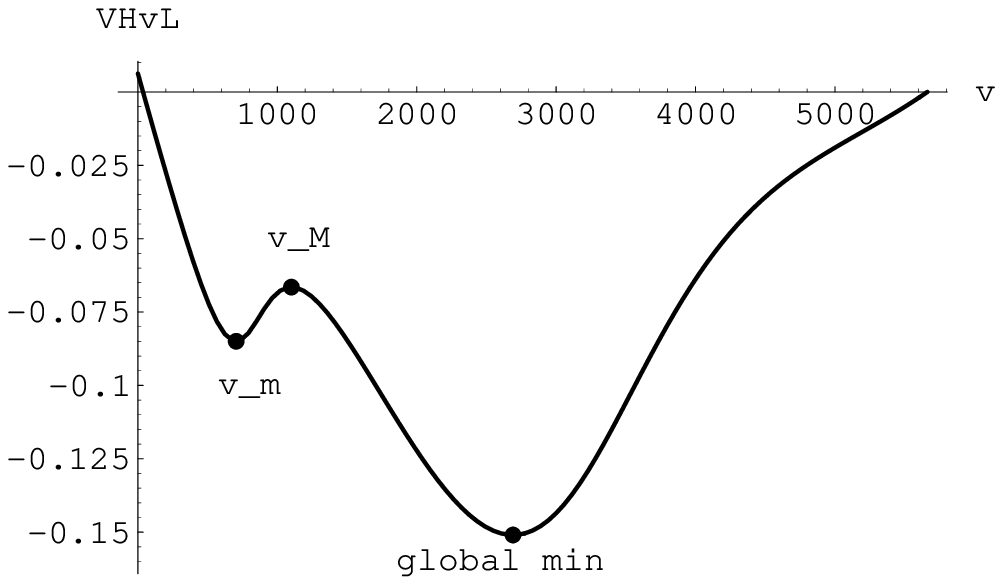}}}
\caption{Tunneling from local to global minimum in the effective potential
 $V(\nu)$, plotted for realistic microlaser parameters with $\Neff=600$.}
\label{fig-tunneling}
\efig

\chapter{Numerical simulations of many-atom microlaser}
\label{chap-simulations}

We present results from quantum trajectory simulations of a
microlaser/micromaser device in which more than one atom at a time may
be present and atom transit times are not required to be much smaller
than the cavity decay time.
%
%
%
%
%
%
For parameters in which the system is well removed from the influence
of trap states
the average photon number is in good agreement with an appropriately
scaled single-atom, weak-decay micromaser theory.  The photon
number variance, however, shows an increase proportional to $N/N_{\rm ex}=
\Gamma_{\rm cav} t_{\rm int}$ for parameters in which the weak-decay theory
predicts sub-Poisson statistics..
This result is similar to that found by Elk in the case of regular
atomic injection \cite{Elk-PRA96}.  By considering the case $N_{\rm
ex}=1$ we show that this broadening of the photon number distribution
is not a many-atom or correlation effect, but is due to
cavity decay during the atom's transit through the cavity.




\section{Introduction}

The microlaser and micromaser differ from conventional lasers and
masers not only in that they use single atoms as the gain medium, but
in the controlled nature of the atom-cavity interaction.  Uniformity
of the coupling strength between atom and cavity \cite{An-OL97b} and
interaction time $t_{\rm int}$ of the atom through the cavity, in
addition to the long lifetime of the atom upper level state compared
to the transit time, create an unusual situation in which different
atoms undergo nearly the same interaction with the cavity.  That is,
interaction is almost independent of decay and other random effects.
This is in contrast to descriptions of a conventional laser
\cite{Scully-Lamb-LP} in which ensemble averages are performed to
account for the lifetimes of the excited atoms, inhomogeneous effects,
etc.

It is natural to ask which properties of the microlaser/micromaser
persist even when the restriction of a single atom is lifted.
Dynamics based on single-atom quantum Rabi oscillations are not
necessarily expected to hold; in addition, correlation effects may
cause unknown differences from single-atom theory.

A related problem is to determine the effect of abandoning an
assumption made in analytical treatments of the microlaser/micromaser
\cite{Filipowicz-PRA86} that the cavity decay time is much longer than
the interaction time, i.e.  $\Gamma_{\rm cav} t_{\rm int} << 1,$ where
$\Gamma_{\rm cav}$ is the cavity linewidth.  In this paper we will
call this the condition of {\it weak decay}.

The present study was motivated by the realization that with the
current experimental parameters, more than one atom at a time must be
present for laser oscillation to occur.  In going from the standing
wave to the tilted atomic beam, uniform-coupling, traveling-wave
interaction \cite{An-OL97b} the coupling strength $g$ is decreased by
a factor of 2 relative to the peak standing-wave value.  (This factor
of 2 was incorrectly reported as a factor of $\sqrt 2$ in
\cite{An-OL97b}.)  The threshold number of atoms is given by
\cite{Filipowicz-PRA86}:

\[ N_{\rm th} = {\Gamma_{\rm cav} \over {g^2 t_{\rm int}}} \sim 4, \]

\noindent
This is to be compared with $N_{\rm th} \sim 1$ for the original
(standing wave) microlaser experiment \cite{An-PRL94} if $g$ is taken
at an antinode.  Therefore the study of multiple atom effects is
necessary for understanding the microlaser with traveling-wave
interaction.

Effects relating to finite $\Gamma_{\rm cav} t_{\rm int}$ are also of
interest in the microlaser, since with present parameters
\cite{An-OL97b} we have $\Gamma_{\rm cav} t_{\rm int} \sim 0.1$.

\section{Previous approaches}

The effect of 2-atom events was studied in \cite{trap-state} in the
context of micromaser trap states.  It was shown that collective
events destroy trap state resonances with great efficiency.

More general treatments of multi-atom behavior have involved strong
simplifications to the atom injection scheme.

D'Ariano studied a micromaser system pumped by clusters of up to $N=100$
atoms which all enter and leave the cavity at the same time
\cite{DAriano-PRL95}. It was shown that for $N < N_{\rm ex}$ the system
behaves similarly to the one-atom maser, and for $N > N_{\rm ex}$ the system
exhibits multiple thresholds.

Elk \cite{Elk-PRA96} considered an injection scheme in which atom
arrivals are equally spaced in time.  Results for small $\theta$ and
$N < N_{\rm ex}$ showed a nearly constant average photon number but a
linear increase in photon number distribution width $\sigma$ with $N /
N_{\rm ex}$.  However, it is difficult to determine whether the
results are related to regular injection of atoms, which in general
causes a reduction of noise (i.e. smaller $\sigma$).  We note that the
results in the case $N=1$ do not agree with single-atom micromaser
theory \cite{Filipowicz-PRA86}.

In this paper we study the case of realistic, Poisson atomic
injection, and interpret the results based on finite transit-time
rather than many-atom effects.  

%
%


\section{Quantum trajectory analysis}

To simulate the microlaser with random (Poisson) injection of atoms,
we used a quantum trajectory algorithm
\cite{Molmer-1991},\cite{Carmichael-1993}.  This technique was
previously used to analyze the threshold-like transition, and the
possibility of many-atom effects, in the original standing-wave
microlaser in \cite{Yang-PRA97}.  This paper contains for a detailed
description of the extended quantum trajectory technique also employed
in the present study.

Due to limitations of computer memory and processing time, for each
simulation run it was necessary to set a maximum value $N_{\rm max}$
for the number of atoms that could be present in the cavity at a given
time.  
In each case $N_{\rm max}$ was adjusted such that the ``overflow''
probability was 1\% or less, ensuring that the statistics are very
close to Poissonian.

In single-atom micromaser theory \cite{Filipowicz-PRA86}, the behavior
of the system is determined by three parameters: (i) the expected photon number
$N_{\rm ex} = r / \Gamma_{\rm cav}$
where $r$ is the atom injection rate; (ii) the pump parameter
$\theta = \sqrt{N_{\rm ex}} \ g t_{\rm int}
$ where $g$ is the atom-field coupling constant; and (iii) the
thermal photon number $n_b$.

It is then shown that, with the assumptions of single-atom events and
$\Gamma_{\rm cav} t_{\rm int} << 1$ (weak decay), \cite{Filipowicz-PRA86} the
steady-state photon number distribution in the cavity is given by

\begin{equation}
%
\label{micromaser-theory-eqn}
p_n = p_0{\left[\frac{n_b}{1+n_b}\right]}^n \prod_{k=1}^{n}\left[1+
\frac{N_{\rm ex} \sin ^2(\sqrt{k+1} g t_{\rm int})}{n_b k}\right]
\end{equation}

\noindent where $p_0$ is determined by normalization.  In this paper
Eq.~\ref{micromaser-theory-eqn} will be referred to as the result from
micromaser theory, although it is also applicable to the microlaser.

In this study we set thermal photon number $n_b = 0$

We performed runs of 10,000 -- 100,000 atom injections, for varying
values of $N_{\rm ex}$, $\theta$ and number of intracavity atoms $N
\equiv r t_{\rm int}$.  We scaled $g$ and $\Gamma_c$ as necessary to
obtain the correct values for $N_{\rm ex}$ and $\theta$ as $N$ was
varied.  At the beginning of each run, at least 2,000 atom
injections were neglected in order to allow the system to reach
steady state before data collection began.

Most simulations were performed with $N_{\rm ex}=10$.  This value was
chosen for several reasons.  First, a relatively low photon number
allows larger numbers of atoms to be considered.  Second, effects that
scale as $N/N_{\rm ex}$ are more apparent for smaller $N_{\rm ex}$.
Finally, we observed that simulations reached their steady-state
distributions much more rapidly for lower $N_{\rm ex}$.  

For simplicity, and to allow comparison with analytic theory, we
regard $g$ and $t_{\rm int}$ to be perfectly well-defined
(i.e. uniform coupling and monovelocity atoms).  Our QTS program has
also allowed us to simulate the effects of various types of variation
and broadening in a realistic system.  Results of these investigations
will be reported elsewhere.

The simulation was written in C (see Appendix) and all
calculations were performed on a 400 MHz Pentium II workstation
running Linux.  The calculations presented here required roughly 2000
hours of CPU time in total.

\section{Results}

\subsection{Single atom limit; Trap states}

\bfig 
\centerline{
\resizebox{\textwidth}{!}{\includegraphics{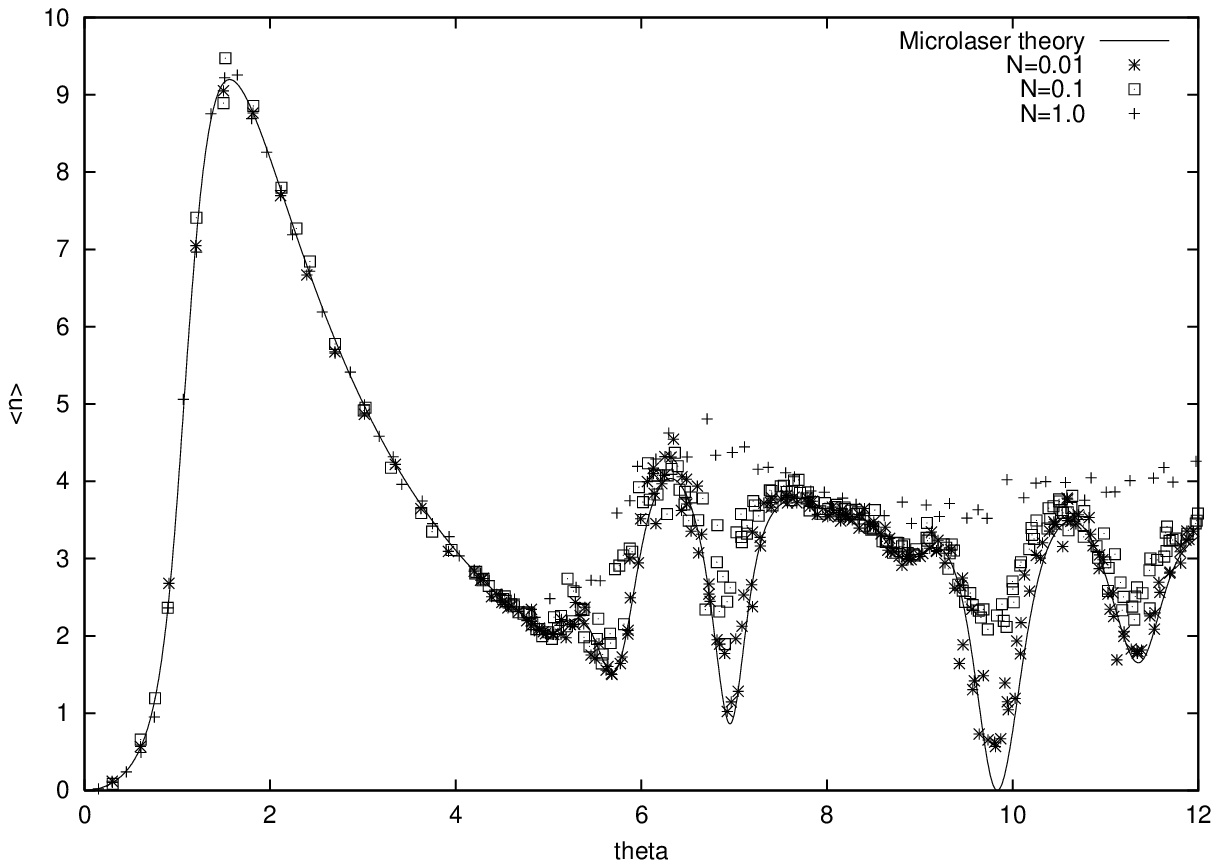}}
}
\caption{Two-atom events destroy the trap state resonances}
\label{fig-trap-state}
\efig

Fig.~\ref{fig-trap-state} shows the average photon number as a
function of $\theta$, for $N_{\rm ex}=10$.  Note the deep depressions
due to trap states resonances in which atoms perform integral numbers
of Rabi oscillations.

With $N=0.1$ the QTS results for average photon number $\langle
n\rangle$ show significantly smaller modulation due to trap states than
the result of Eq.~\ref{micromaser-theory-eqn} although the probability
of the cavity containing 2 or more atoms at any given time is $\sim 2
\%$.  It is only with $N=0.01$ that QTS results give good agreement
with micromaser theory.  This illustrates the sensitivity of the trap
state resonances to 2-atom events as described in \cite{trap-state}.

The photon number distributions in the low-$N$ limit were verified to
be in excellent agreement with Eq.~\ref{micromaser-theory-eqn},
further confirming that the simulation was working properly.


\subsection{Broadening of photon number distribution}

Figs.~\ref{fig-qts-n} and \ref{fig-qts-q-plot-main} show
values of average photon number $n$ and Mandel parameter $Q \equiv
({\langle n^2\rangle-\langle n\rangle^2})/{\langle n\rangle} - 1$ as a
function of $\theta$ for a range of atom numbers $N$, with $N_{\rm
ex}=10$.  Note that for larger values of $\theta$, $Q$ is large and
may be difficult to interpret due to the presence of 2 or more peaks
in the photon number distribution.

\bfig 
\centerline{
\resizebox{\textwidth}{!}{\includegraphics{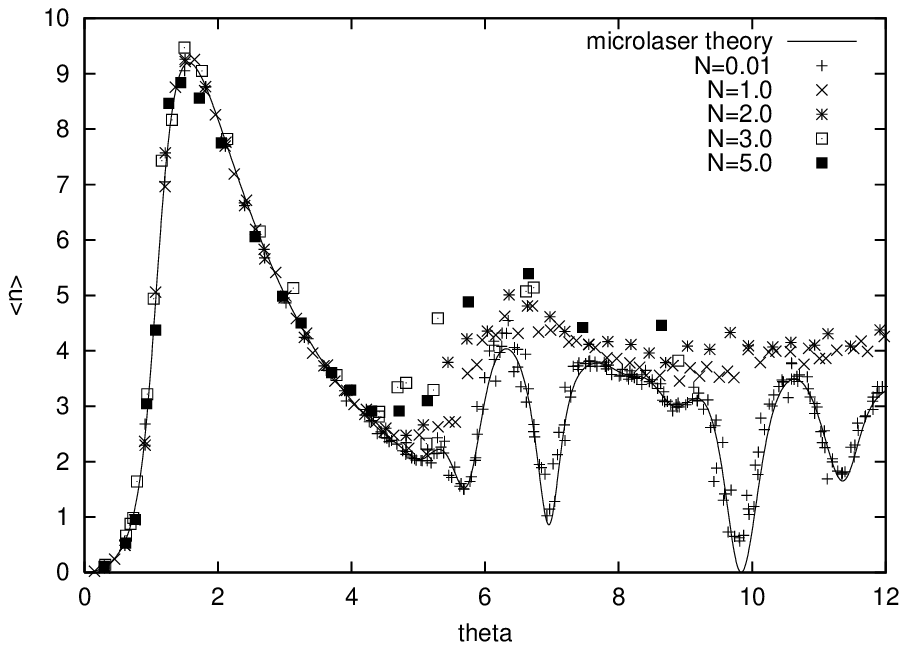}}
}
\caption{Quantum trajectory results for average photon number}
\label{fig-qts-n}
\efig

\bfig 
\centerline{
\resizebox{\textwidth}{!}{\includegraphics{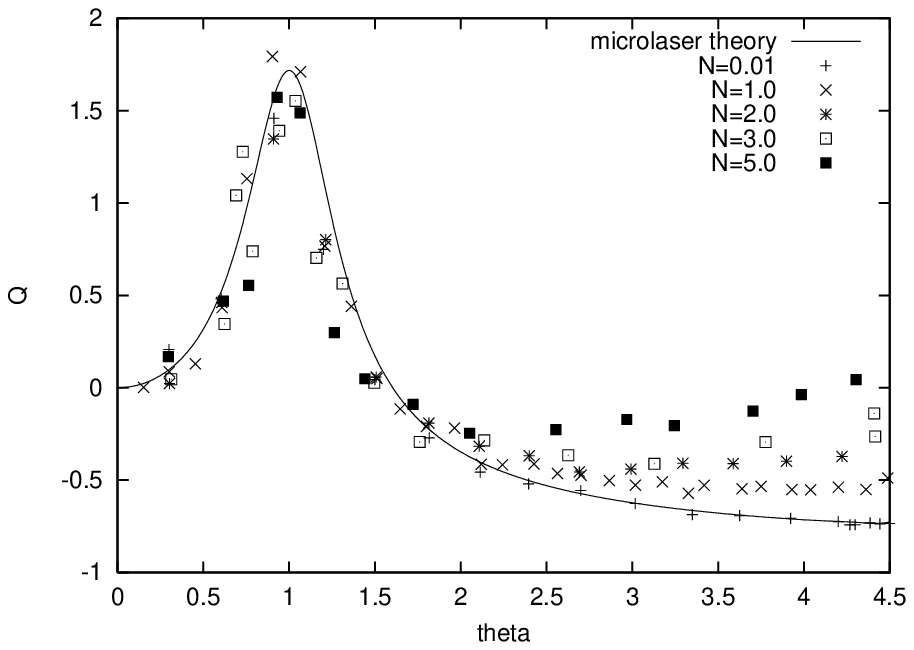}}
}
\caption{Quantum trajectory results for Mandel $Q$ parameter}
\label{fig-qts-q-plot-main}
\efig

For $\theta < 5$, before the trap state resonances occur, the results
for average photon number remain remarkably close to the prediction of
micromaser theory.  The values for $Q$, however, show a significant
increase with $N$ in the range $2 < \theta < 5$.  This
interval corresponds to the region in which according to micromaser
theory the system displays sub-Poisson statistics ($Q<0$).


Similar increases in $Q$ are observed in a small interval near $\theta
= 8$, between two trap states.  Simulations with $N_{ex} = $30 and $50$,
in which trap state resonances are more narrow, showed similar
behavior over wider intervals between trap states.

\bfig 
\centerline{
\resizebox{\textwidth}{!}{\includegraphics{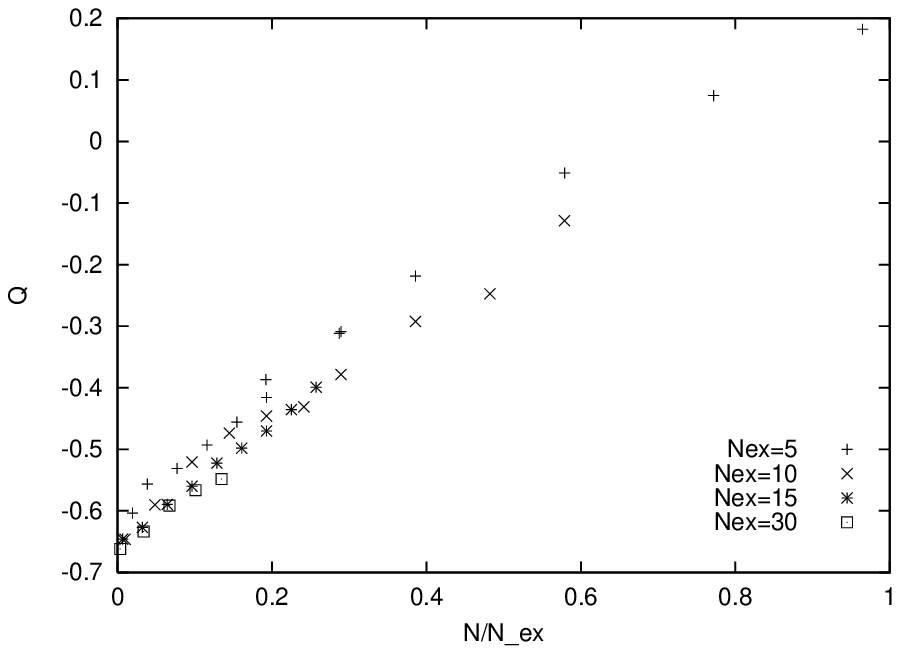}}
}
\caption{Universal curve showing an increase in $Q$ which scales with $N/\Nex$}
\label{fig-qts-universal-curve}
\efig

The parameter $Q$ was calculated as a function of $N/N_{ex}$, with
$\theta=3$, for $N_{\rm ex} = 5$, $10$, $15$, $30$.  The increase of
$Q$ is seen to be linear with $N/N_{ex}$; for $N_{ex} \geq 5$ we find
(Fig.\ref{fig-qts-universal-curve}) that all points may be
approximately described by

\begin{equation}
Q = Q_0(N_{ex}, \theta) + \alpha(\theta) (N/N_{\rm ex})
\label{eq:extra-variance}
\end{equation}
\noindent
where $Q_0$ is the value of the Mandel parameter from micromaser theory and
$\alpha(\theta=3) \approx 0.80 \pm .05$.  

\bfig 
\centerline{
\resizebox{\textwidth}{!}{\includegraphics{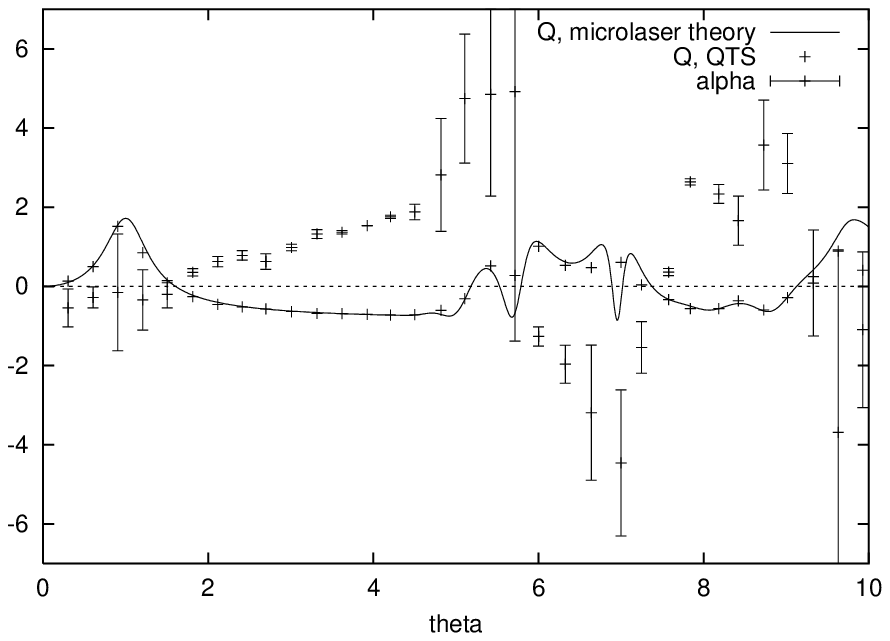}}
}
\caption{Slope  $\alpha(\theta)$ of the increase in $Q$ with $N/\Nex$, as a
function of pump parameter}
\label{fig-qts-alpha-plot}
\efig

Fig.~\ref{fig-qts-alpha-plot} shows $\alpha(\theta)$ calculated via a
least-squares fit with $N_{\rm ex}=10$ and $0 < N < 2$.  Error bars
indicate the degree of linearity.  It is seen that $\alpha$ exhibits
oscillation with $\theta$ and is positive for $\theta$ where the
single-atom, weak-decay theory predicts sub-Poisson statistics
($Q<0$).

\subsection{Many-atom vs. finite transit time effects}

An increase of variance proportional to $N/N_{\rm ex}$ was previously
found by Elk, who interpreted it as being due to many-atom
interactions \cite{Elk-PRA96}.

We note, however, that $N/N_{\rm ex}$ is equal to simply $t_{\rm int}
\Gamma_c$, a parameter describing cavity decay during the transit
time, and is not directly related to the number of atoms.
The extra variance observed in this study and by Elk can therefore be
seen as a consequence not of multiple atoms, but of cavity decay
during an atom's transit time.  For example, we may have $N \gg 1$ but
virtually no extra variance if $N_{\rm ex} \gg N$.  More importantly,
there are examples in which $N \ll 1$ but excess variance still
occurs.  Fig.~\ref{fig-nex1} shows $Q$ as a function of $N$ for $N_{\rm ex}=1$,
$\theta=2$.  A linear increase in variance is seen even for $N = 0.1$.

\bfig 
\centerline{
\resizebox{4in}{!}{\includegraphics{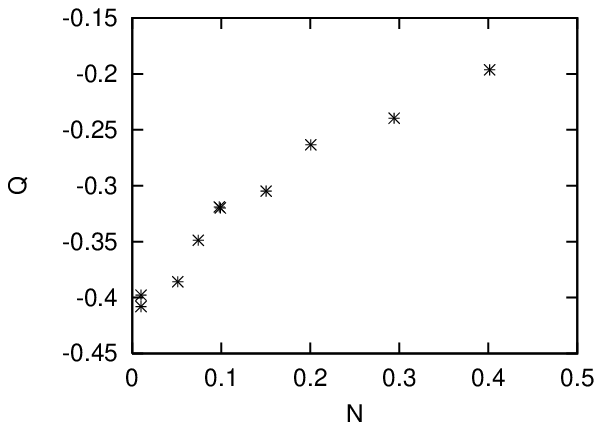}}
}
\caption{Increase of variance with $N/\Nex$ for $\Nex=1$}
\label{fig-nex1}
\efig

\section{Predictions for experiment}

To apply our quantum trajectory results to our microlaser system, we
make a seemingly precarious assumption that simulations for $N \leq 5$
atoms represent the behavior for up to $N\sim 1000$.  If true then the
predictions for the experiment are as follows.

No significant deviation from the single-atom theory was found for the
few-atoms microlaser; therefore we would expect the cavity photon
number to follow the prediction of the microlaser with realistic
parameters described in Sec.~\ref{theory-realistic-effects}, as
plotted in Fig.~\ref{fig-re-mt-realistic}.


For the photon statistics, we have in the microlaser $N/\Nex \approx
0.1$, and therefore the perturbation to photon statistics is fairly
small; it would not differ strongly from that given in
Fig.~\ref{fig-qplot02}.

An upper-bound prediction for the $Q$ parameter is given in
Fig.~\ref{fig-qts-microlaser-q}.


\bfig
\centerline{$Q$ \hspace*{4.5in}}
\centerline{\resizebox{5in}{!}{\includegraphics{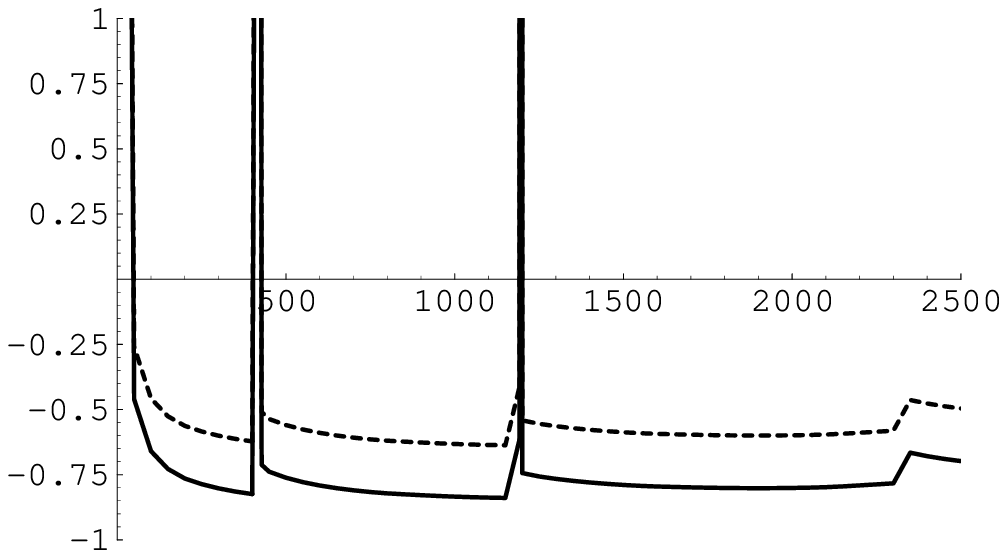}}}
\centerline{$N$}
\caption{Dashed line: upper bound of $Q$ parameter, including finite decay time
effect.  Solid line: $Q$ parameter from microlaser theory without
without finite decay time effects, as in Fig.~\ref{fig-qplot02}.}
\label{fig-qts-microlaser-q}
\efig

\section{Discussion}

Our calculations have clarified the distinction between many-atom
effects and effects due to cavity decay during transit-time.
Many-atom events destroy the trap state resonances and therefore have
strong effects for parameters for which these resonances are
prominent.  Away from these resonances, the system shows remarkable
agreement with single-atom micromaser theory.  However, for values of
$\theta$ for which micromaser theory predicts sub-Poisson statistics,
cavity decay during transit time cause an increase in variance
relative to micromaser theory, by an amount proportional to $N /
N_{\rm ex} = t_{\rm int} \Gamma_c$.




Lack of collective effects away from trap states may be a consequence
of the noncorrelated Rabi phases of atoms which have entered the
cavity at different times.  Increase of variance represents an
absolute decrease in the negative differential gain in the atom-cavity
system.  Future work will include development of models to explain the
observed effects quantitatively.




\chapter{Apparatus and methods of microlaser experiment}
\label{chap-methods}



In this chapter we describe the components of the microlaser and
techniques used in its study.  In particular we describe how a
well-defined interaction between atoms and cavity can be achieved.




\section{Two-level atom}


\subsection{Barium energy levels}

\bfig
\resizebox{\textwidth}{!}{\includegraphics{fig-791-553-levels.epsi}}
\caption{Relevant energy levels for \ba.}
\label{fig-ba-energy-levels}
\efig

Figure~\ref{fig-ba-energy-levels} shows the levels of \ba\ relevant to
the microlaser experiment.  The microlaser uses the
\transitiontriplet\ transition of wavelength $\lambda = 791.1$ nm and
linewidth approximately 50 kHz.  The excited state decays to the
ground state and two metastable D states with a branching ratio of
0.43:0.41:0.16.  We use this intercombination line due to its
relatively long lifetime $\ga^{-1} = 1.3\ \us\ $.

The most frequently studied transition in \ba is \transitionsinglet\
at 553 nm.  It has a linewidth of about 19 MHz.  The fluorescence
spectrum of a naturally occuring barium sample is shown, with isotope
shifts, in Fig.~\ref{fig-ba-553-spectrum}.

The 553 nm transition was used to measure atomic density in the
cavity, determine pump beam efficiency, and measure the atomic
velocity distribution.

\bfig[t]
\centerline{\resizebox{4in}{!}{\includegraphics{fig-Ba-isotopes-sm.epsi}}}
\caption{Isotope spectrum of the \transitionsinglet\ transition at 553 nm.}
\label{fig-ba-553-spectrum}
\efig

Natural occuring barium, of which 71.7\% is \ba, is relatively
inexpensive and readily available.  We purchased the metal in rod
form, packed in argon, from Alfa Aesar.

\section{Optical resonator}

The microlaser cavity mirrors consist of superpolished fused silica
substrates coated with a multilayer, ion-sputtered dielectric coating.
The substrates and coatings were purchased from Research
Electro-Optics (Boulder, Colorado).  The coating consists of 45
alternating layers of ${\rm SiO}_2$ and ${\rm Ta}_2{\rm O}_5$.

\subsection{Characteristics}

The microlaser cavity consists of two mirrors in a symmetrical, near-planar
Fabry-Perot configuration.  The mirror separation, measured via
transverse mode spacing, is 1.1 mm.  The radius of curvature of the
coated surfaces is 10 cm.



The mirrors have an overall diameter of 7?mm and length of ? mm.  The
actual reflective surfaces are 45-degree beveled to a diameter of 3 mm
in order to facilitate alignment of the tilted atomic beam and
cleaning of the coated surfaces.

The mirrors are glued into stainless steel holders
(Fig.~\ref{fig-cav-holder}) designed to hold the mirrors without
birefringence-inducing stress.  The cavity is aligned outside the
chamber, then the holders are glued on the inside of a cylindrical
piezo transducer (PZT) used to modulate the cavity mirror spacing.

\bfig
\centerline{\resizebox{4in}{!}{\includegraphics{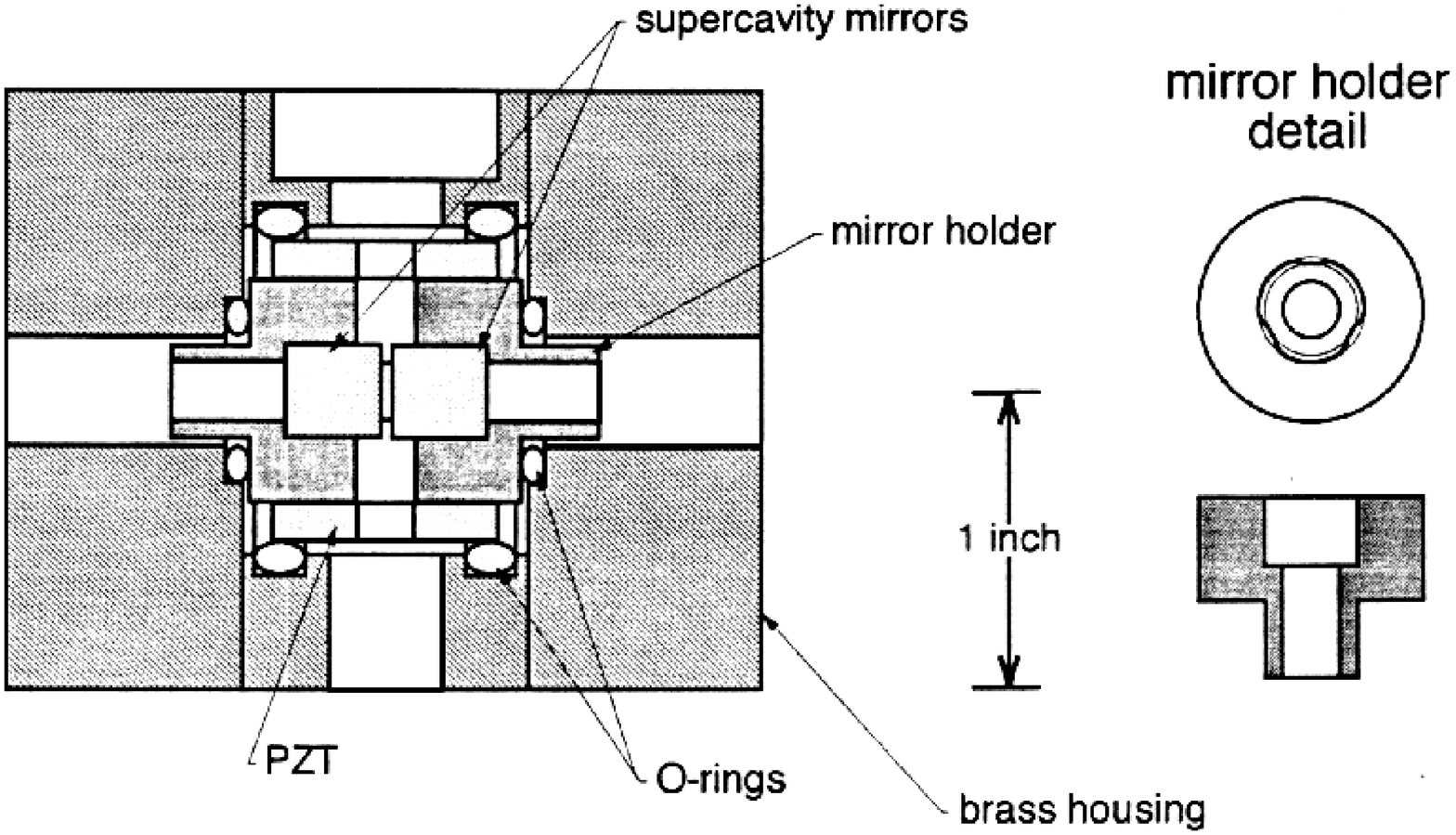}}}
\caption{Brass cavity housing assembly and stainless steel
cavity holders, from \cite{an-thesis}. }
\label{fig-cav-holder}
\efig
\clearpage

\bfig
\center
\resizebox{4in}{!}{\includegraphics{cavity-assembled.epsi}}
\caption{Microlaser resonator assembled with PZT}
\label{}
\efig

\bfig

\centerline{\resizebox{6in}{!}{\includegraphics{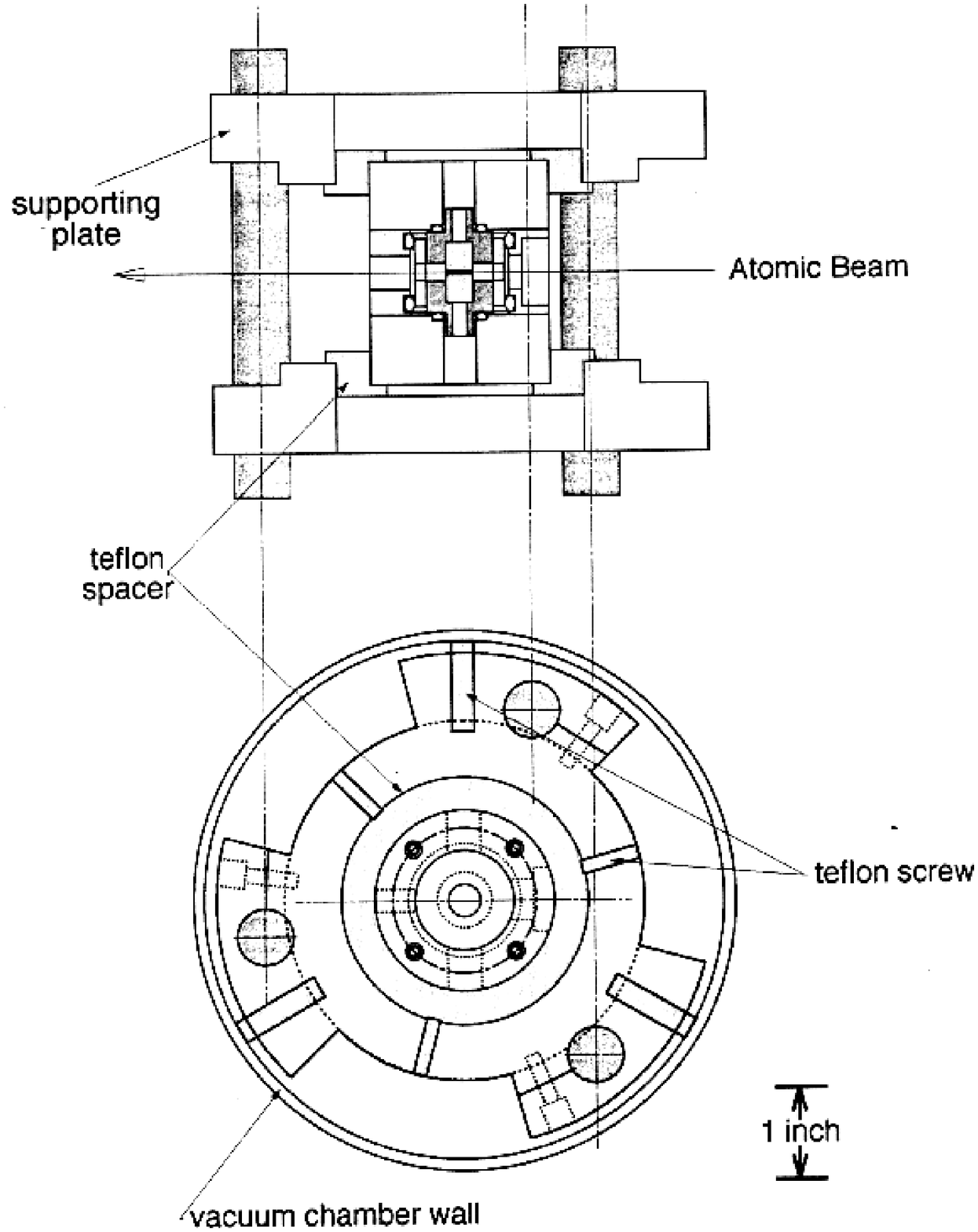}}}
\caption{Support system for cavity assembly in chamber.}
\label{fig-cav-mount}
\efig

The microlaser experiment is performed with the fundamental
(TEM$_{00}$) mode of the cavity (it is also possible to use
higher-order transverse modes).  The mode waist is given by 
\be
w_0 = \left({r_0 L \lambda^2} \over {2 \pi^2} \right)^{1/4} \approx 41.1 \um
\ee
\noindent where $r_0$ is the mirror radius of curvature, $L$ is the cavity
spacing, and $\lambda$ is the wavelength.

\bfig
\centerline{\resizebox{4in}{!}{\includegraphics{fig-tem00.epsi}}}
\caption{Cavity transmission of an earlier cavity (from \cite{an-thesis})
showing TEM00 and higher order transverse modes.  Transverse mode spacing
in the current cavity is 7.11 GHz.}
\label{fig-tem00}
\efig

The cavity mode volume is given by
\be V = {\pi \over 4} L w_0^2 \ee
\noindent and the peak atom-cavity coupling for the standing-wave case is
\be 
g_0 = {\mu \over \hbar}\sqrt{{2 \pi \hbar \omega}\over V} = 385\ {\rm kHz}
\ee

The atom-cavity interaction time for a particular atom is defined as
\be
\tint = {{\sqrt{\pi} w_0} \over v}
\ee
\noindent where $v$ is the atom's velocity.  The $\sqrt{\pi}$ factor
accounts for the Gaussian shape of the cavity mode: the integral of
$g$ with time for a moving atom is the same as an atom of the same
velocity traversing a ``top hat'' mode of constant coupling $g_0$ for
time \tint:
\be
\int_{-\infty}^{+\infty} g_0 \exp\left({{-(vt)^2}\over{w_0^2}}\right) dt = 
{{\sqrt{\pi} w_0 g_0} \over v} =  g_0 \tint 
\ee

The cavity spacing $L$ was determined via measurement of the
transverse mode spacing.  In a near-planar cavity,

\be \Delta \nu_{\rm t} = {c \over {2 \pi \sqrt{L R / 2}}}, \ \  L \gg r_0 \ee

\noindent where higher-order corrections can be shown to be  $\sim 10^{-3}$
for our cavity \cite[ch.\,19]{siegman-lasers}.

We determined $\Delta \nu_{\rm t}$ by measuring the difference in PZT
voltages between the fundamental and first transverse mode resonances.
The scan was calibrated by measuring the PZT voltage shift between two
sidebands of the 24.67 MHz FM modulation used for laser frequency
locking (see Sec.~X).  We find the PZT scan calibration is 171.6 MHz/V
and verified that it does not vary measurably over the PZT voltage
range in question.  The transverse mode spacing was measured to be
\be \Delta
\nu_{\rm t} = 7.11\ \rm{GHz} \ee
\noindent giving a cavity length \be L = 0.90\ {\rm mm}. \ee

\bfig[t]
\centerline{\resizebox{\textwidth}{!}{\includegraphics{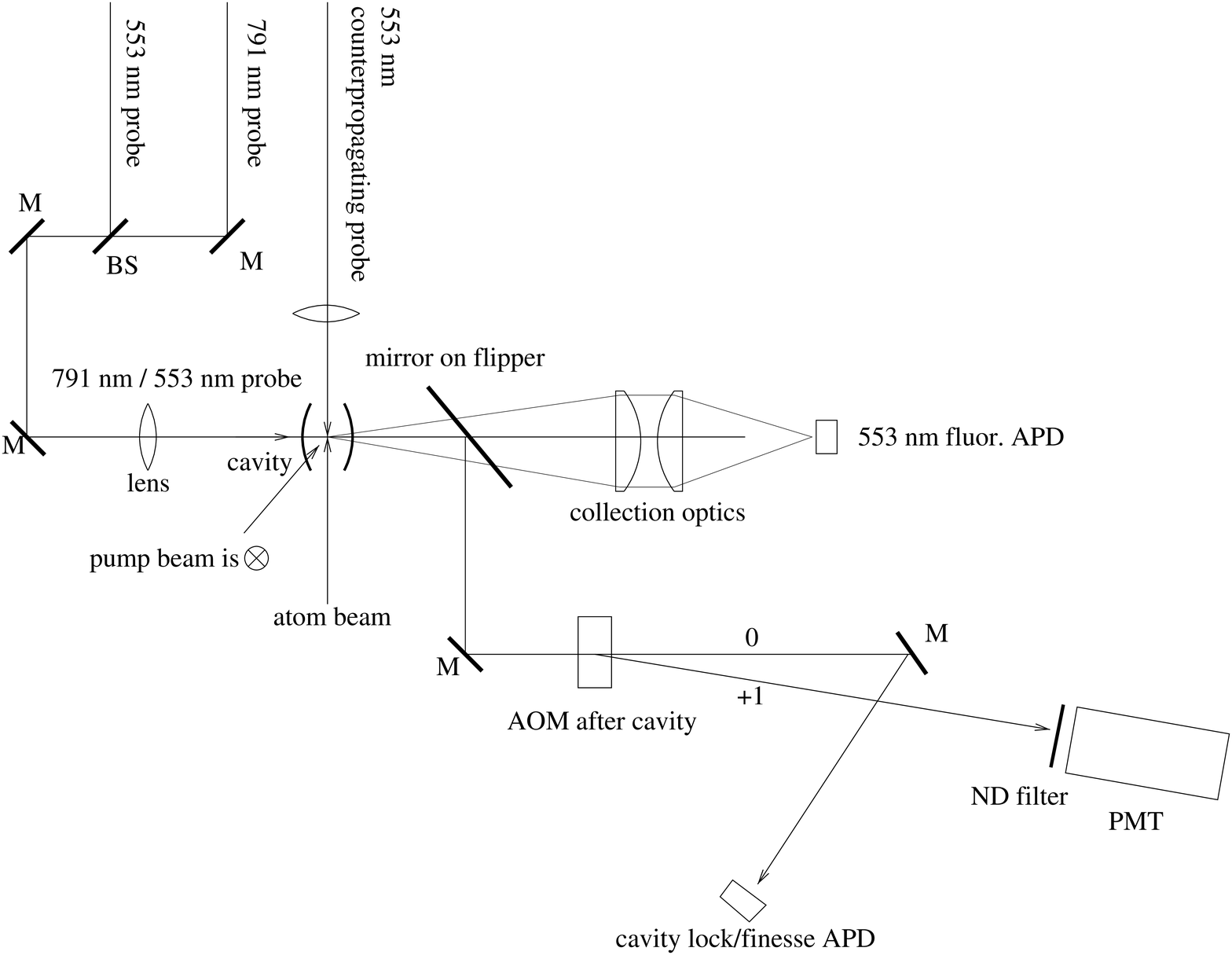}}}
\caption{Schematic of optics for microlaser experiment.  To the right
of the cavity a mirror on a ``flipper'' mount is used to switch
between cavity fluorescence detection for velocity distributions and
detection of cavity transmission.}
\label{fig-mlsetup}
\efig

\subsection{Cavity PZT}

The PZT controlling the cavity separation is a tube (Vernitron PZT-5A)
of height 0.75'' and inner diameter 0.75'', and wall thickness
0.125''.  Four holes are drilled in the cylinder wall to allow passage
of the atom and pump beams.  

Isolation mechanisms are used to isolate the cavity from vibrations.
Viton o-rings hold the PZT inside a heavy brass housing
(Fig.~\ref{fig-cav-holder}).  The housing is placed in a supporting
structure by fittings made of teflon (Fig.~\ref{fig-cav-mount}); the
support in turn is secured inside the vacuum chamber with teflon screws.

\subsection{Finesse measurement by ringdown technique}

\bfig 
\center
\resizebox{5in}{!}{\includegraphics{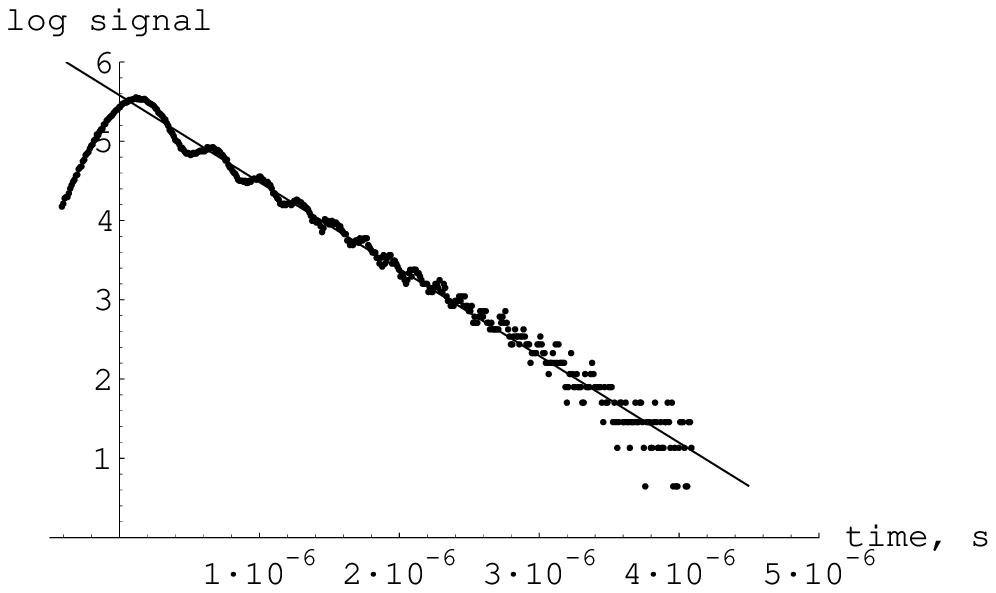}}
\caption{Log of ringdown signal (a.u.) vs.\ time.  Line shows least-squares 
linear fit. Here $T_{\rm cav} = 1.02\ \us$ and finesse $F~=~1.0\times~10^6$.}
\label{fig-ringdown}
\efig

The finesse of a Fabry-Perot cavity is defined as the ratio of free
spectral range (FSR) to the FWHM of a cavity transmission peak, or equivalently.
\be \f = { {\sqrt{R} \pi } \over {1-R}} \approx  {{ \pi } \over {1-R}} 
\ee
\noindent where $R$ is the geometric average of the intensity 
reflectivity of the two mirrors, i.e.~$R=\sqrt{R_1 R_2}$ and we assume
$R$ very close to unity.

To measure finesse we scan the cavity PZT through resonance with a
probe laser beam.  The cavity decay time is measured by fitting a
curve to the exponential decay of the cavity field
(Fig.~\ref{fig-ringdown}).  The signal shows a sinusoidal modulation
due to interference between the probe beam and the Doppler-shifted
cavity field.  This is described further in Ref.~\cite{An-OL95}.

\subsection{Absorption measurement via thermally-induced optical bistability}

The finesse is given by $\pi/(1-R)$, with $R \approx 1$ the mirror
reflectivity and $1-R$ the total loss, equal to the sum of three loss
parameters, the absorption $A$, scatter $S$ and transmission $T$.  R,
hence the total loss, can be measured by cavity ringdown
\cite{An-OL95}.  However, determining how the loss is distributed
among $A$, $S$ and $T$ is difficult.  In particular, few techniques
exist for measuring very small absorption coefficients.  Methods
such as photothermal deflection \cite{Commandre-AO96} are capable of
measuring $A$ down to 1 ppm.


\begin{figure}
\centerline{\resizebox{4in}{!}{\includegraphics*{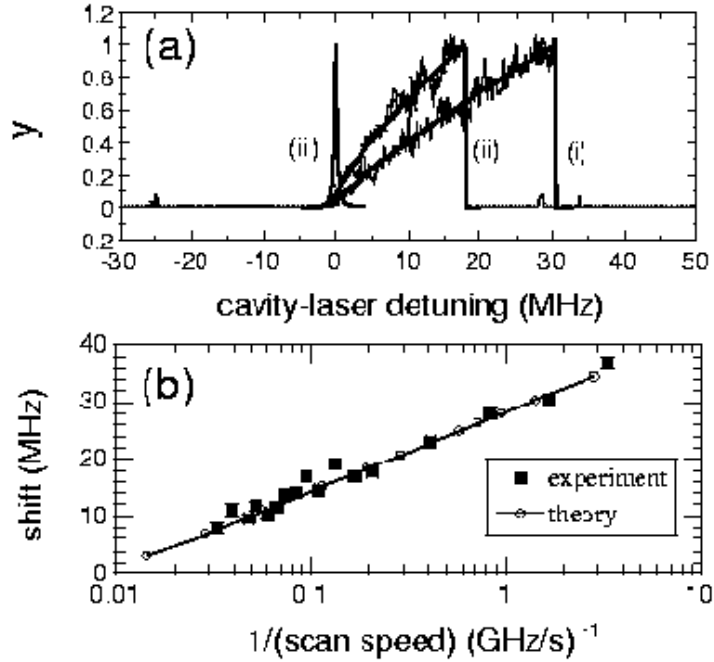}}}

\caption{Figure 4.  Anomalous cavity transmission.  Scan speeds: (i)
0.6, (ii) 4.8 GHz/s, compared with theory; (iii) unmodified cavity
lineshape.  (b) Frequency shift vs.\ inverse cavity scan speed.  The
best fit is for $A = 0.2$ ppm.}
\end{figure}

Recently we have developed a new technique to measure $A$ at the
sub-ppm level \cite{An-OL97a,Bryndol-thesis}.  This technique relies
on the intense circulating field created when a resonant probe beam is
incident on a high finesse cavity.  Any absorption of the intense
internal field causes heating in the coating layers, and the
subsequent thermal expansion of the mirror substrate reduces the
mirror separation distance and leads to a thermally-induced optical
bistability of the empty resonator.  This optical bistability causes
distortion in the measured cavity lineshape, as observed by slowly
scanning the cavity length.  The lineshape exhibits an extended
resonance with a slow rise and rapid drop (Fig.~4), the duration of
the resonance being directly related to the absorption coefficient $A$
\cite{An-OL97a}.  

We have characterized all three parameters and established that $A
\approx 0.2$ ppm and $S \approx 2.5$ ppm, thus establishing that
scattering is the predominant loss mechanism in our supercavity
mirrors \cite{Bryndol-thesis}.



\newpage

%
%
%

\subsection{Uniform coupling by traveling-wave interaction}

In the original microlaser experiment, the atom beam is incident
normal to the cavity axis and atoms interact with the standing wave of
the Fabry-Perot cavity.  Since the coupling $g$ is proportional to the
local field amplitude, which varies as $\sin^2(kz)$, atoms experience
a wide range of coupling strengths.

We eliminated this problem by introducing a small tilt $\theta \ll 1$
between the atom beam direction and the normal to the cavity axis.
Due to the Doppler shift, each atom then experiences not one standing
wave but two traveling waves of frequencies $\omega \pm kv\theta$.  If
$\theta$ is sufficiently large, the atom will be resonant with only
one traveling wave component at a time.  This eliminates the fast
spatial variation due to the standing wave.  It should be noted that
in the traveling-wave configuration the peak coupling constant, $g_0$,
is reduced by a factor of $2$ compared to the peak standing-wave value
(see Appendix~\ref{app-tw}).  (This factor of 2 was incorrectly
reported as $\sqrt{2}$ in \cite{An-OL97b}.)

\begin{figure}
\centerline{\resizebox{3.5in}{!}{\includegraphics{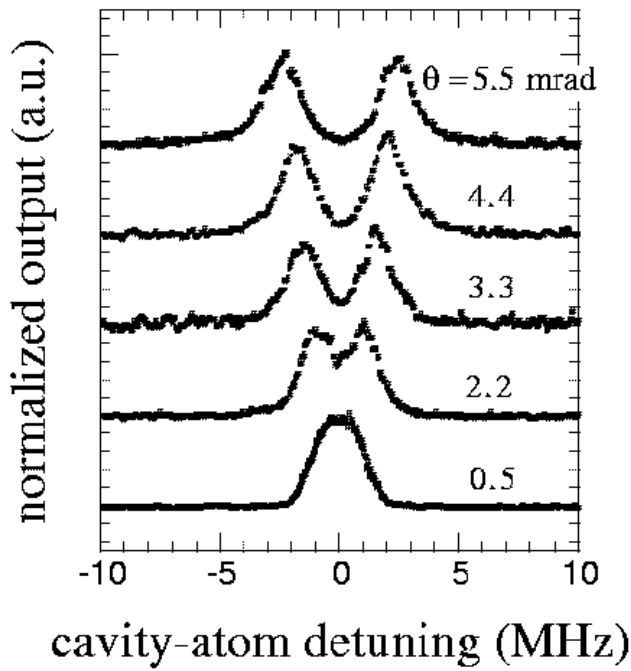}}}
\caption{Microlaser output as a function of
cavity-atom detuning, for varying atomic beam tilt angles $\theta$.
From \cite{An-OL97b}}
\label{TW}
\end{figure}


\section{Atomic beam}

A typical atom beam apparatus consists of a container in which the
sample is heated in vacuum to form a vapor; a small nozzle allows
vapor atoms to escape.  Apertures are placed downstream to collimate
the beam.

We have two main requirements for our barium atomic beam: high beam
density and narrow velocity distribution in the ground state.

\subsection{Velocity selection: mechanical, optical, or supersonic}

An atomic beam with narrow velocity spread is needed to provide a
well-defined atom-cavity interaction time (i.e. uniform transit time),
and to ensure that all atoms traversing the pump laser field prior to
entering the cavity are excited to a state of complete inversion
($\pi$ pulse condition).

The ``classical'' technique of mechanical velocity selection (via
e.g. Fizeau wheel) has several serious drawbacks: the beam created is
pulsed, not continuous, and the overall efficiency is low.

Two new methods for creating narrow-velocity ground state atom barium
beams have been developed in our laboratory.

\subsection{Atomic velocity selection via optical pumping}
\begin{figure}
\hspace{-.3in}\noindent \resizebox{!}{3in}{\includegraphics*[1in,7.1in][8in,10in]{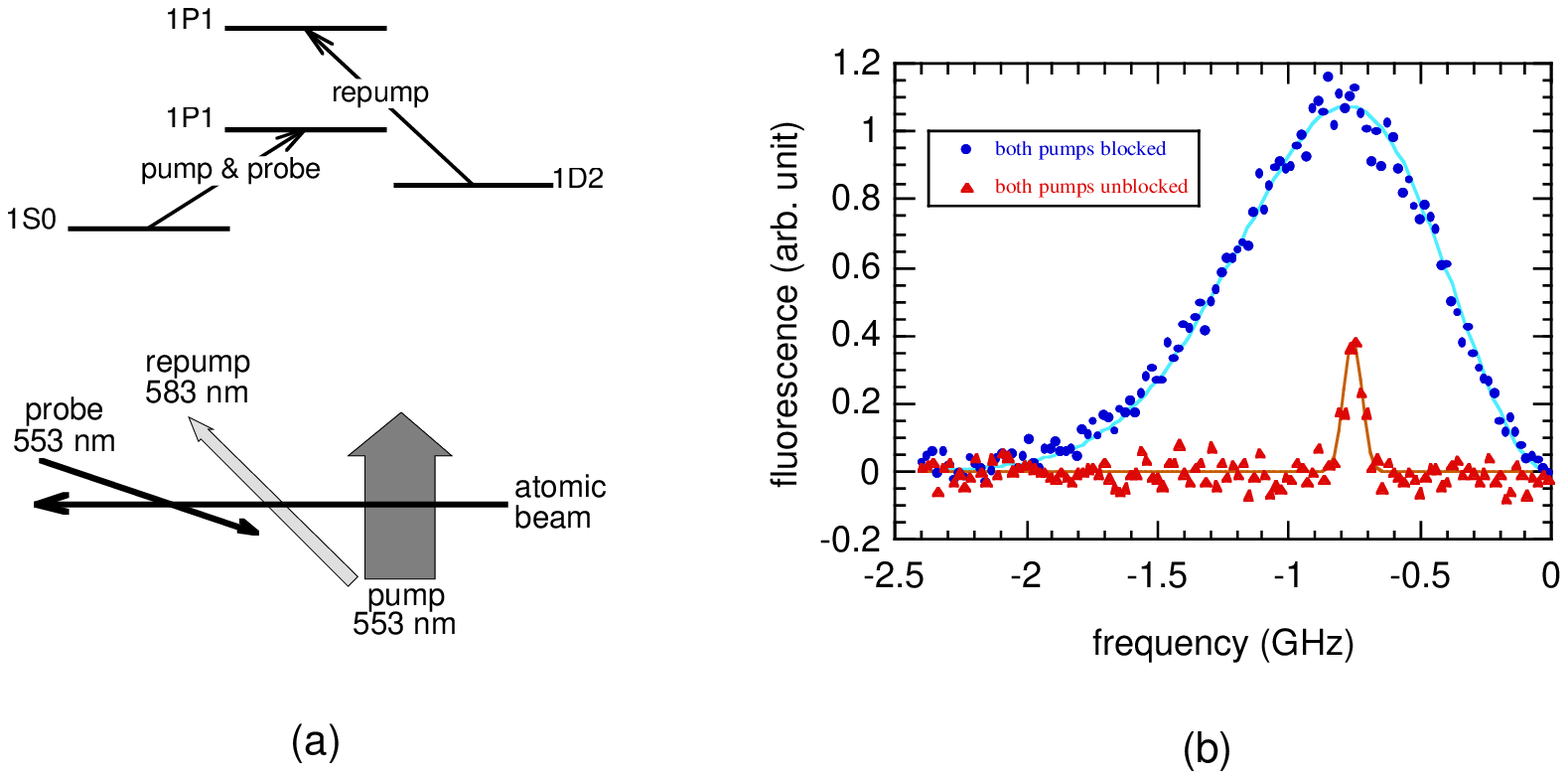}}
\vspace{-.3in}
\caption{Left: Twocolor optical pumping scheme.  Right: Velocity
profile of the selected atoms.  In this experiment the pump laser is
tuned to resonance, wherease the repump laser is tuned to resonance
with the atoms with most probable velocity while the probe laser is
scanned.  In this example $\delta v$(FW)$/v\approx$ 13\%.}
\label{vel-sel-result}
\end{figure}

We have developed a two-color optical pumping method for selectively
preparing atoms in the ground state at a certain velocity, in an
effusive beam.  The beam is created in a cylindrical, externally
heated oven made of Inconel.  Continuous-wave lasers are used to pump
tbe barium atoms, first into the metastable $^1$D$_2$ state via the
6s$^2$ $^1$S$_0\leftrightarrow$ 6s6p $^1$P$_1$ cycling transition
($\lambda$=553 nm), and then back to the 6s$^2$ $^1$S$_0$ ground state
with velocity selectivity via a tilted laser tuned to the 583 nm 6s5d
$^1$D$_2\leftrightarrow$ 5d6p $^1$P$_1$ transition (Fig.\
\ref{vel-sel-result}).  When the most probable velocity $v_0$ is
selected we obtain $\Delta v_{\rm FWMM}/v_0\approx 10\%$  with approximately
50\% repumping efficiency (Fig.~\ref{vel-sel-result}).

\bfig
\resizebox{\textwidth}{!}{\includegraphics{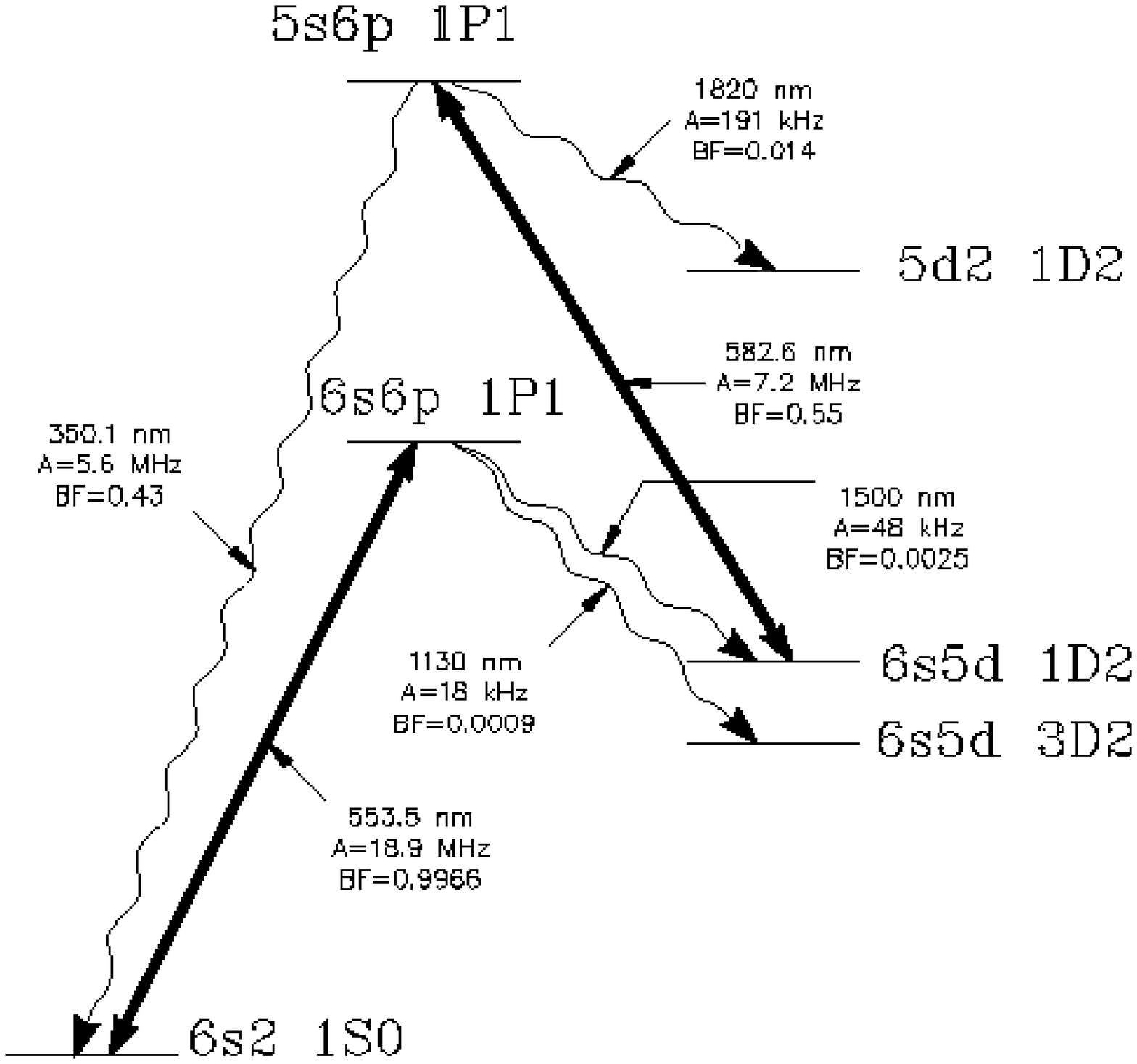}}
\caption{Relevant \ba\ energy levels for optical velocity selection scheme.  Dark arrows represent transitions used for pump (553.5 nm) and repump (582.6 nm).  A = Einstein A coefficient; BF = branching fraction.}
\efig

%
%
%
\subsection{Supersonic oven}

\subsubsection{Background}

Supersonic beam techniques have found many applications in
spectroscopy and atom interferometry since their invention in 1951
\cite{kantrowitz-rsi51,kistiakowsky-rsi51,anderson-pf65}.  
In a supersonic beam, adiabatic expansion as the beam exits the oven
nozzle causes a dramatic cooling in the frame of the moving atoms or
molecules; random thermal energy is converted into directed kinetic
energy.  In addition, the net flow rate is generally much higher than
for an effusive beam.  This solves the two problems associated with
conventional effusive beams: low intensity and a wide distribution of
velocities.

\subsubsection{Supersonic barium oven: apparatus}

We have developed a supersonic beam oven for barium closely following
the design of Ref.~\cite{thomas-ol89}, which was based on an effusive
beam source developed in our laboratory and previously used for the
microlaser experiment \cite{An-PRL94}.

The atomic beam oven consists of resistively heated, barium-filled
piece of tantalum tubing, in which a small nozzle hole has been
drilled.  A diagram in shown in
Fig.~\ref{fig-supersonic-oven-diagram}.  The tubing is 6.5'' long,
0.5'' in diameter, with 0.015'' wall thickness.  The nozzle diameter
varied from 150 \um\ to 406 \um.

Like the atoms strontium and ytterbium used in
Ref.~\cite{thomas-ol89}, barium has two valence electrons and
negligible dimer formation.  

\bfig
\centerline{\resizebox{\textwidth}{!}{\includegraphics{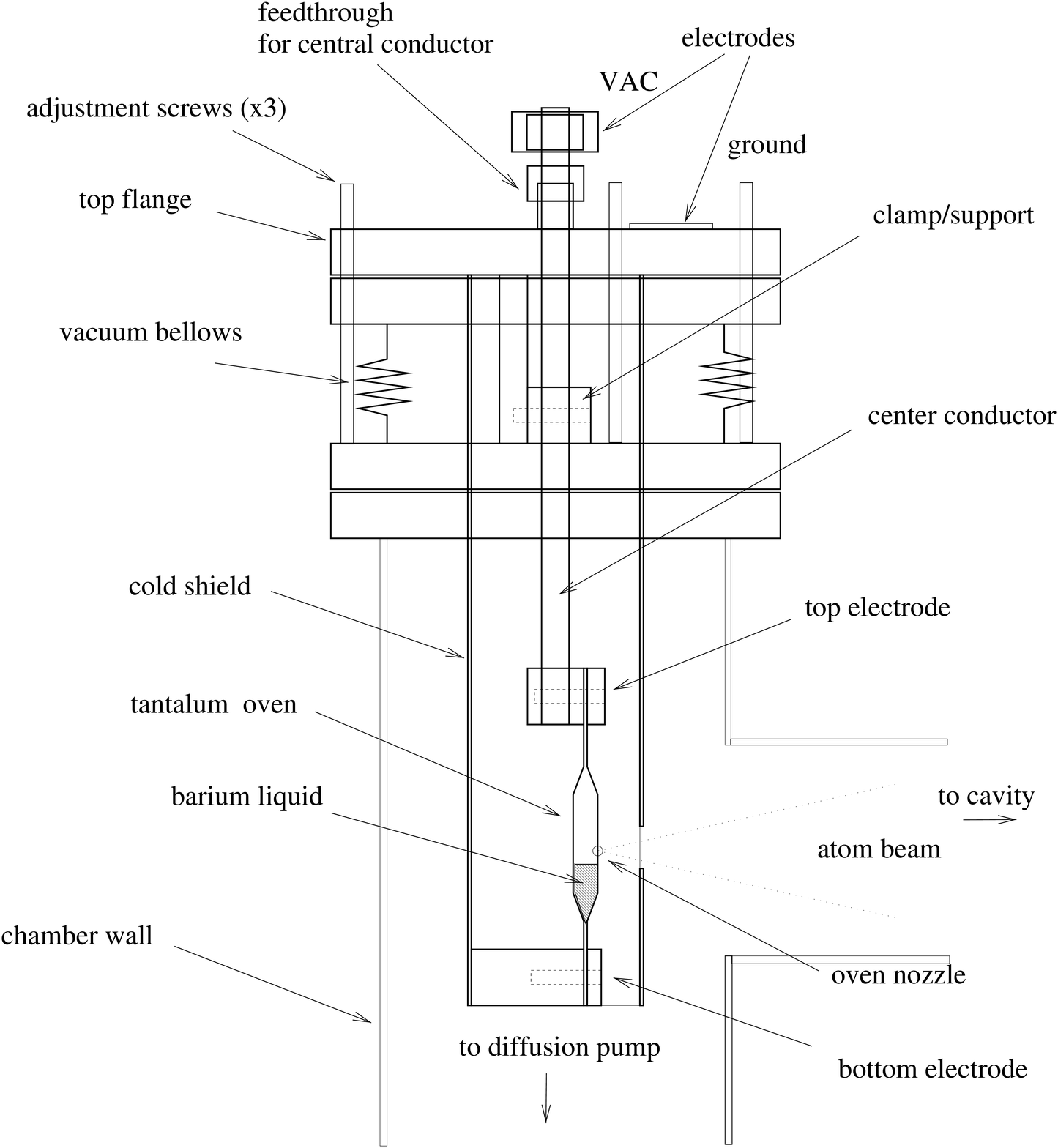}}}
\caption{Supersonic oven design}
\label{fig-supersonic-oven-diagram}
\efig


The oven is created by the following procedure.  A length of tubing is
measured and cut with a pipe cutter.  The bottom 1.125'' is pinched
flat in a vise between two aluminum blocks.  One quarter of a 2 cm
diameter barium rod (cut by bandsaw lengthwise) is inserted into the
tantalum tube.  The typical mass of barium in one oven is 12-16 g.
The top 1.5'' of the oven is then flattened in a vise, parallel with
the bottom of the oven.  The oven is mounted in a specially designed
holder to position it with flat parts horizontal.  It is clamped in a
milling machine where a 1/8'' end mill removes about 0.007'' from the
future location of the nozzle.  This gives a shorter ``channel'' for
the atom flow and creates a flat surface for drilling the nozzle,
which is done in a miniature drill press.

In the experiment, the oven is clamped by two screws at each end
between copper electrodes.  The upper electrode is connected to a
copper tube which extends through an 8'' Conflat flange via a 3/4''
diameter Ultra-Torr fitting.  The bottom electrode is soldered to a
4'' diameter copper cold shield tube which both absorbs the oven's
large thermal radiation and conducts current back up to the flange
coaxially with the central conductor, to minimize stray magnetic
fields from the oven.  The upper and bottom electrodes and the cold
shield are cooled by water flowing through 1/4'' copper tubing which
has been soldered to the copper pieces.

A Proteus flowmeter placed on the return path of the oven cooling
water is connected to a lamp which provides visual feedback that the
water is flowing.

The oven drive current is controlled manually using a variable Variac
autotransformer with 208 VAC line input and 0-230 VAC output.  The
Variac output is connected to a step-down transformer with 1:8 turn
ratio, and the resulting output is connected to the center conductor
and top oven flange (ground) via 2 pairs of 1'' diameter welding
cables.  The oven voltage and current are monitored by AC voltmeter
and a clamp-on ammeter.  The voltage varies from 0-5VAC and current
0-1000 A rms?.  The resistance of the entire circuit varies from
approximately 4 m$\Omega$ to 6 m$\Omega$ depending on oven
temperature.

\subsubsection{Supersonic barium oven: results}

As oven power is increased, the beam density increases and the beam
undergoes a transition from Maxwell-Boltzmann (thermal) velocity
distribution to supersonic velocity distribution.

We measured the atomic density and velocity distribution via \fft\
\transitionsinglet\ fluorescence.  To measure density, the \fft\ beam
was aligned to overlap the \sno\ probe beam along the cavity axis and
the fluorescence was detected by CCD (see section \ref{sec-imaging}).
Velocity distribution was measured by aligning a focused \fft\ beam in
the cavity in a direction opposite to atom beam direction, and
detecting the Doppler-dependent fluorescence via a photon counting APD
(EG\&G C30902TE) previously aligned with the cavity transmission.  A
500
\um\ diameter pinhole located in front of the APD and mounted on an XY
translation stage blocks scattered light from the ground glass beveled
edges of the cavity mirrors.

Figure~\ref{fig-supersonic-veldist} shows the velocity distribution
for the supersonic beam used in the experiments.  This oven had a
nozzle size of 406 \um\ and was driven by about an rms current of
about 600 amperes.

The degree of velocity compression can be characterized by the ratio
between FWHM of the {\it measured} velocity distribution and most
probable velocity ${\Delta v} / {v_0}$.


The measured velocity distribution $f_m(v)$ is equal to the actual
velocity distribution $f(v)$ convoluted with a lineshape function
accounting for atom linewidth and isotope shifts.  Laser power
broadening effects can be minimized by using a probe laser beam
intensity well below saturation.



\bfig
\centerline{\resizebox{4in}{!}{\includegraphics{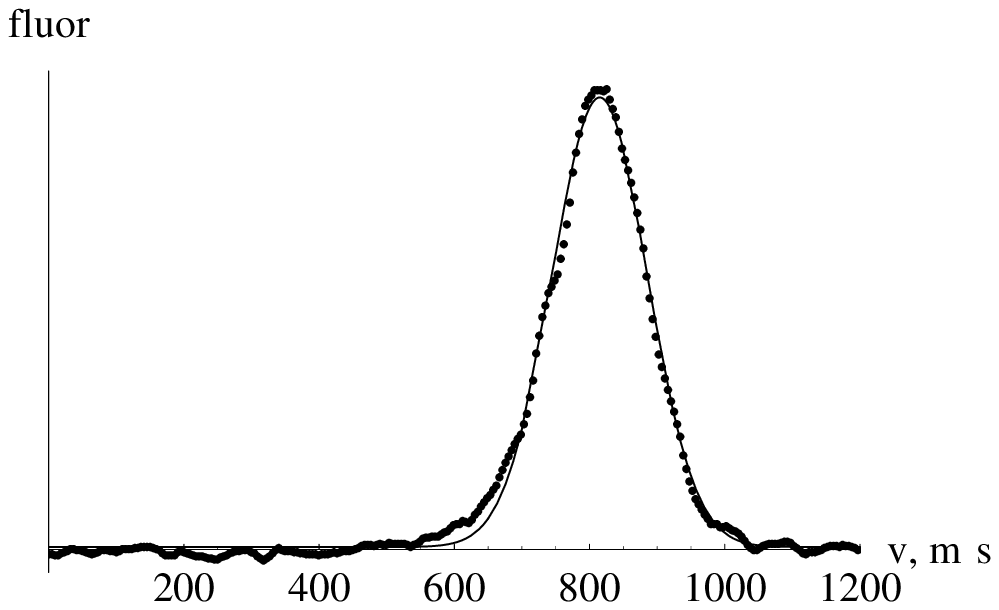}}}
\caption{Fluorescence signal of a counterpropagating \fft\ probe beam as
measurement of supersonic beam velocity distribution.  Zero of distribution
determined by simulataneous measurement of Lamb dip transmission.
Curve: Gaussian fit with peak velocity $v_0 = 816$ m/s, $\sigma =
70.2$ m/s.  When natural linewidth and isotope shifts are taken into
account, we find $\Delta v_{\rm FWHM}/v_0 = 16.5\%$.}
\label{fig-supersonic-veldist}
\efig


Although the supersonic expansion results in a higher velocity and
thus smaller interaction time (roughly by a factor of 2), there are
several advantages of the supersonic oven.  First, and most
importantly, velocity {\it narrowing} gives higher densities than
velocity {\it selection} because non-selected atoms are lost in the
latter.  In addition, the supersonic beam gives far higher densities
than even a non-selected thermal beam, due to the higher oven
temperatures.  Experiments with the supersonic beam are simpler
because the additional lasers and locking systems are not needed.  Due
to these advantages, all of the microlaser experiments in this thesis
were done with the supersonic beam.

\subsection{Final aperture}

The small dimensions of the microlaser cavity mode place strong
constraints on the size and location of the atomic beam.  It must be
small and closely centered on the mode if reasonably uniform coupling
is to be obtained.

Variation due to the Gaussian shape of the cavity mode can be made
relatively small by limiting the dimensions of the atomic beam inside
the cavity.  For an atomic beam of height $h$ centered in the mode,
the ratio of minimum to maximum coupling is
\be {{g_{\mathrm{min}}} \over {g_{\mathrm{max}}}} = 
{{e^{-\left({h / {2 w_0}}\right)^2}}} \ee 
This value is plotted in Figure
\ref{fig-coupling-variation}.  For a beam size of 25 \um\ we have
${{g_{\rm min}}/{g_{\rm max}}} \approx 0.90$

\bfig 
\centerline{\resizebox{4in}{!}{\includegraphics{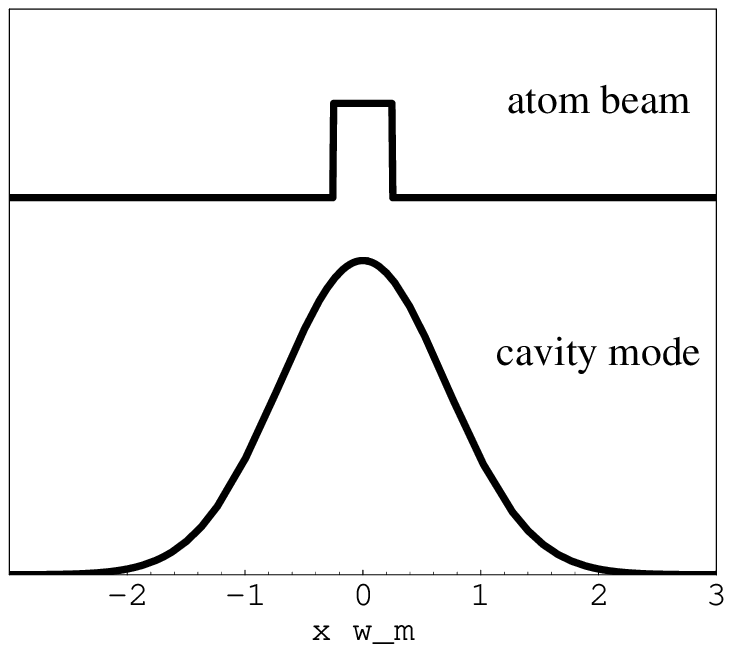}}
}
\caption{Atom beam width compared with cavity mode.  Atom beam should ``sample'' only the peak of the mode to minimize coupling variation.  Here the aperture width is set equal to one-half the mode waist $w_m$.}
\label{fig-coupling-diagram}
\efig

\bfig 
\centerline{\resizebox{4in}{!}{\includegraphics{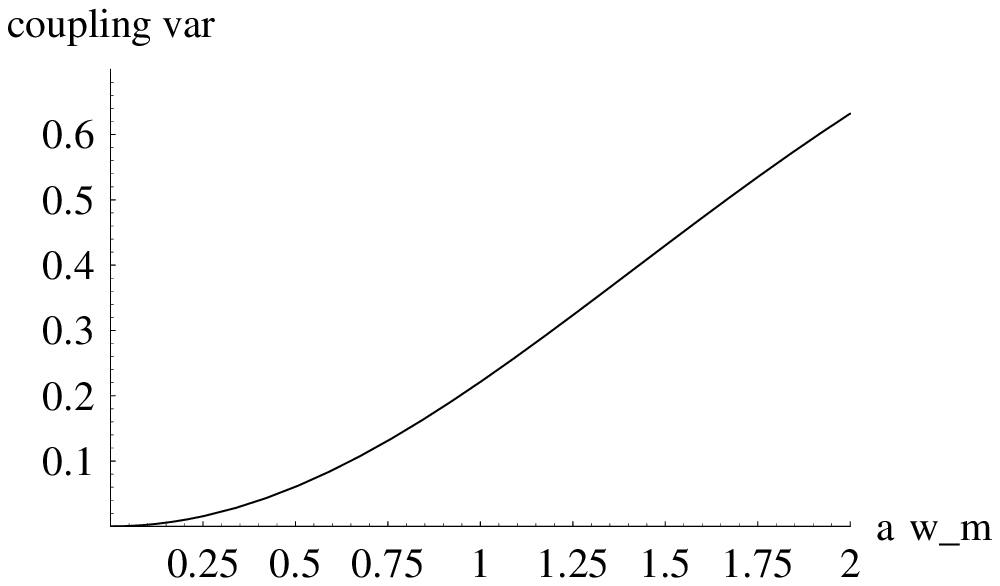}}}
\caption{Coupling variation $1-{{g_{\mathrm{min}}} /{g_{\mathrm{max}}}}$ due to finite aperture size $a$}
\label{fig-coupling-variation}
\efig

In order to create a small beam in the cavity we collimated the atomic
beam with a 24.5 \um\ diameter aperture placed approximately 3 mm from
the cavity axis.  The distance between aperture and cavity is
important because a larger distance increases the broadening of the
atom beam due to the finite solid angle of the atom beam source
(i.e. oven nozzle).  The aperture was mounted onto a thin, hollow,
tapered ``pencil'' which fit into the hole drilled into the PZT to
allow passage of the atomic beam.  The pencil is attached to the
center of a 1/2'' stainless-steel compact mirror mount 
which allows the orientation of the pencil to be modified precisely.
The mirror mount was mounted on a miniature motorized rotation stage
(National Aperture MM--3M--R).  A stepper driver (National Aperture
MC-II) located outside the chamber is connected to the rotation stage
by a 10-pin electrical feedthrough.  The mirror mount adjustment
screws were tweaked so that center of rotation is located at the
center of the aperture.
Finally, the rotation stage and aperture assembly were connected to an
XY miniature translation stage (National Aperture ST1XY) which could
be manipulated from outside the chamber through two $1/8$ inch
Ultra-Torr feedthroughs.  The feedthroughs couple to the stage via
flexible couplings.  

The pencil has a tapering design in order to be small enough at the
tip to approach the cavity closely, and yet large enough to accomodate
a range of atomic beam angles and avoid clipping the beam.  The pencil
was made from several pieces of brass telescoping square tubing, glued
together with Varian Torr-Seal vacuum epoxy.  The aperture (12.7 \um\
thick stainless steel, from National Aperture) was attached to the
pencil tip with Torr-Seal.

\bfig 
\centerline{
\centerline{\resizebox{\textwidth}{!}{\includegraphics*[0in,0in][6in,3in]{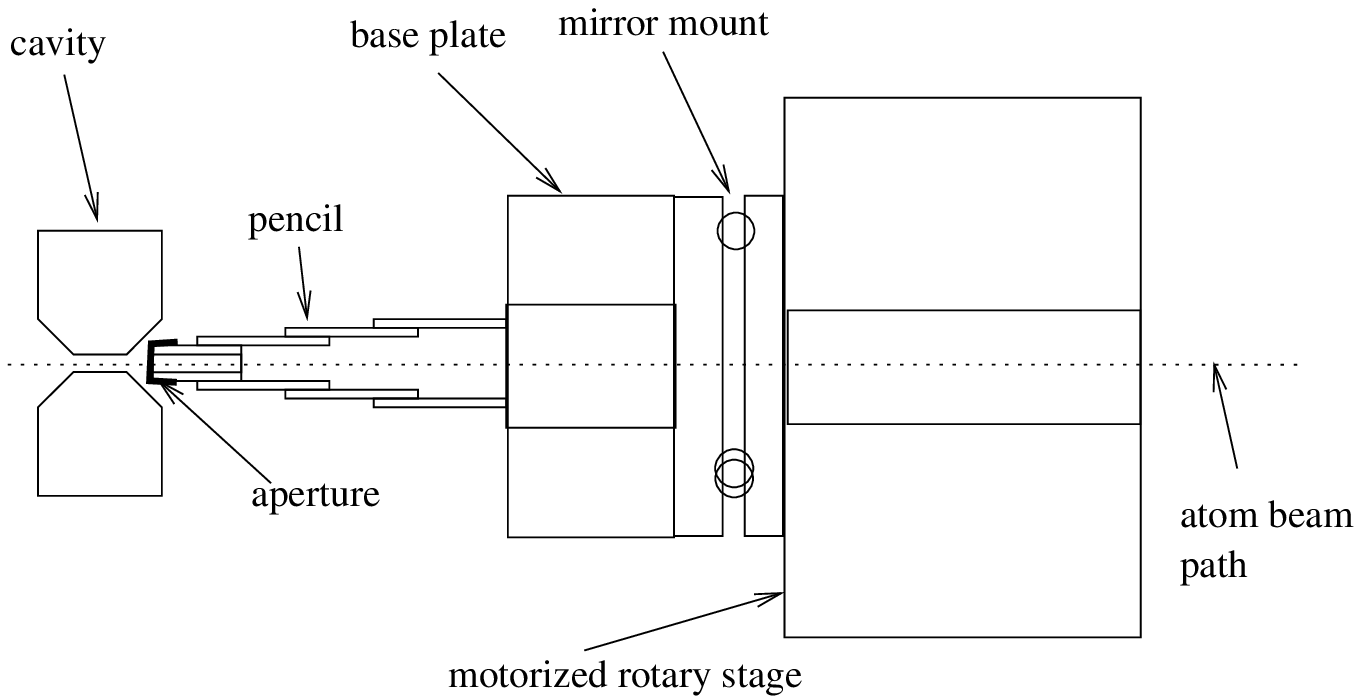}}}
}
\caption{Top view of aperture assembly including motorized rotary 
stage, mirror mount for angular adjustments, tapering ``pencil'', and
final aperture in relation to optical cavity.  Not shown: X-Y
translation stages attached to rotation stage.}
\label{fig-pencil-design}
\efig

\bfig 
\centerline{\resizebox{5in}{!}{\includegraphics*[0in,0in][6.5in,8in]{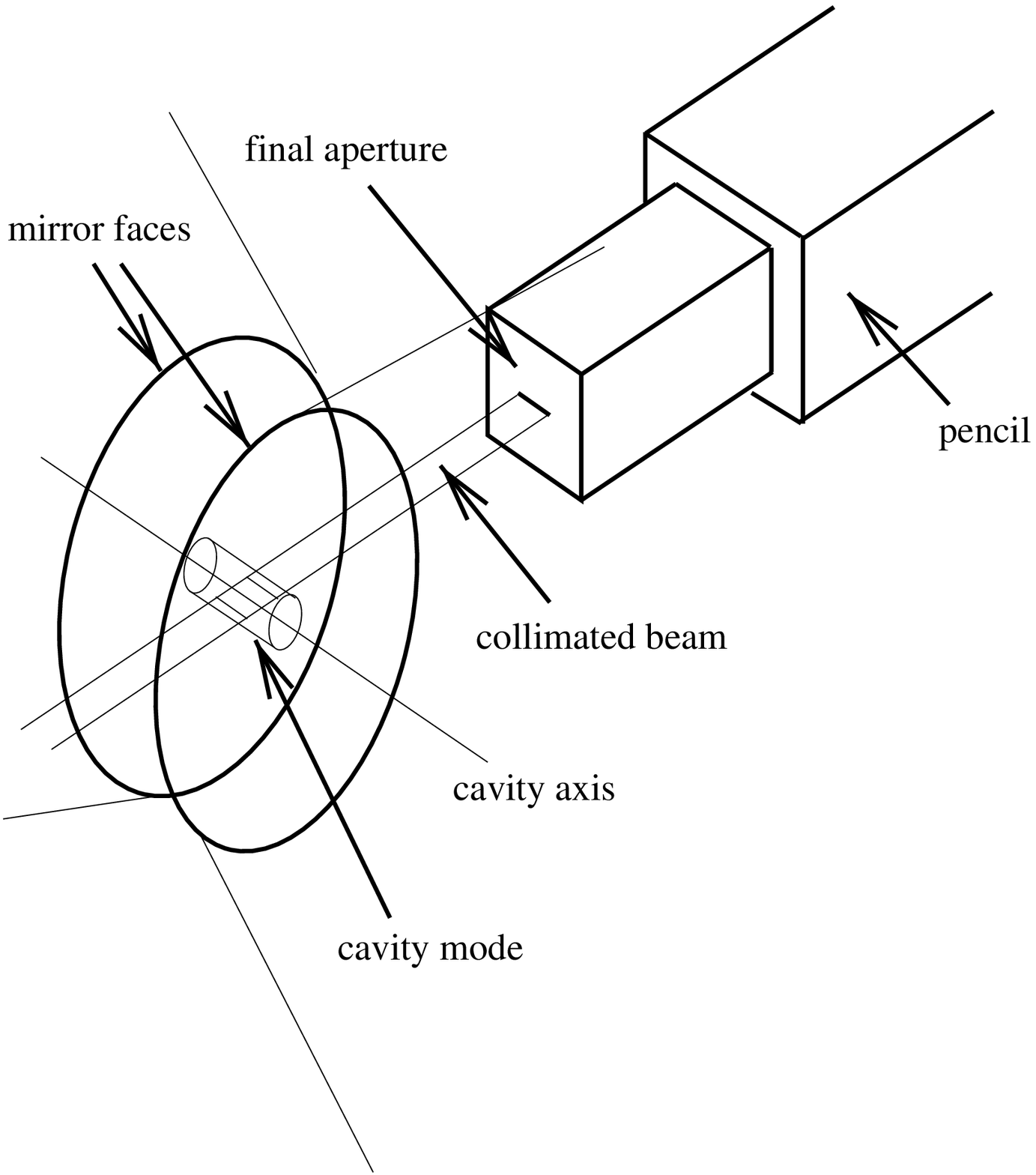}}}
\caption{Close-up of tip of aperture assembly and cavity.  
Atomic beam is collimated by aperture slit and passes through the
center of the cavity mode.}
\label{fig-aperture-design}
\efig

To center the atomic beam on the cavity mode axis, we monitored the
microlaser emission power as a function of atom beam vertical position
(measured by imaging the aperture with the imaging system described in
Sec.~\ref{sec-imaging}).  By this method we are able to find the
center of the mode to a few \um\ .  A typical result is shown in
Fig.~\ref{fig-aperture-translation}.  Repeated measurements are taken
as the aperture position is varied; low points are due to fluctuations
in chamber pressure when the feedthrough is manipulated.  Aperture
position is measured by the CCD pixel number corresponding to the
maximum intensity in an image of the oven blackbody radiation passing
through the aperture.

\bfig
\centerline{\resizebox{4in}{!}{\includegraphics{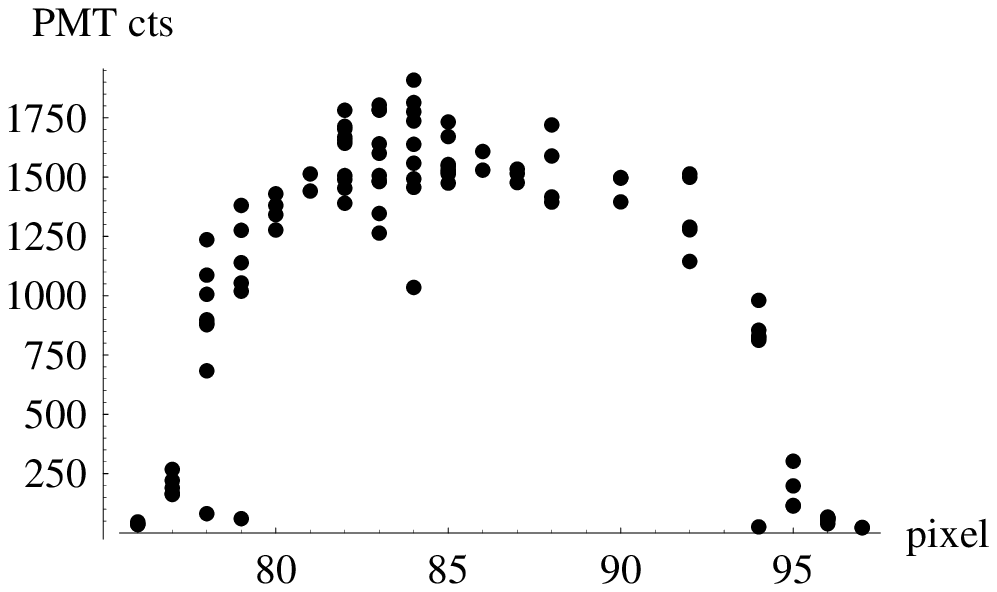}}}
\caption{Microlaser output for microlaser for constant density in initial 
threshold region, as a function of aperture position in vertical
direction.  1 pixel corresponds to about 4.5 \um.  }
\label{fig-aperture-translation}
\efig

\subsection{Atom density modulator}

The density, peak velocity, and velocity compression of a supersonic
beam are all functions of the oven current (i.e.\ temperature) and the
amount of barium remaining in the oven.  During our experiment, we
would like to vary the density only.  Therefore the oven current alone
cannot be used to modulate density in the cavity, as it was for the
original microlaser experiment \cite{An-PRL94} or some recent
experiments \cite{aljalal-thesis}

\bfig 
\centerline{
\resizebox{4in}{!}{\includegraphics{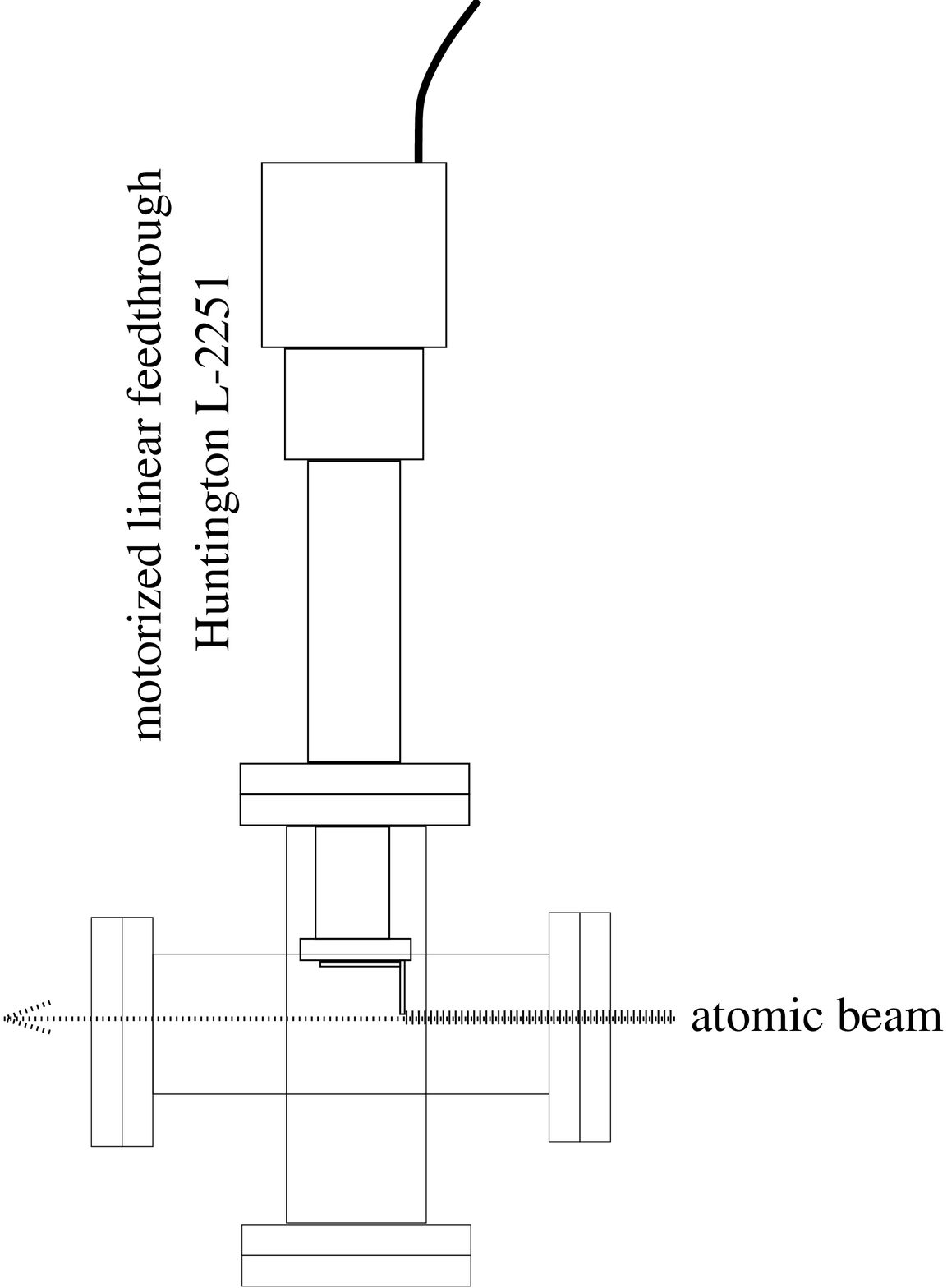}}
}
\caption{Atomic beam density modulator}
\label{fig-beam-blocker}
\efig

We have introduced a variable beam density modulator which intercepts
the atomic beam between the oven and cavity and prevents some or all
of the atoms from entering the cavity.  The modulator consists of a
piece of sheet aluminum attached to a motorized linear feedthrough
(Huntington L-2251-2) mounted at the 6-way cross between oven and
cavity.  This is shown in Fig.~\ref{fig-beam-blocker}.  The aluminum
piece was initially cut in a wedge shape in order to give a longer
modulation range as it is moved vertically.  This was later found to
be unnecessary as sufficiently high resolution was obtained with a
flat piece of aluminum.  
The modulator is located 21 cm from the
oven; the cavity is 42 cm from the oven.

The linear feedthrough has a lead of $0.05$'' per revolution.  The
stepper motor turns one revolution per 200 steps, giving a linear
resolution of $6.35 \um$ per step.  Higher resolution is possible via
microstepping.  Full linear travel is 2''.

The stepper motor is controlled by an Intelligent Motions Systems
Panther LI2 (aka Huntington MLC-11) Driver.  This driver was found to
create a great deal of electromagnetic interference in the laboratory,
which was reduced using a set of opto-isolators and line filters.  The
driver is controlled by a PC via a standard RS232 serial connection.
The driver is equipped with 3 logic-level auxiliary outputs which were
used to control other experimental parameters (described in section
\ref{sec-automation}).

\bfig 
\centerline{
\resizebox{\textwidth}{!}{\includegraphics{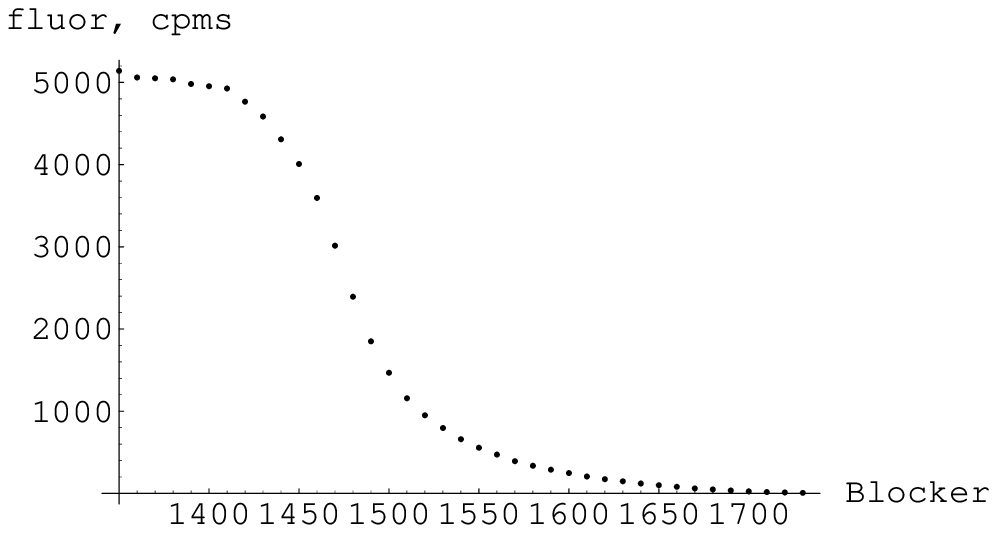}}
}
\caption{Beam density (fluorescence CCD counts per ms) as a function of 
beam modulator position; 1 step = 6.35 \um.}
\label{fig-beam-blocker-result}
\efig

The cavity fluorescence signal as a function of modulator position is
shown in Fig.~\ref{fig-beam-blocker-result}


\section{Laser systems, locking, and stabilization}

\subsection{Laser equipment}
Two tunable lasers were used in the microlaser experiments.  To
provide the 791nm microlaser pump beam and cavity probe, we used a
Coherent 899-21 Titanium-Sapphire ring laser pumped by an Innova 310
argon ion laser operating at 8W in multiline visible mode.

A Coherent 699-21 dye laser tuned to the 553 nm \transitionsinglet\
transition was used to monitor atomic density in the cavity and
measure the atomic velocity distribution.  The dye used in this laser
was Rhodamine 110.  The laser was pumped by an 8W multiline visible
beam from a Coherent Innova 200 Argon ion laser.

\subsection{Frequency stabilization of 791 nm beam by Doppler-free 
FM spectroscopy}

Frequency locking of the Ti:Sapphire laser to the \transitiontriplet\
transition was done by a standard FM modulation technique.  This
locking system is described in detail in \cite{an-apl95}.

An evacuated stainless-steel cylindrical cell containing about 2g
barium is heated to $600^\circ$ C by external heating elements to
produce a barium vapor.  Pump and probe laser beams pass through the
cell from opposite directions, through two glass windows.  The probe
beam first passes through an electro-optic phase modulator (EOM)
driven at 24.6 MHz.  After passing through the cell the probe beam is
incident on a photodiode (Thorlabs DET110).  The signal is amplified
and demodulated via mixing with the 24.6 MHz local oscillator signal
in a double-balanced mixer (Minicircuits).  The relative phase of the
two inputs to the mixer is adjusted by varying the modulation
frequency and the lengths of BNC cables to produce a maximum
dispersion signal, shown in \ref{fig-fmlock-sig}.  The dispersion
signal is amplified and shifted to remove any DC offset from the
mixer.  It has a zero crossing at the peak of the Lamb dip (atom
resonance).  To lock the laser to the zero crossing, the dispersion
signal is fed back to to Ti:Sapphire laser as an error signal in place
of the signal from the reference cavity.  The linewidth of the lock
is estimated as approximately 300 \khz\ \cite{an-apl95}.

\bfig
\centerline{\resizebox{4in}{!}{\includegraphics{fig-791lambdip.epsi}}}
\caption{Lamb dip used to lock Ti:Sapphire laser to 
\sno\ \transitiontriplet\ transition.}
\label{fig-791lambdip}
\efig

\bfig
\centerline{\resizebox{4in}{!}{\includegraphics*[1.5in,6in][7in,10in]{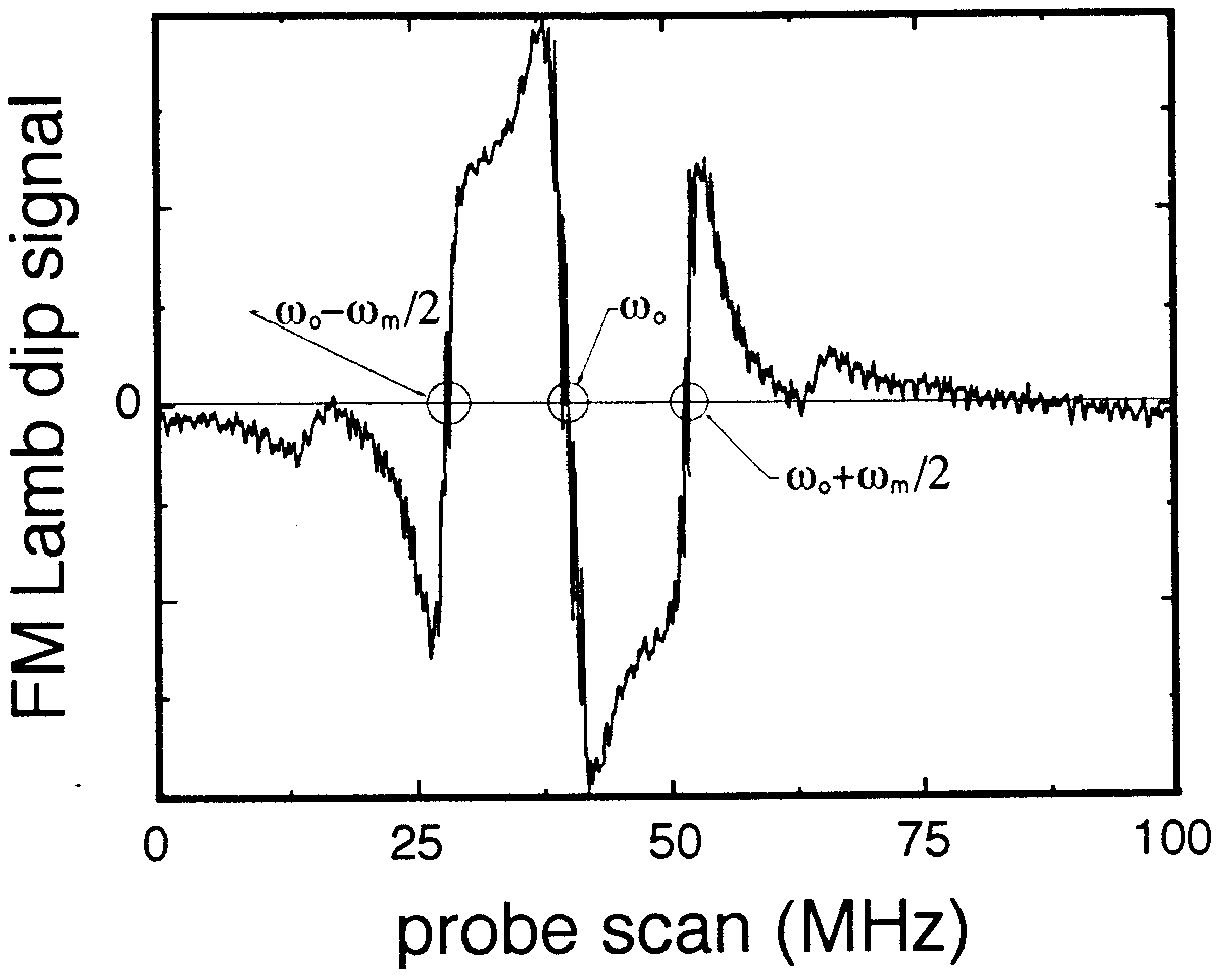}}}
\caption{Dispersion signal used for locking of Ti:Sapphire laser to
\transitiontriplet\ transition at \sno.}
\label{fig-fmlock-sig}
\efig


\subsection{Locking of 553 nm beam to atomic beam}

The 699-21 dye laser was locked to the \transitionsinglet\ transition
via a different Lamb dip technique.  After reflection through a
beamsplitter, a strong 553 nm pump laser beam is incident on a barium
cell heated to 490 C.  The transmission through the cell is apertured
then retroreflected back into the cell, overlapped by the original
beam.  The reflected beam forms the probe, which is detected by a
photodiode (Thorlabs DET110) after the beamsplitter.  The resulting
absorption signal shows a Lamb dip (Fig.~\ref{fig-lamb-dip}).

\bfig 
\centerline{
\resizebox{4in}{!}{\includegraphics*[2in,0.5in][5.5in,3in]{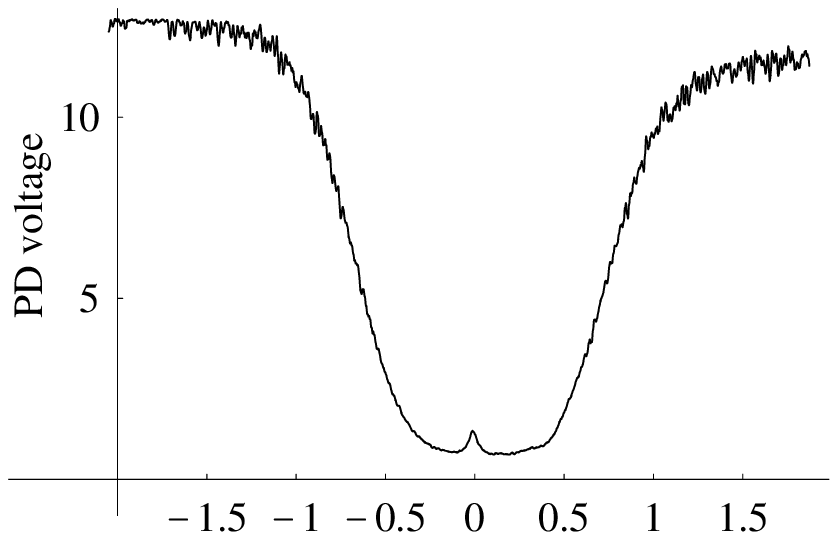}}
}
\vspace{-1em}
\centerline{Detuning, GHz}
\caption{Lamb dip signal used for dye laser locking to the \fft\ \transitionsinglet\ transition}
\label{fig-lamb-dip}
\efig

%


\bfig
\centerline{\resizebox{\textwidth}{!}{\includegraphics{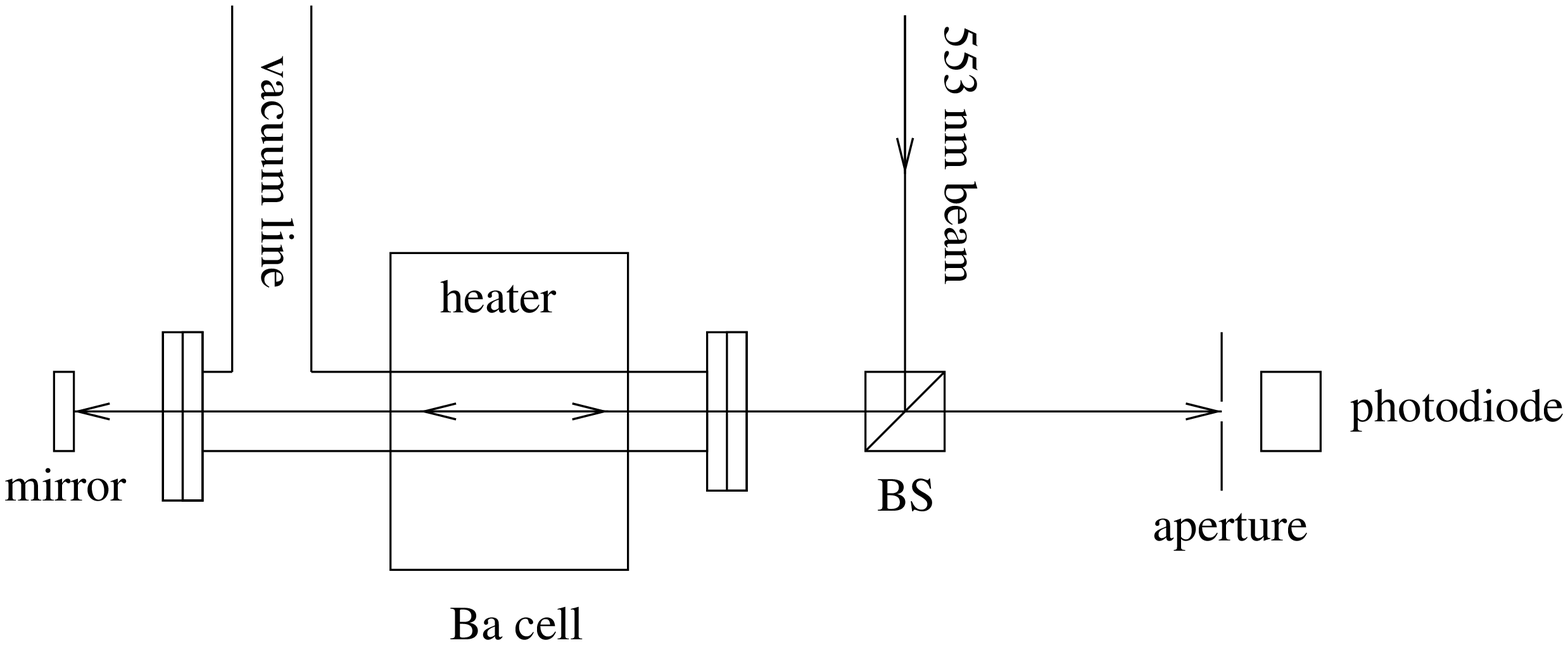}}}
\caption{\fft\ Lamb dip setup for 699 dye laser locking.}
\label{fig-553cell}
\efig

The laser locking system consists of an integrating amplifier and
summing amplifier.  A local oscillator at about 300 Hz frequency, 100
mV peak to peak amplitude dithers the frequency of the dye laser
through the external horizontal input of the laser control box.  The
photodiode signal and local oscillator are connected to a lock-in
amplifier.  After adjusting the phase, the output signal is of
dispersion type, with zero crossing at the peak of the Lamb dip.  This
signal is fed to an integrating amplifier, which controls the DC
offset of the laser external input.  The resulting lock stays within
approximately 2 MHz of resonance.  This is much smaller than the 19
MHz natural linewidth of the transition and was judged to be adequate.

\subsection{Intensity and position stabilization}

Both the 791nm and 553nm lasers experience some short-term and
long-term intensity fluctuations.  The dye laser has considerable (
$\sim 20\%$) noise in the kHz range and the average intensity
decreases gradually over hours as the dye decays.  The Ti:Sapphire
laser has little short-term noise but does have slow intensity
fluctuations, most likely due to pump beam wander.

Pump beam wander, thermal effects, daily power peaking, and other
effects cause substantial variations in laser beam direction from day
to day and within single days.  Often these effects are corrected for
by aligning a laser beam through two adjustable irises.  However,
since many aspects of the microlaser experiment require very precise
beam positions (micron scale) and angles (milliradian scale), irises
do not provide adequate repeatability.

\bfig 
\centerline{
\resizebox{\textwidth}{!}{\includegraphics*[0in,0in][10in,6in]{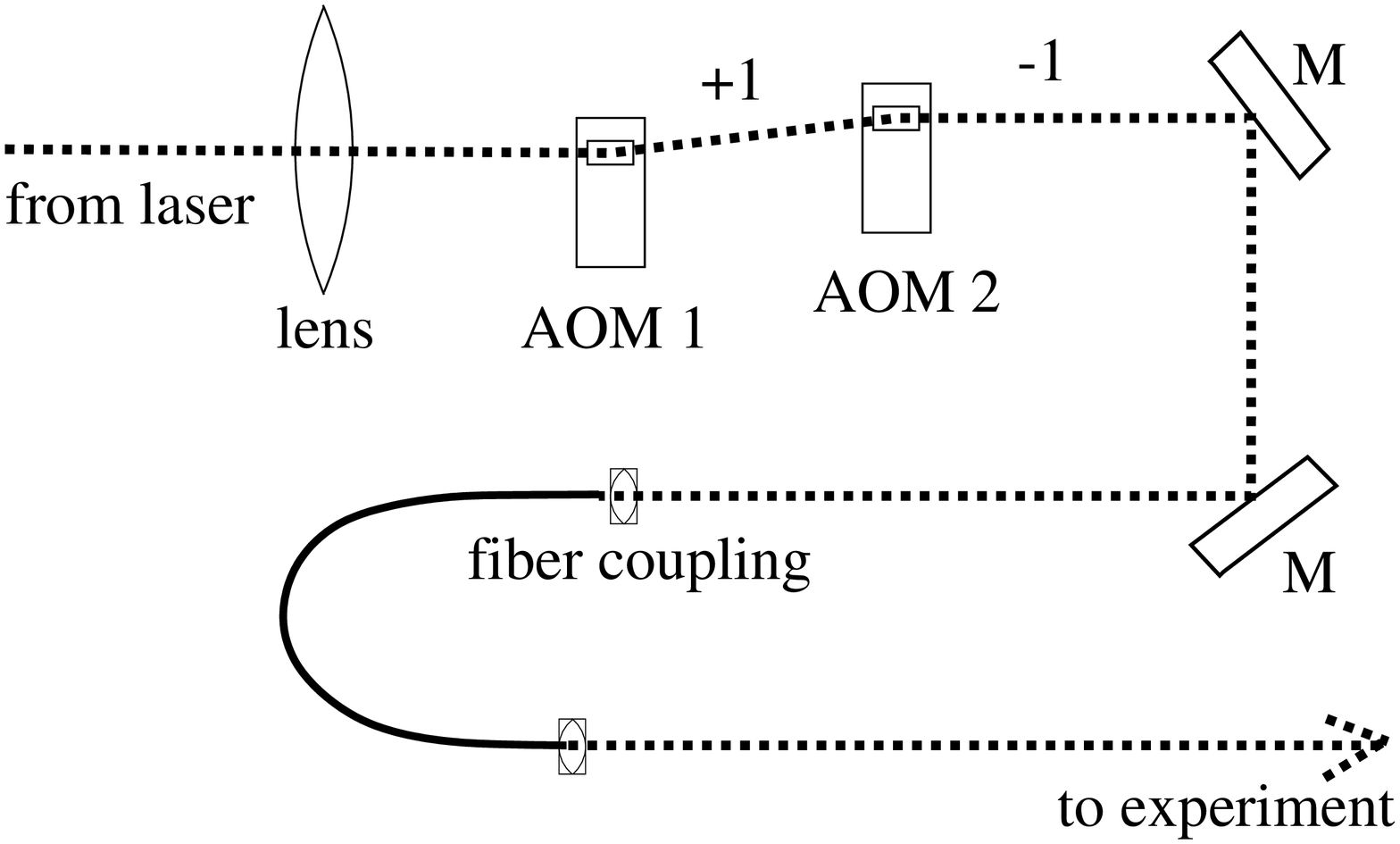}}
}
\caption{Laser intensity and position stabilization system for 791 nm and 553 nm beams}
\label{fig-intensity-stab-setup}
\efig

To reduce laser intensity noise and stabilize long-term power, and to
stabilize laser positions, we have developed a laser stabilization
system.  Acousto-optic modulators (AOMs) are used to cancel intensity
fluctuations.  Coupling through single-mode optical fibers allows very
consistent optical alignment of beams with respect to the setup.

\bfig
\resizebox{\textwidth}{!}{\includegraphics{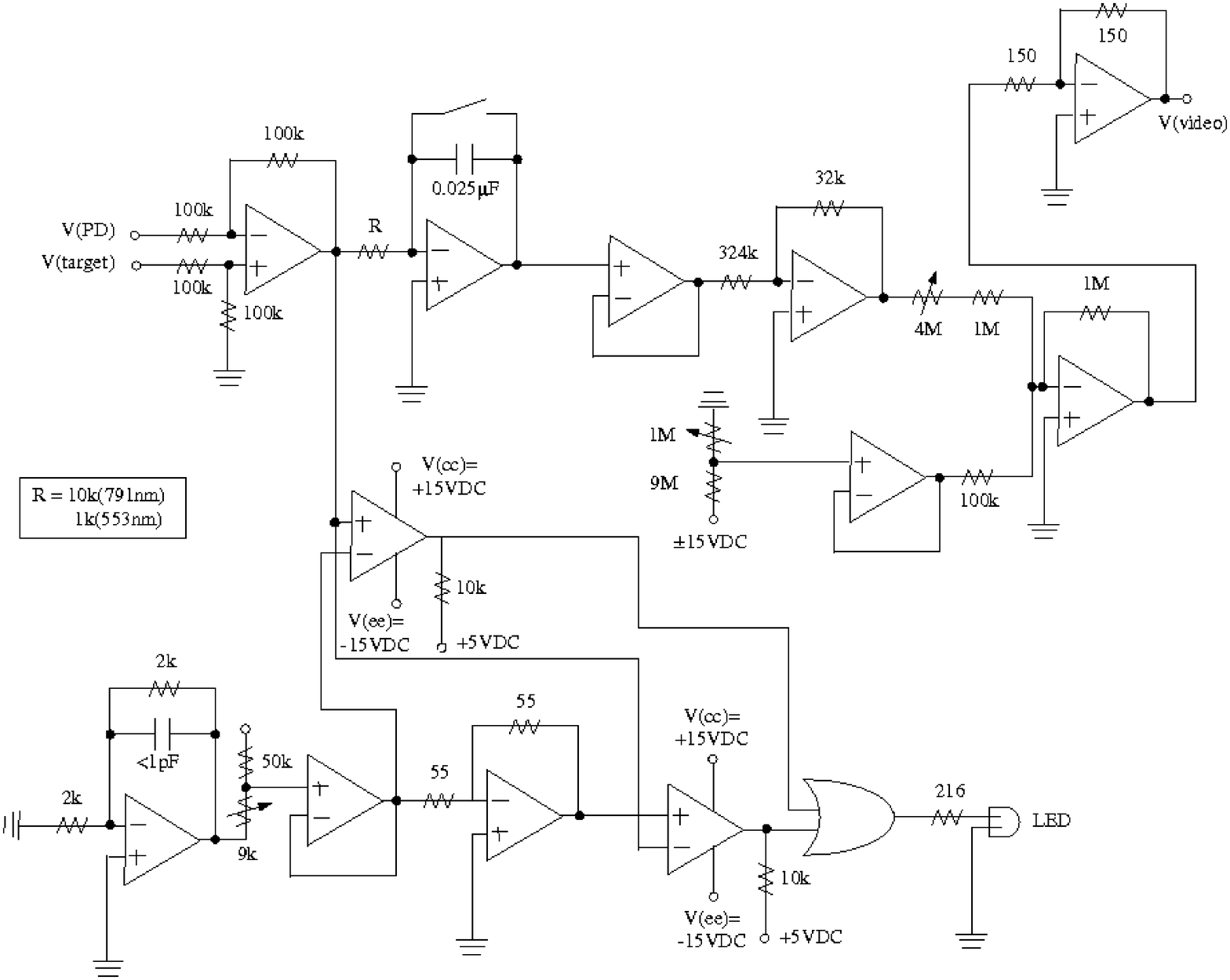}}
\caption{Circuit diagram for intensity stabilization system}
\efig

The intensity stabilization system uses a photodiode and integrating
amplifier to provide feedback through an AOM VIDEO signal.  A
photodiode connected to a variable-termination, transconductance
amplifier monitors the laser power after the single mode fiber as
shown in Fig.\ref{fig-intensity-stab-setup}.  The photodiode signal
$V_{PD}$ is compared with a target voltage $V_T$ and the difference is
passed to an integrating amplifier, the output of which is connected
to the AOM VIDEO input (amplitude modulation).  The output voltage
range is trimmed by potentiometers to correspond to a range from
minimum to maximum AOM transmission.

Polarization maintaining (PM) single-mode fibers are used to stabilize
the position of the 791 pump beam and the 553nm probe laser beams.  We
use 1m long Newport SP-V fiber for the 791nm beam and a 5m Newport
SP-?  fiber for the 553nm beam.  A schematic is shown in
Fig.~\ref{fig-intensity-stab-setup}.

\section{Pump beam}

The pump beam is supplied by the FM-locked Ti:Sapphire laser.  The
problem of how to pump the atoms correctly proved to be more
complicated than expected, and led to the rediscovery of a method for
inverting atoms by a defocused laser beam.

The pump beam intersection with the atomic beam must be sufficiently
small that upper state spontaneous decay does not significantly affect
the final degree of inversion.  For the same reason, the pump beam
spot must be placed close to the cavity mode, but without overlap.

Other critical parameters include the angle between the pump beam and
atom beam direction (ideally exactly 90 degrees), total pump power,
location of pump focus, uniformity of the pump beam intensity over the
width of the atomic beam, and laser frequency jitter about the atom
resonance.



\subsection{Optical setup}

As described in Sec.X the pump beam is coupled through a single-mode
polarization-maintaining (PM) fiber for intensity and position
stabilization.  The emerging beam is collimated by an aspheric lens
and passes through a beam steering apparatus
to raise the beam height above the vacuum chamber.

The pump beam steering apparatus is illustrated in
Fig.~\ref{fig-pumpbeamsetup}.  It consists of mirrors and a lens
mounted on translation stages such that one stage controls the pump
beam focus and one controls the pump beam position (in the z
direction), nearly independently.

The lens is a single element, BK7 plano-cylindrical lens with focal
length 400? mm.  Tests using the CCD camera as a beam analyzer showed
that a nearly diffraction-limited spot size of 35 \um\ could be
achieved at the focus.

The pump beam is reflected down into the cavity by an 2'' diameter,
45-degree-angled dichroic beamsplitter which has reflectivity
approximately 90\% at \sno\ and 10\% at \fft.  A dichroic beamsplitter
is used instead of a mirror in order to allow imaging of \fft\
fluorescence from above the beam splitter.  In addition, the weak
\sno\ transmitted beam can be used to analyze the focus and beam
quality of the pump.

\bfig
\centerline{\resizebox{5in}{!}{\includegraphics{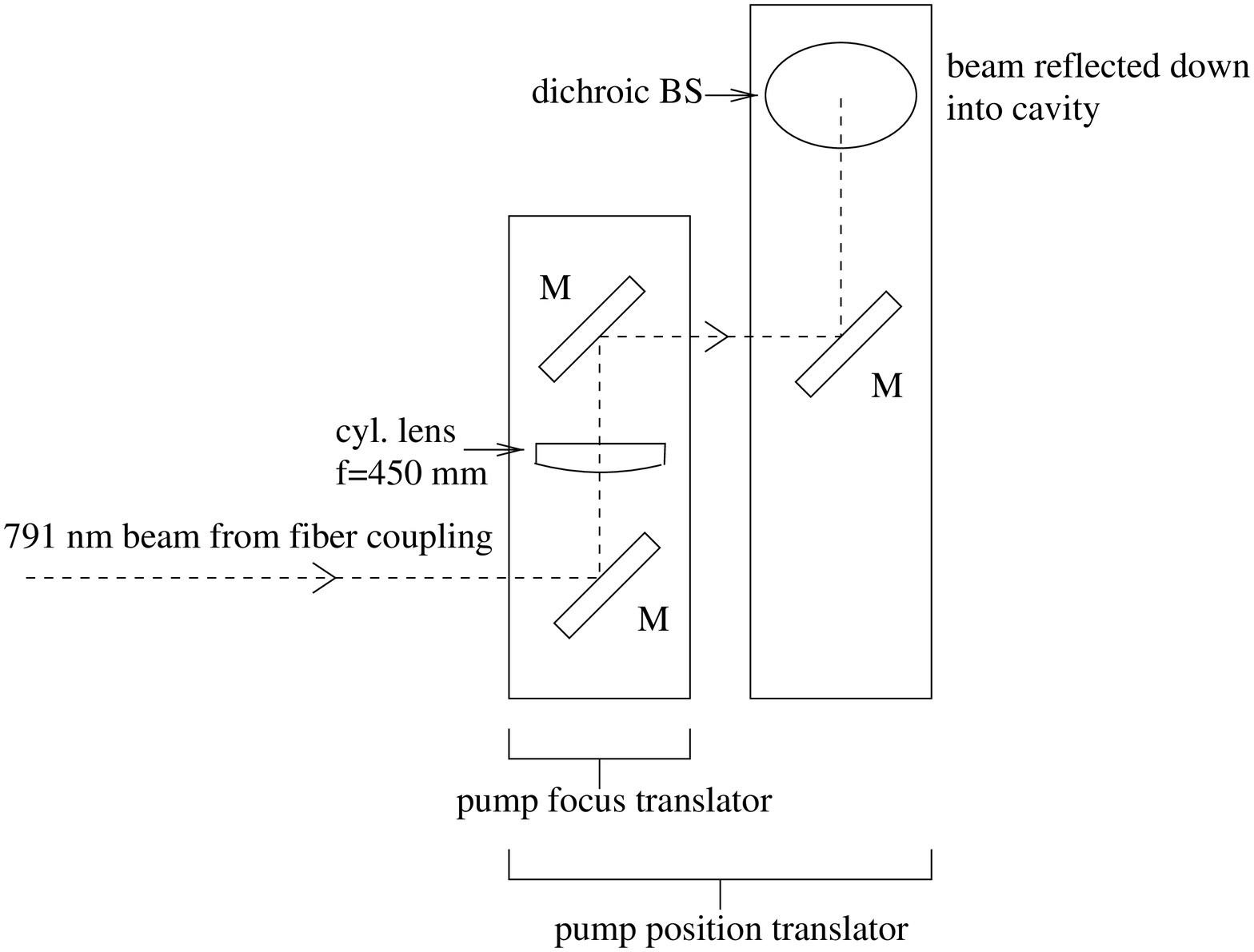}}}
\caption{Pump beam optical setup.  Optics are fixed on two small
rectangular breadboards attached to separate translation stages to
control pump focus and translation independently.}
\label{fig-pumpbeamsetup}
\efig

The window on the top of the chamber is an anti-reflection coated
glass disc held in place by a custom metal flange and Viton O-rings.

A coil-generated magnetic field of approximately 50 gauss parallel to
the direction of pump polarization (perpendicular to the cavity axis)
splits the $m$-states by about 50 \mhz\ and ensures that only
transitions between sublevels $m=0$ and $m=0$ occur.

\subsection{Pump beam parameters}


Here we describe the optimization of the pump beam parameters in more
detail.

\subsubsection{Pump beam focus}

The pump focusing lens is placed approximately one focal length away
from the cavity.  To find the focus more precisely, we overlap the
\sno\ pump with a resonant \fft\ probe beam.  
With the imaging system (Sec.~\ref{sec-imaging}) in the top viewing
configuration, we take pictures of the fluorescence of atoms in the
probe field.  By measuring the width of the fluorescence spot as a
function of the pump focus translation stage micrometer, we can find
the focus of the \fft\ beam.  However, this is in general displaced
from the \sno\ focus due to lens dispersion and different laser beam
divergences of the beam before entering the pump beam setup.  To
address this problem we used the CCD camera, without lenses, as a
laser beam analyzer for the \sno\ and \fft\ beams {\it transmitted} by
the dichroic beamsplitter.  The CCD position was positioned the same
distance from the beamsplitter as the cavity is from the beamsplitter.
The camera was moved by a translation stage to locate the positions of the
\fft\ and \sno\  beam foci.  This gave the displacements of the two
beam waists, typically about 1 cm apart.  The pump focus was then
adjusted by this amount to place the \sno\ focus as close as possible
to the atoms in the cavity.

This procedure was not perfect: as described in
Sec.~\ref{sec-adiabatic}, significant pump defocus was found to occur.

\subsubsection{Pump position}

Pump position can be monitored by a number of methods involving the
\fft\ transition and the imaging system.  However, the most effective
methods is to monitor the microlaser output power as a function of
pump position.  When the pump is placed downstream of the cavity mode,
no emission is seen.  For pump beam overlapping the mode, a signal
with unequal traveling-wave peak heights is usually observed.  The
reason for the unequal traveling-wave peaks is not clear.  For pump
beam upstream of the cavity mode, a signal is observed that is nearly
flat, with a small slope corresponding to increasing atomic decay with
pump-mode distance.

In our experiments we aim for about 100 \um\ separation between the
pump field and cavity mode.  At this distance, which is admittedly
somewhat arbitrary, atom decay between pump and mode is quite small,
$< 10\%.$


\subsubsection{Pump angle}

To minimize Doppler broadening it is important for the pump beam to be
as close as possible to perpendicular to the atomic beam direction.
Two methods for ensuring this condition were explored.  First, we
overlapped a \fft\ probe beam over the \sno\ pump as closely as
possible, and blocked the \sno\ beam.  Then we measured the linewidth
of the \sno\ fluorescence as the dye laser was scanned in frequency,
and repeated this process as the angle between laser and atom beam was
varied.  The linewidth increased linearly with any angular deviation
from a right angle.  This technique was very time-consuming and gave a
pump beam perpendicular to the atom beam only to the accuracy that the
\sno\ and \fft\ beams were parallel in the cavity.  The second technique
was to monitor the microlaser output for a constant atom density a few
times over threshold, while the frequency of the \sno\ pump laser was
varied over a 50 \mhz\ range via the external input of the Ti:Sapphire
laser.  In this manner, it could be determined if the microlaser power
was maximized for positive or negative detunings relative to the atom
frequency (i.e.\ the atoms experienced a blueshifted or redshifted
pump field).  The pump angle was adjusted until the the microlaser
power was maximum with the laser on exact resonance.

\subsection{Shelving experiment}

In order to test the efficiency of the pump process, we perform a
``shelving'' experiment as follows.

A \fft\ probe beam is aligned with the \sno\ probe which had been
matched with the $TEM_{00}$ mode of the cavity.  Therefore the \fft\
probe field is close to centered on the cavity mode (the beam size may
be different, but it will be close to the focus since the
\fft\ probe passes through the same achromatic lens).  Using
the CCD camera, we measure the fluorescence from the atoms in the
cavity with and without the pump beam.  Since the fluorescence is a
measure of the ground state population, the ratio between the
fluorescence intensities with and without the pump beam is equal to
the average upper state population.  If the pump beam were perfectly
efficient, no fluorescence at \fft\ would be seen, because all atoms
are in the upper state of the \transitiontriplet\ transition.

Shelving ratios as low as 10\% (i.e. 90\% of atoms in the upper state)
have been measured.  

\subsection{Pump power dependence of inversion}

When the first shelving data was collected, a puzzle presented itself.
Atoms interact with the pump beam for a certain time dependiing on
their velocity $v$,
\be 
t_{\rm pump} = {{\sqrt\pi\;w_{\rm pump}}\over v}
\ee
where $w_{\rm pump}$ is the waist of the pump beam, assumed to be
Gaussian. The excited-state probability for a given atom after leaving
the pump beam is expected to be
\be
P_e^{(1)} = \sin^2\left({\Omega_R t_{\rm pump}}\over 2\right)
\ee
Averaging over atom velocity distribution $f(v)$ gives
\be
P_e = \int_0^\infty \sin^2\left({\Omega_R \sqrt \pi w_{\rm pump}}\over 2 v\right) f(v)\; dv
\ee

%

We expect to observe an oscillatory excited state probability damped
by the velocity distributions.  In the limit of strong field the
excited state population should be $1/2$.

\bfig[t]
\centerline{\resizebox{4in}{!}{\includegraphics{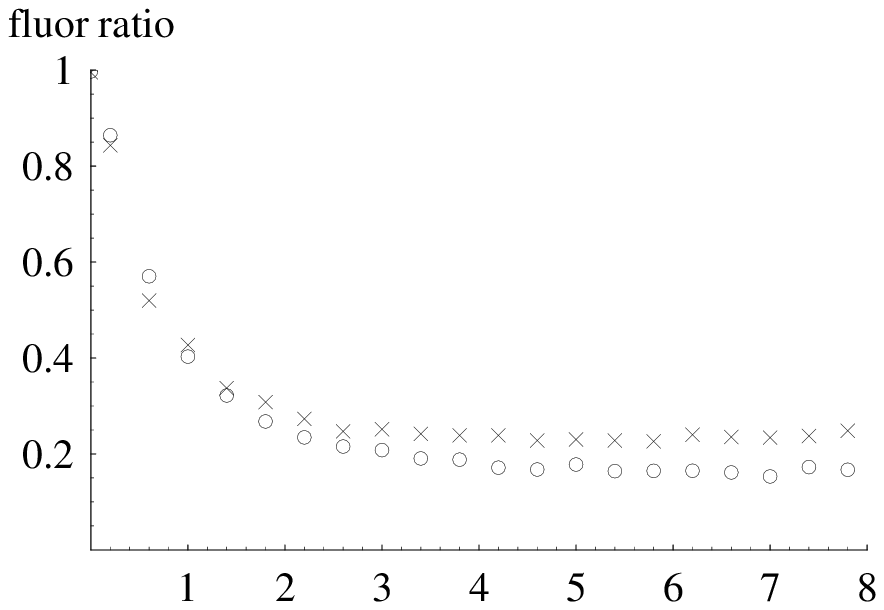}}}
\centerline{Pump power, \uw}
\caption{Ground state amplitude of atoms, measured by \fft\ fluorescence:
monotonic dependence on pump power.  Crosses ($\times$): with B field.
Circles($\circ$): without B field.}
\label{fig-shelving-monotonic}
\efig

Oscillations as a function of pump power were initially {\it not}
observed.  Instead, we observed the dependence drawn in
Fig.~\ref{fig-shelving-monotonic}.  The excited state population
increases monotonically to approximately 80\% and does not approach
1/2.

\subsection{Adiabatic inversion by defocused pump beam}
\label{sec-adiabatic}
The unusual behavior was explained by the following effect.  Suppose
there is some distance $y$ between the atoms and the focus (minimum
waist) of the pump beam -- for example, the beam comes to a focus
before reaching the atoms.  Then the atoms pass through a laser beam
which is diverging in the direction of their propagation.  As a
result, the angle between atom velocity $\bf v$ and the wave vector of
the pump field $\bf k$ varies approximately linearly with position,
causing a Doppler-induced frequency chirp through resonance.  This
leads to an {\it adiabatic inversion} process as the atom's Bloch
vector precesses about a relatively slowly varying ``torque'' which
tends to align the atomic Bloch vectors with the inversion axis.  A
quantitative analysis follows.

Without loss of generality we restrict our attention to the $x-y$
plane containing both the atomic beam and Gaussian beam axis.
The phase function of a Gaussian beam depicted in Fig.~\ref{fig-gaussian-beam}
is (see e.g.~\cite{yariv-qe})
\be
\Phi(x,y) = -k y + \tan^{-1}(y/y_0) - \frac{k x^2}{2R(y)}
\ee
\noindent where $k$ is the wave vector, $y_0 = \pi w_p^2/\lambda$ is 
the Rayleigh range and the wavefront radius of curvature
\be
R(y) = y\left(1+\frac{y_0^2}{y^2}\right)
\ee

\bfig
\centerline{\resizebox{\textwidth}{!}{\includegraphics{fig-gaussian.epsi}}}
\caption{Atoms displaced from pump beam focus experience frequency chirp
and adiabatic inversion.}
\label{fig-gaussian-beam}
\efig

The angle $\theta$ between the Gaussian beam axis and the normal to
the phase front can be found via $\theta = \tan^{-1}(dy/dx)$ where
$dy/dx$ is the slope of the wavefront (i.e. line of constant $\Phi$):
\be
\theta = \tan^{-1}\left[\frac{|y|}{-(y^2+y_0^2)+\frac{y_0}{k}+\frac{x^2}{2}
\frac{y^2-y_0^2}{y^2+y_0^2}} x\right]
\ee
The sign of the angle is chosen to be positive when the x component of
the vector normal to the phase front is opposite the direction of the
atom velocity.  For simplicity, we consider only the case of atomic
velocity perpendicular to Gaussian beam axis.

The equation for $\theta$ simplifies assuming $k \gg y_0$ and $x \ll y$,
both valid easily in our case:
\be
\theta \approx - \frac{|y|}{y^2+y_0^2} x
\ee
The Gaussian waist of the beam as a function of $y$ is given by 
\be
w(y) = w_p \sqrt{1 + (y/y_0)^2}
\ee
An atom traversing the pump beam at vertical position $y$ experiences,
as a function of position, the following Doppler shift:
\be
\Delta_d(x) = k v \sin \theta \approx k v \frac{|y|}{y^2+y_0^2} x
\ee
The excited state may decay back to the ground state or to a
metastable D state.  The decay rate to the ground state is $\Gamma_g =
50 \khz$ and to the metastable state is $\Gamma_m = 70 \khz$.  The
optical Bloch equations describing the pumping process including loss are
\be
\dot{\rho}_{ee}(t) = - \frac{\Omega_0(t)}{2}R_2(t) - 
\Gamma_g+\Gamma_m)\rho_{ee}(t)
\ee
\be
\dot{\rho}_{gg}(t) = + \frac{\Omega_0(t)}{2}R_2(t) - \Gamma_g \rho_{ee}(t)
\ee
\be
\dot{\rho}_{mm}(t) = + \Gamma_m \rho_{ee}(t)
\ee
\be
\dot{R}_{2}(t) = \Delta(t) R_1(t) + \Omega_0(t)(\rho_{ee}(t)-\rho_{gg}(t)) - 
\frac{\Gamma_g}{2} R_2(t)
\ee
\be
\dot{R}_{1}(t) = - \Delta(t) R_2(t) - \frac{\Gamma_g}{2} R_1(t)
\ee
\noindent where $\rho_{ee}$, $\rho_{gg}$, and $\rho_{mm}$ are  excited state,
ground state, and metastable state probabilities;
\be
R_1(t) \equiv \rho_{eg} + \rho_{ge}
\ee
\be
R_2(t) \equiv i (\rho_{ge} - \rho_{eg})
\ee
and $\Delta $ is the atom-laser detuning
\be
\Delta = \omega_l + \Delta_d - \omega_a
\ee
\noindent where $\omega_l$ and $\omega_a$ are laser frequency and atom 
resonance frequency.  The on-resonant Rabi frequency is given by
\be
\Omega_0(t) = \Omega_0 \exp[-(vt/w(y))^2] = \sqrt{\frac{I}{2I_s}} \Gamma_g 
\exp[-(vt/w(y))^2] 
\ee
\noindent where $\Omega_0$ and $I$ are the Rabi frequency and on-axis 
laser intensity and $I_s$ is the saturation intensity
\be
I_s = \frac{\pi h c}{3 \lambda^3} \Gamma_g \approx 13 \uw/{\rm cm}^2
\ee
Fig.~\ref{fig-pump-sim} shows the results of numerical integration of the 
Bloch equations for different displacements $y$.
  For
$y=3y_0$ the Rabi oscillations are almost completely absent, and inversion
occurs almost independently of pump intensity for $I>2I_s$.

\bfig
\centerline{\resizebox{5in}{!}{\includegraphics{fig-pump-sim.epsi}}}
\caption{Results of adiabatic inversion pump simulations: final ground state
probabilities as functions of pump power.}
\label{fig-pump-sim}
\efig
\clearpage

If adiabatic inversion is responsible for the pump power dependence,
we should be able to observe Rabi oscillations by focusing the pump
beam closer to the atoms (i.e.\ decrease $y$).  The results are shown
in Fig.~\ref{fig-rabi-pump}.  Note that peaks and valleys of the
oscillations occur at integer values of $\sqrt{p/p_{\pi/2}}$.  The
curve is not asymptotic to $1/2$ for large pump fields; due to
uncertainties in the experiment this data may reflect a intermediate
state between pure Rabi oscillation and adiabatic inversion.

\bfig
\centerline{\resizebox{5in}{!}{\includegraphics*[1.3in,0in][4.8in,3in]{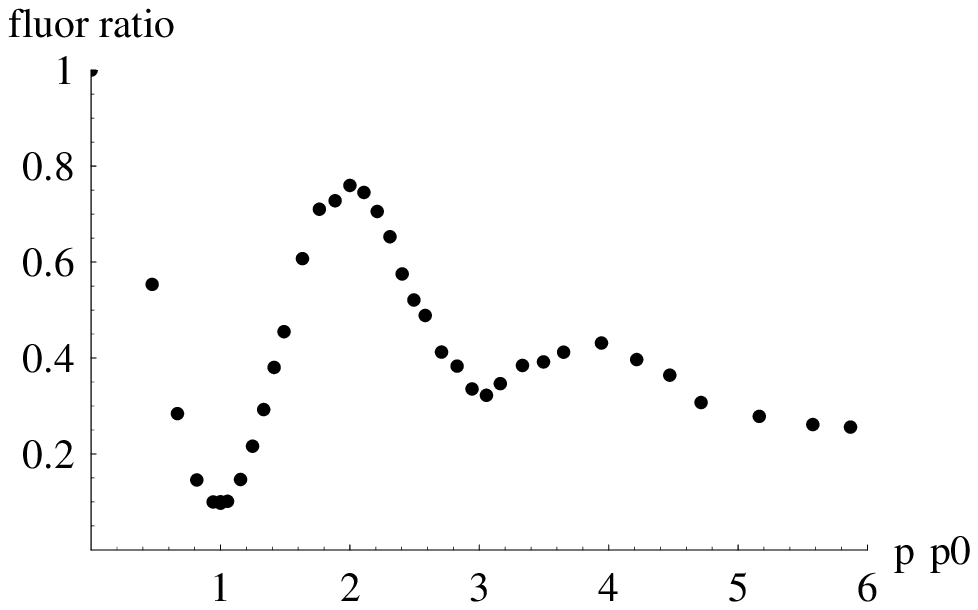}}}
\centerline{$\sqrt{p/p_{\pi/2}}$}
\caption{Ground state probability as a function of $\sqrt{p/p_{\pi/2}}$, where
$p$ is the pump power and $p_{\pi/2} \approx 1.8 \uw$.}
\label{fig-rabi-pump}
\efig


A quantitative comparison between experiment and theory will be
performed in a future study.  It has come to our attention that
adiabatic following due to a diverging laser beam was previously
observed in \cite{nguyen-pra} and \cite{kroon-pra85}.  These experiments
involve atoms with more complex level structures than the present one.

\section{Imaging system}
\label{sec-imaging}
A cavity imaging microscope was developed to perform several functions,
described below and in following sections.

\subsection{Setup}
The optical design of the imaging system is shown in
Fig. \ref{fig-imaging-setup}.  It can be thought of as a relay lens
system combined with a CCD microscope.  A 2:1 lens system consisting
of two 1'' achromatic doublets creates a real image of the cavity.  A
microscope objective relays the image to a CCD camera.  The CCD is
positioned at the field focus plane of the objective.  The use of
microscope objectives gives well-corrected imaging and allows the
magnification of the system to be varied easily without realignment.
In our experiments we use the 5x objective for the side imaging and
10x for the top imaging (see below for descriptions of the two imaging
configurations); the 10x and 20x objectives can also be used in the
side imaging configuration when the position of the final collimation
aperture needs to be determined precisely.

\bfig 
\centerline{\resizebox{\textwidth}{!}{\includegraphics*[0in,0in][5.7in,3in]{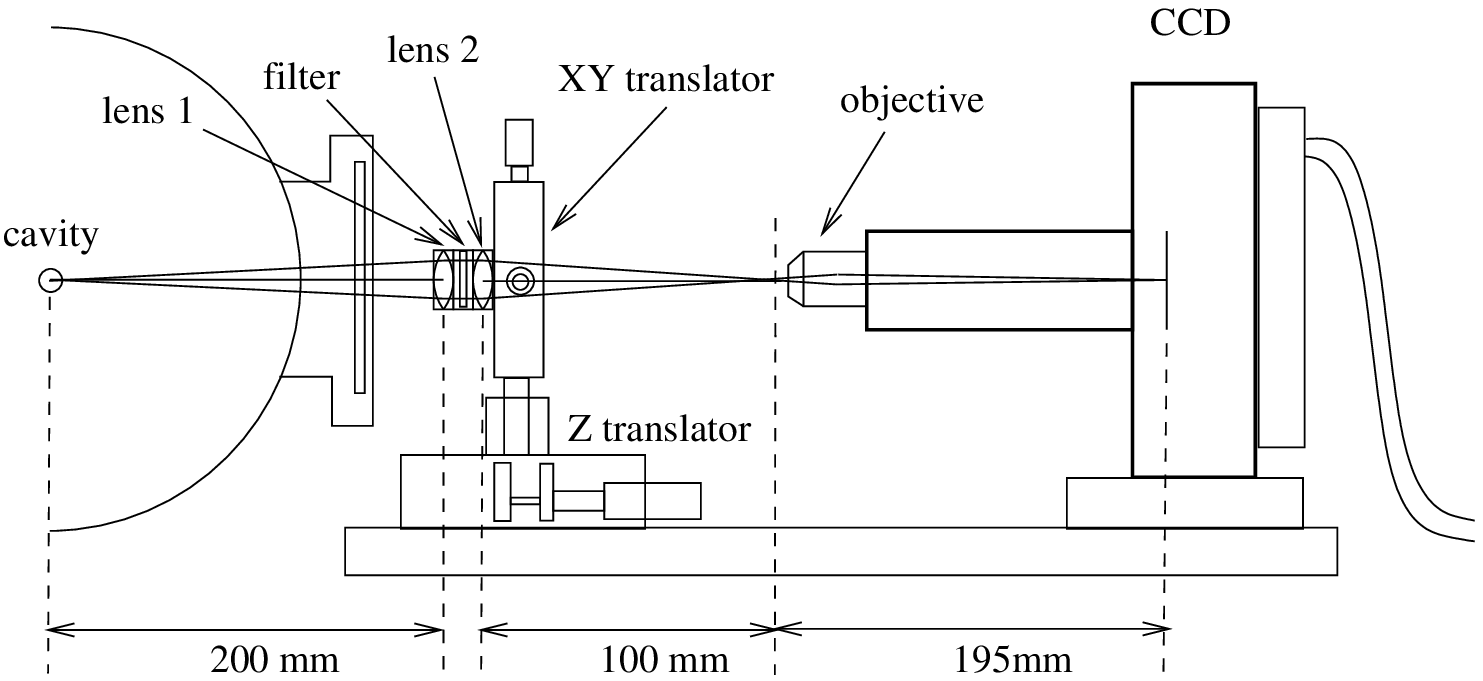}}}
\caption{Schematic of imaging system}
\label{fig-imaging-setup}
\efig
\clearpage

\bfig 
\centerline{\resizebox{\textwidth}{!}{\includegraphics*[0in,0in][7in,5in]{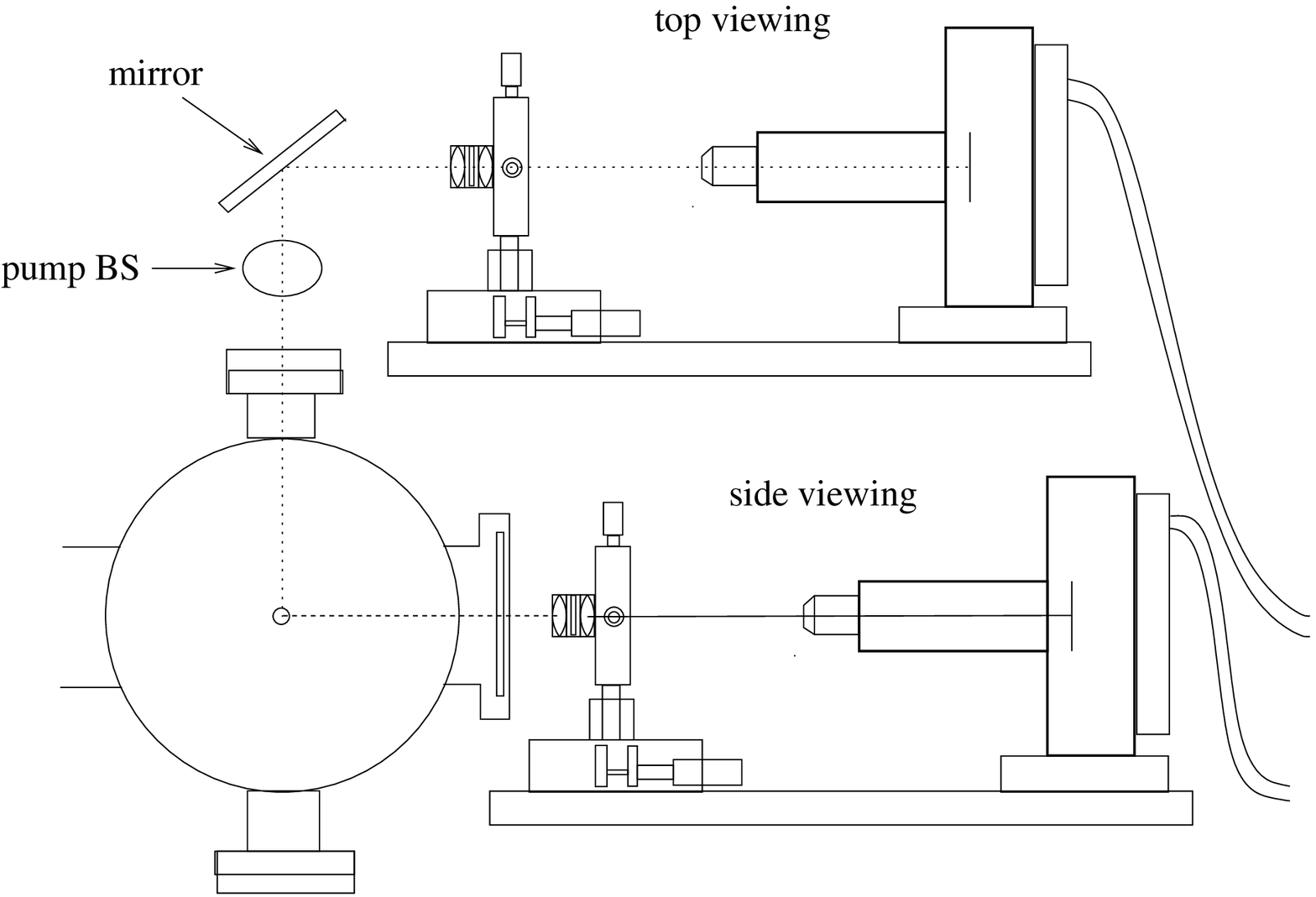}}}
\caption{Side and top viewing imaging system configurations.  
Lab jacks and supporting structures not shown.}
\label{fig-camera-configs}
\efig

The achromatic lenses are translatable in 3 directions by translation
stages for centering and focusing of the image.  The optics,
translators, and camera are mounted on an aluminum plate which is
fixed to a laboratory jack.

For the side imaging configuration, a 1 inch diameter interference
filter (Oriel 53890, $550 \pm 5$ nm) is placed between the two
achromats to block transmission of the oven's blackbody radiation.
For the top imaging configuration, a color glass filter 
is used to eliminate pump beam reflection and scatter into the imaging
system.  This filter has transmission $T > 95$\% at 553 nm and $T
<$0.1\% at 791 nm.

Background light is blocked by a plastic C-mount coupled tube connecting
the CCD camera and objective lens, and a sliding cardboard tube covering
the optical path between the objective and relay lens.

The CCD used is a Roper Scientific / Photometrics Sensys 400.  This
camera uses a thermoelectrically cooled, 1/2-inch format, 768 $\times$ 512
pixel Kodak KAF 0400 CCD.  Readout digitization is 12 bits.

\begin{table}
\begin{center}
\begin{tabular}{|c|c|c|c|c|} \hline
configuration&lens 1 f.l.&lens 2 f.l.&objective (typ.)&magnification \\ \hline 
side view     & 20 cm & 10 cm & 5x & 2.5 \\ 
top view     & 45 cm & 10 cm & 10x & 2.2 \\ \hline
\end{tabular}
\end{center}
\caption{Optics for two imaging system configurations}
\label{optics-table}
\end{table}

The camera can be used in two configurations
(Fig.~\ref{fig-camera-configs}.  In the first configuration, the
camera views the cavity from the side, antiparallel to the atomic beam
direction.  In this view the aperture can be imaged and fluorescence
of the atom beam appears as a small spot.  In the second configuration
the camera images the cavity from above.  A mirror is mounted at a
45-degree angle above the top window of the vacuum chamber, bending
the optical path so that the optics and CCD can be mounted
horizontally, above the optical path for the side-viewing
configuration.  The optics used for the two configurations is
summarized in Table~\ref{optics-table}.  The CCD mount is designed so
that the CCD can be swapped between the two configurations easily.

\subsection{Uses of imaging system}

\subsubsection{Atom-cavity alignment}
To achieve the most uniform atom-cavity coupling possible, it is
essential that the atomic beam be centered on the cavity mode.  This
can be done by adjusting the final aperture vertical translation stage
to maximize output power.  However, the miniature translation stage is
manipulated via a flexible coupling through an Ultra-torr vacuum
fitting and exhibits considerable backlash (hysteresis).  In addition,
the microlaser output power is not a highly sensitive measure of the
centering of the aperture on the mode, especially when the photon
number stabilization effect is taken into account.  Therefore is it
useful to be able to monitor the aperture position accurately.  This
can be done by focusing the imaging system on the oven's blackbody
radiation emitted from the aperture.  A computer program monitors the
microlaser output and blackbody peak vertical position as the latter
is varied manually.  


\subsubsection{Density calibration}

\bfig
\centerline{\resizebox{4in}{!}{\includegraphics{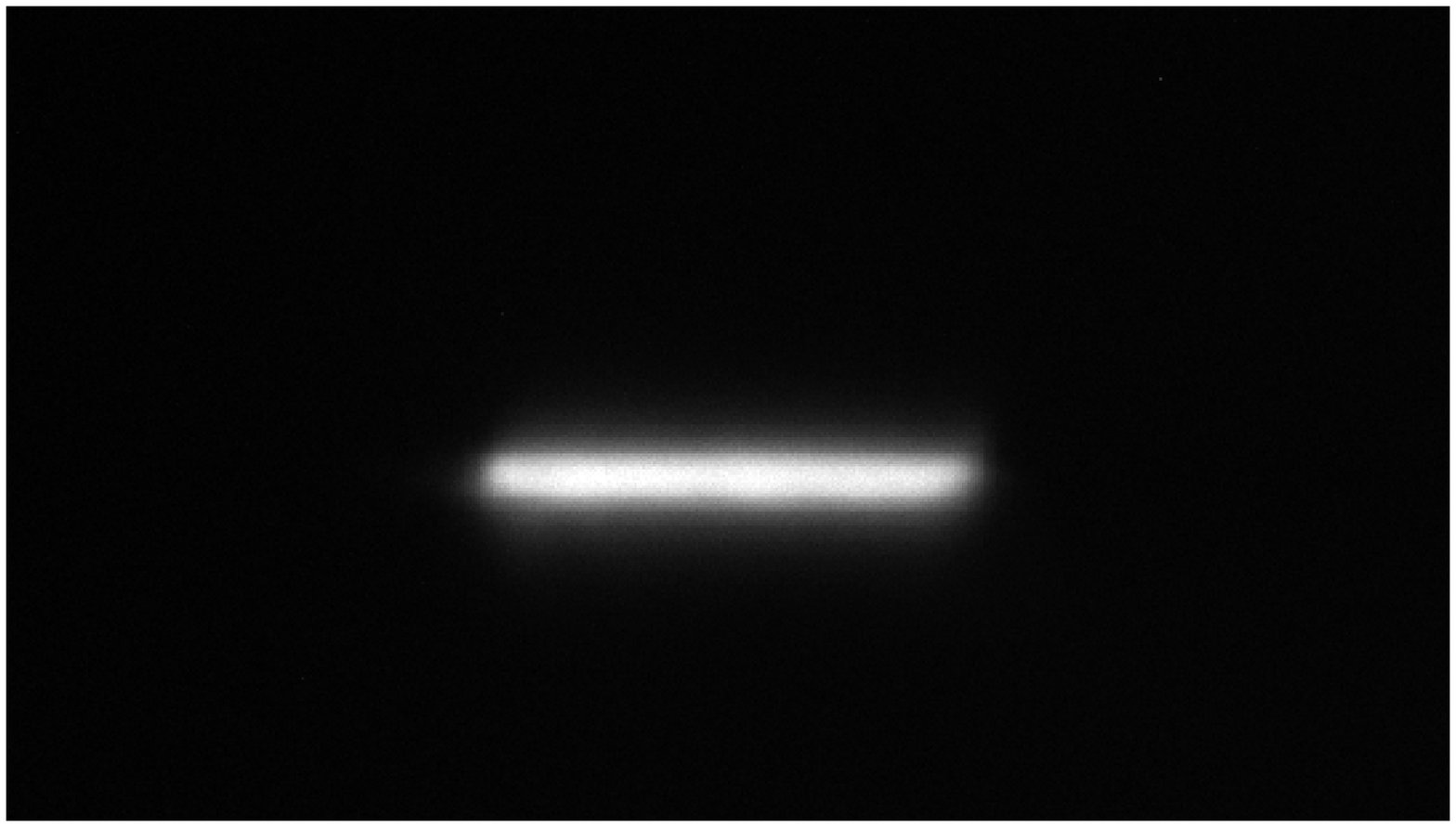}}}
\caption{Atom fluorescence imaged by microscope in side imaging configuration.
Similar but smaller images were used to measure atomic density during the 
experiment.}
%
\label{fig-atoms}
\efig

\bfig
\centerline{\resizebox{4in}{!}{\includegraphics{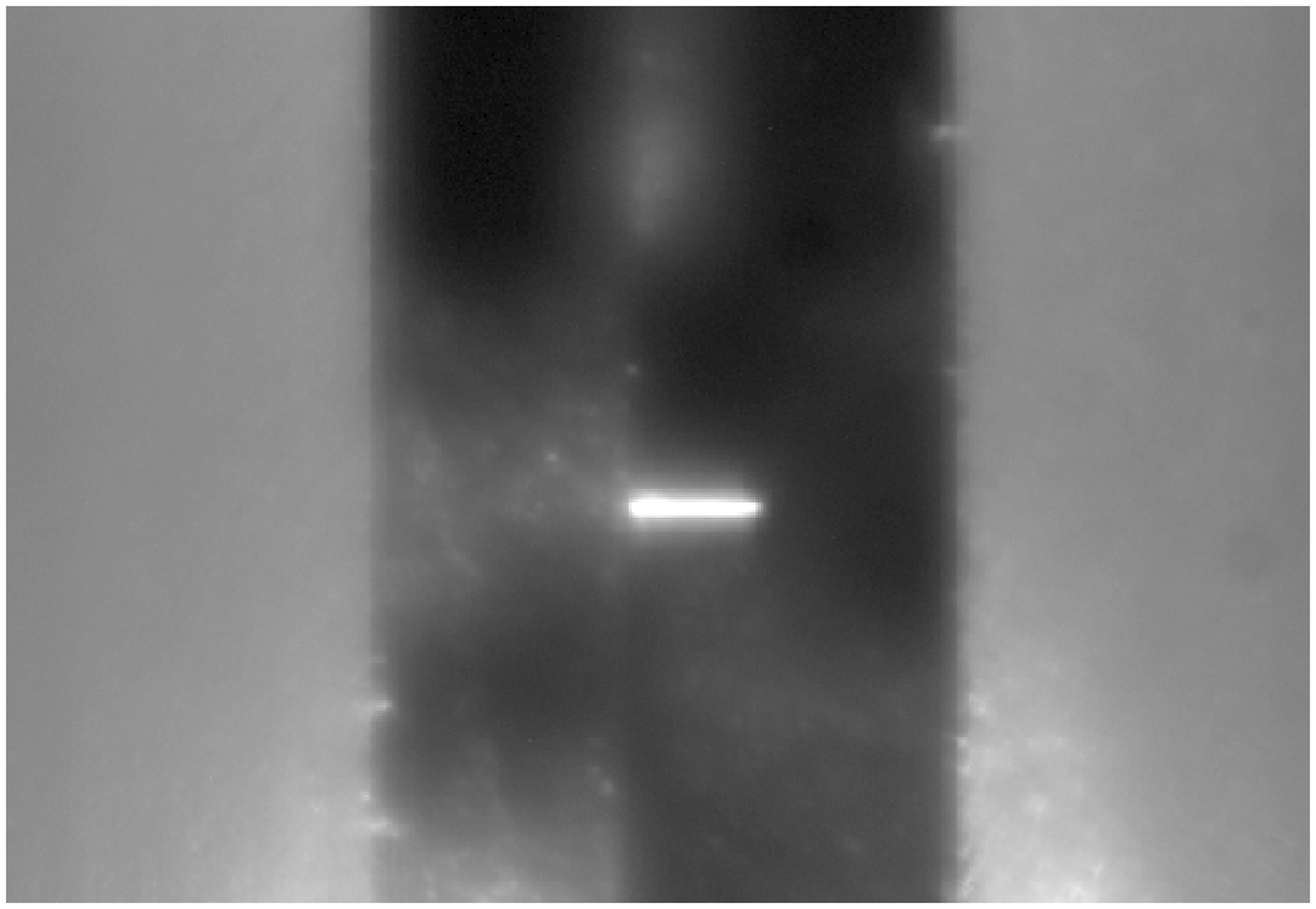}}}
\caption{Microscope image of center of cavity taken from side imaging
configuration.  Left and right: cavity mirrors.  Center: final
collimating aperture.  For this image, cavity was illuminated by
flashlight and oven blackbody radiation was used to image aperture
slit.}
\label{fig-cavbw}
\efig

The density of atoms in the atomic beam, and therefore the average
number of atoms in the cavity, can be monitored quantitatively by
detecting fluorescence of atoms in the cavity.  Details of this
experiment are given in Sec.~\ref{sec-density-cal}.

In addition to density calibration, the imaging system can be used to
monitor the condition of the final atom beam aperture.  Over time this
25 \um\ aperture becomes coated and eventually clogged with barium.
By monitoring the fluorescence and blackbody radiation, we can tell
when the aperture needs to be removed and cleaned.

\subsubsection{Characterization of laser and atomic beam}

In order to calibrate density, focus the pump and probe beams, and
position the pump beam correctly, it is necessary to determine the
position and focus properties of the 791nm and 553nm laser beams.  The
imaging system can be used to measure the focus and position of these
beams.

\subsubsection{Shelving measurement}

The imaging system was used to detect \fft\ fluorescence in the cavity
while the \sno\ pump beam was present.  This allows us to measure the 
degree of excitation due to the pump beam.

\section{Detectors}
\subsection{Avalanche photodiodes}


A photon-counting, thermoelectrically-cooled avalanche photodiode
(EG\&G 10902TC) is used to detect the \fft\ fluorescence in the cavity
when measuring beam velocity distributions.  An uncooled APD (EG\&G
10902) detects the cavity transmission for cavity locking and finesse
measurements.

\subsection{Photomultiplier tubes}

A photomultiplier tube was used to detect microlaser emission.  Our
PMT is a Hamamatsu R943-2.  It is cooled to less than $-20^\circ$~C by
a thermoelectric cooler (Pacific Instruments 3470).  At this
temperature it has a dark count rate of about 20 cps.  The bias
voltage used was -1500V.  The signal from the PMT is terminated into
50 $\Omega$ and discriminated/counted by a Stanford Research SR400
photon counter.

The PMT count rate was verified to be linear for the count rate used
in the experiment.  The saturation count rate was not measured and may
be significantly higher than the damage threshold of the PMT.


\section{Vacuum system}

A vacuum chamber houses the cavity and atomic beam oven.  It consists
of two large cylindrical sections, the first containing the oven and
the second containing the cavity.  The two chambers are connected by
flexible welded bellows couplings.  The atomic beam passes through the
flexible segment and into a 6-way cross before entering the cavity
chamber.  
%
A diagram is shown in
Fig.\ref{fig-chamberdraw}.  

\bfig
\centerline{\resizebox{\textwidth}{!}{\includegraphics{chamber4.epsi}}}
\caption{Photo of the microlaser experiment.  Atom beam oven is attached
to top flange of vertical cylinder portion of chamber at right.  
Chamber containing cavity is at center.}
\label{chamber4-photo}
\efig

\bfig[t]
\centerline{\resizebox{\textwidth}{!}{\includegraphics{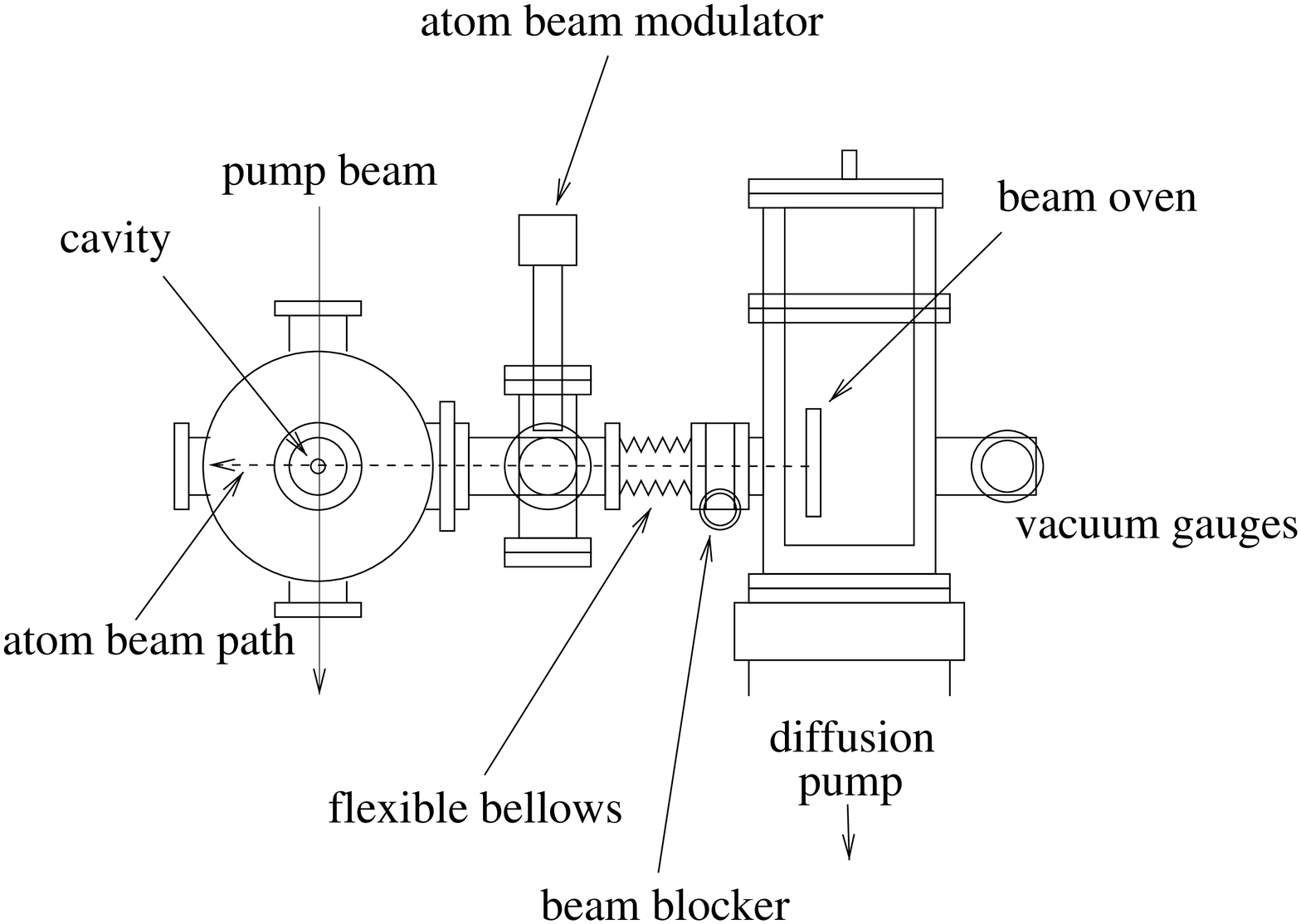}}}
\caption{Diagram of major chamber components showing atom beam path.}
\label{fig-chamberdraw}
\efig


All ports are faced with Conflat-type flanges.  OFHC copper and Viton
gaskets are used to create seals.
%
%
The oven chamber is connected to a diffusion pump (Varian VHS-4) with
pumping speed 1000 l/s.  A pneumatically driven 8'' gate valve (MDC
GV-50000) separates the oven from the chamber.  An interlock circuit
ensures that the diffusion pump can only be operated when the chamber
and foreline pressures (measured by two thermocouple gauges) are
sufficiently low, and the diffusion pump cooling water is flowing.
The foreline of the diffusion pump is connected by hoses to a roughing
pump  located in an adjoining room to minimize vibrations
and noise.  The foreline pressure was about 5 millitorr, measured by a
thermocouple vacuum gauge.  A tubulated ion gauge measures residual
pressure in the chamber, which during the experiment is typically
$\sim 10^{-6}$ Torr.

The vacuum system was fabricated by MDC Vacuum Products, and several
modifications were performed by Sharon Vacuum and the MIT Central
Machine Shop.

From time to time it was necessary to bring to chamber to atmospheric
pressure, in order to change the oven, reload the barium cell, etc.
In these cases we filled the chamber with argon to avoid oxidizing the
barium in the chamber.  This also avoids introducing water vapor into
the chamber which is difficult to pump out.


\section{Density calibration}
\label{sec-density-cal}
We now describe the experiments and calculations used to determine the
number of atoms present in the cavity.  First, let us clarify the
definition of the intracavity atom number $N$.  Since the cavity mode
is Gaussian, it is not clear where a boundary between the inside and
outside of the mode can be drawn.  We define $N$ to be the average
number of atoms within the boundary of a top-hat distribution
(Fig.~\ref{fig-tophat}) with height equal to that of the Gaussian and
with equal area (i.e. total interaction). This condition is met for a
top hat of width $w_{\rm tophat}= \sqrt{\pi} w_m$.

\bfig 
\centerline{\resizebox{3.5in}{!}{\includegraphics{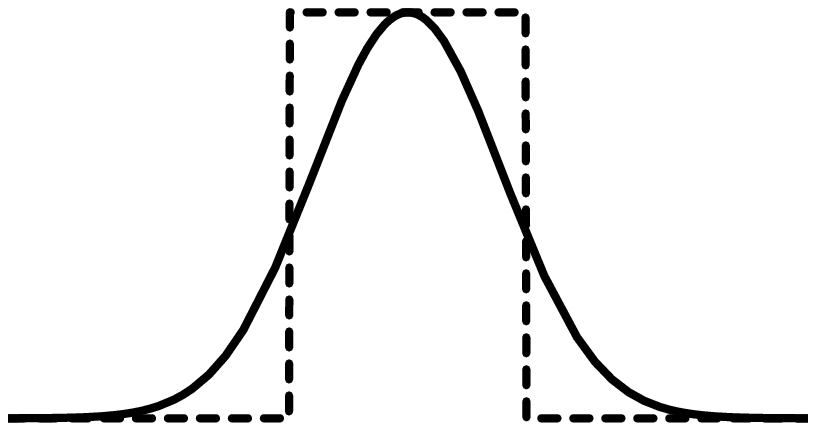}}}
\caption{Gaussian mode and equivalent tophat profile}
\label{fig-tophat}
\efig

Note that we assume that each atom in the atom beam passes through the
mode at some time; this is true because the atomic beam size in the
direction perpendicular to the cavity axis is smaller than the mode
waist.  By contrast, in the original microlaser experiment the atomic
beam was 300 \um\ in diameter, much larger than the mode.

To measure the atomic density we overlap a \fft\ probe beam onto a
\sno\ beam which has been focused to match as close as possible
only the $TEM_{00}$ mode of the cavity.  The \fft\ beam then
intersects the atomic beam perpendicularly except for any cavity tilt
that is introduced.  The resulting \transitionsinglet\ fluorescence is
collected by the camera in the side viewing configuration.

\bfig
\centerline{\resizebox{4in}{!}{\includegraphics{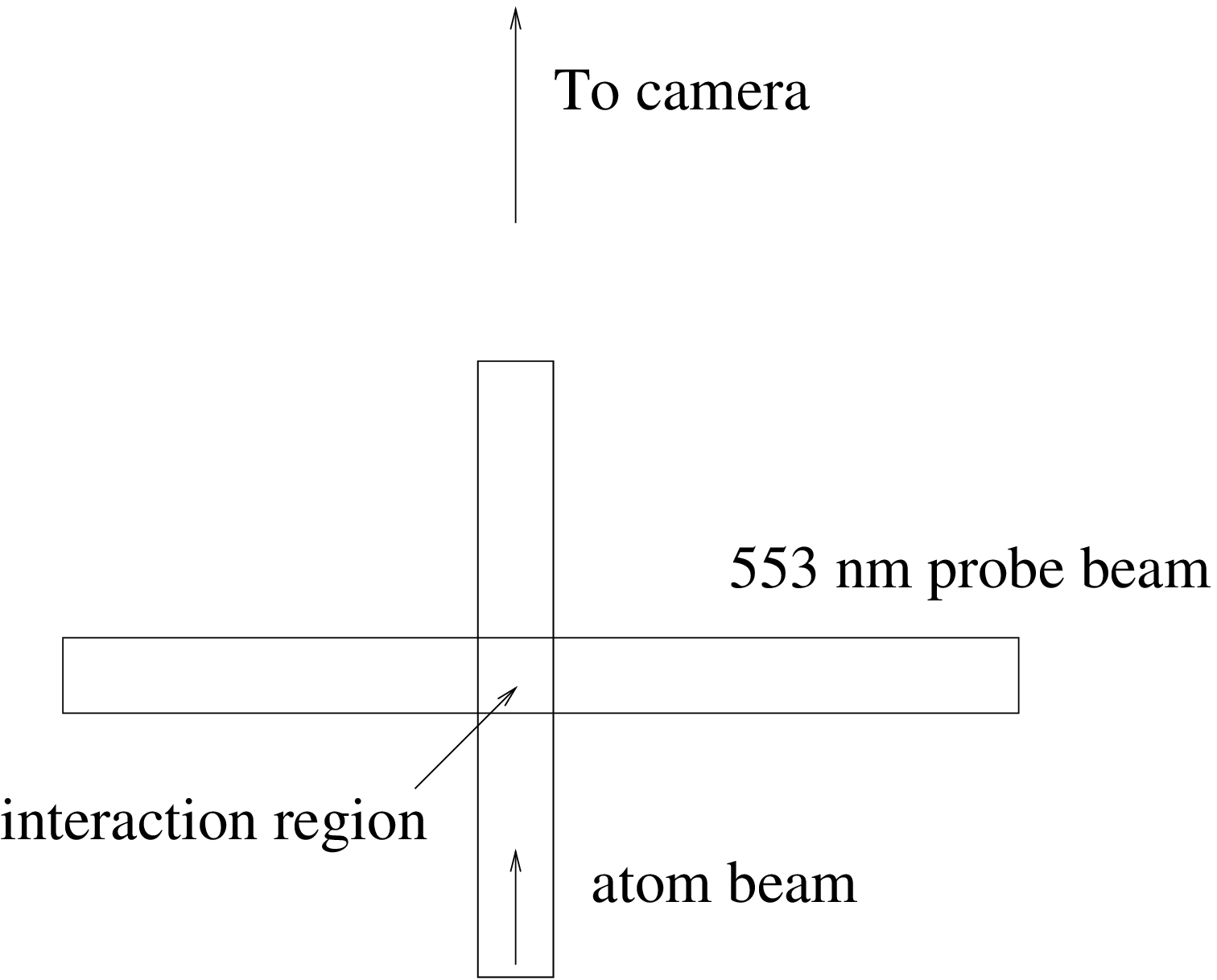}}}
\caption{density-measurement-setup}
\label{density-measurement-setup}
\efig

The total number of fluorescence photons incident on the CCD is given by

\be
n_p = \eta_{\rm optics} F_\Omega W F_\delta T_{\rm ex}
\ee

where $\eta_{\rm optics}$ is the transmission of the optical system
including chamber window, $ F_\Omega$ is the fraction of scattered
light which travels in the direction of the collection optics, and $W$
is the total scattering rate.  $T_{\rm ex}$ is the exposure time.
$ F_\delta < 1$ is a detuning factor reflecting the Doppler broadening due
to the tilted cavity, and is approximately equal to the ratio between
the homogeneous and inhomogeneous atom linewidths.

The transmission $\eta_{\rm optics}$ is the product of the
transmission of the chamber window, two lenses, \fft\ interference
filter, and microscope objective.  This was measured by passing a
laser beam through these optics:
\be
\eta_{\rm optics} \approx 22\%
\ee

The parameter $ F_\Omega$ involves the solid angle of collection and
the dipole radiation pattern of the atoms.  The polarization of the
probe field is set to be vertical so that the camera collects light
scattered at the optical 90 degrees from the dipole direction.  The
fraction of power emitted into the solid angle subtended by the camera
optics is

\be
F_\Omega = \frac{\int_{\rm optics} \sin^2(\theta) d\Omega}{\int_{\rm
all} \sin^2(\theta) d\Omega} \approx \frac{\Omega_{\rm optics} }{2 \pi 
\int_0^\pi d\theta \sin^3(\theta)} = {3 \over 2} \frac{\Omega_{\rm optics}}
{4\pi}
\ee
\noindent where we have assumed $\theta$ does not deviate significantly from
$\pi/2$ over the solid angle of the collection optics.  This result
simply means that the fluorescence detected at 90 degrees from the
dipole direction is 50\% than that expected from an isotropic
radiation pattern.  

The solid angle of the optics is that of a 1'' lens located 20 cm from
the fluorescence source
\be
\Omega_{\rm optics} \approx \frac{\pi(0.5)(2.54 cm)^2}{(20 cm)^2}=0.019 sr
\ee

The total rate of photon scattering from the source is
\be
W = \int d{\bf r} \rho({\bf r}) R({\bf r})
\ee
where $\rho({\bf r})$ is the atom density and
$R({\bf r})$ is the scattering rate per atom at point $\rm{r}$:
\be
R({\bf r}) = \frac{\ga}{2} \frac{I(x)/I_{\rm sat}}{1+I(x)/I_{\rm sat}}
= \frac{\ga}{2} \frac{ (p/p_s) \exp(-2(x/w_p)^2) }
{1+(p/p_s) \exp(-2(x/w_p)^2)}
\ee
Here $\ga\ \approx 19 MHz$ is the natural linewidth of the
\transitionsinglet\ transition; $w_p$ is the mode waist of the probe beam; 
$I_s$ is the saturation intensity; $p_s$ is the power of the \fft\
probe, measured outside the chamber, such that the laser intensity
center of the focused probe beam is $I_s$.  We then have
\be
W = \int_{-\infty}^{\infty}dx \lambda' \frac{\ga}{2} \frac{ (p/p_s) \exp(-2(x/w_p)^2) }
{1+(p/p_s) \exp(-2(x/w_p)^2)}
\ee
\noindent where $\lambda' = \int dy\ dz\ \rho$ is the constant linear density along the atom beam direction.

The parameter $p_s$ is measured by using the camera in the top viewing
configuration and taking exposures for different probe powers.  The
peak of the measured fluorescence distribution corresponds to the most
intense (axial) part of the probe beam; by plotting this against the
probe power and fitting a saturation curve we obtain a saturation
power (Fig.~\ref{fig-553sat}) of
\be
p_s = 11.2 \uw
\ee


\bfig
\hspace{1in}fluor.~cts\\ \mbox{}\\
\centerline{\resizebox{4in}{!}{\includegraphics{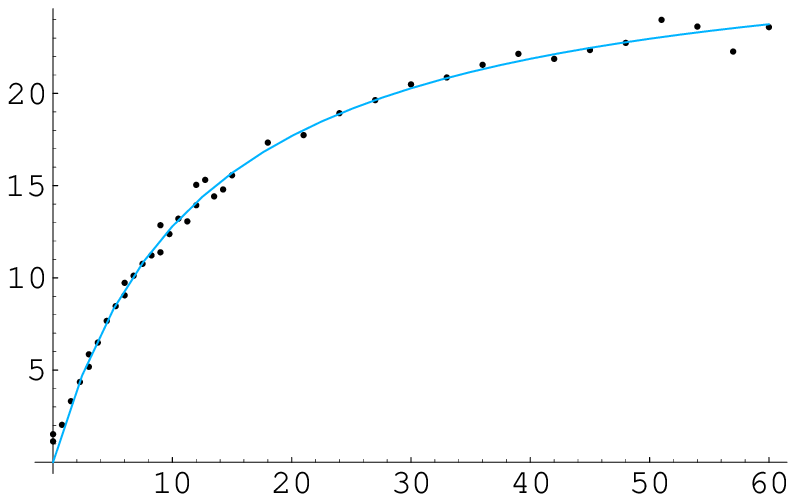}}}
\centerline{probe power, \uw}
\caption{Saturation curve of peak of \fft\ fluorescence: fluorescence counts vs.\ input probe power, \uw.}
\label{fig-553sat}
\efig

\bfig
\hspace{1in}fluor.~cts\\ \mbox{}\\
\centerline{\resizebox{4in}{!}{\includegraphics{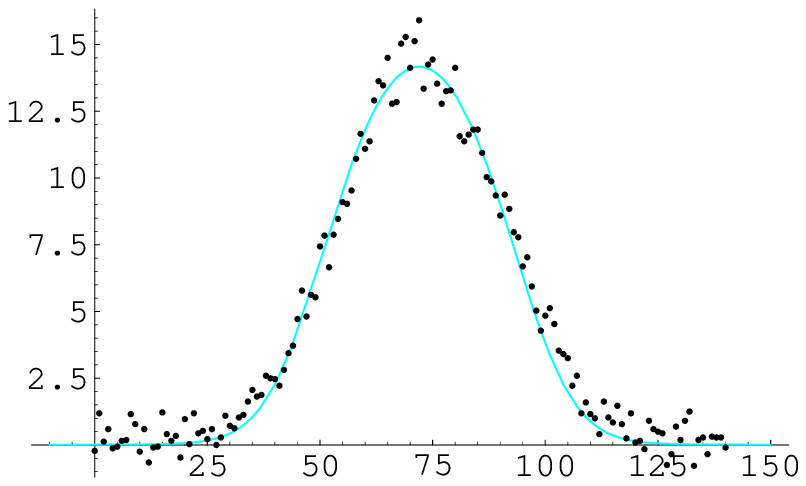}}}
\centerline{Pixel number}
\caption{\fft\ fluorescence distribution from CCD.  Fit includes saturation
effects and optical aberration function. }
\label{fig-553waist}
\efig

The probe beam waist is measured with the same camera data by imaging
the fluorescence distribution directly (Fig.~\ref{fig-553waist}).  After
accounting for atom saturation and camera defocus/aberration effects
we obtain
\be
w_p = 139 \um.
\ee
The total integrated counts recorded on the CCD is given by
\be
C = n_p \cdot \eta_{\rm CCD} / G \approx 0.0625 n_p
\ee
where $\eta_{\rm CCD} \approx 0.35$ is the CCD quantum efficiency at
\fft\, according to manufacturer specifications, and the gain is $G
\approx 5.6$ electrons per ADU (analog-to-digital unit) at the camera's 
gain index 3 setting.

The CCD exposure time was typically set in software to be 300 msec.
However, it was discovered (Fig.~\ref{fig-ccd-nonuniform}) that the
total fluorescence measured was not proportional to the exposure time
when those exposure times were small; this is probably due to finite
shutter opening and closing times.  An extra factor was used to
compensates for this effect: $T_{\rm ex} = 142$ ms for a set time of
300 ms.

\bfig
\centerline{\resizebox{4in}{!}{\includegraphics{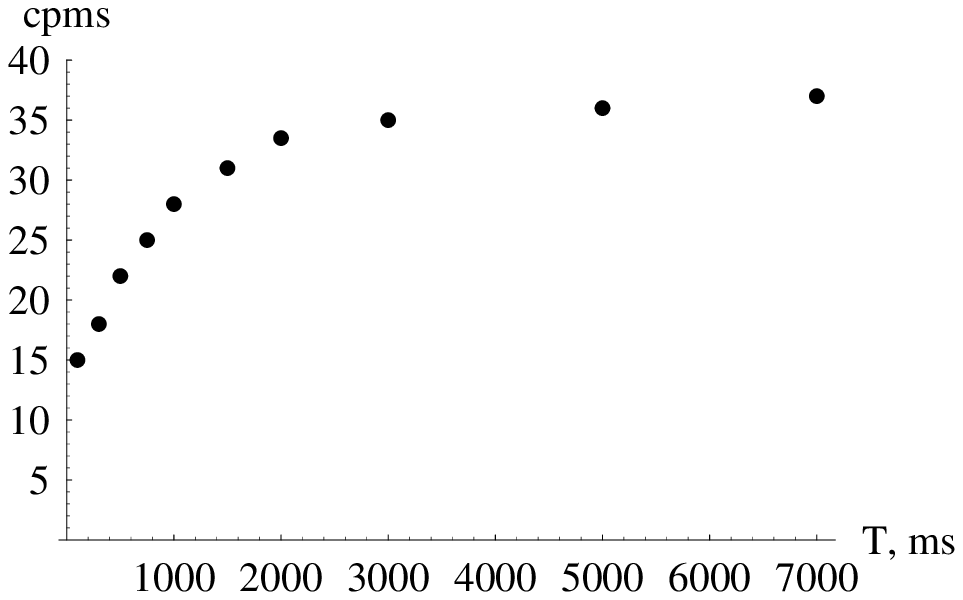}}}
\caption{CCD counts per millisecond (cpms) for a constant atom density
when the exposure time $T$ is varied.  Drop-off for short exposures is
probably due to opening and closing times of mechanical shutter.}
\label{fig-ccd-nonuniform}
\efig

The lineshape of the \fft\ transition was measured by varying the
laser detuning from the computer and recording the total fluorescence.
The resulting curve was fit accurately by a Lorentzian with FWHM of 34
MHz.  The detuning correction factor is then $ F_\delta = 19 {\rm MHz}/34
{\rm MHz} = 0.559$ 

Finally we calculate the intracavity atom number:
\be
N = \sqrt{\pi} w_m \lambda'  
\ee

The ratio between CCD counts and atom number is most conveniently
expressed as the number of counts per millisecond of exposure time,
per atom, with probe power 10 \uw.  We find that $N=1$ atom
corresponds to 1.16
counts per ms on the CCD; 1 cpms corresponds to 0.862 intracavity
atoms.

\section{Photon number calibration}

The PMT count rate $n_d$ as a function of cavity photon number $n$ is
\be
n_d = n (\frac{T}{2(T+A+S)}) \gc \eta_{\rm optics}' \eta_{\rm AOM} \eta_{\rm PMT}
\ee
\noindent where $T=0.5 \times 10^{-6}$, $A = 0.2 \times 10^{-6}$, 
and $S = 3.0 \times 10^{-6}$ are estimates of the transmission,
absorption, and scattering coefficients of the cavity mirrors; the factor
of 2 accounts for the collection of light from only one side of cavity.
\gc\ is the cavity loss rate; $\eta_{\rm AOM}$ is the fraction of light
switched by the AOM after the cavity into the PMT (i.e. diffraction
efficiency, about 70\%); $\eta_{\rm optics}$ is the transmission of
optics including the rear mirror surface, chamber window, several
mirrors and lenses guiding the beam to the PMT, and the PMT window.
$\eta_{\rm PMT}$ is the quantum efficiency of the PMT at \sno, about 12\%
according to manufacturer specifications.  

Our estimate for the number of photons in the cavity is $n = 0.16$ per
$10^3$ counts per second on the PMT, with no filter in front of the PMT.

Both the estimate for atom and photon number are subject to fairly
large systematic error; each may have errors as large as a factor of
2.

\section{Data acquisition and computer control}

The experiment is controlled by a Gateway 200 MHz PC running
Windows 95.  The sequencing and data collection routines are written
in VPascal, the language used by the camera control software, Digital
Optics V++.  The sequencing programs used in this experiment can be
found in Appendix \ref{app-vpascal}.

\subsection{Experiment automation}
\label{sec-automation}
Table~\ref{table-automation} summarizes the computer automation of
various devices.  These are further described and commented on below.

\mbox{}

\noindent
\begin{table}[h]
\begin{tabular}{|l|c|c|} \hline
Device & Level & Control \\ \hline
SR400 photon counter & digital data & PC serial port 2 \\
IMS atom beam mod.\ driver & digital data & PC serial port 2 \\
Laser intensity mod.\,(791 nm pump)& 0-1 V & SR400 PORT1 output \\
Laser intensity mod.\,(553 nm probe)& 0-1 V & SR400 PORT2 output \\
Analog multiplexer (MUX) & TTL $\times$ 3 & IMS TTL OUT 1-3 \\
Cavity lock disable or scan trigger& TTL & MUX output 1\\
791nm probe beam AOM $V_T$ & 10V(on), 4V(off) & MUX output 2\\
AOM after cavity $V_T$& 10V(on), 4V(off) &  MUX output 3\\ \hline
\end{tabular}
\caption{Summary of computer automation setup}
\label{table-automation}
\end{table}

\subsubsection{Laser intensity modulation}

The two analog outputs of the SR400 photon counter are used to control
the intensities of the 553 nm probe and 791 nm pump.  During the
``off'' states the target voltage is set to -0.1 volt in order to
ensure that the minimum intensity is reached.

\subsubsection{Atom density modulation}

The stepper motor driving the density modulator is controlled by the
computer through the serial port.  The stepper rate was 400 steps per
second.  No microstepping was necessary.

\subsubsection{AOM switching}

Two acousto-optic modulators required binary control: the 791 nm probe
switching between cavity locking and data collection, and the AOM
deflecting the cavity emission to the \gtwo\ setup during data
collection.  Both AOMs were controlled by TTL signals from the stepper
driver.



\section{Cavity PZT control}

The cavity frequency may be scanned through the microlaser resonance
or locked onto it.  In some sense a scanning experiment is more
general since it includes variation in atom-cavity detuning.  Other
experiments require the laser to be kept on resonance while other
parameters are varied.


\subsubsection{Cavity scanning}
\label{sec-cavscan-sequence}
For cavity scanning, the PZT is simply driven by a $\approx 1$ V ramp
signal generated from an oscilloscope, which sweeps in 'AUTO' mode or
is triggered by a pulse from the TTL output from the computer.  To create
bidirectional scans we designed an additional circuit which
outputs $|V_{\rm in}(t)-V_0|$ given an input voltage $V_{
\rm in}(t)$ from the oscilloscope or other ramp source.  $V_0$ is equal to
half the peak-peak height of the ramp signal.  The scan and sample
PMT signal is shown in Fig.~\ref{fig-pztscan}.

The experimental sequence for cavity scanning experiments is the following:

\begin{enumerate}
\item Move atom density modulator for the desired density
\item  Turn off the \sno\ pump beam, 
\item Turn on the \fft\ probe, with power 10 \uw
\item Measure atomic density by exposing the CCD in the side viewing
configuration for a set time, usually 300 msec,
\item Turn off \fft\
probe and turn on \sno\ pump beam 
\item Trigger oscilloscope to drive
cavity PZT with a bidirectional ramp of duration about 4 seconds 
\item Start photon
counter to collect data from PMT during cavity scan 
\item Download
PMT count data to computer via serial connection
\end{enumerate}

It was occasionally necessary to adjust the HV PZT offset voltage
manually to compensate for a slow PZT drift which would otherwise
cause the microlaser resonances to drift out of the scan range.  These
adjustments were made between data collection intervals so that the
linearity of the PZT scans was preserved.

\bfig
\centerline{\resizebox{4in}{!}{\includegraphics{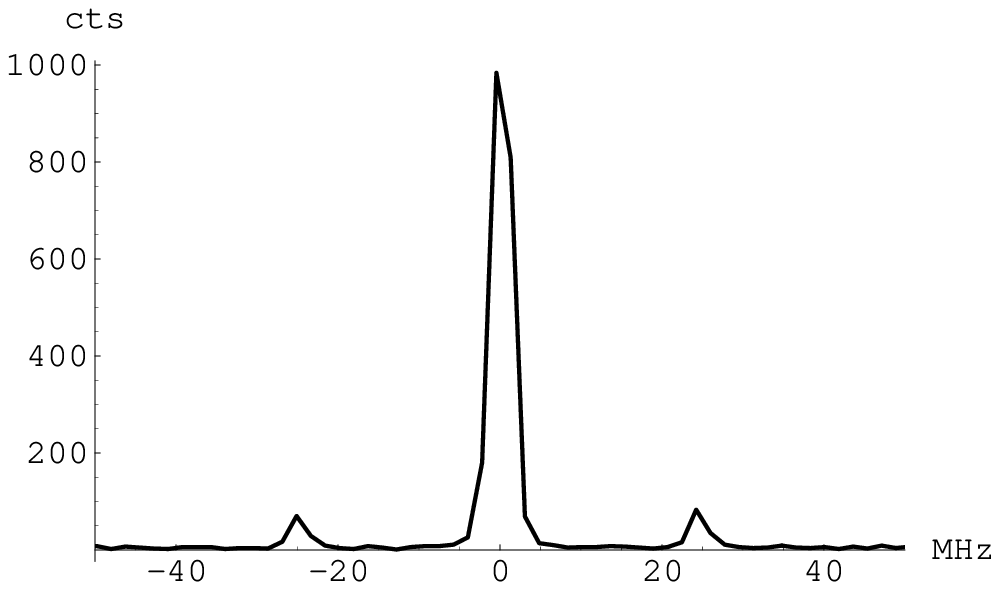}}}
\caption{Probe transmission spectrum for frequency calibration of PZT 
scans.  Sideband frequency shift from carrier is 24.67 MHz.}
\label{calibscan2}
\efig

To calibrate the frequency of the scans we measured the cavity
transmission of a very low power probe beam with no atoms present; the
result is shown in Fig.~\ref{calibscan2}.

The background counts level, due primarily to pump scattering from
cavity mirrors and stray room light, is subtracted by computer.


\subsubsection{Cavity locking}

To maintain the cavity on resonance, we use the following cavity lock
technique.

\bfig
\centerline{\resizebox{\textwidth}{!}
{\includegraphics{fig-cavlock-circuit.epsi}}}
\caption{Cavity locking circuit}
\label{fig-cavlock-circuit}
\efig

A probe beam from the Ti:Sapphire laser is upshifted $100$ MHz and
then downshifted about $113$ MHz by a pair of AOMs.  The frequency of
the second AOM adjusted via the input tuning voltage $V_T$ to match
the peak of one of the microlaser resonances, in our case the lower
frequency resonance at $\omega_a - k v_0 \theta$.  The AOMs are also
used to switch the probe beam on and off.

For cavity locking, a probe beam of about 1 \uw\ is incident on the
resonator.  The cavity transmission is focused onto an APD.  A circuit
compares the APD voltage with a fixed reference voltage adjusted to a
value approximately one-third of the peak signal height.  The voltage
difference is integrated and the result fed back to the PZT control
voltage.  If no strong mechanical or acoustical disturbances occur,
this system can remain locked for more than 1 hour.  


Experiments which involve cavity locking require us to switch between
locking and data collection.  The cavity can be unlocked for a maximum
of about 1 second before the cavity drifts too far from resonance for
lock to be regained.  During cavity non-lock, a TTL signal applied to
the lock box causes the integrator to hold its value by grounding its
input.


\chapter{Results and Analysis}
\label{chap-results}



We describe the results of microlaser experiments with a high-density
supersonic beam and nearly uniform atom-cavity coupling.  

Experiments were performed with the goal of answering the following
questions: (i) Does the theory of Chapter~\ref{chap-theory} apply to
the microlaser with $\Neff \sim 1000$?  (ii) Do multiple thresholds
occur, and if so, where?  (iii) What are the time scales of the
transitions, and can hysteresis be observed?




\section{Study of microlaser with variable detuning}

We first measured the microlaser output as a function of atom-cavity
detuning by applying a bidirectional ramp signal to the cavity PZT.
The experiment sequence is given in Section~\ref{sec-cavscan-sequence}.
%
The atom density modulator position was varied between 1350 and 1730
with a step size of 10, and 1450 to 1500 with a step size of 2 (c.f.\
Fig~\ref{fig-beam-blocker}).  

\bfig
\centerline{\resizebox{4in}{!}{\includegraphics{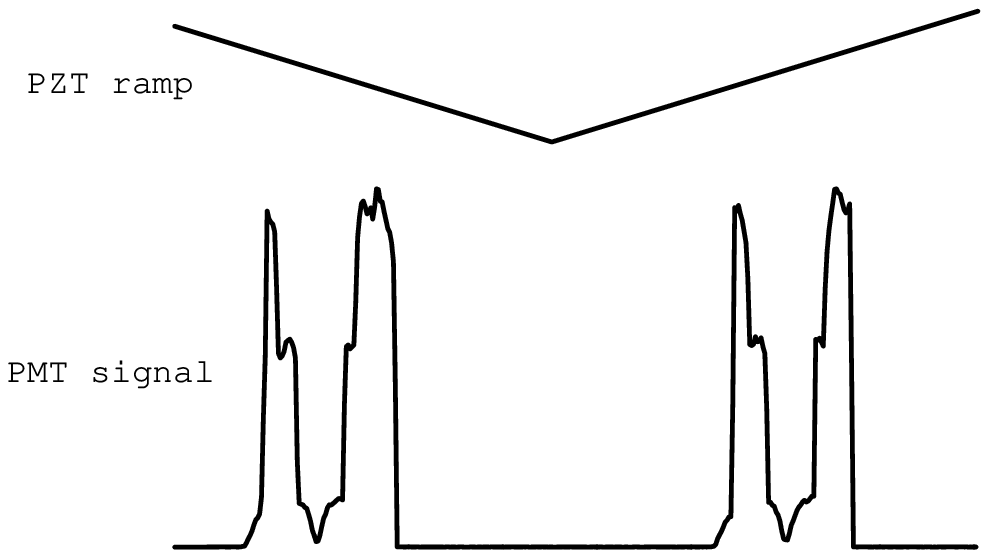}} \hspace*{.5in}}
\centerline{time~$\rightarrow$}
\caption{Bidirectional PZT voltage scan shown with typical PMT data for
a high atom density. Horizontal range corresponds to 4 seconds.}
\label{fig-pztscan}
\efig

%

\section{Cavity scanning data}

Data for the detuning curves for a range of atom densities are shown
in Figs.~\ref{data2index30} to \ref{data2index0}.  
The PMT data is folded back onto itself to show both results from the
PZT scan in both directions.  The solid and dashed lines represent
scans in the positive and negative detuning directions, respectively.
PMT counts are measured in time bins of 0.01 seconds.

Effective atom numbers are following the estimates given later in
this chapter.

\bfig
\centerline{\resizebox{4in}{!}{\includegraphics{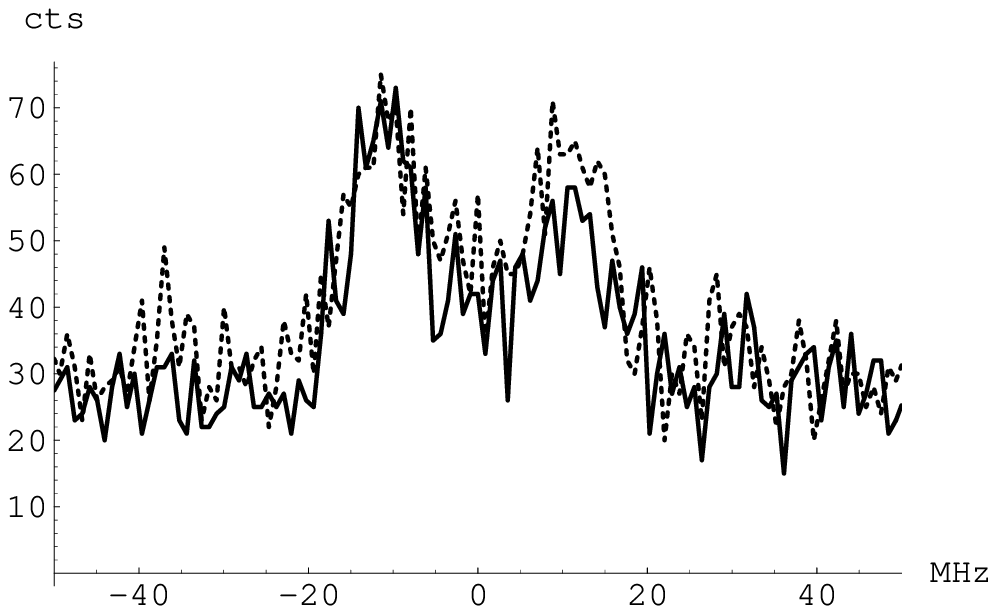}}}
\caption{PMT counts vs.\ cavity-atom detuning.  
Fluorescence 48.5 cpms ($\Neff = 10.7$) }
\label{data2index33}
\efig

\bfig
\centerline{\resizebox{4in}{!}{\includegraphics{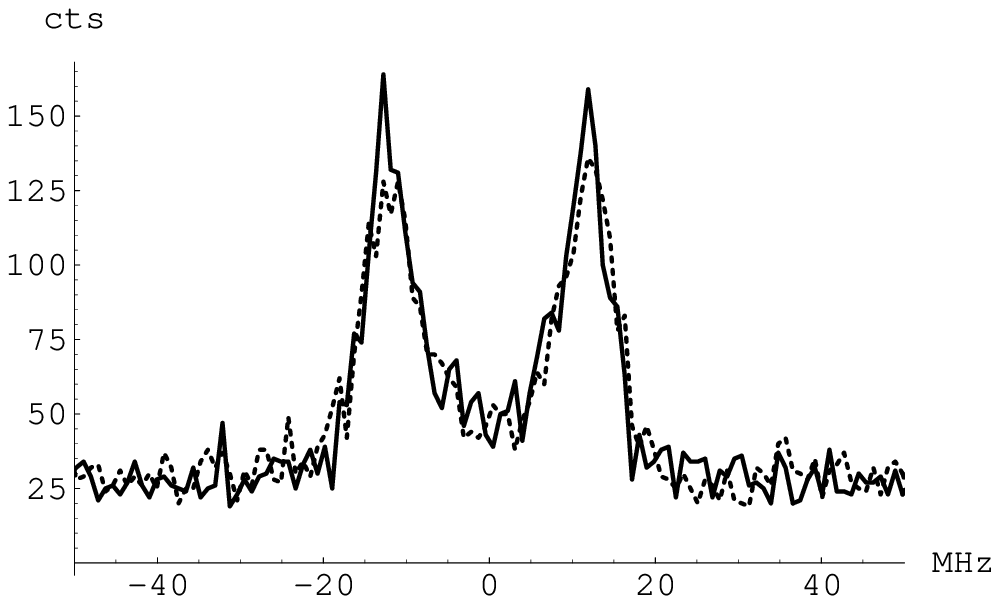}}}
\caption{Fluorescence 60.9 cpms ($\Neff = 13.4$)}
\label{data2index32}
\efig

\bfig
\centerline{\resizebox{4in}{!}{\includegraphics{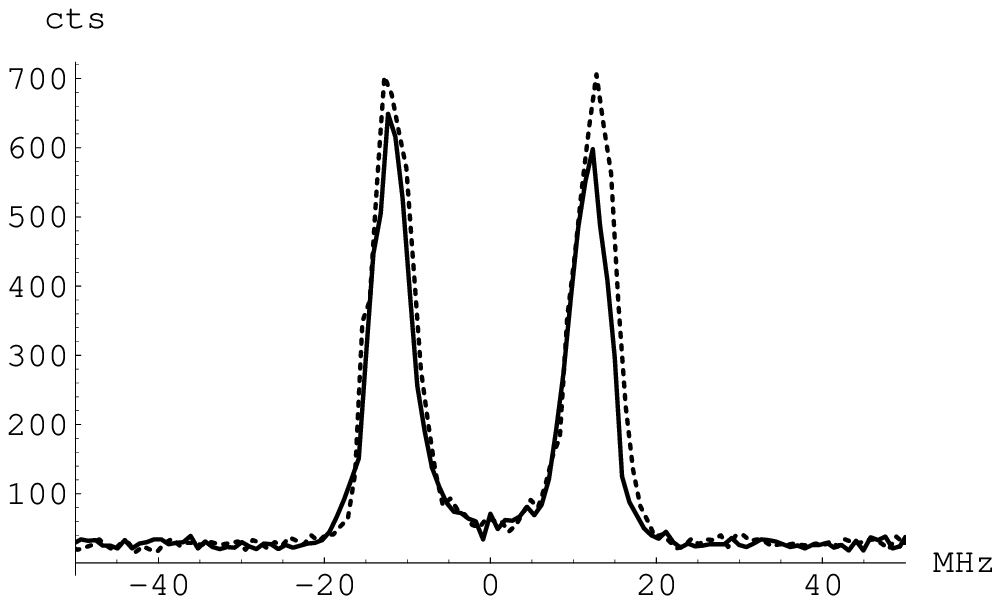}}}
\caption{ Fluorescence 80.2 cpms ($\Neff = 17.6$) }
\label{data2index31}
\efig

\bfig
\centerline{\resizebox{4in}{!}{\includegraphics{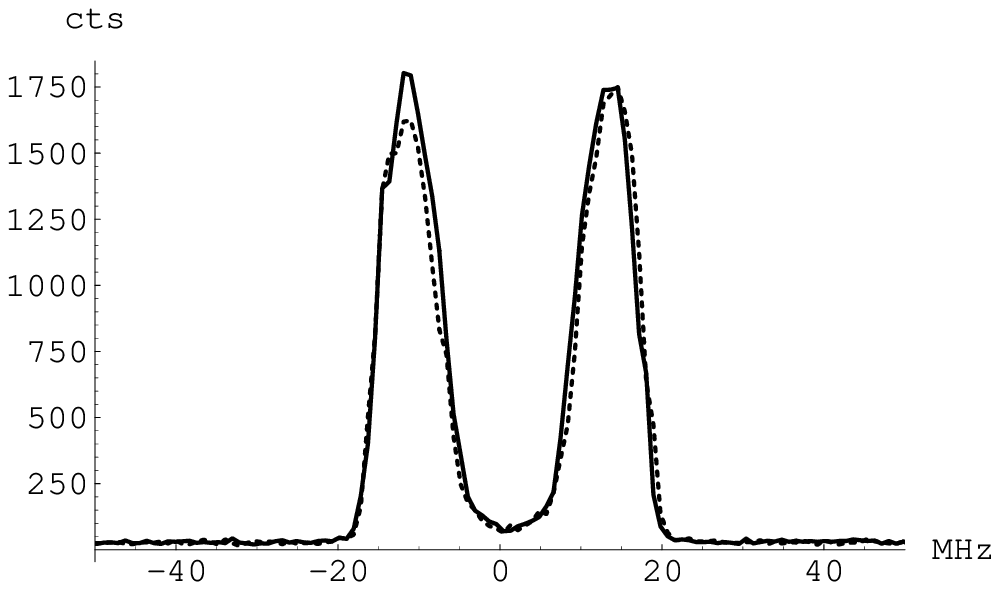}}}
\caption{Fluorescence 100 cpms ($\Neff = 22$) }
\label{data2index30}
\efig


\bfig
\centerline{\resizebox{4in}{!}{\includegraphics{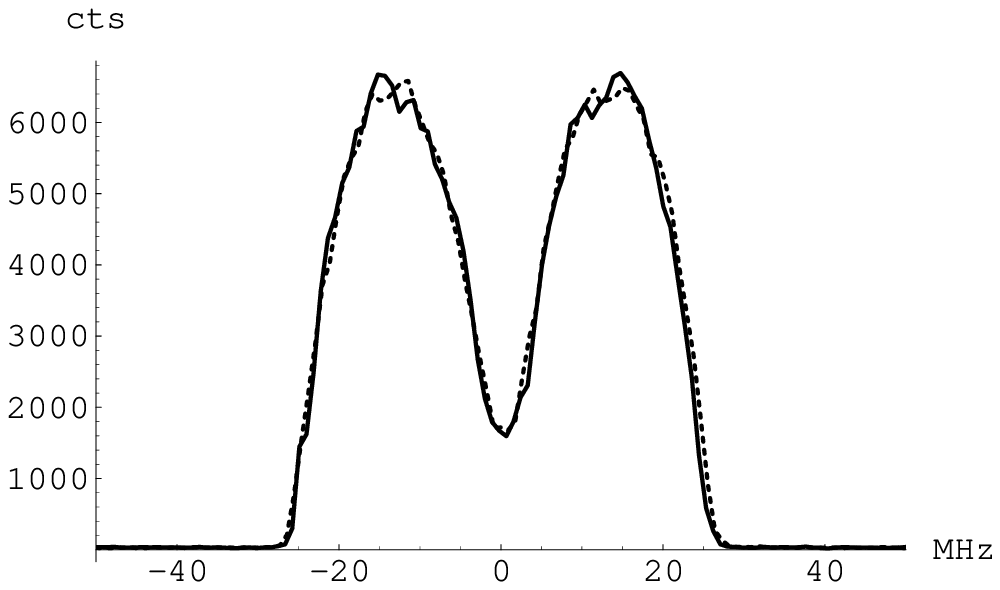}}}
\caption{Fluorescence 659.1 cpms ($\Neff = 145$) }
\label{data2index19}
\efig

\bfig
\centerline{\resizebox{4in}{!}{\includegraphics{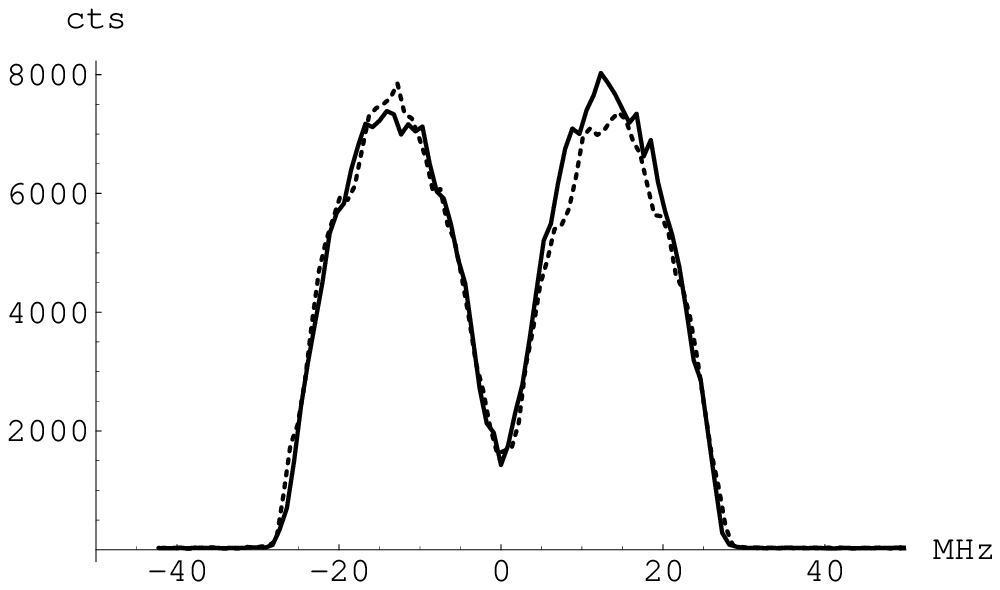}}}
\caption{Fluorescence 950 cpms ($\Neff = 209$) }
\label{data2index17}
\efig

\bfig
\centerline{\resizebox{4in}{!}{\includegraphics{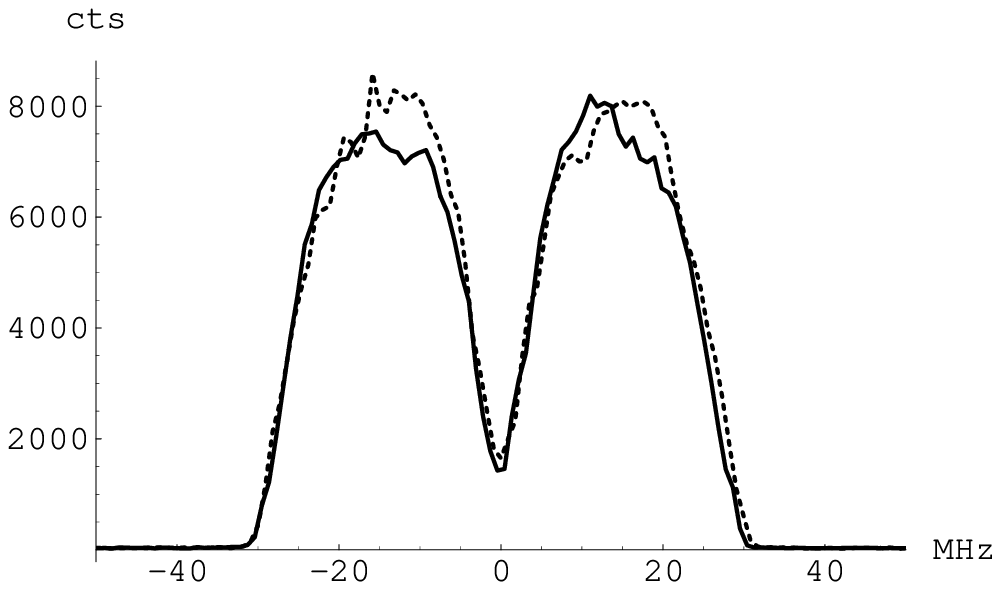}}}
\caption{Fluorescence 1850 cpms ($\Neff = 407$) }
\label{data2index14}
\efig

\bfig
\centerline{\resizebox{4in}{!}{\includegraphics{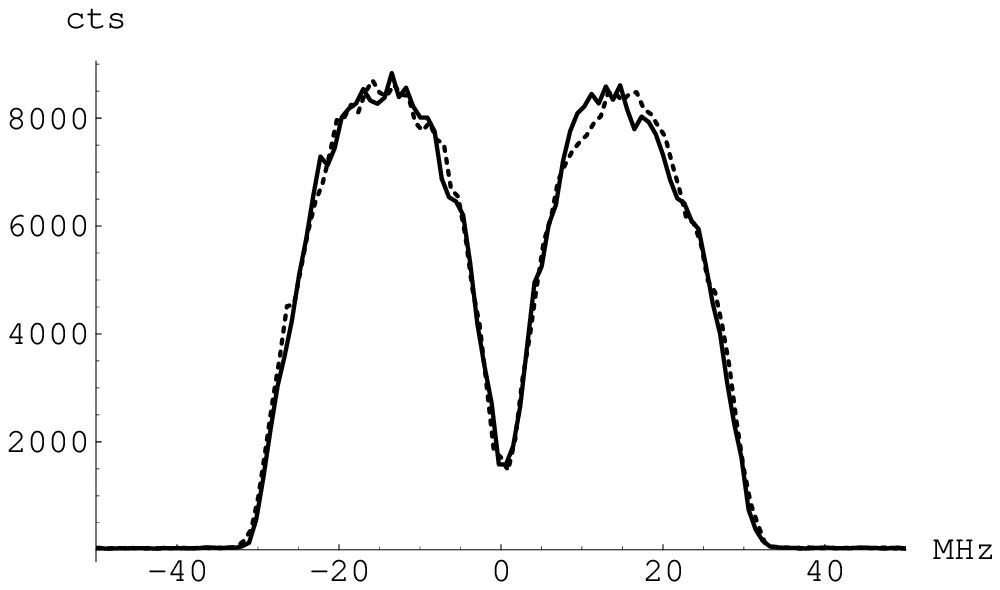}}}
\caption{Fluorescence 2393 cpms ($\Neff = 526$) }
\label{data2index13}
\efig

\bfig
\centerline{\resizebox{4in}{!}{\includegraphics{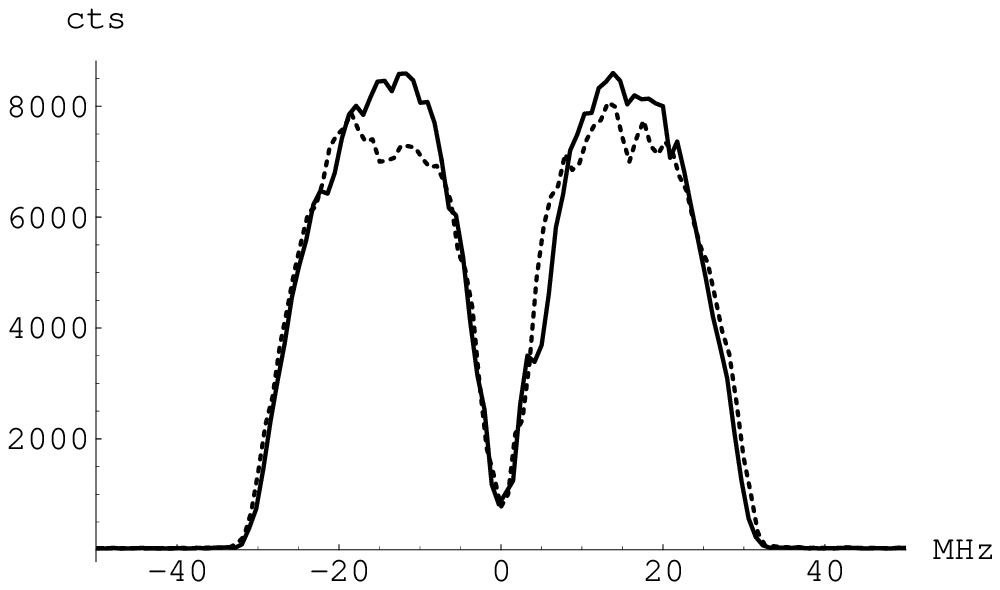}}}
\caption{Fluorescence 2748 cpms ($\Neff = 605$) }
\label{data3index10}
\efig

\bfig
\centerline{\resizebox{4in}{!}{\includegraphics{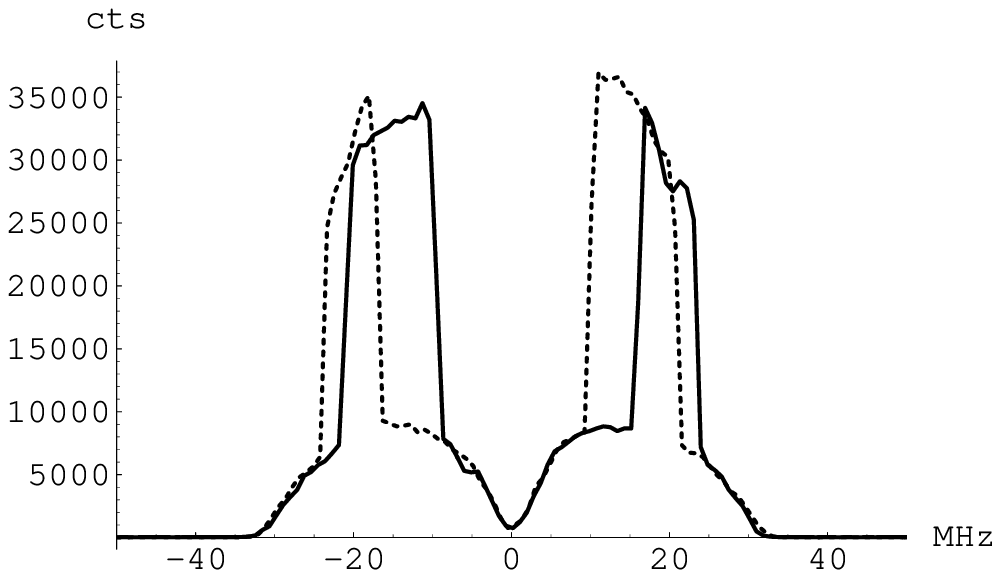}}}
\caption{Fluorescence 2889 cpms ($\Neff = 636$).  Second threshold is
observed. }
\label{data3index9}
\efig

\bfig
\centerline{\resizebox{4in}{!}{\includegraphics{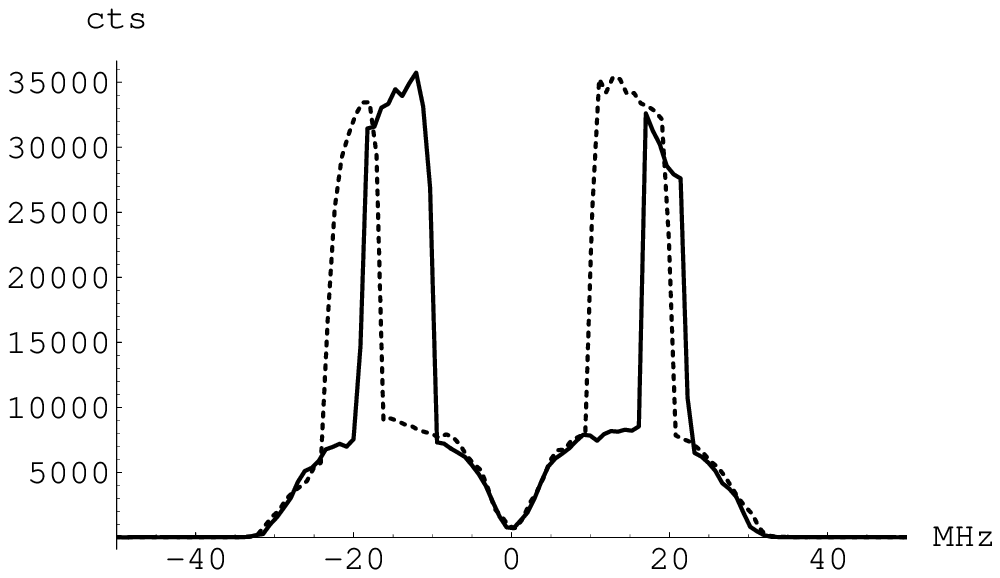}}}
\caption{Fluorescence 2986 cpms ($\Neff = 657$) }
\label{data3index8}
\efig

\bfig
\centerline{\resizebox{4in}{!}{\includegraphics{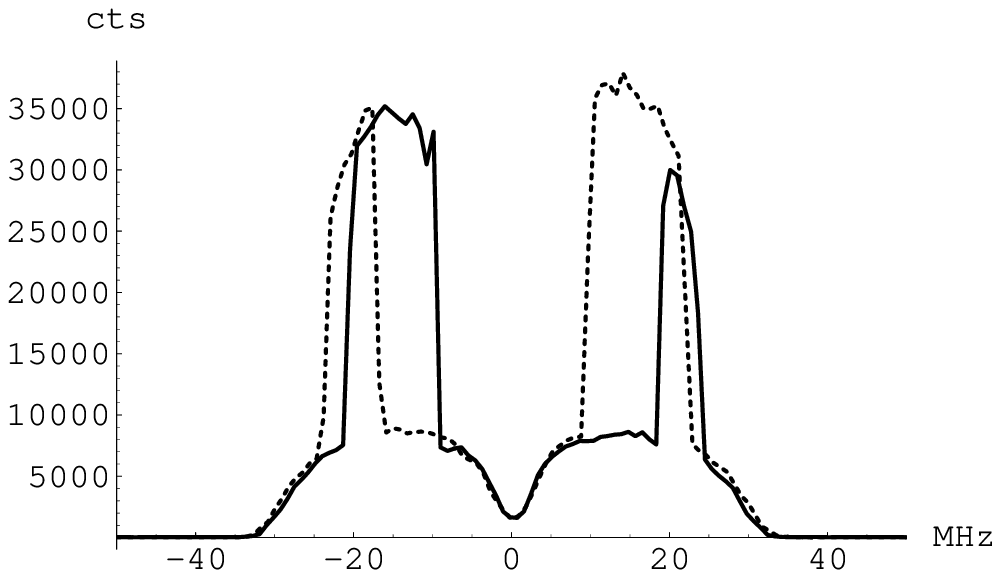}}}
\caption{Fluorescence 3014 cpms ($\Neff = 663$) }
\label{data2index12}
\efig

\bfig
\centerline{\resizebox{4in}{!}{\includegraphics{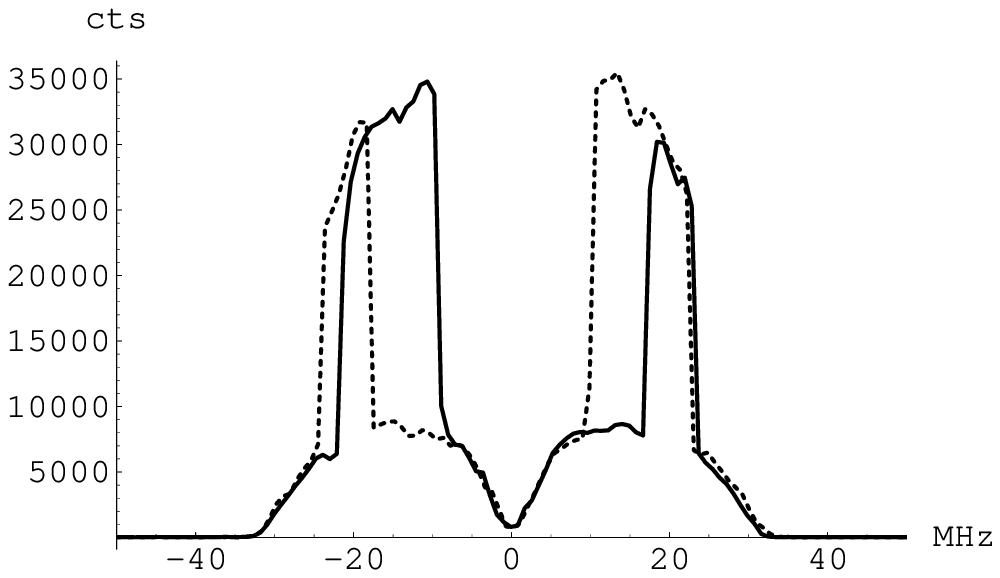}}}
\caption{Fluorescence 3075 cpms ($\Neff = 677$) }
\label{data3index7}
\efig

\bfig
\centerline{\resizebox{4in}{!}{\includegraphics{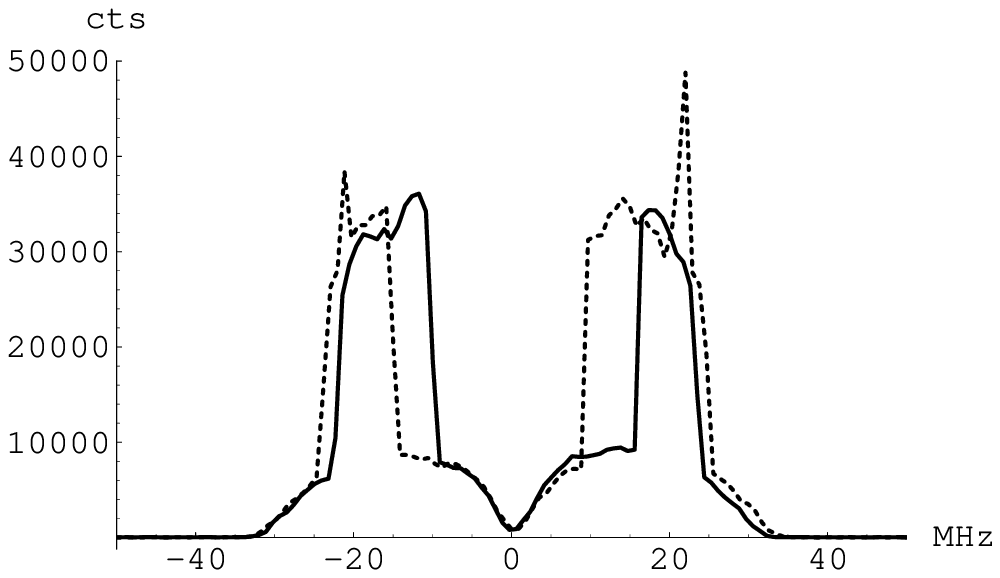}}}
\caption{Fluorescence 3252 cpms ($\Neff = 715$).  Spikes are the first signs of third threshold.}
\label{data3index6}
\efig

\bfig
\centerline{\resizebox{4in}{!}{\includegraphics{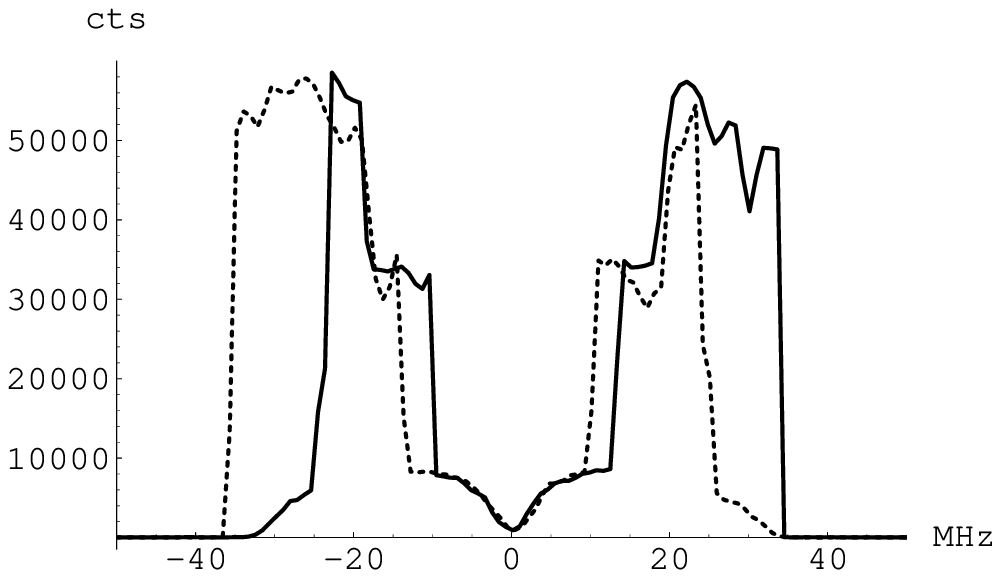}}}
\caption{Fluorescence 3432 cpms ($\Neff = 755$) }
\label{data3index5}
\efig

\clearpage 

\bfig
\centerline{\resizebox{4in}{!}{\includegraphics{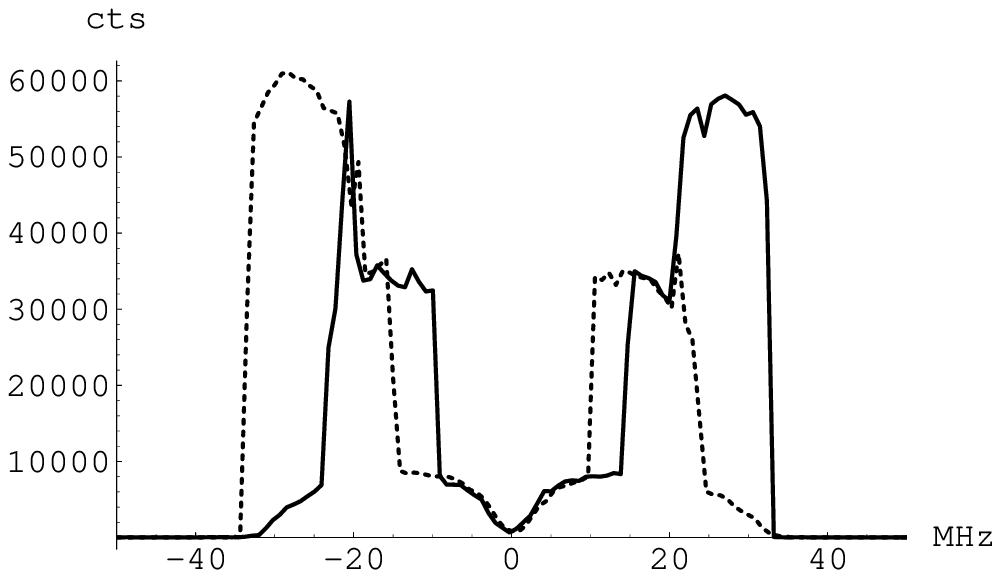}}}
\caption{Fluorescence 3556 cpms ($\Neff = 782$) }
\label{data3index4}
\efig

\bfig
\centerline{\resizebox{4in}{!}{\includegraphics{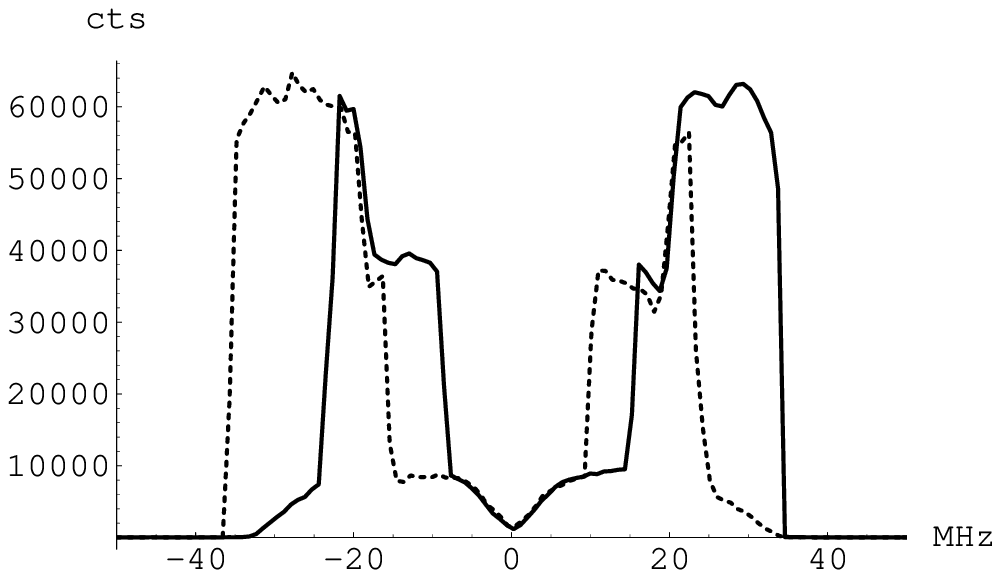}}}
\caption{Fluorescence 3595 cpms ($\Neff = 791$) }
\label{data2index11}
\efig

\bfig
\centerline{\resizebox{4in}{!}{\includegraphics{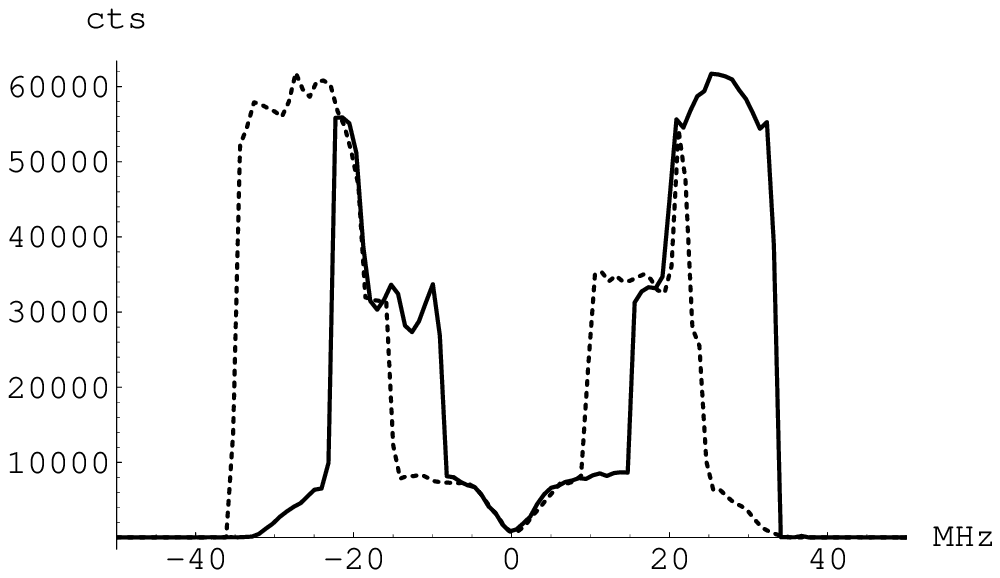}}}
\caption{Fluorescence 3628 cpms ($\Neff = 798$) }
\label{data3index3}
\efig

\bfig
\centerline{\resizebox{4in}{!}{\includegraphics{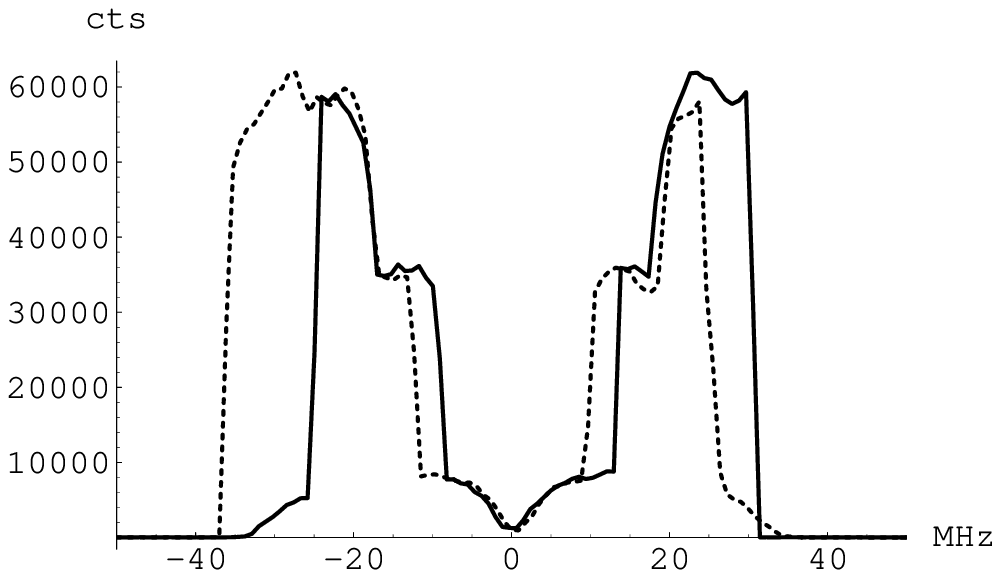}}}
\caption{Fluorescence 3690 cpms ($\Neff = 812$) }
\label{data3index2}
\efig

\bfig
\centerline{\resizebox{4in}{!}{\includegraphics{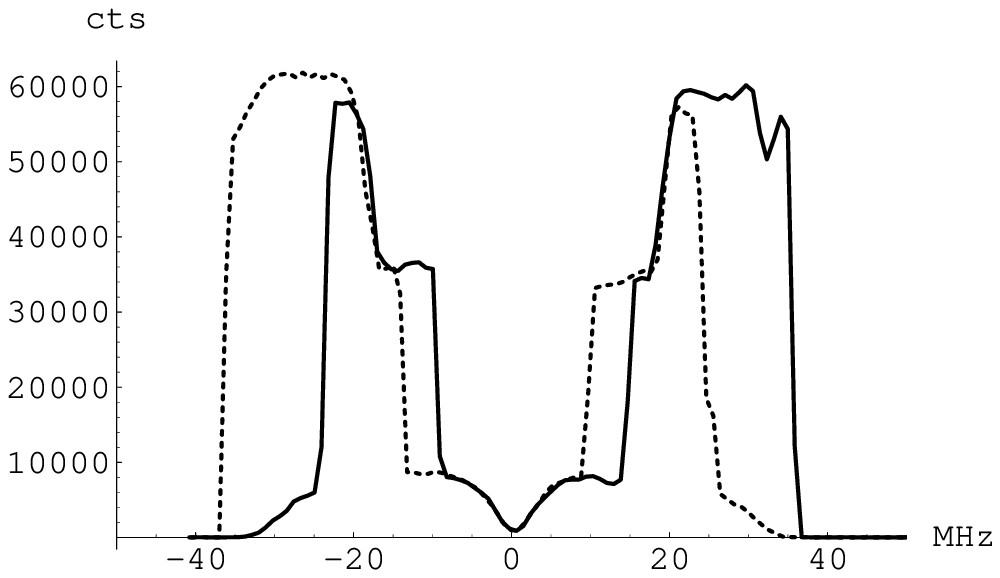}}}
\caption{Fluorescence 3996 cpms ($\Neff = 879$) }
\label{data3index0}
\efig

\bfig
\centerline{\resizebox{4in}{!}{\includegraphics{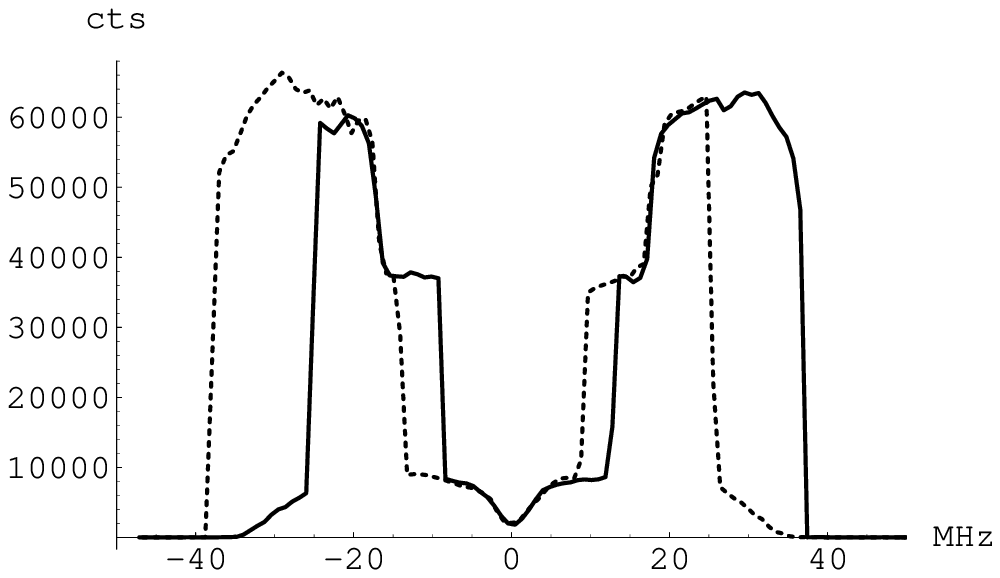}}}
\caption{Fluorescence 4008 cpms ($\Neff = 882$) }
\label{data2index10}
\efig

\bfig
\centerline{\resizebox{4in}{!}{\includegraphics{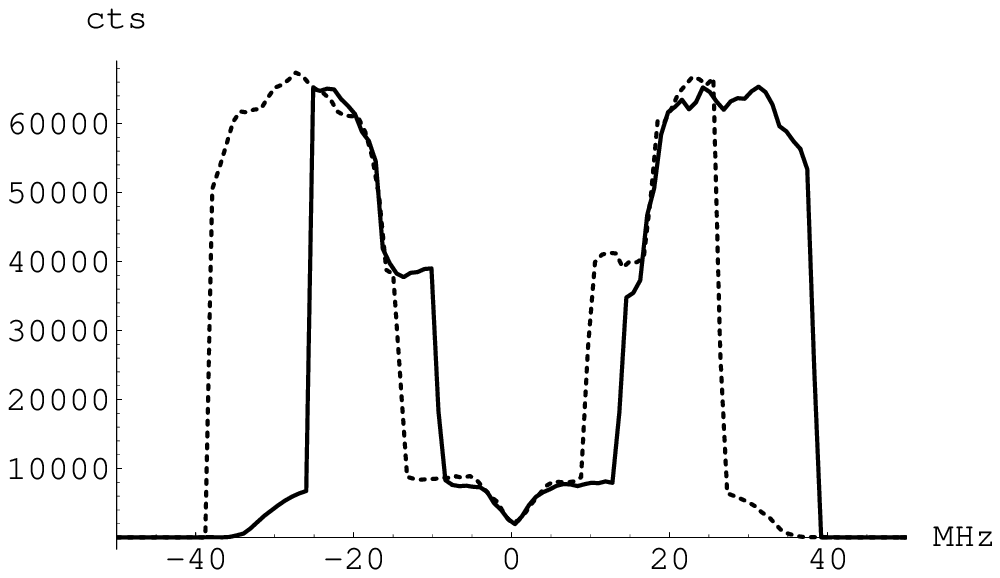}}}
\caption{Fluorescence 4585 cpms ($\Neff = 1009$) }
\label{data2index8}
\efig

\bfig
\centerline{\resizebox{4in}{!}{\includegraphics{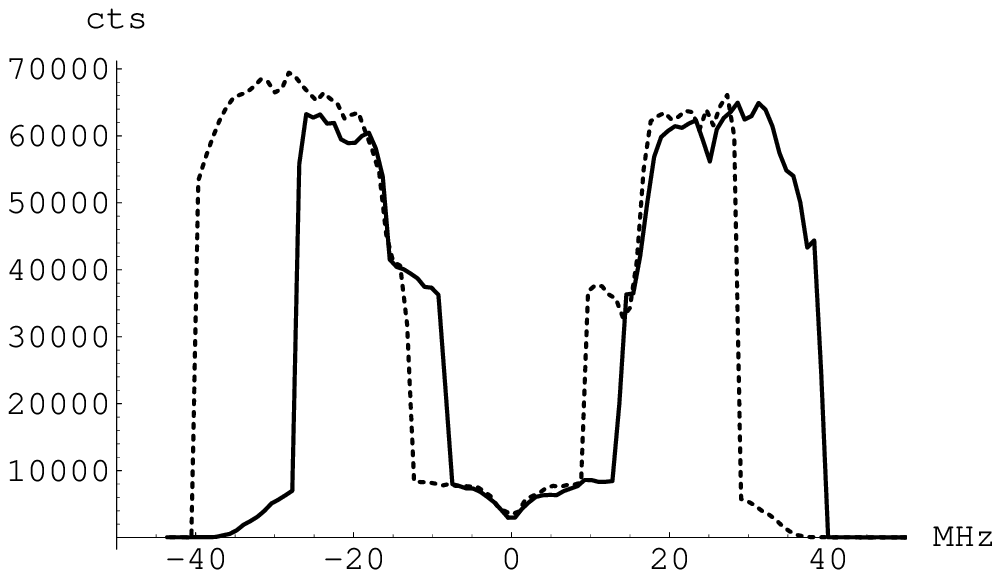}}}
\caption{Fluorescence 5140 cpms ($\Neff = 1131$) }
\label{data2index0}
\efig

\clearpage


As density is increased from zero, the two resonances broaden and
increase in amplitude.  At about 3000 cpms (Fig.~\ref{data3index9}) a
``spike'' appears near the peak of each resonance.  We will show below
that this is a transition to the second branch.  With an further
increase in density to 3250 cpms (Fig.~\ref{data3index6}) a jump to
another, third branch appears.


Above second threshold, the positive and negative direction scans are
in general quite different in shape, although they overlap over some
intervals.  Transitions in the positive-detuning direction occur at
equal or greater (more positive) detunings than the corresponding
transitions in the negative-detuning direction.  This is as expected
from hysteresis behavior.





To measure the frequency splitting between the two traveling-wave
peaks in the microlaser, we found the frequency shift between the two
maxima for each scan below the second threshold; the average value was
27.1 \mhz.  The measured peak of the velocity distribution was $v_0=
815$ m/s; therefore we infer the cavity tilt angle is $\theta 
\approx 13$ mrad.

\subsection{Photon number vs.\ atom number}

The detuning curve data was analyzed by computer to give, for a fixed
detuning, the microlaser emission rate as a function of atomic
density.


Data for center of the peaks $\Delta=\pm13.5 \mhz$ is given in
Fig.~\ref{fig-plot23det135} in terms of an estimated cavity photon
number $n$ and effective atom number $\Neff$.  The curve is the result
of the rate equation model of Chapter
\ref{chap-theory} with modifications for velocity distribution,
nonuniform coupling, and detuning effects.  


Calibrations for the atom density and photon number were adjusted
to give the best fit between theory and experiment as follows.

The photon number calibration from Chapter~\ref{chap-methods} was $n =
0.16$ for every $10^3$ cps on the PMT, with no filters.  With the
OD~1.0 filter used in this experiment, this becomes $n = 1.6$ for
every $10^3$ cps on the PMT.  The experimental best fit is with $n =
0.25 \times $ PMT counts per 0.01 second bin, or $n = 2.5$ for every
$10^3$ cps.  The experimentally derived calibration is 56\% higher
than the original estimate.

The estimate of atom number from Chapter~\ref{chap-methods} was $N =
1.16$ for every count per millisecond (cpms) on the CCD.  For the 300
msec exposures in this experiment, the measured fluorescence count
rate is underestimated by a factor of 2.11 due to the nonuniform
exposure effect (Fig.~\ref{fig-ccd-nonuniform}).  Therefore the
expected value is $N = 0.55$ atoms per cpms.  The effective atom
number is $\Neff = (0.80 - 0.20) N = 0.33$ per cpms.  The experimental
fit gives $\Neff = 0.22$ per cpms, 33\% lower than the estimate.
%
%
This fitting procedure is justified by the large uncertainties in our
estimates for absolute photon and atom numbers.

Note that in the detuning curves, the positive-direction data and
negative-direction data are in most cases nearly mirror images of one
another.  This is expected from the symmetry between the two
traveling-wave components.  We therefore choose to display the photon
number vs.\ atom number data in following way: Circles ($\circ$)
represent data from (i) positive-direction detuning scans with
positive detunings and (ii) negative-direction scans with negative
detunings; crosses ($\times$) give data from (iii) positive-direction
detuning scans with negative detunings and (iv) negative-direction
scans with positive detunings.

\bfig
\centerline{\resizebox{5in}{!}{\includegraphics{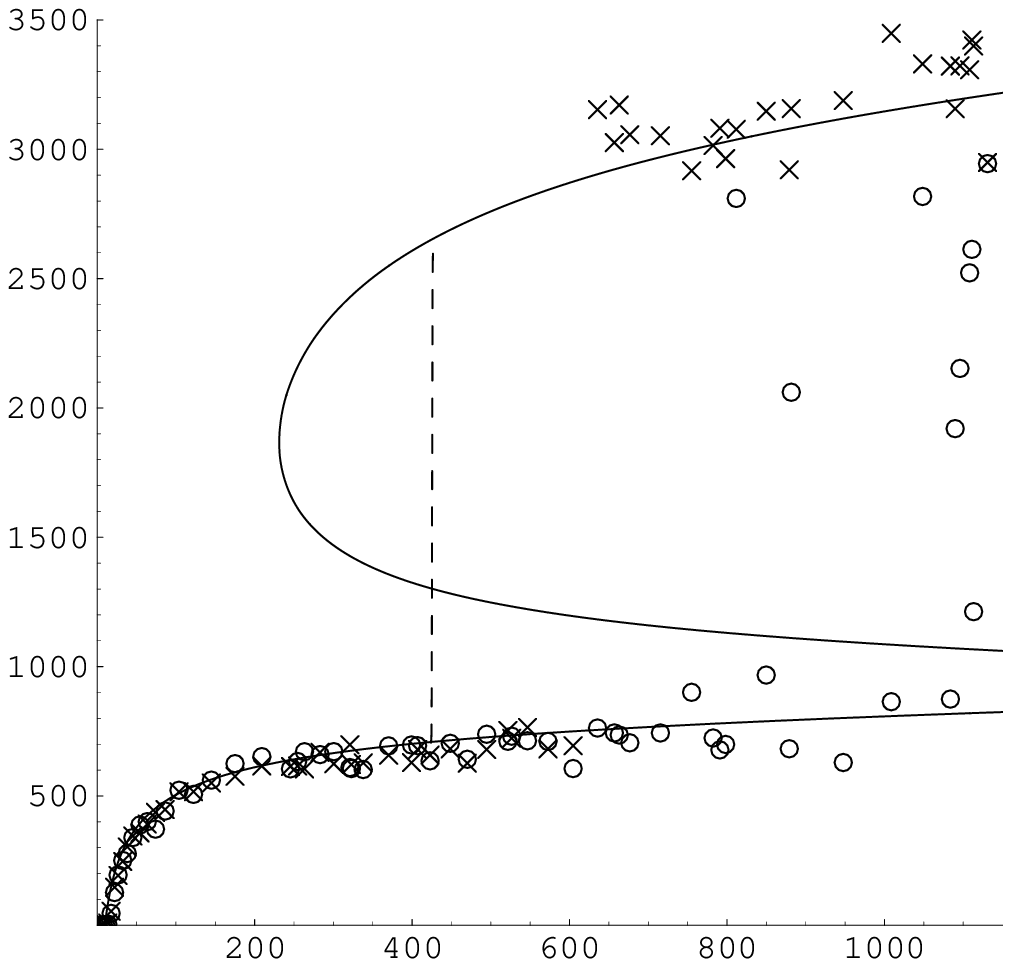}}}
\caption{Photon number $n$ vs.\ effective atom number \Neff\ for center of 
peak $\Delta=\pm 13.5 \mhz$ (cavity on resonant with atoms of most
probable velocity).  Circles ($\circ$): Data from positive-direction
detuning scans with positive detunings, negative-direction scans with
negative detunings.  Crosses ($\times$): Data from positive-direction
detuning scans with negative detunings, negative-direction scans with
positive detunings.  Solid line: Rate equation model from
Chap.~\ref{chap-theory}.  Dashed line: location of second threshold
from corresponding quantum microlaser theory.}
\label{fig-plot23det135}
\efig


The agreement between theory and experiment is, overall, quite good.
A fairly large number of ``outliers'' are present.  These points
represent the system undergoing transitions between the first and
second branches and reflect the fact that the second threshold occurs
very close in detuning to the central point.  To support this point,
Figs.~\ref{fig-plot23det120} and \ref{fig-plot23det150} show the
microlaser output curves for nearby detunings 12 \mhz\ and 15
\mhz.  Dramatic changes occur for very small changes in detuning.  The
line is the same as in Fig.~\ref{fig-plot23det135}.  
%
In the next section, a ``cleaner'' way to do the experiment via cavity
locking finds no intermediate states exist between the first and
second branches.

In these plots, the point of onset of the second branch points reflect
the atom numbers at which the second threshold appears for some
detuning (not necessarily the detuning being considered).

\bfig
\centerline{\resizebox{5in}{!}{\includegraphics{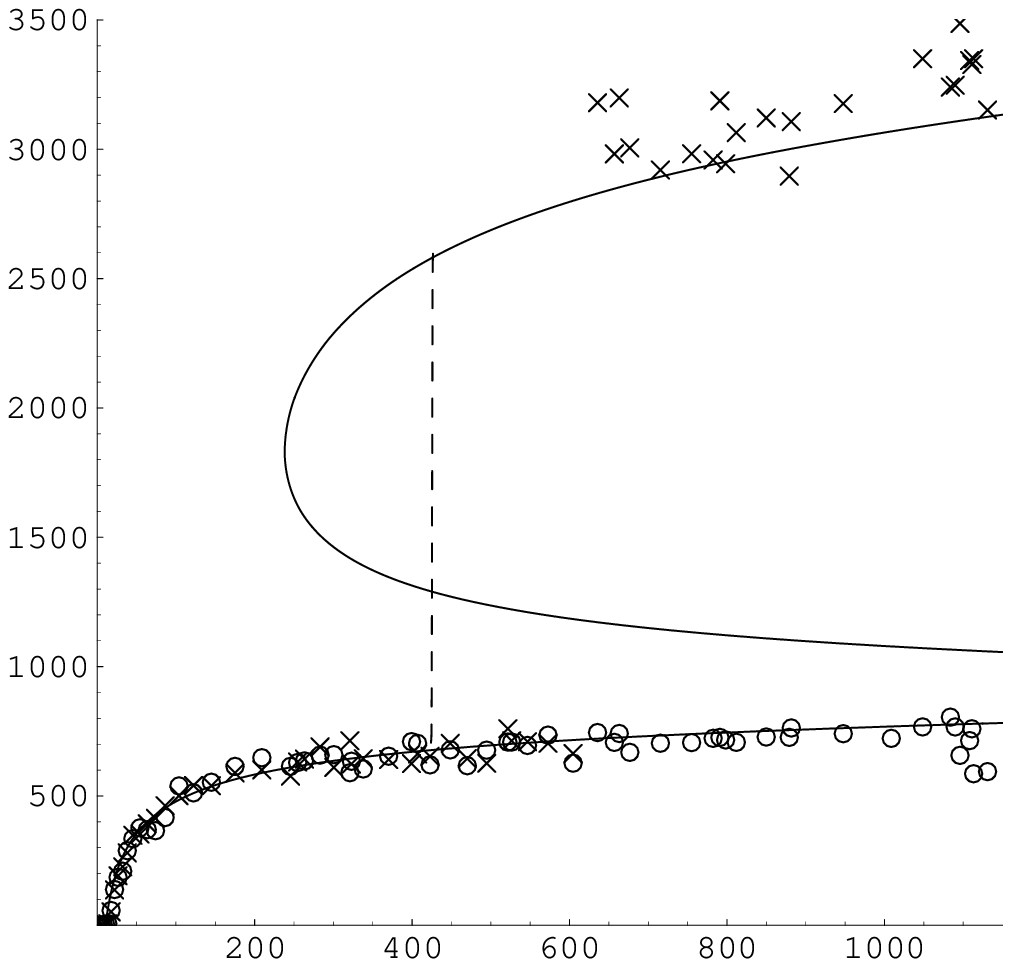}}}
\caption{Photon number $n$ vs.\ effective atom number \Neff\ for 
$\Delta=\pm 12.0 \mhz$.  Circles ($\circ$) and crosses ($\times$)
are as in Fig.~\ref{fig-plot23det135}.  Line: rate equation model}
\label{fig-plot23det120}
\efig


\bfig
\centerline{\resizebox{5in}{!}{\includegraphics{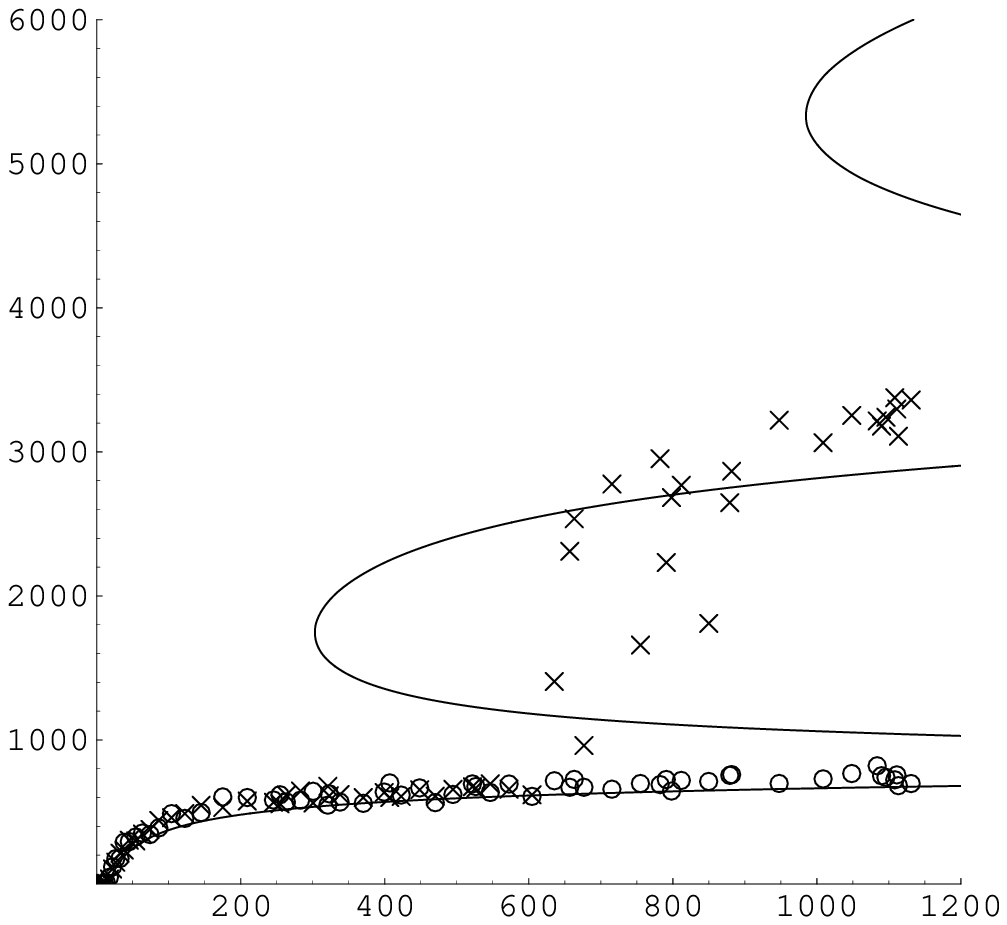}}}
\caption{Photon number $n$ vs.\ effective atom number \Neff\ for 
$\Delta=\pm 10.0 \mhz$.  Circles ($\circ$) and crosses ($\times$)
are as in Fig.~\ref{fig-plot23det135}.}
\label{fig-plot23det100}
\efig


\bfig
\centerline{\resizebox{5in}{!}{\includegraphics{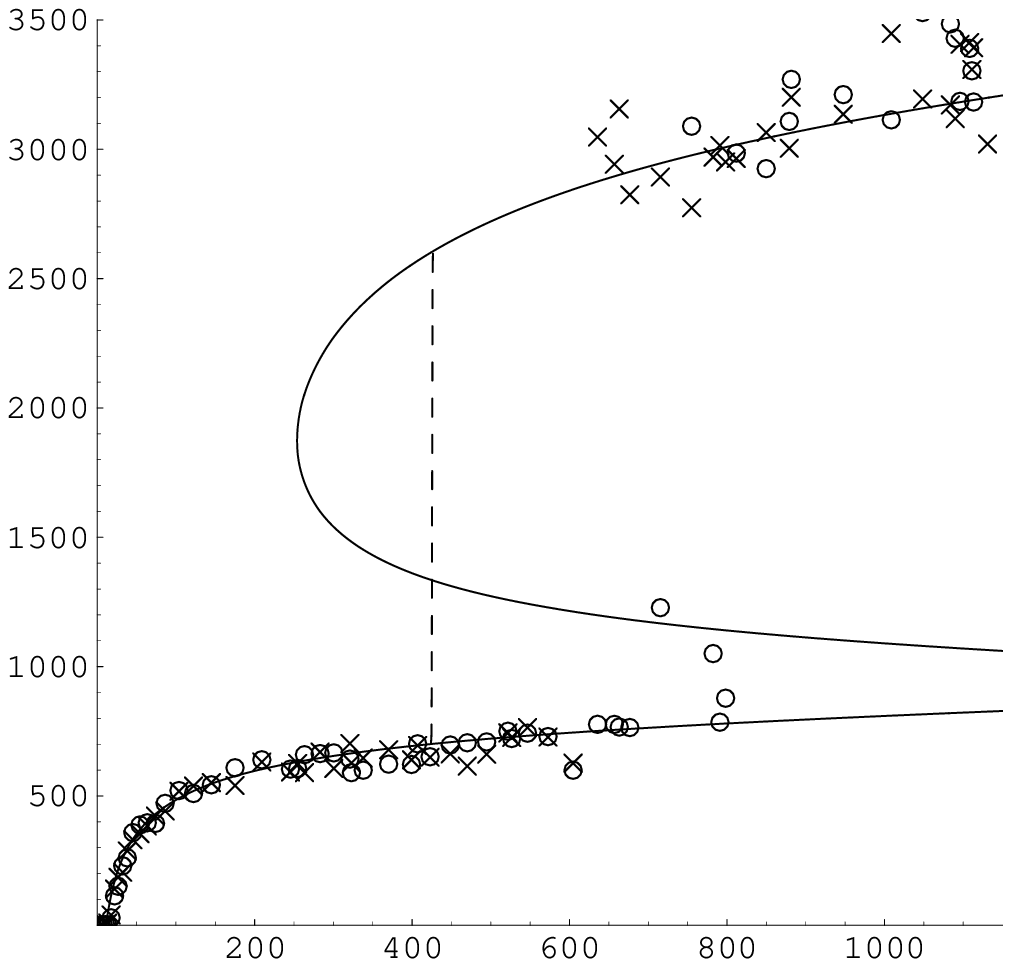}}}
\caption{Photon number $n$ vs.\ effective atom number \Neff\ for 
$\Delta=\pm 15.0 \mhz$.  Circles ($\circ$) and crosses ($\times$)
are as in Fig.~\ref{fig-plot23det150}.}
\label{fig-plot23det150}
\efig






\bfig
\centerline{\resizebox{5in}{!}{\includegraphics{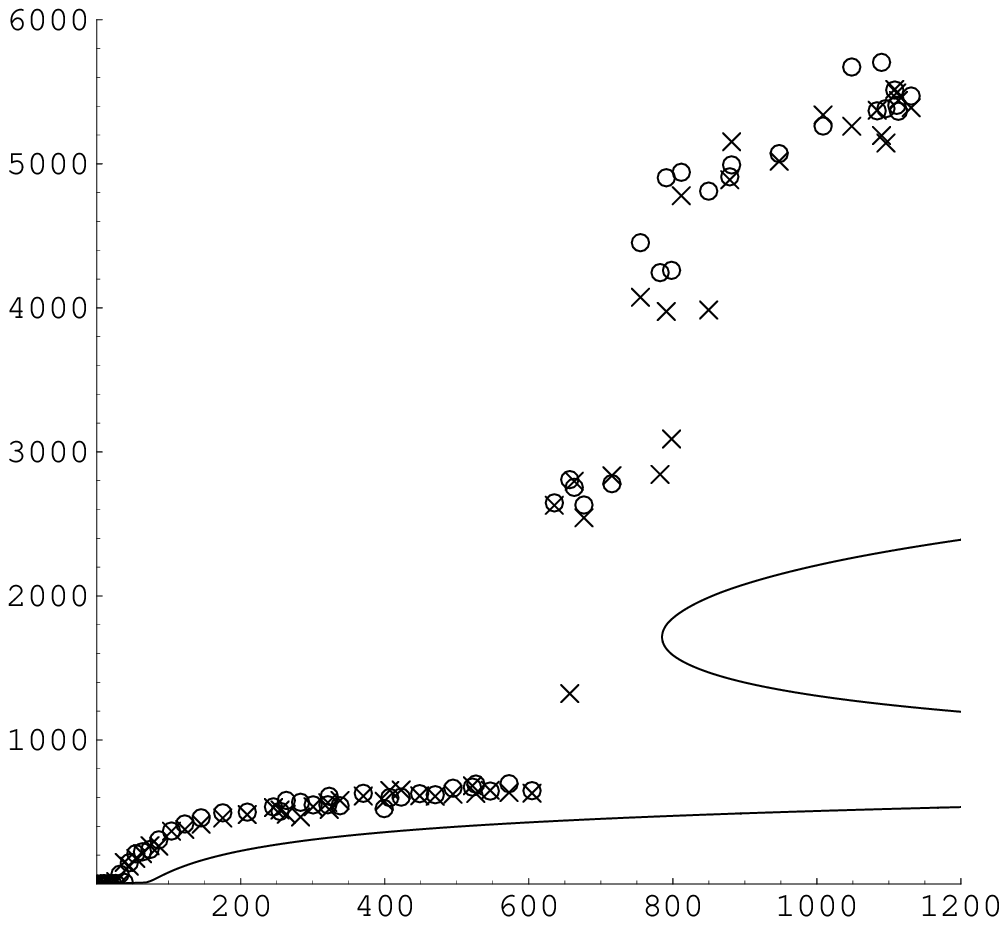}}}
\caption{Photon number $n$ vs.\ effective atom number \Neff\ for 
$\Delta=\pm 20.0 \mhz$.  Circles ($\circ$) and crosses ($\times$) are
as in Fig.~\ref{fig-plot23det135}.  Line: rate equation model.  Third
threshold is apparent.}
\label{fig-plot23det200}
\efig

\bfig
\centerline{\resizebox{5in}{!}{\includegraphics{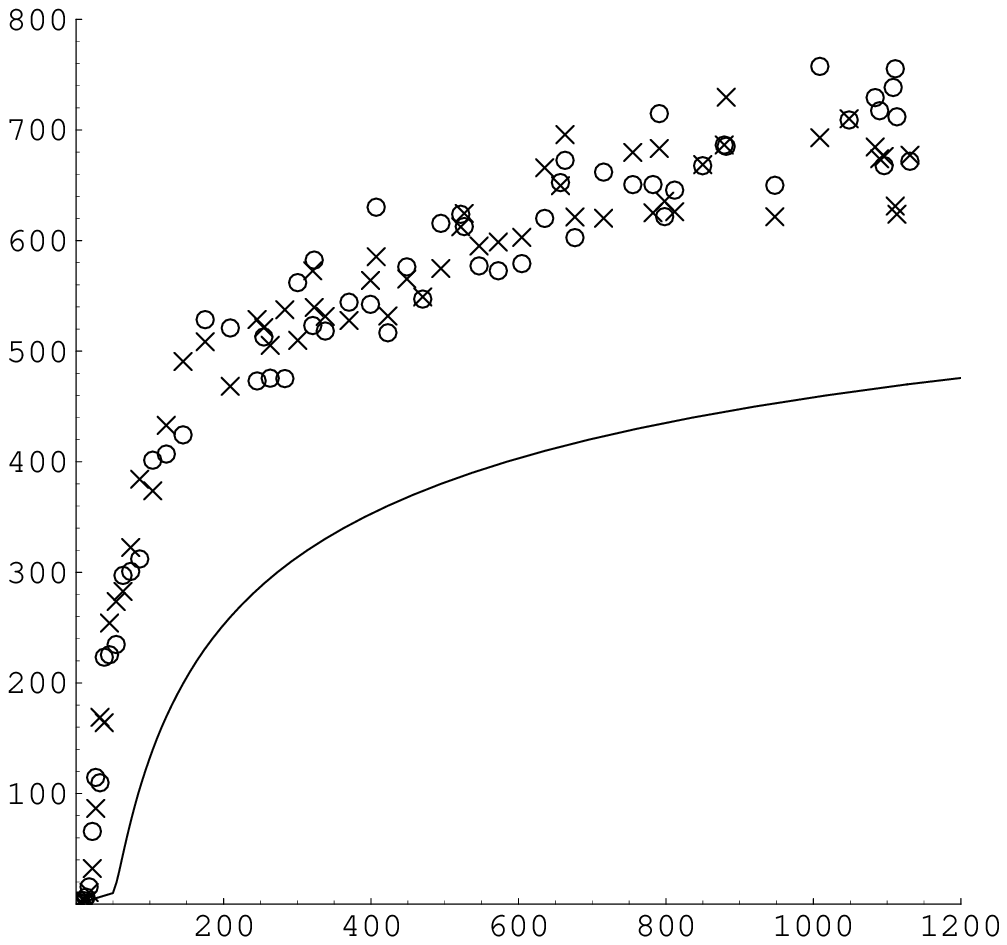}}}
\caption{Photon number $n$ vs.\ effective atom number \Neff\ for 
$\Delta=\pm 7.5 \mhz$.  Circles ($\circ$) and crosses ($\times$)
are as in Fig.~\ref{fig-plot23det135}.  Line: rate equation model}
\label{fig-plot23det075}
\efig

\subsection{Disagreement with theory}

The data does not agree with theoretical curves in several ways.
First, the theory underestimates the photon number in the case of
cavity detuned significantly from resonance with the atoms.  For
example, see Fig.~\ref{fig-plot23det075}, Fig.~\ref{fig-plot23det100},
and Fig.~\ref{fig-plot23det200}.  For a given density, the theory
predicts a narrower first-branch resonance than is observed.

Second, the peaks for the second and third branches occur at greater
absolute detunings than for the initial branch.  The third branch, in
particular, exhibits a maximum at approximately $\pm 28\ \mhz$, more than
twice the detuning of the first branch peak; the third threshold does
not appear at all for absolute detunings less than about $17.5\
\mhz$.  From the model of Chapter \ref{chap-theory} we would expect
all branches to be approximately centered on zero detuning relative to
atoms of the most probable velocity.


In Sec.~\ref{sec-detuning-disagree} we discuss possible reasons for
the disagreement.

\subsection{Peak heights of detuning curves}

Suppose we consider the points of maximum photon number in the
detuning curves as representing the condition of the cavity resonant
with atoms of the most probable velocity.  

In Figure \ref{fig-peakheights} we plot the peak in each detuning
curve against the effective atom number.  Transitions to the second
and third branches occur very close to eachother.  The agreement
between theory and experiment is quite good except for the third
branch, for which data points are considerably smaller than the theory
values.  Note also that the transitions to the third branch occur at a
lower atom number the than predicted by the fully quantized theory.

\bfig
\centerline{\resizebox{5in}{!}{\includegraphics{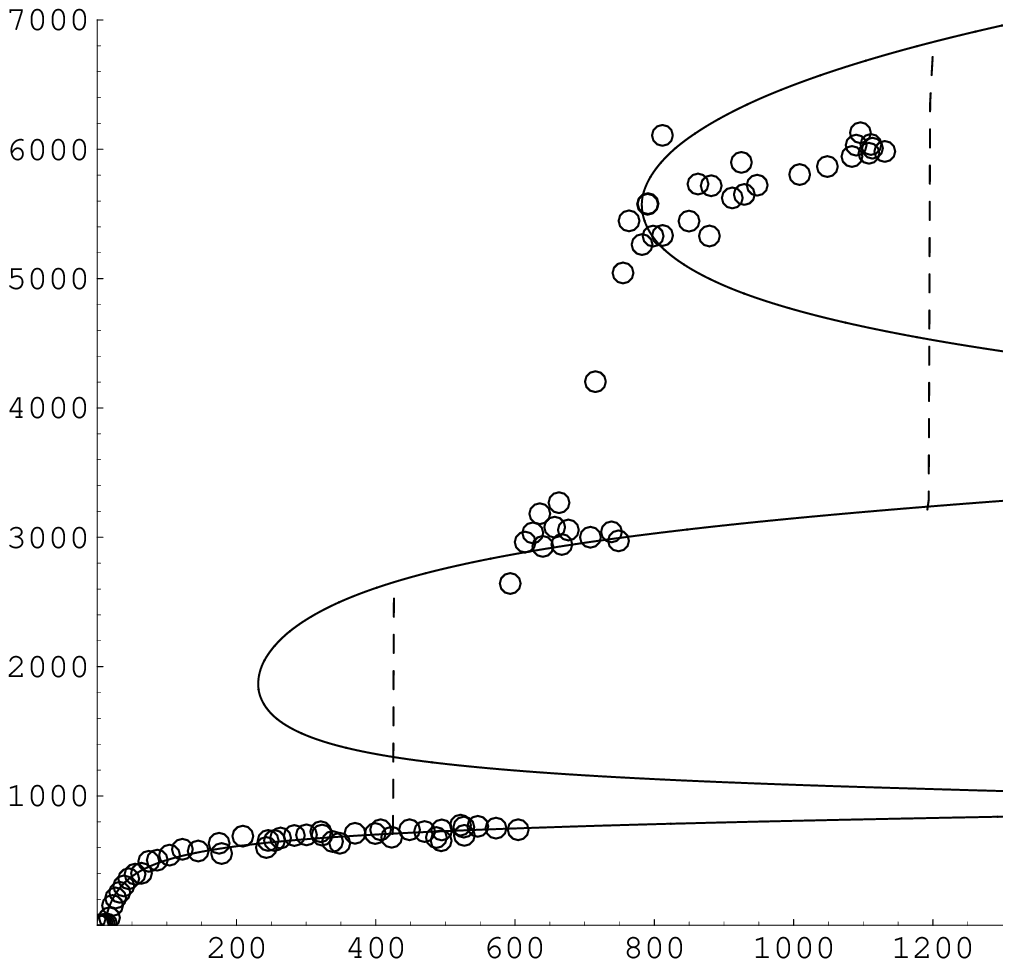}}}
\centerline{\Nex}
\caption{Peak photon numbers of cavity scanning data vs.\ effective atom 
number.  Solid line: rate equation solution for atom-cavity resonance
($\Delta=\pm13.5 \mhz$).  Dotted line: location of thresholds by
quantum theory.}
\label{fig-peakheights}
\efig

\section{Cavity locking}

In order to study the system at a single detuning corresponding to
resonance with the most probably velocity atoms, and to consider the
effect of different initial conditions of the cavity field more
directly, we conducted experiments with the cavity locked on
resonance.  


Cavity locking experiments used the following sequence:
\begin{singlespace}
\begin{enumerate}
\item Lock cavity with a probe beam which has frequency corresponding to most
probable velocity atoms, as described in Chapter~\ref{chap-methods}
\item Unlock cavity (remove probe beam and disengage cavity lock circuit)
\item Move atom density modulator by a certain step size
\item Measure atomic density by \fft\ probe and CCD (no \sno\ pump or probe)
\item Lock cavity
\item Unlock cavity
\item Turn on the \sno\ pump beam, 
\item Trigger photon counter to measure microlaser emission for 100 msec (``unseeded'' data); download data to computer
\item With the \sno\ pump remaining on, lock cavity again
\item Unlock cavity
\item Trigger photon counter to measure microlaser emission for 100 msec (``seeded'' data); download data to computer
\item Turn off \sno\ pump and repeat
\end{enumerate}
\end{singlespace}
The repeated locking and unlocking of the cavity was necessary to
refresh the locking in between other operations.  A locking time
of 300 msec was used.

The frequency shift of the probe beam relative to the atomic line
center was found to be +12.94 \mhz.

Note that for each step in density, two microlaser data samples are
collected: (1) ``unseeded'', in which the cavity is empty of photons
when the pump beam is turned on, and (2) ``seeded'' in which the pump
beam enters while the cavity has a very large ($n_0 \sim 10^{12}$)
number of photons due to the cavity locking beam.  When the cavity
lock is turned off, the microlaser field develops from this
large-$n$ initial condition.


\subsection{Cavity locking results}

\bfig
\centerline{\resizebox{5in}{!}{\includegraphics{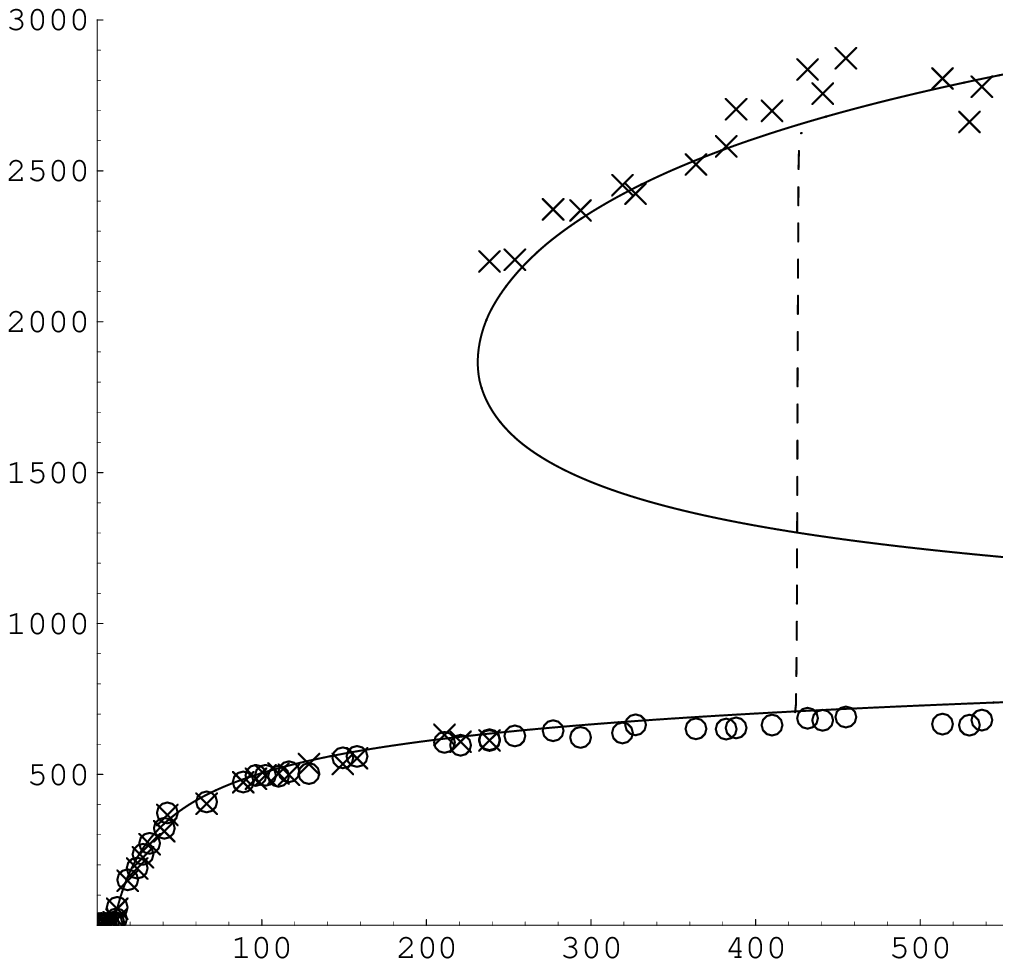}}}
\caption{Photon number $n$ vs.\ effective atom number \Neff\ for cavity
locking experiment.  Circles ($\circ$): ``unseeded'' data; Crosses
($\times$): ``seeded'' data.  Solid line: Rate equation model from
Chap.~\ref{chap-theory}.  Dashed line: location of second threshold
from microlaser quantum theory.}
\label{fig-cavlock}
\efig

\bfig
\centerline{\resizebox{5in}{!}{\includegraphics{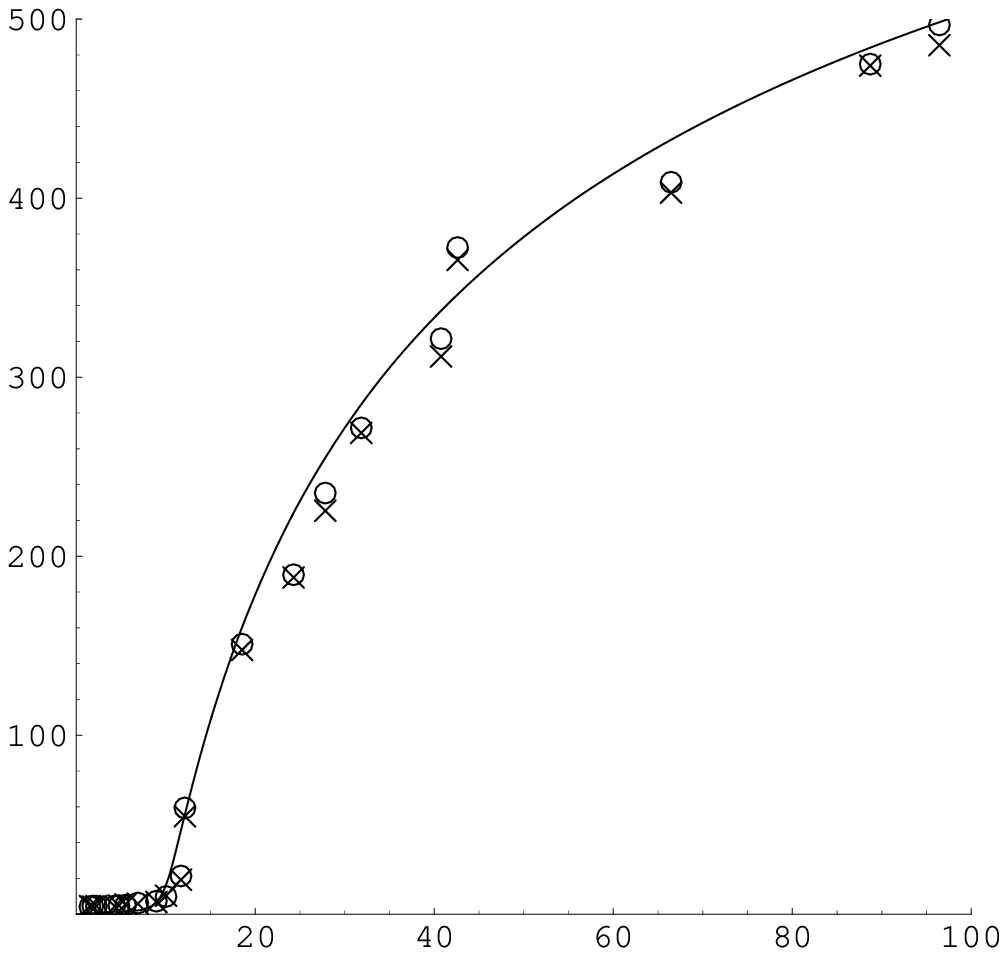}}}
\caption{Detail of Fig.~\ref{fig-cavlock} in the initial threshold region}
\label{fig-cavlock-zoom}
\efig

The results of the cavity locking experiment are plotted in
Figs.~\ref{fig-cavlock} and \ref{fig-cavlock-zoom}.  The unseeded
results agree with data from the cavity scanning experiment.  The
seeded results agree with the first branch for densities up to the
onset of bistability; above this point the results agree with the
second stable branch.

Several outlying points in which the photon counts recorded were
extremely high, $n \sim 10^5$, were removed.  These points were
clearly artifacts because they occasionally appeared even when no
atoms were present.  They were most likely due to a problem in the
sequencing electronics which on occasion allowed some of the locking
probe beam to leak into the cavity during the microlaser experiment.


\section{Discussion}

The interpretation of the cavity locking results seems
straightforward.  The microlaser is becoming trapped in metastable states.
The transition rate from a metastable to a stable solution is longer
than the time of the experiment, which is about 100 msec and limited
to that order of magnitude or smaller by the cavity re-locking
requirement.  Note that 100 msec $\approx 10^5\ t_{\rm cav}$ is very
long by the time scale of the cavity decay time.  
The estimate of a typical tunneling time between metastable and stable
states is given by the Fokker-Planck analysis is
\cite{Filipowicz-PRA86}:
\be
t_g \sim t_{\rm cav} \exp(\alpha \Nex)
\ee
\noindent where $\alpha$ is on the order of unity.  In our experiment we
have $\Nex \approx 10 \Neff \sim 5000$ in the region where the
transition would be expected to occur spontaneously; this gives a
transition time scale that is extraordinarily long.  A calculation of
Eq.~\ref{eq-transition-rate} gives a transition rate $W \sim 10^{-91}\
\gc$ for $\Neff=600$ and zero detuning.  By this calculation,
spontaneous jumps will occur on time scales of 1 second or faster only
for points extremely close (less than $\sim 10^{-3}$ in relative
distance) to points of marginal stability.  Then as far as our
experiments are concerned, the system would effectively behave as if
no spontaneous transitions at all occur.

Some evidence suggests the situation may not be as simple as this.  In
the cavity scanning experiments, transitions from the first branch to
and from the second often occurs with a abrupt bend in the curve; this
occurs in both scan directions.  This seems to suggest this transition
occurs spontaneously rather than as a result of a branch becoming
marginally stable and then disappearing.  Note that the curve of
solutions, as a function of detuning, must have a divergent slope at a
marginally stable point.  (see Fig.~\ref{fig-transitions}).  It may be
that other sources of noise (classical or quantum) drive fluctuations
in the system faster than would be predicted by the Fokker-Planck
analysis.  We note that studies of bistability in the micromaser
\cite{benson-prl94} found transition rates much higher than predicted
by a similar Fokker-Planck analysis.



A similar analysis of hysteresis applies to the cavity scanning
experiments.  When plotted as photon number versus detuning for a
fixed $\Neff$, rate equation solutions appear as closed loops or
curves as discussed in Chapter~\ref{chap-theory}.




Transitions from the first branch to and from the second often occurs
with a abrupt bend in the curve; this occurs in both scan directions.
This seems to suggest this transition occurs spontaneously rather than
as a result of a branch becoming marginally stable and then
disappearing.  Note that the curve of solutions, as a function of
detuning or atom number, has a divergent slope at a marginally stable
point.  (Fig.~\ref{fig-transitions}).

\bfig
\centerline{\resizebox{4in}{!}{\includegraphics{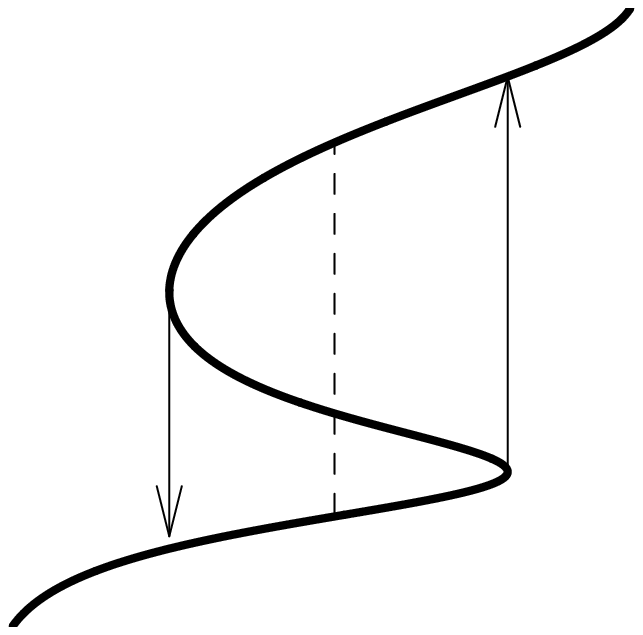}}}
\caption{Arrows: transitions due to disappearance of a stable point.  
Dotted line: spontaneous transition from a metastable to stable state.
$x$-axis may represent atom number or detuning}
\label{fig-transitions}
\efig



\section{Summary}

We have observed second and third thresholds in the microlaser.
Theory and experiment are in remarkably good agreement for the
resonant case.  Highly history-dependent behavior is observed which
suggests that the system becomes trapped in long-lived metastable
states.

Possible reasons for disagreement in the nonresonant case will be
discussed further in the next chapter.

\chapter{Discussion and Conclusions}
\label{chap-discussion}


\section{Multiple thresholds and the phase transition analogy}


The discontinuous nature of the higher thresholds and the existence of
hysteresis support the identification of these jumps as first-order
transitions.  Due to long metastable lifetimes, however, the observed
jumps in the field do not occur at points predicted from the quantum
microlaser theory.  

We may ask just how fruitful the phase transition analogy will be in
this context.  What insights from the study of many-particle systems
can be applied to the microlaser, or vice versa?  Ideas from
nonequilbrium statistical dynamics might be applied to the microlaser
in describing the approach to steady-state and the transitions between
stable states.  One theoretical paper \cite{balconi96} suggests that
under some variations a many-atom ``mesolaser'' with injected signal
will exhibit not only multistability but temporal instabilities,
period doubling, self-pulsing, quasiperiodicity, and chaos.  

We note that the microlaser is itself a many-particle system, but the
particles are not strongly interacting; they may interact only through
the common cavity field. 




\section{Many-atom effects}

The microlaser output in the resonant case agrees remarkably well with
the rate equation solutions of Chapter \ref{chap-theory}.  Both the
quantum microlaser theory and the rate equation were derived by
considering the influence of a single atoms on the field and assuming
all atoms act independently.  These experiments confirm the validity
of this independent approximation.  At first glance this may seem
surprising, since the eigenvalues for the atom-cavity system with even
two atoms are different from the single atom case.  On the other hand,
note that in general the photon number distribution in the cavity is
strongly peaked (in fact, often more strongly peaked than a classical
beam).  The effect of the other atoms in the cavity can be neglected
so long as their perturbation $\delta n$ is small compared to $n$.
Moreover, the effect of the other atoms will tend to cancel eachother
if the phases of their own Rabi oscillations are randomly or broadly
distributed.

\section{Microlasers, lasers, and randomness}

We now comment on the relation between the microlaser and a
conventional laser.  The microlaser's unique properties stem from its
fully coherent atom-cavity interaction, which leads to an emission
probability per atom (in the case of no broadening) of
\be
P_e = \sin^2(\sqrt{n+1} g \tint)
\ee
To model a conventional laser we simulate atomic decay by averaging over
a transit time distribution weighted by $(1/\tau_1)\exp(-t/\tau_a)$:
\be
P_e = \frac{1}{\tau_a}\int_0^\infty d\tint \exp(-t/\tau_a) 
\sin^2(\sqrt{n+1} g \tint) \\
= \frac{1}{2}\left(
\frac{(n+1)g^2\tau_a^2}{1+(n+1)g^2\tau_a^2}\right)
\ee
Solving the resulting gain-loss rate equation (Fig.~\ref{fig-conv-re})
gives a single solution for the photon number which increases linearly
over threshold as a function of injection rate.
Substituting this equation into the
quantum microlaser theory gives steady-state photon statistics which
are Poissonian far above threshold.

\bfig
\centerline{\resizebox{4in}{!}{\includegraphics{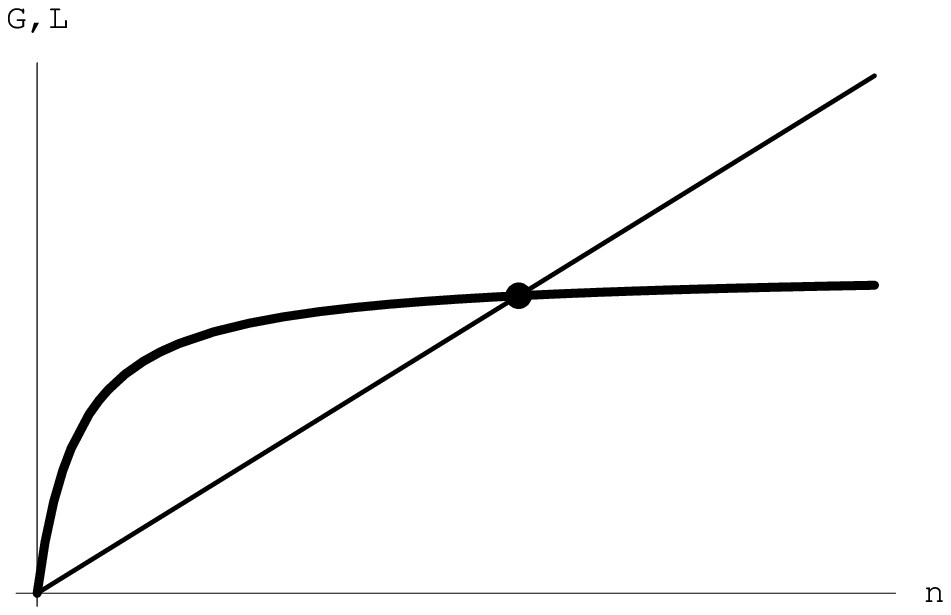}}}
\caption{Gain-loss graphical solution for conventional laser as modeled
in the text.  Gain function follows saturation curve.}
\label{fig-conv-re}
\efig

\bfig
\centerline{\resizebox{4in}{!}{\includegraphics{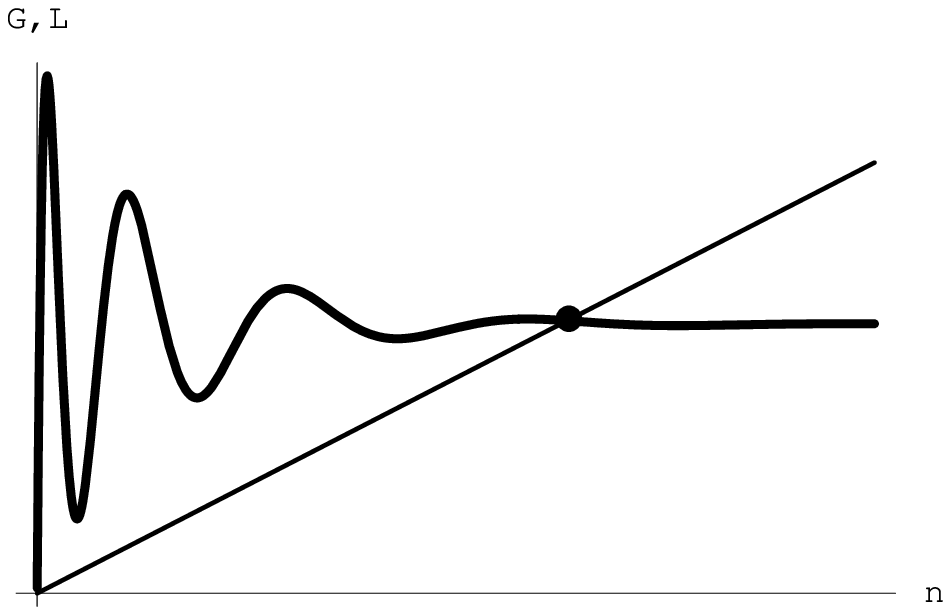}}}
\caption{Gain-loss graphical solution for microlaser with some broadening.
For sufficiently large $n$, gain line is flat and solution is similar to
that of conventional laser.}
\label{fig-rate-eqn-ul}
\efig


From this perspective, the microlaser can be seen as a laser with an
insufficient amount of noise or randomness to be a conventional laser.
Conversely, a laser must possess a certain amount of noise to behave
in the conventional way with monostability and a single threshold.

We note that any realistic microlaser will contain some averaging due
to velocity distribution, coupling variation, etc.  For a sufficiently
large photon number, roughly $n > (\pi/\delta(g\tint))^2$ where
$\delta(g\tint)$ is uncertainty in $g\tint$ the averaging will cause
wash out oscillations in the gain function (Fig.\ref{fig-rate-eqn-ul})
to a value of one-half.  Therefore the conventional laser is
invariably recovered in the limit of large photon number.

\section{Value of quantum theory}

Our quantum theory predicts the steady-state photon number
distribution in the microlaser cavity.  However, long metastable
lifetimes prevent the true steady-state from being reached on an
experimental time scale.  Since the quantum theory otherwise agrees
with the rate equation solutions, we might ask what value the quantum
theory has in our case.  One answer is that it is a more rigorous
treatment than obtained by the rate equation, and therefore serves as
a validation for it on theoretical grounds.  Another consideration is
that the quantum theory also describes the statistics of the photon
field.  Experiments have been done to investigate the microlaser's
photon statistics in the threshold region \cite{aljalal-thesis} and an
efforts to measure nonclassical statistics in the current microlaser
are now under way.

\section{Detuning curves}
\label{sec-detuning-disagree}
How do we explain the discrepancy between theory and experiment for
the cavity nonresonant with the most probable velocity atoms?

If we confine our attention to the resonance with a single
traveling-wave component, the asymmetry in the detuning curves between
positive and negative detunings (relative to the most probable
velocities) is particularly puzzling.  For a truly monovelocity atom
beam one would expect perfect symmetry between positive and negative
detunings, by virtue of the symmetry in the Bloch equations when final
state probabilities are measured.  In our case we have a quite narrow
beam: $\Delta v_{\rm FWMM}/v_0\approx 17\% $, corresponding to a FWHM
Doppler broadening of only $17\% \times 13.5 \mhz \approx 2.3 \mhz$.
And yet the center of the third branch (on the positive detuning side)
is found at a detuning of about $+25 \mhz$ relative to atomic line
center, or $+12
\mhz$ relative to the resonance with the peak of the velocity
distribution.  It seems unlikely that the velocity distribution would
lead to such a large effect.

We now suggest a two ideas which are being explored in connection with
the microlaser's detuning dependence.

\subsection{Influence of nonresonant traveling-wave field: atom in a 
bichromatic field}

The microlaser's traveling-wave (TW) interaction scheme requires a
Doppler splitting larger than the widths of the microlaser resonances
(i.e.\ no significant overlap between the two peaks).  It was
initially believed that if the two resonances are separated in this
way, the interaction with the nonresonant TW component can be
neglected.  This may not be true: the nonresonant field may exert a
significant influence on the resonant interaction even it would be
directly responsible for an insignificant number of photons in the
absence of the resonant interaction.

To treat this problem exactly, we note that in the atom frame of
reference, the cavity field consists of two components of frequency
$\ocav \pm k v \theta$.  The equations of motion for an atom in a
bichromatic field can be derived, analogous to Eqns.~\ref{eq-bloch1}
and \ref{eq-bloch2}:

\be
\dot{c}_a = \left[ \frac{i \Omega_1}{2} e^{-i \phi_1} 
e^{i(\omega_0 - \omega_1)t} + \frac{i \Omega_2}{2} e^{-i \phi_2} 
e^{i(\omega_0 - \omega_2)t} \right] c_b 
\ee

\be
\dot{c}_b = \left[ \frac{i \Omega_1}{2} e^{i \phi_1} 
e^{-i(\omega_0 - \omega_1)t} + \frac{i \Omega_2}{2} e^{i \phi_2} 
e^{-i(\omega_0 - \omega_2)t} \right] c_a 
\ee

\noindent where $\omega_1 = \omega_{\rm cav} + kv \theta$, 
$\omega_2 = \omega_{\rm cav} + kv \theta$ are the field frequencies,
$\omega_0$ is the atom resonance frequency, and $\phi_1$, $\phi_2$ are
the field phases.  The solutions strongly depends on the choice of
initial phase difference between the two fields.  In a simulations, an
averaging over phase difference must be performed.


\subsection{Dispersive effects}

A related question is that of dispersive effects of the resonant and
nonresonant atoms in the cavity.  The susceptibility of a collection
of $N$ atoms in a light field of frequency $\omega$ (see
\cite{atom-photon-interactions} ) is given by $\chi = \chi' + i \chi''$,
with

\be
\chi' = - \frac{N' | \mu|^2}{\epsilon_0 \hbar} \frac{\delta_L}{\delta_L^2 + \Gamma_a^2/4 + \Omega_R^2/2}
\ee

\be
\chi'' =  \frac{N' | \mu|^2}{\epsilon_0 \hbar} \frac{\Gamma_a/2}{\delta_L^2 + \Gamma_a^2/4 + \Omega_R^2/2}
\ee

\noindent where $N'$ is the atom density, 
$\Omega_R$ is the Rabi frequency of atoms in the field, $\delta_L$ is
the laser-atom detunig, and $\Gamma_a$ is the atom linewidth.

This equation must be modified for the microlaser system, in which
transit time is much shorter than atom lifetimes.  A first
approximation is to replace $\Gamma_a$ with $\tint^{-1}$.  

The refractive index of the atoms is
\be
n = \sqrt{1+\chi'} \approx 1 + \chi' / 2
\ee

The shift in cavity frequency due to atom dispersion will be on the
order of $\delta_{\rm cav} \sim \chi'/2 \omega_{\rm cav}$ which may be
as large as $\sim 10\ \mhz$ for 1000 atoms in the cavity.  The shift
is a function of the Rabi frequency $2\sqrt{n}g$ and atom-cavity
detuning.
In general there will be two such frequency shifts, one for each
traveling-wave component, which add in effect.

Frequency shifts from the dispersive effects may lead to another form
of optical bistability, due to a nonlinear medium.  The hysteresis
observed in the microlaser may in fact be a combination of the
microlaser's oscillatory gain function and ``classical'' optical
bistability effects.  Unraveling the two forms of bistabilities will
be an interesting challenge.

\section{Future directions}

\subsection{Theory of microlaser}

Clearly, the next step is to develop a more complete description of
the microlaser, particularly for the nonresonant case.  Additional
data for high densities, and results from cavity locking experiments
as a function of detuning as well as atom number, should provide
important clues to the source of discrepancy between experiments and
current theory.  The possibility of many-atom effects playing a
significant role cannot be ruled out, despite what the quantum
trajectory results of Chapter \ref{chap-simulations} might suggest.


In parallel, a number of other microlaser experiments are being
considered.

\subsubsection{Second-order correlation measurements}

We have performed several experiments to measure the second-order
correlation function $g^{(2)}(t)$ of the microlaser emission
\cite{aljalal-thesis}.  Experiments have shown an reduction of amplitude of
fluctuations, and a local maximum of the time scale of fluctuations,
during the initial (second-order) threshold region.  Our primary goal
however has been to observe antibunching in the output field.
Although the photon statistics inside the cavity are highly squeezed
in amplitude (with a Mandel $Q$ parameter as low as -0.8), the
resulting nonclassical effect in the output light is limited by the
relatively large number of photons in the cavity.  The second-order
correlation function at zero time delay is related to the average
cavity photon number and $Q$ parameter by $g^{(2)}(0) = 1 + Q/\langle
n \rangle$; for $\langle n
\rangle \approx 500 $ we expect $g^{(2)}(0) - 1 \approx -0.002$.
Classical fluctuations and other systematic effects may obscure this
small effect.

\subsubsection{Internal statistics: Cavity dumping experiment}

One method of more directly investigating photon statisics of the
cavity field is via a ``cavity dumping'' scheme.  The cavity-atom
interaction could suddenly be disabled, by detuning or by optical
mean, after which the internal photons are collected as they exit the
cavity.  By repeating this process many times, a photon count
distribution $P_m^{({\rm PCD})}$ is obtained which is related to the
cavity photon number distribution $P_n^{({\rm PND})} $ by the
partition formula
\be
P_m^{({\rm PCD})} = \sum_{n=m}^{\infty} 
\left(\begin{array}{@{}c@{}}n\\m\end{array}\right) 
\eta^m (1-\eta)^n P_n^{({\rm PND})},
\ee
\noindent where $\eta$ is the overall counting efficiency.  

\subsubsection{Microlaser lineshape measurement}

Lineshape measurements in the many atom regime is a potentially rich
field of study.  Number-phase uncertainty requires sub-Poissonian
statistics to be associated with increased phase diffusion, and thus a
larger linewidth.  For this reason we expect to observe dramatic
changes in microlaser linewidth as the number of atoms and other
parameters are varied.
Such behavior has been predicted in \cite{Scully-PRA91} in the
context of the micromaser, but has yet to be observed, either directly
or indirectly.  

\section{Summary}

We have developed a laser system in which atom-field interaction is
truly coherent.  Rabi oscillations of the atoms lead to multistability
in the laser field intensity, and rapid jumps occur between different
stable points.  Theory and experiment are in good agreement for the
case of the cavity resonant with atoms of the most probable velocity,
but long lifetime of metastable states prevent verification of true
steady-state transition points.  Hysteresis is observed as a function
of atom-cavity detuning and initial cavity photon number.
The detuning behavior of the microlaser is not yet well-understood;
interactions with the nonresonant field and dispersive effects of the
resonant and nonresonant atoms are being investigated.


Quantum trajectory simulations support the application of single-atom
theory to the many-atom case.  In addition, simulations predict an
increase in photon number variance proportional to the product of the
interaction time and cavity decay rate, for parameters such that the
single-atom theory predicts sub-Poisson statistics.

\appendix
\chapter{Microlaser theory calculations}
\label{app-calcs}
\begin{singlespace}
These programs were written and executed using Mathematica 3.0 and 4.0
by Wolfram Software.

\section{Micromaser theory}

\begin{verbatim}

(* All distances in microns   *)
(* All times in microseconds  *)
(* All frequencies in 10^6 /s *)

us = 1;
mhz = 2. Pi;

lambda = 0.7911;
wm = 41.1 ;
v0 = 815; (* most probable velocity, m/s *)
deltapeak = +13.5 mhz;
tiltangle = (deltapeak / (2Pi)) * lambda / v0;

gammacav = 1/0.88;

v = 815.;
tint = Sqrt[Pi] * wm / v;
deltaVOverV = 0.165;

gcoupling0 = 2*Pi*192.9 * 10^-3;
gcoupling = gcoupling0 ;

deltacavlowerlimit = 0;
deltacavupperlimit = 50 mhz;

sqnupperlimit = 150;


doMicromaserCalc :=
    	(pOverC = Table[0, {k, 0, kmax}];
      	p = Table[0, {k, 0, kmax}];
      	pOverC[[0 + 1]] = 1; 
      	sumpOverC = 1;
      	Do[
        	pOverC[[k + 1]] = pOverC[[k]] * Nex *
			betaapprox[Sqrt[k], delta]/k;
                 sumpOverC = sumpOverC + pOverC[[k + 1]],
        	{k, 1, kmax}];
      	Do[p[[k]] = pOverC[[k]] / sumpOverC, {k, 1, kmax + 1}];
      	nAvg = Sum[k * p[[k + 1]], {k, 0, kmax}];
                n2Avg = Sum[k^2 * p[[k + 1]], {k, 0, kmax}];
      	mandelQ = (n2Avg - nAvg^2 - nAvg)/nAvg;	
      );

getnQ := (Nex = natoms / (0.10595 * 1); 
      kmax = Floor[Nex + 2]; doMicromaserCalc; Return[{nAvg, mandelQ}]);


\end{verbatim}

\section{Calculation of gain function}

This code includes bichromatic effects (set $\Omega_2 = 0$ for
ordinary monochromatic theory)

\begin{verbatim}


Delta1 = 10;             (* relative to atoms *)
Delta2 = 37 mhz;    (* relative to atoms *)
tstart = -2* tint;
tstop = +2 * tint;

dosoln := NDSolve[{ca'[
            t] == ((I *omegaR1[t]/2) E^(-I *phi1) E^(I* Delta1* t) + (I *
                      omegaR2[t]/2) E^(-I *phi2) E^(I Delta2 * t))* cb[t], 
        cb'[t] == ((I *omegaR1[t]/2) E^(I *phi1) E^(- I* Delta1* t) + (I *
                      omegaR2[t]/2) E^(I *phi2) E^(- I Delta2 * t))*ca[t], 
        ca[tstart] == 1., cb[tstart] == 0.}, {ca, cb}, {t, tstart, tstop}, 
      MaxSteps -> 10000, StartingStepSize -> Automatic, 
      MaxStepSize -> tint/10, WorkingPrecision -> 10] ;

(* Average over relative phases! *)


numphases = 10.;
phase[i_] := 2. Pi * (i - 1)/numphases;
solarray = Table[i, {i, 1, numphases}];

dophaseavg :=
    (
      omega0 = 2* Sqrt[n + 1] * gcoupling;
      omegaR1[t_] := omega0 * Exp[-(v t/wm)^2] ; 
      omegaR2[t_] :=  omega0 * Exp[-(v t/wm)^2] ; 
      
      Do[
        (
           phi1 = 0*phase[i];
          phi2 = phase[i];
          solarray[[i]] = dosoln;), {i, 1, numphases}];
      
caarray = Table[(ca /. solarray[[i]])[[1]], {i, 1, numphases}];
(* cbarray = Table[(cb /. solarray[[i]])[[1]], {i, 1, numphases}];  *)
      
ca2avg2[t_] := Sum[(Abs[caarray[[i]][t]]^2)/numphases, {i, 1, numphases}];
(* cb2avg2[t_] := 
            Sum[(Abs[cbarray[[i]][t]]^2)/numphases, {i, 1, numphases}]; *)
      (* ca2avg[t] = 
            FunctionInterpolation[
              Sum[(Abs[caarray[[i]][t]]^2)/numphases, {i, 1, numphases}], {t, 
                tstart, tstop}]; *)(* 
        cb2avg = 
            FunctionInterpolation[
              Sum[(Abs[cbarray[[i]][t]]^2)/numphases, {i, 1, numphases}], {t, 
                tstart, tstop}]; *)
      Return[1 - ca2avg2[tstop]];
      );

gaussian[x_, xa_, sigma_] := 
    1/(Sqrt[2. Pi]*sigma) E^(-(x - xa)^2/(2.*sigma^2));
veldist[vel_] := gaussian[vel, v0, v0*deltaVOverV / 2.355];

numvels = 11;  (* odd number is better *)
vlowerlimit = v0 (1 - 1.2*deltaVOverV);
vupperlimit = v0 (1 + 1.2*deltaVOverV);
vsample[i_] := 
    N[vlowerlimit + (i - 1)*(vupperlimit - vlowerlimit)/(numvels - 1)];

deltaofv[vel_] := (vel/v0 - 1.)*deltapeak;
vweight[i_] := 
    veldist[vsample[i]] * (vupperlimit - vlowerlimit)/(numvels - 1);
vlist = Table[{vsample[i], vweight[i]}, {i, 1, numvels}];

numgs = 4; 
gvariation = 0.09;
gupperlimit = 0.99*gcoupling0;
glowerlimit = 1.01 * gcoupling0* (1 - gvariation);
gdist[g_] := 
    UnitStep[1 - g/gcoupling0]*UnitStep[g/gcoupling0 - (1 - gvariation)]*
      Sqrt[Log[gcoupling0/g]];

gsample[i_] := 
    N[glowerlimit + (i - 1)*(gupperlimit - glowerlimit)/(numgs - 1)];
gweight2[i_] := gdist[gsample[i]] * (gupperlimit - glowerlimit)/(numgs - 1);
gnorm = Sum[gweight2[i], {i, 1, numgs}];
gweight[i_] := gweight2[i]/gnorm;
glist = Table[{gsample[i], gweight[i]}, {i, 1, numgs}];


(* beta1 : averaged over phase, velocity = detuning *)

dobeta1 := 
    Sum[ vweight[i]*
        ( v = vsample[i];
          Delta1 = deltacav - (v/v0) * deltapeak;
           Delta2 = -(deltacav + (v/v0)*deltapeak );
          dophaseavg
          ), {i, 1, numvels}];

(* beta2 : averaged over phase, velocity = det, gcoupling *)

dobeta2 := 
    Sum[gweight[j]*
        ( gcoupling = gsample[j];
          Sum[ vweight[i]*
              ( v = vsample[i];
                Delta1 = deltacav - (v/v0) * deltapeak;
                 Delta2 = -(deltacav + (v/v0)*deltapeak );
                dophaseavg
                ), {i, 1, numvels}])
      , {j, 1, numgs}];


\end{verbatim}

\end{singlespace}

\chapter{Quantum trajectory program}
\label{app-qts.tex}

\section{Main program: qts.c}
\begin{singlespace}


\section{newatoms2-d.c}

\begin{verbatim}

#define		DELTA_T_MAX		10.		/* in _Delta_t scale */
#define		V_MIN			0.0
#define		V_MAX			4.
#define		A			2.43952

double	get_velocity();

float	get_height()
{
double	a, b, bb, y;

if (!(CIRCULAR_BEAM)) y=REC_HALF_HEIGHT*drand();
if (CIRCULAR_BEAM)	{
	do {
		a=drand();	
		b=sqrt(1.-sqr(a));
		bb=drand();
		} while (bb>b);
	y=_beam_radius*a;
	}
return y;
}	

/* inject_new_atom() returns negative value if not successful to inject an atom
   Otherwise, returns excited state probability */ 

		
double	inject_new_atom(double t)
{
double	v, y, excited_prob;
long	m, max_m;

void	construct_C();

long	two_to(int);
/* amj #ifdef	PUMPING_SIMULATED */
void	get_state_amplitude(double); 
/* amj #endif */

_N_atom++;
/* if(_N_atom == 1) printf("First atom enters at time %g\n", t);  */

/* printf("%d ", _N_atom);  */
if (_N_atom > _max_atoms) {
	fprintf(stderr, "Too many atoms! ");
	fprintf(stderr, "t=%lf  ", t);
	_N_atom--;				/* throw away this new atom */
	excited_prob = -1.;	/* negative means failure */
	goto LAST_LINE;
	}

/* get new atomic parameters */
if (MONO_VELOCITY) v=1.;
else v=get_velocity();

/* amj  #ifdef	PUMPING_SIMULATED *//* amj move this block to this position */

if(PUMPING_SIMULATED)  /* amj */
  {
    /* check overlap, include spont. emiss. 
       PUMP_POS, INTERACTION_LENGTH, PUMPING_LENGTH to be used  */

    /* _omega_Rabi=dimensionless, in 2g unit, pi pulse */
    _omega_Rabi=PI*v/_t_int/PUMP_WAIST; /* amj */

    if((PUMP_POS+PUMPING_LENGTH/2.) <= INTERACTION_LENGTH/2.)
      {
	_A_info[_N_atom][4]=1.;  /* ground state */	
	_A_info[_N_atom][5]=0.;  /* excited state */
      }
    else
      get_state_amplitude(v);
  }

/* amj #else */
else { /* amj */
  
	_A_info[_N_atom][4]=cos(PI/2.*_v0/v);  /* ground state */	
	_A_info[_N_atom][5]=-sin(PI/2.*_v0/v); /* excited state */
      } /*amj  */


if(0)
  {     printf("_A_info[_N_atom][4]=%g\n", _A_info[_N_atom][4] );
        printf("_A_info[_N_atom][5]=%g\n", _A_info[_N_atom][5] );

	exit(0);
  }


/* CFY: change sign on sin and cos  to agree with pumping-d.c */
/* #endif */

excited_prob=sqr(_A_info[_N_atom][5]);

if(excited_prob>99) {
	_N_atom--;				/* throw away this new atom */
	goto LAST_LINE;
}               /* amj add this block; for d state decayexcited_prob=100 */

if (GAUSSIAN_Y) y=get_height();
else y=0.;
_A_info[_N_atom][0]=v; 		/* in u_th */
_A_info[_N_atom][1]=t;		/* arrival time in 1/2g */

_A_info[_N_atom][2]=t+INTERACTION_LENGTH*_w0/(v*_u_th)
						*2.*_g_over_2pi*2.*PI;  
							/* in 1/2g */
_A_info[_N_atom][3]=1.;			
if (STANDING_WAVE) 
	_A_info[_N_atom][3] *=sin((drand()-0.5)*PI);
 
if (GAUSSIAN_Y)	{
	if (sqr(y)>10.) _A_info[_N_atom][3]=0.;
	else _A_info[_N_atom][3] *=exp(-sqr(y));
	}



_A_info[_N_atom][6]=y;		/* vertical position */

/* Construct new coefficients _C[] */
max_m=two_to(_N_atom-1)*_max_photon_size;

for (m=0; m<max_m; ++m) {
	_C[m+max_m]=_C[m]*_A_info[_N_atom][5];
	_C[m]*=_A_info[_N_atom][4];
	_D[m+max_m]=_D[m]*_A_info[_N_atom][5];
	_D[m]*=_A_info[_N_atom][4];
	/* 
	printf("_C[%ld]=%lf  _C[%ld]=%lf\n", 
					m, _C[m], m+max_m, _C[m+max_m]);
	*/
	}
LAST_LINE:
return excited_prob; 
}



/*
double	get_arrival_time_old()
{
double	x, y, yy;
do {
	x=drand()*DELTA_T_MAX;
	y=drand();
	if (x<10.) yy=exp(-x);
	else yy=0.;
	} while (y>yy);                 
return x*_Delta_t;
}
*/


double	get_velocity()
{
double	x, x2, y, yy;

if (USE_MAXWELL_BOLTZMANN) do {
	x=(V_MAX-V_MIN)*drand()+V_MIN;
	y=drand();
	x2=sqr(x);
	if (x2<10.) yy=exp(-x2);
	else	yy=0.;
	yy*=A*(x*x2);
	} while (y>yy);

else do {
  x=(V_MAX-V_MIN)*drand()+V_MIN;
  y=drand();
  x2=sqr(x);
  yy=A*x*x2*exp(-x2)/(1+sqr(2*(x-_v0)/_delta_v));
} while (y>yy); /* amj changes this block */

return x;
}


long	two_to(int n)
{
long	x=1L;
int		i;

if (n<0) {
	fprintf(stderr, "Error in two_to()\n");
	exit(1);
	}
for (i=0; i<n; ++i) x *=2L;
return x;
#undef	V_MIN
#undef	V_MAX

\end{verbatim}

\section{exitatom2-d.c}
\begin{verbatim}

void	exchange_atom_with_last_one(int p)
{
long	max_k, mask, k, new_k, new_m0, old_m0;
int		j, bits[MAX_NUM_ATOM];
double	C_tmp, D_tmp;

/* exchange p-th atom with the last atom */
if (p<_N_atom) { 
	/*
	printf("First exchanging %d-th atom with %d-th one\n", 
				p, _N_atom);
	*/
 
	/* re-ordering of _C[]*/
	max_k=two_to(_N_atom-1);
	
	for (k=0; k<max_k; ++k) {
		/* evaluate bit patter of k */
		mask=k;
		for (j=1; j<=_N_atom; ++j) {
			bits[j]=mask%2;	
			/* j-th atom state: 0=ground  1=excited */
			mask/=2;
			}
		/*
		printf("k=%ld --> ", k);
		for (j=1; j<=_N_atom; ++j) {
			printf("bits[%d]=%d  ", j, bits[j]);
			}
		printf("\n"); 
		*/

				
		/* exchange bits[p] with bits[_N_atom]=0 */
		bits[_N_atom]=bits[p];
		bits[p]=0;

		/* evaluate new k */
		new_k=0;
		for (j=1; j<=_N_atom; ++j) new_k+=bits[j]*two_to(j-1);

		/*
		printf("new_k=%ld --> ", new_k);
		for (j=1; j<=_N_atom; ++j) {
			printf("bits[%d]=%d  ", j, bits[j]);
			}
		printf("\n"); 
		*/
		
		/* rename indices of _C[] */
		if (!(new_k==k)) {
			new_m0=new_k*_max_photon_size;
			old_m0=k*_max_photon_size;
		
			for (j=0; j<_max_photon_size; ++j) {
				C_tmp=_C[new_m0+j];
				_C[new_m0+j]=_C[old_m0+j];
				_C[old_m0+j]=C_tmp;
				
				D_tmp=_D[new_m0+j];
				_D[new_m0+j]=_D[old_m0+j];
				_D[old_m0+j]=D_tmp;
				}
			
			/*
			printf("_C[%ld]=%lf   _C[%ld]=%lf\n\n",
				new_m0, _C[new_m0], old_m0, _C[old_m0]);
			*/
			}
		else {
			/* 
			printf("Same k and new_k. Not exchanged.\n\n"); 
			*/
			}		
		}
	/* re-order _A_info[][]*/
	for (j=0; j<A_INFO_ITEMS; ++j) 
		_A_info[p][j]=_A_info[_N_atom][j];
	}
}


int	remove_atom(int p)
{
double	A0;
long	m, max_m;
int		exiting_excited=0;

double	normalize_C();

exchange_atom_with_last_one(p);

/* 	decide whether p-th atom, which is now _N_atom-th atom, 
	leaves the cavity in excited or ground state */

/* If in ground state, do nothing. If in excited state, 
take last half of _C[] */
if (is_excited(&A0)) {
	max_m=two_to(_N_atom-1)*_max_photon_size;
	for (m=0; m<max_m; ++m) {
		_C[m]=_C[m+max_m];
		_D[m]=_D[m+max_m];
		/* if ((m%(_max_photon_size))==0)  */
		/* 
		printf("_C[%ld]=%lf  _C[%ld]=%lf\n", m, _C[m], m+max_m, _C[m+max_m]);
		*/
		}
	
	/*printf("exiting in excited state (P_ex=%.3lf) with 
resulting _C[] array\n\n", 
				1.-A0);*/
	
	exiting_excited=1;
	}

/* reduce _N_atom by 1 */
_N_atom--;

	
/* normalize _C[] */
normalize_C();

/*printf("exiting_excited=%d\n", exiting_excited);*/
return exiting_excited;
}


int		is_excited(double *A0)
{
long	max_m, m;
double	X;

max_m=two_to(_N_atom-1)*_max_photon_size;
*A0=0.;
for (m=0; m<max_m; ++m) *A0+=(sqr(_C[m])+sqr(_D[m]));
	/* ground state probability */
X=drand();
/*printf("A0=%f, X=%f\n", *A0, X);*/
if (X> *A0) return 1;
else {
	/*printf("Exiting in ground state\n");*/
	return 0;	
	}						
}


double	normalize_C()
{
long	m, max_m;
double	norm;

max_m=two_to(_N_atom)*_max_photon_size;
norm=0.;
for (m=0; m<max_m; ++m) norm+=(sqr(_C[m])+sqr(_D[m]));
for (m=0; m<max_m; ++m) {
	_C[m]=_C[m]/sqrt(norm);
	_D[m]=_D[m]/sqrt(norm);
	}	
return norm;
}

\end{verbatim}

\section{decay2-d.c}
\begin{verbatim}
int	cavity_decay_occurs(double dt, double *n_mean,  double *n2_mean)
{
double	mean, sqrt_j;
long	k, max_k;
int	j;
double	mult_factor;

double	calculate_n_mean();
double	calculate_n2_mean(); /* amj */
int	does_cavity_decay(double, double);

/* calculate n_mean */
*n_mean=calculate_n_mean();
*n2_mean=calculate_n2_mean();
max_k=two_to(_N_atom);
if (does_cavity_decay(*n_mean, dt)) {
	/* collapse wave function */
	for (j=1; j<_max_photon_size; ++j) {
		sqrt_j=sqrt((double)j);
		for (k=0; k<max_k; ++k) { 
			_C[k*_max_photon_size+j-1]
				=_C[k*_max_photon_size+j]*sqrt_j;
			_D[k*_max_photon_size+j-1]
				=_D[k*_max_photon_size+j]*sqrt_j;
			}
		}
	for (k=1; k<=max_k; ++k) {
		_C[k*_max_photon_size-1]=0.;
		_D[k*_max_photon_size-1]=0.;
		}
	
	/* normalize it */
	mean=normalize_C();

	/*
	printf("Cavity decay occurred from n_mean=%f\n\n", 
				n_mean);
	*/
	return 1;
	}
else {
	/* decrease non-zero photon amplitudes */
	for (j=0; j<_max_photon_size; ++j) {
		mult_factor=1.-(double)j*_Gc_over_2pi/_g_over_2pi/4.*dt;
		for (k=0; k<max_k; ++k) {
			_C[k*_max_photon_size+j]*=mult_factor;
			_D[k*_max_photon_size+j]*=mult_factor;
			}
		}
	mean=normalize_C();
 	return 0;
	}
}

double	calculate_n_mean()
/* calculate <n> mean photon number as well as _Pn_dist[] 
photon number distribution */
{
double	mean;
long	k, max_k;
int	j;

mean=0.;
max_k=two_to(_N_atom);
for (j=0; j<_max_photon_size; ++j) {
	_Pn_dist[j]=0.;
	for (k=0; k<max_k; ++k) 
		_Pn_dist[j]+=(sqr(_C[k*_max_photon_size+j])
			+sqr(_D[k*_max_photon_size+j]));
	_Pn_dist_mean[j]+=_Pn_dist[j];
	mean+=_Pn_dist[j]*(double)j;
	}
return mean;
}

/*  amj add the fununction calculate_n2_mean() */
double	calculate_n2_mean()
/* calculate <n*n> */
{
double	mean2;
int	j;

mean2=0.;

for (j=0; j<_max_photon_size; ++j) 
	mean2+=_Pn_dist[j]*((double)j)*((double)j);
	
return mean2;
}


#define	SMALL_PROB	0.3

int		does_cavity_decay(double mean, double dt)
{
double	prob;

/* calculate decay rate */
prob=_Gc_over_2pi/_g_over_2pi/2.*dt*mean;
if ((prob > 0.3)&&(prob < 10.)) prob=1.-exp(-prob);
if (prob>10.) prob=1.;
if (drand()<prob) return 1;
else return 0;
}

\end{verbatim}

\section{evolve2-2-d.c}
\begin{verbatim}
void	evolve_wave_function(double dt, double t)
/* Runge-Kutta 4th order algorithm */
{
double	dC, dD, x, y, g[MAX_NUM_ATOM];
long	m0, m0j, k_new, k, mask;
int		j, i;
double	xp, t0, omega[MAX_NUM_ATOM];


for (k=0; k<_max_k; ++k) {
	/* evaluate bit pattern of k */
	mask=k;
	_detuning_sum[k]=0.; /* freq/2g unit */
	for (j=1; j<=_N_atom; ++j) {
		_bits[k][j]=mask%2;	
		/* j-th atom state: 0=ground  1=excited */
		if (_bits[k][j]>0) 
			_detuning_sum[k]+=(_CA_detuning-
		_A_info[j][0]*_mean_shift_over_2g/_v0);
		else 
			_detuning_sum[k]-=(_CA_detuning-
		_A_info[j][0]*_mean_shift_over_2g/_v0);
		mask/=2;
		}  /* amj add _v0 jun 16,98 */
	_detuning_sum[k]*=0.5;	
	}
	
/* calculate coupling constant of i-th atom */
for (i=1; i<=_N_atom; ++i) {
	g[i]=_A_info[i][3]; /* of i-th atom, already including 
exp(-sqr(y)) factor */
	
	x=(2.*(t-_A_info[i][1]) /* calculate its x position at t */
		/(_A_info[i][2]-_A_info[i][1])-1.)
			*INTERACTION_LENGTH/2.;
	/* -INTERACTION_LENGTH/2 < x < +INTERACTION_LENGTH/2 */

	if (GAUSSIAN_X) {
		if (sqr(x)<10.) g[i]*=exp(-sqr(x));
		else g[i]=0.;
		}
	else { /* Top hat in X */
		if (fabs(x)>INTERACTION_LENGTH/2.) g[i]=0.;
		}
	/*
	if(i==1 ) printf("t=%g: g[1]=%g\n", t, g[1]);
	*/

	if ((PUMPING_SIMULATED)&&(PUMPING_EFFECT_ON)) {
	  /* check if the atom is still in the pump beam */
	  if (fabs(x+PUMP_POS)<PUMPING_LENGTH/2.) {
	    xp=(x+PUMP_POS)/PUMP_WAIST;
	    if (sqr(xp)<10.) omega[i]=_omega_Rabi*exp(-sqr(xp));
	    else omega[i]=0.;
	  }	/* atom is still in the pump */
	  else omega[i]=0.;
 
	} /* the pump overlaps with the cavity */
}

	
/************************************************/		
/* First, calculate y1=y0+y'(t,y0)*dt/2 for RK4 */
/* or just y=y0+y'(t,y0)*dt for non-RK4         */
/************************************************/		
for (k=0; k<_max_k; ++k) {
	m0=k*_max_photon_size;
	
	for (j=0; j<_max_photon_size; ++j) {
	
	  /* evaluate two atom-field interaction terms */
		/* first, initialize with detuning term */
		m0j=m0+j;
		dC=-_D[m0j]*_detuning_sum[k];	/* detuning in 2g unit */
		dD=_C[m0j]*_detuning_sum[k];
		
		if (j > 0) {  /* a  S- term */
			for (i=1; i<=_N_atom; ++i) {
				if (_bits[k][i]==0) {	
				/* only when i-th atom in ground state */
					k_new=k+two_to(i-1);

					/* H=hbar*g*(a+S- + aS+) used here */
		dC+=_D[k_new*_max_photon_size+j-1]*g[i]*sqrt((double)j)*0.5;
 		dD-=_C[k_new*_max_photon_size+j-1]*g[i]*sqrt((double)j)*0.5;
					
				/* include pump beam effect on lasing */
		if ((PUMPING_SIMULATED)&&(PUMPING_EFFECT_ON)) {
			dC+=omega[i]*_D[k_new*_max_photon_size+j]*0.5;
			dD-=omega[i]*_C[k_new*_max_photon_size+j]*0.5;
			} /* the pump overlaps with the cavity */
					} /* if i-th atom in ground state */
				} /* for i loop */
			} /* if */

		if (j<_max_photon_size-1) { /* a S+ term */
			for (i=1; i<=_N_atom; ++i) {
				if (_bits[k][i]==1) {	
				/* only when i-th atom in excited state */
					k_new=k-two_to(i-1);
	dC+=_D[k_new*_max_photon_size+j+1]*g[i]*sqrt((double)(j+1))*0.5;
	dD-=_C[k_new*_max_photon_size+j+1]*g[i]*sqrt((double)(j+1))*0.5;
					
				/* include pump beam effect on lasing */
		if ((PUMPING_SIMULATED)&&(PUMPING_EFFECT_ON)) {
				dC+=omega[i]*_D[k_new*_max_photon_size+j]*0.5;
				dD-=omega[i]*_C[k_new*_max_photon_size+j]*0.5;
				} /* the pump overlaps with the cavity */
			} /* if i-th atom in excited state */
		} /* for i loop */
	} /* if */
		
		dC*=dt;	/* dt in 1/2g unit */
		dD*=dt;	/* dt in 1/2g unit */		
		#ifdef USE_RK4_ALGORITHM		
			dC/=2.;
			dD/=2.;	
		#endif
		_C_1[m0j]=_C[m0j]+dC;	/* dt in 1/2g unit */
		_D_1[m0j]=_D[m0j]+dD;	/* dt in 1/2g unit */		
		} /* for j loop */
	} /* for k loop */


#ifndef USE_RK4_ALGORITHM /* not USE_RK4_ALGORITHM */
	for (k=0; k<_max_k; ++k) {
		m0=k*_max_photon_size;
		for (j=0; j<_max_photon_size; ++j) {
			m0j=m0+j;
			_C[m0j]=_C_1[m0j]; 
			_D[m0j]=_D_1[m0j];
			}
		}
#else /* do USE_RK4_ALGORITHM */
 	**************************************************************/
	/* Second, calculate y2=y0+y'(t+dt/2,y1)*dt/2                 */
	/**************************************************************/
t+=dt/2;
	for (k=0; k<_max_k; ++k) {
		m0=k*_max_photon_size;
		
		for (j=0; j<_max_photon_size; ++j) {
			m0j=m0+j;
			/* evaluate two atom-field interaction terms */
			/* first, initialize with detuning term */
		dC=-_D_1[m0j]*_detuning_sum[k];	/* detuning in 2g unit */
		dD=_C_1[m0j]*_detuning_sum[k];
			
			if (j > 0) {  /* a  S- term */
				for (i=1; i<=_N_atom; ++i) {
				if (_bits[k][i]==0) {	
				/* only when i-th atom in ground state */
						k_new=k+two_to(i-1);

	dC+=_D_1[k_new*_max_photon_size+j-1]*g[i]*sqrt((double)j)*0.5;
	dD-=_C_1[k_new*_max_photon_size+j-1]*g[i]*sqrt((double)j)*0.5;
						
				/* include pump beam effect on lasing */
			if ((PUMPING_SIMULATED)&&(PUMPING_EFFECT_ON)) {
			dC+=omega[i]*_D_1[k_new*_max_photon_size+j]*0.5;
			dD-=omega[i]*_C_1[k_new*_max_photon_size+j]*0.5;
				} /* the pump overlaps with the cavity */
			} /* if i-th atom in ground state */
		}/* for loop */
	}

			if (j<_max_photon_size-1) { /* a S+ term */
				for (i=1; i<=_N_atom; ++i) {
					if (_bits[k][i]==1) {	
			/* only when i-th atom in excited state */
						k_new=k-two_to(i-1);
	dC+=_D_1[k_new*_max_photon_size+j+1]*g[i]*sqrt((double)(j+1))*0.5;
	dD-=_C_1[k_new*_max_photon_size+j+1]*g[i]*sqrt((double)(j+1))*0.5;
						
			/* include pump beam effect on lasing */
			if ((PUMPING_SIMULATED)&&(PUMPING_EFFECT_ON)) {
		dC+=omega[i]*_D_1[k_new*_max_photon_size+j]*0.5;
		dD-=omega[i]*_C_1[k_new*_max_photon_size+j]*0.5;
			}	/* atom is still in the pump */
			} /* if i-th atom in excited state */
				}
			}
						
		_C_2[m0j]=_C[m0j]+dC*dt/2.;	/* dt in 1/2g unit */
		_D_2[m0j]=_D[m0j]+dD*dt/2.;	/* dt in 1/2g unit */
			} /* for j loop */
		}/* for k loop */
		
		
	/**************************************************************/
	/* Third, calculate y3=y0+y'(t+dt/2,y2)*dt                         */
	/**************************************************************/
	for (k=0; k<_max_k; ++k) {
		m0=k*_max_photon_size;
		
		for (j=0; j<_max_photon_size; ++j) {
			m0j=m0+j;
			/* evaluate two atom-field interaction terms */
			/* first, initialize with detuning term */
			dC=-_D_2[m0j]*_detuning_sum[k];	/* detuning in 2g unit */
			dD=_C_2[m0j]*_detuning_sum[k];
			
			if (j > 0) {  /* a  S- term */
				for (i=1; i<=_N_atom; ++i) {
					if (_bits[k][i]==0) {	
				/* only when i-th atom in ground state */
						k_new=k+two_to(i-1);
		dC+=_D_2[k_new*_max_photon_size+j-1]*g[i]*sqrt((double)j)*0.5;
		dD-=_C_2[k_new*_max_photon_size+j-1]*g[i]*sqrt((double)j)*0.5;
						
			/* include pump beam effect on lasing */
			if ((PUMPING_SIMULATED)&&(PUMPING_EFFECT_ON)) {
			dC+=omega[i]*_D_2[k_new*_max_photon_size+j]*0.5;
			dD-=omega[i]*_C_2[k_new*_max_photon_size+j]*0.5;
			}	/* atom is still in the pump */
				} /* if i-th atom in ground state */
			} /* for loop */
		}

			if (j<_max_photon_size-1) { /* a S+ term */
				for (i=1; i<=_N_atom; ++i) {
					if (_bits[k][i]==1) {	
			/* only when i-th atom in excited state */
						k_new=k-two_to(i-1);
	dC+=_D_2[k_new*_max_photon_size+j+1]*g[i]*sqrt((double)(j+1))*0.5;
	dD-=_C_2[k_new*_max_photon_size+j+1]*g[i]*sqrt((double)(j+1))*0.5;
						
				/* include pump beam effect on lasing */
			if ((PUMPING_SIMULATED)&&(PUMPING_EFFECT_ON)) {
	dC+=omega[i]*_D_2[k_new*_max_photon_size+j]*0.5;
	dD-=omega[i]*_C_2[k_new*_max_photon_size+j]*0.5;
				}	/* atom is still in the pump */
				} /* if i-th atom in excited state */
			}
		}
			
		_C_3[m0j]=_C[m0j]+dC*dt;	/* dt in 1/2g unit */
		_D_3[m0j]=_D[m0j]+dD*dt;	/* dt in 1/2g unit */		
			} /* for j loop */
		} /* for k loop */
		
	/**************************************************************/		
	/* Fourth, calculate yf=(y1+2*y2+y3-y0)/3+y'(t+dt,y3)*dt/6    */
	/**************************************************************/		
	t+=dt/2;
	for (k=0; k<_max_k; ++k) {
		m0=k*_max_photon_size;
		
		for (j=0; j<_max_photon_size; ++j) {
			m0j=m0+j;
			/* evaluate two atom-field interaction terms */
			/* first, initialize with detuning term */
		dC=-_D_3[m0j]*_detuning_sum[k];	/* detuning in 2g unit */
			dD=_C_3[m0j]*_detuning_sum[k];
			
			if (j > 0) {  /* a  S- term */
				for (i=1; i<=_N_atom; ++i) {
					if (_bits[k][i]==0) {	
			/* only when i-th atom in ground state */
				k_new=k+two_to(i-1);
	dC+=_D_3[k_new*_max_photon_size+j-1]*g[i]*sqrt((double)j)*0.5;
	dD-=_C_3[k_new*_max_photon_size+j-1]*g[i]*sqrt((double)j)*0.5;
						
	/* include pump beam effect on lasing */
	if ((PUMPING_SIMULATED)&&(PUMPING_EFFECT_ON)) {
		dC+=omega[i]*_D_3[k_new*_max_photon_size+j]*0.5;
		dD-=omega[i]*_C_3[k_new*_max_photon_size+j]*0.5;
		}	/* atom is still in the pump */
		} /* if i-th atom in ground state */
		} /* for loop */
		}

			if (j<_max_photon_size-1) { /* a S+ term */
				for (i=1; i<=_N_atom; ++i) {
					if (_bits[k][i]==1) {	
		/* only when i-th atom in excited state */
			k_new=k-two_to(i-1);
	dC+=_D_3[k_new*_max_photon_size+j+1]*g[i]*sqrt((double)(j+1))*0.5;
	dD-=_C_3[k_new*_max_photon_size+j+1]*g[i]*sqrt((double)(j+1))*0.5;
						
			/* include pump beam effect on lasing */
	if ((PUMPING_SIMULATED)&&(PUMPING_EFFECT_ON)) {
		dC+=omega[i]*_D_3[k_new*_max_photon_size+j]*0.5;
		dD-=omega[i]*_C_3[k_new*_max_photon_size+j]*0.5;
				}	/* atom is still in the pump */
			} /* if i-th atom in excited state */
					}
				}
			
_C[m0j]=(_C_1[m0j]+_C_2[m0j]*2.+_C_3[m0j]-_C[m0j])/3.+dC*dt/6.;	
	/* dt in 1/2g unit */
_D[m0j]=(_D_1[m0j]+_D_2[m0j]*2.+_D_3[m0j]-_D[m0j])/3.+dD*dt/6.;	
	/* dt in 1/2g unit */		
			} /* for j loop */
		} /* for k loop */

	t-=dt;	/* unchanged t */

#endif
}


\end{verbatim}
\section{spont2-d.c}
\begin{verbatim}
int	decide_atom_decay(int *does_it_decay, double dt)
{
int	i, n=0;
double	prob;

for (i=1; i<=_N_atom; ++i) {
	prob=_Ga_over_2g*dt*_population[i];
	if ((prob>0.3)&&(prob<10.)) prob=1.-exp(-prob);
	if (prob>10.) prob=1.;
	if (drand()<prob) {
		does_it_decay[i]=1;
		n++;
		/*
		printf("%d-th atom undergoes spontaneous emission.\n", i);
		*/
		}
	else does_it_decay[i]=0;
	}
return n;
}
	
	
int	how_many_atoms_decay(double dt)
{
long	max_k, k, mask, new_k, m, m2;
int		bits[MAX_NUM_ATOM], j, i, does_it_decay[MAX_NUM_ATOM];
double	mean;
int		num_photons;

/* initialize population array */
for (i=0; i<MAX_NUM_ATOM; ++i) _population[i]=0.;

max_k=two_to(_N_atom);

/* calculate excited state population */
for (k=0; k<max_k; ++k) {
	mask=k;
	for (i=1; i<=_N_atom; ++i) {
		/* evaluate i-th bit pattern */
		bits[i]=mask%2;
		mask/=2;
		if (bits[i]==1) {
			for (j=0; j<_max_photon_size; ++j) {
				m=k*_max_photon_size+j;
				_population[i]+=(sqr(_C[m])+sqr(_D[m]));
				}
			}
		}
	}

num_photons=decide_atom_decay(does_it_decay, dt);

for (k=0; k<max_k; ++k) {
	mask=k;
	for (i=1; i<=_N_atom; ++i) {
		/* evaluate i-th bit pattern */
		bits[i]=mask%2;
		mask/=2;
		
		if (does_it_decay[i]==1) {
			/* collapse wave function */
			if (bits[i]==1) {
				new_k=k-two_to(i-1);  /* exist */
				for (j=0; j<_max_photon_size; ++j) {
					m2=new_k*_max_photon_size+j;
					m=k*_max_photon_size+j;
					_C[m2]+=_C[m]; /* down shift */
					_C[m]=0.;
					_D[m2]+=_D[m]; /* down shift */
					_D[m]=0.;
					} /* otherwise no new term */
				}
			else 
				for (j=0; j<_max_photon_size; ++j) {
					_C[k*_max_photon_size+j]=0.;					
					_D[k*_max_photon_size+j]=0.;
					}					
			}
							
		else {/* does not decay */
			if (bits[i]==1) {
				/* decrease excited state amplitude */
				for (j=0; j<_max_photon_size; ++j) {
					m=k*_max_photon_size+j;
					/*
					printf("C[%ld]=%f -->", m, _C[m]); 
					*/
					_C[m]*=(1.-_Ga_over_2g/2.*dt);
					_D[m]*=(1.-_Ga_over_2g/2.*dt);
					/*
					printf("C[%ld]=%f\n", m, _C[m]);
					*/
					}
				}
			else {
				/* do nothing */
				/*
				for (j=0; j<_max_photon_size; ++j) {
					m=k*_max_photon_size+j;
					printf("C[%ld]=%f\n", m, _C[m]); 
					}
				*/
				}
			}
		}
	}
mean=normalize_C();
return num_photons;
}

\end{verbatim}

\section{pumping-d.c}
\begin{verbatim}
/* This part incorporates coherent pumping process */
/* #define	SELF_TESTING  ------ this part removed by CFY because it
/* did not compile correctly - old version still exists in pumping-d.old.c  */

int       TEST_PUMP=0;
double    FUDGE_FACTOR=1;

void	get_state_amplitude(double v)
{
double	dt, Dt, a, b, t, t1, t2, tp, x1, x2, 
		Ga_over_Omega_R, prob;

int i;
char *output_name = "pump.out";
FILE *fp;

/* integration from -PUMPING_LENGTH/2 to PUMP_POS-INTERACTION_LENGTH/2 */

Ga_over_Omega_R=_Ga_over_2g*_t_int*PUMP_WAIST/PI;

x1=-PUMPING_LENGTH/2.;
x2=PUMP_POS-INTERACTION_LENGTH/2.;      /* amj /2 */
/* x1, x2 in w0 unit */

/* printf("*** v=%g, _v0=%g\n", v, _v0);  */


FUDGE_FACTOR = 1.187;
tp=sqrt(PI)/(v/_v0) * FUDGE_FACTOR;
t1=x1/PUMP_WAIST*tp;
t2=x2/PUMP_WAIST*tp;
/* tp, t1, t2 in 1/Omega_R unit */

t=t1;
dt=0.01;
a=1.;	/* ground state amplitude */
b=0.;	/* excited state amplitude */

do {
	Dt=dt*exp(-sqr(t/tp));
	a+=b/2.*Dt;
	b+=-a/2.*Dt;
	
	t+=dt;


  
	/* amj test starts jul 27, 98 */

  if(TEST_PUMP)
  {
	for (i=0; i<2; ++i) 
	  {
	    if (i==0)
	      fp=fopen(output_name, "a");
	    else fp=stderr;
	    fprintf(fp, "%lf   %lf   %lf \n", t, a,  b);
            if(i==0) fclose(fp);
	  }
  }


	/* amj test ends jul 27, 98 */


	/* calculate decay probability */
	/* amj #ifdef	INCLUDE_ATOM_DECAY */
	if(INCLUDE_ATOM_DECAY){
	  prob=Ga_over_Omega_R*dt*sqr(b);
	  if ((prob>0.3)&&(prob<10.)) prob=1.-exp(-prob);
	  if (prob>10.) prob=1.;
	
	/* generate a random number */
	if (drand()<prob) {
		/* atom decays */
		b=0.;
		a=1.;
	}}

	if(INCLUDE_D_STATE_DECAY){
	  prob=D_STATE_DECAY_CORRECTION*Ga_over_Omega_R*dt*sqr(b);
	  if ((prob>0.3)&&(prob<10.)) prob=1.-exp(-prob);
	  if (prob>10.) prob=1.;
	
	/* generate a random number */
	if (drand()<prob) b=10; /* amj by definition b=10 for d state decay */
	} /* amj adds this block */

/* amj else {} */ /* if not decaying, do nothing */
	/* amj #endif */

	} while (t<t2);

/* amj test starts jul 27, 98 */

/*
 for (i=0; i<2; ++i) 
   {
     if (i==0)
       fp=fopen("pumping.out", "a");
     else fp=stderr;
          fprintf(fp, "after this point atom in interaction region\n"); 
     if (i==0) fclose(fp);
   }
 */

if(TEST_PUMP) 
  exit(0);

 /* amj test ends jul 27, 98 */

	_A_info[_N_atom][4]=a;
	_A_info[_N_atom][5]=b;


}

\end{verbatim}

\section{d\_state2-d.c}

\begin{verbatim}
int	decide_d_state_decay(int *does_it_decay, double dt)
{
int	i, n=0;
double	prob;

for (i=1; i<=_N_atom; ++i) {
	/* calculate decay probability to D state */
	prob=_Ga_over_2g*dt*_population[i]*D_STATE_DECAY_CORRECTION;
	if ((prob>0.3)&&(prob<10.)) prob=1.-exp(-prob);
	if (prob>10.) prob=1.;
	if (drand()<prob) {
		does_it_decay[i]=1;
		n++;
		/*
		printf("%d-th atom undergoes d-state decay.\n", i);
		*/
		}
	else does_it_decay[i]=0;
	}
return n;
}
	
	
int	how_many_d_state_decay(double dt)
{
long	max_k, k, mask,  m, max_m;
int	bits_i, j, i, does_it_decay[MAX_NUM_ATOM];
double	mean;
int	num_decays;

/* initialize population array */
for (i=0; i<MAX_NUM_ATOM; ++i) _population[i]=0.;

max_k=two_to(_N_atom);

/* calculate excited state population */
for (k=0; k<max_k; ++k) {
	mask=k;
	for (i=1; i<=_N_atom; ++i) {
		/* evaluate i-th bit pattern */
		bits_i=mask%2;
		mask/=2;
		if (bits_i==1) {
			for (j=0; j<_max_photon_size; ++j) {
				m=k*_max_photon_size+j;
				_population[i]+=(sqr(_C[m])+sqr(_D[m]));
				}
			}
		}
	}

num_decays=decide_d_state_decay(does_it_decay, dt);

for (i=1; i<=_N_atom; ++i) {
	/* evaluate i-th bit pattern */
	if (does_it_decay[i]==1) {
		/* i-th atom decays to d-state --> i-th atom should be in the 
	excited state just before the decay.  When D-state decay occurs 
		for i-th atom, first we exchange the atom with the last one. 
	Now the i-th atom is labeled as the last one.  Since the new last 
		one is in D-state, it is as if it were not in the cavity.  We 
	should remove the atom from the wave function.  Just before the 
	removal, the atom should be in the excited state.  This is equivalent 
		to the atom exciting cavity in the excited state prematurally.  */
		
		exchange_atom_with_last_one(i);
			
		/* the last atom must be in the excited state */
		max_m=two_to(_N_atom-1)*_max_photon_size;
		for (m=0; m<max_m; ++m) {
			_C[m]=_C[m+max_m];
			_D[m]=_D[m+max_m];
			}
		
		_N_atom--;
		}
	else {/* does not decay */
		max_k=two_to(_N_atom-1);
		for (k=0; k<max_k; ++k) {
			mask=k;
			
			/* find the i-th bit pattern of k */
			for (j=1; j<=i; ++j) {
				bits_i=mask%2;	
				mask/=2;
				}
				
			if (bits_i==1) {
				for (j=0; j<_max_photon_size; ++j) {
					m=k*_max_photon_size+j;
				/* decrease excited state amplitude */
	_C[m]*=(1.-_Ga_over_2g/2.*dt*D_STATE_DECAY_CORRECTION);
	_D[m]*=(1.-_Ga_over_2g/2.*dt*D_STATE_DECAY_CORRECTION);
					} /* for j loop */
				} /* if */
			} /* for k loop */
		} /* if */
	} /* for i loop */
		
mean=normalize_C();
return num_decays;
}

\end{verbatim}

\section{Sample input file}
\begin{verbatim}
50.000000 # Ga_over_2pi
791.000000 # lambda
10.000000 # r0
10 # finesse
0.100000 # L
200 3 # _max_photon_size _max_atoms
0.05000 # dt0
10000 # tot_atom_count
0 # DEBUG_LEVEL
0 INCLUDE_ATOM_DECAY
0 # INCLUDE_D_STATE_DECAY
0 # STANDING_WAVE
10.000000 # tilt_angle
1 # num_detuning
1.000000 # detuning[0]
4.000000 # INTERACTION_LENGTH
0 # CIRCULAR_BEAM
0.0001 # REC_HALF_HEIGHT
0.443 # effective_V_to_apparent_V
1 # num_N_eff
.1 # N_eff[0]       *** Don't forget summary filename below ***
1 # num_u_th
400.000000 # u_th[0]
0 # MONO_VELOCITY
0 # USE_MAXWELL_BOLTZMANN
0.100000 # _delta_v
1.000000 # _v0
0 # GENERATE_PHOTON_FILE
0 # PUMPING_SIMULATED
run40.data # summary filename

\end{verbatim}

Note: this program may not treat velocity distributions and detuning
correctly; in this thesis simulation results were only cited for the
monovelocity, resonant case.

\end{singlespace}

\chapter{VPascal sequencing program}
\label{app-vpascal}
\begin{singlespace}
\section{Cavity scanning experiment}
\begin{verbatim}

{ microlaser_experiment.v }
{ C. Fang-Yen }

{ Variables }
var
  Z1 ; plotimage ;  bgimage ; img;
  Plot0 ;  Plot1 ;  Plot2 ;  Plot3 ; exposure_time;
  imagename;  backgroundname;  scatteringname; scatteringimage; 
  rowplotname ;  colplotname ;
  do_dataplot;  dataplotname;  crop_colplot;
  i;  index;  x;    { scanned voltage, etc. }
  x_start;  x_stop;  x_increment;  x_use;
  micp;  mirp;  mox;  moy;
  writedata;  textwin;  textwinname;
  do_serial;  total_counts_x;  total_counts_y;  num_pixels;
  wherexmax;  whereymax;  iylimit;  ixlimit;  sum0;  sum1;  sum2;
  rezero;  rezero_sample;  moymox;  y_third_max_minus;  y_third_max_plus;
  ii;  do_gaussian_fit;  closest;  fitmin; fitmax;
  sigma; logmox; logmoxregion;  a;  line; line100;
  test;  absmox;  data_x;  data_y;  data_x_temp ;  data_y_temp ;
  doing_background_now; signal; previous_signal; ratio; serial_error;
  serial_error_msg; dataline; current_position; background_plot;
  n_periods; data_array; n_frames; data_max; data_bkgnd; where_max;
  data_bkgnd_sample; v_t_553_on; v_t_791_on; v_t_553_off; v_t_791_off; 
  use_background; plotimage_no553; datafilename; do_blackbody; do_microlaser;
  save_images; image_array; image_array_filename; cts_per_ms;

procedure take_picture(roi_index; var img);
		  begin
		  { Set Gain Index 3 }
		  pvcSetGain( 3 );
		  exposure_time := 300;
		  pvcSetExpTime( exposure_time);
		  roi_index := 3;   { override }

		  if(roi_index = 0) then 
		  			   img := pvcSequence( 1,386,228,423,257,1,1 ) ; {small}
		  if(roi_index = 1) then
		  			   img := pvcSequence(  1,304,169,501,323,1,1 ) ; {medium}
		  if(roi_index = 2) then
		  			   img := pvcSequence(  1,0,0,767,511,1,1 ); { full CCD } 
	      if(roi_index = 3) then
		  			   img := pvcSequence( 1,352,236,454,270,1,1 ); { small for 
					   	   	  									   slit}
		  end;

begin

{ *******************  INITIALIZATION   **************** }

datafilename := 'data.tif';  { 'data.tif' = generic }
image_array_filename := 'image.tif';
do_blackbody := False;  { Record location of blackbody peak in col plot }
use_background := False;  { else use scattering subtraction }
do_microlaser := True; { else skip cavity scan }
save_images := False; { save fluorescence images } 

imagename := 'image';
rowplotname := 'rowplot';
colplotname := 'colplot';
dataplotname := 'dataplot';
crop_colplot := 0;
backgroundname := 'background';
scatteringname := 'scattering';
index := 0;
x_use := True;
x_start :=1450;
x_stop := 1500;
x_increment := 2;
x := x_start;
writedata := True;
textwinname := 'output';
do_serial := True;
rezero := False;
rezero_sample := 10;
do_gaussian_fit := False;
data_x := MakeLinear(0.0,0.0,1000);
data_y := MakeLinear(0.0,0.0,1000);
do_dataplot := False;
doing_background_now := 0;
v_t_553_on := 0.65;   { V_t 553nm on voltage } 
v_t_791_on := 1.0;   { V_t 791nm on voltage }
v_t_553_off := -0.1;   { V_t 553nm off voltage } 
v_t_791_off := -0.1;   { V_t 791nm off voltage }


if(writedata) then 		  
	begin
	  	if EditorExists(textwinname) then 
			   GetEditor(textwinname, textwin)
			   	else textwin := CreateEditor( 'output');
		WriteLn(textwin, '');
	end;

{ ******* INITIALIZE SERIAL PORTS ******  }

     SetTxEnd( chr(13) ) ;
	 SetRxEnd( chr(13) ) ;
	 SetRxTimeout(5000);

	 WriteLn('----');
	 
  	 OpenSerial( 1,19200,8,NoParity,1 ) ;  { SR400 photon counter A }
	 serial_error := SerialError;
	
  	 if serial_error <> 0 then 
	 Halt( 'Failed to open serial connection 1: error ', serial_error)
	 else WriteLn('Serial connection 1 ok');

	  SelectPort(1);
	  TxFlush ;
  	  Transmit( chr(13));
  	  Transmit( chr(13));
	  TxFlush ;
	  RxFlush;

	Transmit('pl 1');
	  dataline := RxString;
	  WriteLn('SR400A: PORT1=', dataline);

	Transmit('np');
	  dataline := RxString;
	  n_periods := val(dataline);
	  WriteLn('SR400A: NPERIODS=', n_periods);
	  
	  {
	  CloseSerial(1);	  
	  Halt('');
	  }

	  if x_increment <> 0 then
	  n_frames := Floor((x_stop - x_start)/x_increment)+1
	  else
	  n_frames := 100;
	  
	  data_array := CreateArray(single, n_periods+1, n_frames, 1);
	  
	 OpenSerial( 2,9600,8,NoParity,1 ) ;  { IMS stepper driver }
	 serial_error := SerialError;
	
  	 if serial_error <> 0 then 
	 Halt( 'Failed to open serial connection 2: error ', serial_error )
	 else WriteLn('Serial connection 2 ok');
	 

	 { ****** INITIALIZE BLOCKER ****** }

	  SelectPort(2);

	  Transmit(' ');
	  	  Delay(100);
	  RxFlush;
	  WriteLn('RxWaiting=',RxWaiting);
	  Transmit('Z');
	  Delay(100);
	  dataline := RxString;
	  WriteLn('Current position ', dataline);
	  dataline := StrParse(dataline, 'Z');
	  current_position := Val(StrStrip(dataline));
	  WriteLn('Current position Z=', current_position);
	  Transmit('X');
	  dataline := RxString;
	  WriteLn(dataline);
	  dataline := RxString;
	  WriteLn(dataline);
	  
	  {
	  CloseSerial(1); CloseSerial(2);
	  Halt('done');
	  }

{ ************* TAKE 1 FRAME, SET UP IMAGE AND PLOT WINDOWS ************ }
	  
{ Acquire data }
{ plotimage := pvcSequence( 1,386,228,423,257,1,1 ) ; }
take_picture(0, plotimage);
Show( plotimage) ;

num_pixels := GetXSize(plotimage) * GetYSize(plotimage) ;

if save_images then image_array := CreateArray(GetXSize(plotimage), 
GetYSize(plotimage), n_frames);

WriteLn(FileExpand('background.tif'));
{ Halt; }

if use_background then 
 if not ImageExists(backgroundname) then 
 begin  
  if Query('No background: use this?')=id_Yes then
      begin
	   	   Save(plotimage, 'background.tif');
		   Open('background.tif', background_plot);
		   Show(background_plot, 'background');
		   Delete(plotimage);
	  end;
 	  Halt;
 end;
{
if use_background = False then
    if not ImageExists(scatteringname) then Halt('No scattering file');
}

{ Get the active frame number }
Z1 := GetActiveFrame( plotimage ) ;

{ Create a plot window }
if PlotExists(rowplotname) then
   begin
   GetPlot(rowplotname, Plot0);
 {  Halt('ok'); }
   end
      
else   
  begin
   Plot0 := CreatePlot( rowplotname );
   SetTitle( Plot0,'Row Plot' ) ;
   SetSubTitle( Plot0,GetName( plotimage ) ) ;
   SetXLabel( Plot0,'X Position' ) ;
   SetYLabel( Plot0,'Mean Intensity' ) ;
   SetGrid( Plot0,gs_Both ) ;
  end;

{ Row plot }
SetPlotColor( col_Black ) ;
Plot( Plot0,MeanOverY( plotimage[..,..,Z1] ) ) ;


{ Create data plot window }
if PlotExists(dataplotname) then
   begin
   	GetPlot(dataplotname, Plot2);
	end
else
	Plot2 := CreatePlot(dataplotname);
		

{ Create a plot window }
if PlotExists(colplotname) then
   begin
   GetPlot(colplotname, Plot1);
 {  Halt('ok'); }
   end
else begin
	 Plot1 := CreatePlot( colplotname ) ;
	 SetTitle( Plot1,'Column Plot' ) ;
	 SetSubTitle( Plot1,GetName( plotimage ) ) ;
	 SetXLabel( Plot1,'Y Position' ) ;
	 SetYLabel( Plot1,'Mean Intensity' ) ;
	 SetGrid( Plot1,gs_Both ) ;
end;

{ Col plot }
SetPlotColor( col_Black ) ;
Plot( Plot1,MeanOverX( plotimage[..,..,Z1] ) ) ;

{ *** INITIALIZE SR400 PARAMETERS *** }

{ to be written }


{ ******************* MAIN LOOP BEGINS ************************ }

repeat

SelectPort(2);
Transmit(StrCat('R ', Str(x)));  { SET BLOCKER POSITION }
Delay(1000.0*Abs(current_position - x)/400.0);  { Wait for it to move }
current_position := x;

{ Turn on 553nm probe, turn off 791nm pump}
SelectPort(1);
Transmit(StrCat('pl 1, ', Str(v_t_553_on)));
Transmit(StrCat('pl 2, ', Str(v_t_791_off)));

Delay(1000);

{ Acquire data }
take_picture(0, plotimage);
Show( plotimage ) ;

{ Turn off 553nm probe, turn on 791 pump }
SelectPort(1);
Transmit(StrCat('pl 1, ', Str(v_t_553_off)));
Transmit(StrCat('pl 2, ', Str(v_t_791_on)));

mox := MeanOverX( plotimage[..,..,Z1] );
{ plotimage[..,0..crop_colplot,..] := mox[0,crop_colplot,0]; }


GetImage( backgroundname, bgimage ) ;

GetImage( scatteringname, scatteringimage );

{ Perform an image subtraction }

if use_background then plotimage := plotimage - bgimage ;

if not use_background then
   begin
   take_picture(0, plotimage_no553);
   plotimage := plotimage - plotimage_no553 - 0;

   end;
   
{ Rezero: subtract avg of first 10 pixels in col plot }

if rezero then begin
mox := MeanOverX( plotimage[..,..,Z1] );
moymox := MeanOverY(mox[..,0..10,..]);
plotimage := plotimage - moymox[0,0,0];
{ mox[0,crop_colplot,0]; }
end;

if save_images then
begin
image_array[..,..,index] := plotimage;  { store plotimage in array }
   if Save(image_array, image_array_filename) <> file_Ok 
      then Halt('Save error');
end;

{ Row plot }
ClearPlot(Plot0);
SetPlotColor( col_Black ) ;
moy := MeanOverY( plotimage[..,..,Z1] );

Plot( Plot0,moy ) ;
mirp := MaxOf(moy);

wherexmax := 0;
ixlimit := GetXSize(moy)-1;
for i := 0 to ixlimit	do
	begin
	if moy[i,0,0] = mirp
	   then wherexmax := i;
{	WriteLn(str(i));    }
    end;
	
sum0 := 0; sum1 := 0; sum2 := 0;
for i :=0 to ixlimit do
	begin
	sum0 := sum0 + moy[i,0,0];
	sum1 := sum1 + moy[i,0,0] * i;
	sum2 := sum2 + moy[i,0,0] * i * i;	
	end;

WriteLn('x: avg=', sum1/sum0, ' sigma=', 
				 Sqrt(sum2/sum0 - sum1*sum1/(sum0*sum0)), ' sum=', sum0);
 			
WriteLn(x, ' ', sum0);

{ WriteLn(textwin, 'max at x=', wherexmax); }

{ WriteLn(textwin, MaxOf(moy[0..10])); }
total_counts_y := MeanOf(moy) * num_pixels;
SetTitle( Plot0, StrCat(StrCat('Row Plot, max=', Str(mirp)), 
		  StrCat(' at ', Str(wherexmax))) ) ;

{ Col plot }
ClearPlot(Plot1);
SetPlotColor( col_Black ) ;
mox := MeanOverX( plotimage[..,..,Z1] );
{ WriteLn(textwin, GetYSize(mox)); }
mox[..,0..crop_colplot,..] := mox[..,crop_colplot,..]  ;
{ if crop_colplot>0 then 
   for i := 1 to crop_colplot-1 do
   	    begin
		WriteLn(i);
   		mox[i] := mox[crop_colplot];
		end; }
		
Plot( Plot1,mox ) ;
micp := MaxOf(mox);
total_counts_x := MeanOf(mox) * num_pixels;

whereymax := 0;
iylimit := GetYSize(mox)-1;
for i := 0 to iylimit	do
	begin
	if mox[0,i,0] = micp
	   then whereymax := i;
{	WriteLn(str(i));    }
    end;

{ Find 1/3 max points }
	
closest := 10000;	
for ii := whereymax to iylimit do
    begin
	if Abs(mox[0,ii,0] - micp/3.0) < closest then 
		 begin
		 closest := Abs(mox[0,ii,0] - micp/3.0);
		 y_third_max_plus := ii;
		 end;
	end;

closest := 10000;	
for ii:=0 to whereymax do
    begin
	if Abs(mox[0,ii,0] - micp/3.0) < closest then 
		 begin
		 closest := Abs(mox[0,ii,0] - micp/3.0);
		 y_third_max_minus := ii;
		 end;
	end;	 
{
	WriteLn('y_half_max_plus ', str(y_third_max_plus));	
	WriteLn('y_half_max_minus ', str(y_third_max_minus));	
}

{ Gaussian fitting }

sigma:=0;
if do_gaussian_fit then begin

absmox := sqrt(sqr(mox))+0.01;
logmox := Ln(absmox);
fitmin := y_third_max_minus;
fitmax := y_third_max_plus;
line := MakeLinear(fitmin -0.0 , fitmax - 0.0, fitmax - fitmin + 1);
line100 := MakeLinear(0,iylimit+0.0,(iylimit+1.0));
a := PolyFit(line, logmox[0,fitmin..fitmax,0],2);

{ WriteLn(a);  }
{
WriteLn(str(line));
WriteLn(str(logmox[0,fitmin..fitmax,0]));
WriteLn(str(a[0,0,0]));
WriteLn(str(a[1,0,0]));
WriteLn(str(a[2,0,0]));
WriteLn(' '); }

{ logmoxregion := logmox[0,fitmin..fitmax,0];
WriteLn(logmoxregion);
WriteLn(str(logmoxregion[0,0,0])); }

{
  Plot3 := CreatePlot('logplot');
  Plot(Plot3, logmoxregion);
}
{  Plot(Plot3, line, Polyvalue(a, line)); }
  
sigma := Sqrt(Abs(-1.0 / (2.0 * a[2,0,0]))); 
{ SetTitle( Plot1, str(2*sigma)); }

WriteLn('gaussian_fit 2*sigma= ', str(2*sigma));

SetPlotStyle( ps_Line ) ;
SetPlotColor( col_Blue ) ;
PlotXY(Plot1, line100, Exp(Polyvalue(a, line100)));
SetPlotColor( col_Red ) ;
PlotXY(Plot1, line, Exp(Polyvalue(a, line)));
{SetPlotColor( col_Blue ) ;
PlotXY(Plot2, line100, Polyvalue(a, line100));}
end;

{ Take moments }

sum0 := 0; sum1 := 0; sum2 := 0;
for i :=0 to iylimit do
	begin
	sum0 := sum0 + mox[0,i,0];
	sum1 := sum1 + mox[0,i,0] * i;
	sum2 := sum2 + mox[0,i,0] * i * i;	
	end;

WriteLn('y: avg=', sum1/sum0, ' sigma=', 
				 Sqrt(sum2/sum0 - sum1*sum1/(sum0*sum0)), ' sum=', sum0);
 		
{ WriteLn(textwin, 'max at y=', whereymax); }


SetTitle( Plot1, StrCat(StrCat(StrCat(StrCat('y:Colplot, max=', Str(micp)), 
		  StrCat(' at ', Str(whereymax)) ), ' sigma*2= '), Str(2*sigma)) );

  	  signal := micp;

	  cts_per_ms := total_counts_x / exposure_time;

	  WriteLn('blocker=',x,' micp=', signal, ' total_cts_x=', total_counts_x, 
	  'cts_per_ms = ', cts_per_ms);
	
	
	data_x[index+1,0,0] := x;
	data_y[index+1,0,0] := signal;
	data_x[index,0,0] :=x;
	data_y[index,0,0] := signal;
	

	if do_dataplot then
	begin
	ClearPlot(Plot2);
	data_x_temp := data_x[0..(index),0,0];
	data_y_temp := data_y[0..(index),0,0];

	{ WriteLn(data_x[0..index+1,0,0], data_y[0..index+1,0,0]); }
	PlotXY(Plot2, data_x_temp, data_y_temp);		  
	end;   

WriteLn('frame=', index, ' x=',x, ' micp=', micp, ' mirp=', mirp, ' tot_x=', total_counts_x, 
				 ' tot_y', total_counts_y);

{ ***** END OF FLUORESCENCE  ***** }				 
				 
{ ***** TRIGGER RAMP AND COLLECT DATA ***** }				 

if do_microlaser then
begin
				  
   SelectPort(2);
   Transmit('A 8');   { OUTPUT 1 -> HIGH }
   Transmit('A 0');   { OUTPUT 1 -> LOW  }


   SelectPort(1);
   RxFlush;
   Transmit('CR');  { SR400 Counter Reset }
   Transmit('FB');  { Start with continuous data output }

  for i := 1 to n_periods do
	begin
	dataline := RxString;
	data_array[i,index,0] := val(dataline);
{	WriteLn(data_array[i,index,0]);  }
	end;

   data_max := 0; data_bkgnd := 0;
   data_bkgnd_sample := 10;

   for i :=1 to n_periods do
  	begin
	{ WriteLn('i, index=', i, ' ', index); }
	{ WriteLn(i, ' ', data_array); }
	if data_array[i,index,0] > data_max then 
	   		begin
	   		data_max := data_array[i,index,0];
			where_max := i;
			end;
	end;

   data_bkgnd := MeanOf(data_array[1..data_bkgnd_sample,index,0]);

   WriteLn('data_max=', data_max, ' at ', where_max, '. data_bkgnd=', 
					 data_bkgnd, ' signal=', data_max-data_bkgnd);	

   if (do_blackbody) then
      WriteLn(textwin, whereymax, ' ', data_max - data_bkgnd)					 
   else 
	 WriteLn(textwin, index, ' ', x, ' ', cts_per_ms, ' ', data_max - data_bkgnd);
	 { WriteLn(textwin, index, ' ', micp,  }

   Clear(plot2);
   Plot(plot2, data_array[1..n_periods,index,0]);

   if Save(data_array, datafilename) <> file_Ok 
      then Halt('Save error');
  
    x := x + x_increment;
    index := index + 1; 

end;  { end of 'if do_microlaser' }
	
until x > x_stop;

{ *************** END OF MAIN LOOP ************** }

CloseSerial(1); CloseSerial(2);

end


\end{verbatim}

\section{Cavity locking experiment}
\begin{verbatim}

{ microlaser_experiment.v }
{ C. Fang-Yen }

{ Variables }
var
  Z1 ; plotimage ;  bgimage ; img;
  Plot0 ;  Plot1 ;  Plot2 ;  Plot3 ; exposure_time;
  imagename;  backgroundname;  scatteringname; scatteringimage; 
  rowplotname ;  colplotname ;
  do_dataplot;  dataplotname;  crop_colplot;
  i;  index;  x;    { scanned voltage, etc. }
  x_start;  x_stop;  x_increment;  x_use;
  micp;  mirp;  mox;  moy;
  writedata;  textwin;  textwinname;
  do_serial;  total_counts_x;  total_counts_y;  num_pixels;
  wherexmax;  whereymax;  iylimit;  ixlimit;  sum0;  sum1;  sum2;
  rezero;  rezero_sample;  moymox;  y_third_max_minus;  y_third_max_plus;
  ii;  do_gaussian_fit;  closest;  fitmin; fitmax;
  sigma; logmox; logmoxregion;  a;  line; line100;
  test;  absmox;  data_x;  data_y;  data_x_temp ;  data_y_temp ;
  doing_background_now; signal; previous_signal; ratio; serial_error;
  serial_error_msg; dataline; current_position; background_plot;
  n_periods; data_array; data_array2; n_frames; data_max; data_bkgnd; 
  where_max;
  data_bkgnd_sample; v_t_553_on; v_t_791_on; v_t_553_off; v_t_791_off; 
  use_background; plotimage_no553; datafilename; do_blackbody; do_microlaser;
  save_images; image_array; image_array_filename; log_filename; do_blocker;
  lock_time; do_553_ttl; cts_per_ms;

procedure ttl_outputs(out1, out2, out3);
	begin
	SelectPort(2);
	Transmit('A '+  Str(8*(1-out1) + 16*(1-out2) + 32*(1-out3)));
	end;

procedure lock_cavity;
	begin
	ttl_outputs(0 , 0, 0);  { turn off AOM after cavity }
	ttl_outputs(1 , 1, 0);  { turn on 791 probe; no cav lock disable}
	end;

procedure unlock_cavity;
	begin
	ttl_outputs(0, 0, 0);  { turn off 791 probe; cav lock disable }
	ttl_outputs(0, 0, 1);  { turn on AOM after cavity }
	end;

procedure take_picture(roi_index; var img);
		  begin
		  { Set Gain Index 3 }
		  pvcSetGain( 3 );
		  exposure_time := 300;
		  pvcSetExpTime( exposure_time);
		  roi_index := 0;   { override }

		  if(roi_index = 0) then 
		  			   img := pvcSequence( 1,386,228,423,257,1,1 ) ; {small}
		  if(roi_index = 1) then
		  			   img := pvcSequence(  1,304,169,501,323,1,1 ) ; {medium}
		  if(roi_index = 2) then
		  			   img := pvcSequence(  1,0,0,767,511,1,1 ); { full CCD } 
	      if(roi_index = 3) then
		  			   img := pvcSequence( 1,352,236,454,270,1,1 ); { small for 
					   	   	  									   slit}
		  end;

begin

{ *******************  INITIALIZATION   **************** }

datafilename := 'data.tif';  { 'data.tif' = generic }
image_array_filename := 'image.tif';
log_filename := 'log.txt';
do_blackbody := False;  { Record location of blackbody peak in col plot }
use_background := False;  { False = use scattering subtraction }
do_microlaser := False; { False = skip cavity scan }
save_images := False; { save fluorescence images } 
do_blocker := True;

imagename := 'image';
rowplotname := 'rowplot';
colplotname := 'colplot';
dataplotname := 'dataplot';
crop_colplot := 0;
backgroundname := 'background';
scatteringname := 'scattering';
index := 0;
x_use := True;
x_start := 20;
x_stop := 60;
x_increment := 1;
x := x_start;
writedata := True;
textwinname := 'output';
do_serial := True;
rezero := False;
rezero_sample := 10;
do_gaussian_fit := False;
data_x := MakeLinear(0.0,0.0,1000);
data_y := MakeLinear(0.0,0.0,1000);
do_dataplot := False;
doing_background_now := 0;
v_t_553_on := 0.605;   { V_t 553nm on voltage } 
v_t_791_on := 1.0;   { V_t 791nm on voltage }
v_t_553_off := -0.1;   { V_t 553nm off voltage } 
v_t_791_off := -0.1;   { V_t 791nm off voltage }

lock_time := 1000;


if(writedata) then 		  
	begin
	  	if EditorExists(textwinname) then 
			   GetEditor(textwinname, textwin)
			   	else textwin := CreateEditor( 'output' );
		WriteLn(textwin, '');
	end;

{ ******* INITIALIZE SERIAL PORTS ******  }

     SetTxEnd( chr(13) ) ;
	 SetRxEnd( chr(13) ) ;
	 SetRxTimeout(1000);

	 WriteLn('----');
	 
  	 OpenSerial( 1,19200,8,NoParity,1 ) ;  { SR400 photon counter A }
	 serial_error := SerialError;
	
  	 if serial_error <> 0 then 
	 Halt( 'Failed to open serial connection 1: error ', serial_error)
	 else WriteLn('Serial connection 1 ok');

	  SelectPort(1);
	  TxFlush ;
  	  Transmit( chr(13));
  	  Transmit( chr(13));
	  TxFlush ;
	  RxFlush;

	Transmit('pl 1');
	  dataline := RxString;
	  WriteLn('SR400A: PORT1=', dataline);

	Transmit('np');
	  dataline := RxString;
	  n_periods := val(dataline);
	  WriteLn('SR400A: NPERIODS=', n_periods);
	  
	  {
	  CloseSerial(1);	  
	  Halt('');
	  }

	  if x_increment <> 0 then
	  n_frames := Floor((x_stop - x_start)/x_increment)+1
	  else
	  n_frames := 100;
	  
	  data_array := CreateArray(single, n_frames, 1);
	  data_array2 := CreateArray(single,  n_frames, 1);
	  
	 OpenSerial( 2,9600,8,NoParity,1 ) ;  { IMS stepper driver }
	 serial_error := SerialError;
	
  	 if serial_error <> 0 then 
	 Halt( 'Failed to open serial connection 2: error ', serial_error )
	 else WriteLn('Serial connection 2 ok');
	 

	 { ****** INITIALIZE BLOCKER ****** }

	  SelectPort(2);

	  Transmit(' ');
	  	  Delay(100);
	  RxFlush;
{	  WriteLn('RxWaiting=',RxWaiting); }
	  Transmit('Z');
	  Delay(100);
	  dataline := RxString;
	  WriteLn('Current position ', dataline);
	  dataline := StrParse(dataline, 'Z');
	  current_position := Val(StrStrip(dataline));
	  WriteLn('Current position Z=', current_position);
	  Transmit('X');
	  dataline := RxString;
	  WriteLn(dataline);
	  dataline := RxString;
	  WriteLn(dataline);
	  
	
{ ************* TAKE 1 FRAME, SET UP IMAGE AND PLOT WINDOWS ************ }
	  
unlock_cavity;  { unlock cavity }

take_picture(0, plotimage);

lock_cavity;  { re-lock cavity}

Show( plotimage) ;

num_pixels := GetXSize(plotimage) * GetYSize(plotimage) ;

WriteLn('n_frames=', n_frames);

if save_images then 
image_array := CreateArray(GetXSize(plotimage), 
GetYSize(plotimage), n_frames);

WriteLn(FileExpand('background.tif'));
{ Halt; }

if use_background then 
 if not ImageExists(backgroundname) then 
 begin  
  if Query('No background: use this?')=id_Yes then
      begin
	   	   Save(plotimage, 'background.tif');
		   Open('background.tif', background_plot);
		   Show(background_plot, 'background');
		   Delete(plotimage);
	  end;
 	  Halt;
 end;

 {
if use_background = False then
    if not ImageExists(scatteringname) then Halt('No scattering file');
}
 
{ Get the active frame number }
Z1 := GetActiveFrame( plotimage ) ;

{ Create a plot window }
if PlotExists(rowplotname) then
   begin
   GetPlot(rowplotname, Plot0);
 {  Halt('ok'); }
   end
      
else   
  begin
   Plot0 := CreatePlot( rowplotname );
   SetTitle( Plot0,'Row Plot' ) ;
   SetSubTitle( Plot0,GetName( plotimage ) ) ;
   SetXLabel( Plot0,'X Position' ) ;
   SetYLabel( Plot0,'Mean Intensity' ) ;
   SetGrid( Plot0,gs_Both ) ;
  end;

{ Row plot }
SetPlotColor( col_Black ) ;
Plot( Plot0,MeanOverY( plotimage[..,..,Z1] ) ) ;


{ Create data plot window }
if PlotExists(dataplotname) then
   begin
   	GetPlot(dataplotname, Plot2);
	end
else
	Plot2 := CreatePlot(dataplotname);
		

{ Create a plot window }
if PlotExists(colplotname) then
   begin
   GetPlot(colplotname, Plot1);
 {  Halt('ok'); }
   end
else begin
	 Plot1 := CreatePlot( colplotname ) ;
	 SetTitle( Plot1,'Column Plot' ) ;
	 SetSubTitle( Plot1,GetName( plotimage ) ) ;
	 SetXLabel( Plot1,'Y Position' ) ;
	 SetYLabel( Plot1,'Mean Intensity' ) ;
	 SetGrid( Plot1,gs_Both ) ;
end;

{ Col plot }
SetPlotColor( col_Black ) ;
Plot( Plot1,MeanOverX( plotimage[..,..,Z1] ) ) ;

{ *** INITIALIZE SR400 PARAMETERS *** }
{ to be written }


{ ******************* MAIN LOOP BEGINS ************************ }

repeat

if(do_blocker) then
begin
SelectPort(2);
Transmit(StrCat('R ', Str(x)));  { SET BLOCKER POSITION }
Delay(1000.0*Abs(current_position - x)/400.0);  { Wait for it to move }
current_position := x;
end;

{ Turn on 553nm probe, turn off 791nm pump}
SelectPort(1);

Transmit(StrCat('pl 1, ', Str(v_t_553_on)));
Transmit(StrCat('pl 2, ', Str(v_t_791_off)));

unlock_cavity; {unlock}

{ Acquire data }
take_picture(0, plotimage);
Show( plotimage ) ;

lock_cavity; {relock + Turn off 553nm probe}

{ Turn off 553nm probe }
SelectPort(1);
Transmit(StrCat('pl 1, ', Str(v_t_553_off)));
{ Transmit(StrCat('pl 2, ', Str(v_t_791_on))); }

mox := MeanOverX( plotimage[..,..,Z1] );
{ plotimage[..,0..crop_colplot,..] := mox[0,crop_colplot,0]; }


GetImage( backgroundname, bgimage ) ;

{
GetImage( scatteringname, scatteringimage );
}

{ Perform an image subtraction }

if use_background then plotimage := plotimage - bgimage ;

if not use_background then
   begin
   Delay(lock_time);
   unlock_cavity; { unlock }
   take_picture(0, plotimage_no553);
   lock_cavity; { relock }
  { Halt('should be locked');}
   Delay(lock_time);
  { plotimage := plotimage - plotimage_no553 - scatteringimage;}
   plotimage := plotimage - plotimage_no553;
   end;
   
{ Rezero: subtract avg of first 10 pixels in col plot }

if rezero then begin
mox := MeanOverX( plotimage[..,..,Z1] );
moymox := MeanOverY(mox[..,0..10,..]);
plotimage := plotimage - moymox[0,0,0];
{ mox[0,crop_colplot,0]; }
end;

if(save_images) then
begin
image_array[..,..,index] := plotimage;  { store plotimage in array }
   if Save(image_array, image_array_filename) <> file_Ok 
      then Halt('Save error');
end;

{ Row plot }
ClearPlot(Plot0);
SetPlotColor( col_Black ) ;
moy := MeanOverY( plotimage[..,..,Z1] );

Plot( Plot0,moy ) ;
mirp := MaxOf(moy);

wherexmax := 0;
ixlimit := GetXSize(moy)-1;
for i := 0 to ixlimit	do
	begin
	if moy[i,0,0] = mirp
	   then wherexmax := i;
{	WriteLn(str(i));    }
    end;
{	
sum0 := 0; sum1 := 0; sum2 := 0;
for i :=0 to ixlimit do
	begin
	sum0 := sum0 + moy[i,0,0];
	sum1 := sum1 + moy[i,0,0] * i;
	sum2 := sum2 + moy[i,0,0] * i * i;	
	end;

WriteLn('x: avg=', sum1/sum0, ' sigma=', 
				 Sqrt(sum2/sum0 - sum1*sum1/(sum0*sum0)), ' sum=', sum0);
 			
WriteLn(x, ' ', sum0);
}
{ WriteLn(textwin, 'max at x=', wherexmax); }

{ WriteLn(textwin, MaxOf(moy[0..10])); }
total_counts_y := MeanOf(moy) * num_pixels;
SetTitle( Plot0, StrCat(StrCat('Row Plot, max=', Str(mirp)), 
		  StrCat(' at ', Str(wherexmax))) ) ;

{ Col plot }
ClearPlot(Plot1);
SetPlotColor( col_Black ) ;
mox := MeanOverX( plotimage[..,..,Z1] );
{ WriteLn(textwin, GetYSize(mox)); }
mox[..,0..crop_colplot,..] := mox[..,crop_colplot,..]  ;
{ if crop_colplot>0 then 
   for i := 1 to crop_colplot-1 do
   	    begin
		WriteLn(i);
   		mox[i] := mox[crop_colplot];
		end; }
		
Plot( Plot1,mox ) ;
micp := MaxOf(mox);
total_counts_x := MeanOf(mox) * num_pixels;

whereymax := 0;
iylimit := GetYSize(mox)-1;
for i := 0 to iylimit	do
	begin
	if mox[0,i,0] = micp
	   then whereymax := i;
{	WriteLn(str(i));    }
    end;

{ Find 1/3 max points }
	
closest := 10000;	
for ii := whereymax to iylimit do
    begin
	if Abs(mox[0,ii,0] - micp/3.0) < closest then 
		 begin
		 closest := Abs(mox[0,ii,0] - micp/3.0);
		 y_third_max_plus := ii;
		 end;
	end;

closest := 10000;	
for ii:=0 to whereymax do
    begin
	if Abs(mox[0,ii,0] - micp/3.0) < closest then 
		 begin
		 closest := Abs(mox[0,ii,0] - micp/3.0);
		 y_third_max_minus := ii;
		 end;
	end;	 
{
	WriteLn('y_half_max_plus ', str(y_third_max_plus));	
	WriteLn('y_half_max_minus ', str(y_third_max_minus));	
}

{ Gaussian fitting }

sigma:=0;
if do_gaussian_fit then begin

absmox := sqrt(sqr(mox))+0.01;
logmox := Ln(absmox);
fitmin := y_third_max_minus;
fitmax := y_third_max_plus;
line := MakeLinear(fitmin -0.0 , fitmax - 0.0, fitmax - fitmin + 1);
line100 := MakeLinear(0,iylimit+0.0,(iylimit+1.0));
a := PolyFit(line, logmox[0,fitmin..fitmax,0],2);

{ WriteLn(a);  }
{
WriteLn(str(line));
WriteLn(str(logmox[0,fitmin..fitmax,0]));
WriteLn(str(a[0,0,0]));
WriteLn(str(a[1,0,0]));
WriteLn(str(a[2,0,0]));
WriteLn(' '); }

{ logmoxregion := logmox[0,fitmin..fitmax,0];
WriteLn(logmoxregion);
WriteLn(str(logmoxregion[0,0,0])); }

{
  Plot3 := CreatePlot('logplot');
  Plot(Plot3, logmoxregion);
}
{  Plot(Plot3, line, Polyvalue(a, line)); }
  
sigma := Sqrt(Abs(-1.0 / (2.0 * a[2,0,0]))); 
{ SetTitle( Plot1, str(2*sigma)); }

WriteLn('gaussian_fit 2*sigma= ', str(2*sigma));

SetPlotStyle( ps_Line ) ;
SetPlotColor( col_Blue ) ;
PlotXY(Plot1, line100, Exp(Polyvalue(a, line100)));
SetPlotColor( col_Red ) ;
PlotXY(Plot1, line, Exp(Polyvalue(a, line)));
{SetPlotColor( col_Blue ) ;
PlotXY(Plot2, line100, Polyvalue(a, line100));}
end;

{ Take moments }
{
sum0 := 0; sum1 := 0; sum2 := 0;
for i :=0 to iylimit do
	begin
	sum0 := sum0 + mox[0,i,0];
	sum1 := sum1 + mox[0,i,0] * i;
	sum2 := sum2 + mox[0,i,0] * i * i;	
	end;

WriteLn('y: avg=', sum1/sum0, ' sigma=', 
				 Sqrt(sum2/sum0 - sum1*sum1/(sum0*sum0)), ' sum=', sum0);
}	
{ WriteLn(textwin, 'max at y=', whereymax); }

SetTitle( Plot1, StrCat(StrCat(StrCat(StrCat('y:Colplot, max=', Str(micp)), 
		  StrCat(' at ', Str(whereymax)) ), ' sigma*2= '), Str(2*sigma)) );

  	  signal := micp;

	  cts_per_ms := total_counts_x / exposure_time;

	  WriteLn('blocker=',x,' micp=', signal, ' total_cts_x=', total_counts_x, 
	  'cts_per_ms = ', cts_per_ms);
	

if do_dataplot then
	begin
	data_x[index+1,0,0] := x;
	data_y[index+1,0,0] := signal;
	data_x[index,0,0] :=x;
	data_y[index,0,0] := signal;

	ClearPlot(Plot2);
	data_x_temp := data_x[0..(index),0,0];
	data_y_temp := data_y[0..(index),0,0];

	{ WriteLn(data_x[0..index+1,0,0], data_y[0..index+1,0,0]); }
	PlotXY(Plot2, data_x_temp, data_y_temp);		  
	end;   

WriteLn('frame=', index, ' x=',x, ' micp=', micp, ' mirp=', mirp, ' tot_x=', total_counts_x, 
				 ' tot_y', total_counts_y);

{ ***** END OF FLUORESCENCE  ***** }				 
				 
{ ****** MICROLASER I ******* }

unlock_cavity; { unlock }
{ Delay(100); } { Wait a bit }

SelectPort(1);
Transmit(StrCat('pl 2, ', Str(v_t_791_on))); { turn on pump AFTER unlock }
		  
   SelectPort(1);
   RxFlush;
{   Transmit('CR');  }{ SR400 Counter Reset }
{  Transmit('FB');  }{ Start with continuous data output } 
   Delay(100);
   Transmit('QB');  { Read most recent complete data point }

	dataline := RxString;
	data_array[index,0] := val(dataline);

    data_bkgnd := 0;  { Enter correct value here }
	
   if (do_blackbody) then
      WriteLn(textwin, whereymax, ' ', data_array[index,0] - data_bkgnd)					 
   else 
	 WriteLn(textwin, index, ' ', x, ' ', micp, ' ', 
	 				  total_counts_x/exposure_time, ' ',
	 				  Str(data_array[index,0] - data_bkgnd));

{   Clear(plot2);
   Plot(plot2, data_array[1..index,0]);   }

{   if Save(data_array, datafilename) <> file_Ok 
      then Halt('Save error');    }

	lock_cavity; { relock }
	Delay(lock_time);
	unlock_cavity; { unlock, pump still on! }

{ *********** MICROLASER II ******** }

   SelectPort(1);
   RxFlush;
{   Transmit('CR');  }{ SR400 Counter Reset }
{  Transmit('FB');  }{ Start with continuous data output } 
   Delay(100);
   Transmit('QB');  { Read most recent complete data point }
 
	dataline := RxString;
	data_array2[index,0] := val(dataline);

    data_bkgnd := 0;  { Enter correct value here }																							   
																								   
if (do_blackbody) then
      WriteLn(textwin, whereymax, ' ', data_array[index,0] - data_bkgnd)					 
   else 
	 WriteLn(textwin, index, ' ', x, ' ', micp, ' ', 
	 				  total_counts_x/exposure_time, ' ',
	 				  data_array2[index,0] - data_bkgnd);

lock_cavity;
Delay(lock_time);

{
   Clear(plot2);
   Plot(plot2, data_array[1..index,0]);   }

{   if Save(data_array2, datafilename) <> file_Ok 
      then Halt('Save error');
}
  
    x := x + x_increment;
    index := index + 1; 

until x > x_stop;

{ *************** END OF MAIN LOOP ************** }

CloseSerial(1); CloseSerial(2);

end


\end{verbatim}
\end{singlespace}

\chapter{Standing-wave versus traveling-wave coupling}
\label{app-tw}

We now show that the peak atom-cavity coupling $g$ in the
traveling-wave case is equal to one-half the peak coupling of the
standing-wave case.

Consider one atom in a cavity, located along the cavity's axis
at $z=z(t)$ where
\be z(t) = z_0 + vt \ee
We can write the operator for the total electric field seen by the atom:
\be \hat{E} = {1 \over 2}\left\{ \sqrt{{2 \pi \hbar \omega} \over {V}}
\left[{\hat{a} e^{i(kz(t) - \omega t)} + \hat{a}^\dag e^{-i(kz(t) - \omega t)}}
\right] \right. \ee
\be
+ \left. { \sqrt{{2 \pi \hbar \omega} \over {V}}
\left[{\hat{a} e^{i(-kz(t) - \omega t)} + \hat{a}^\dag e^{-i(-kz(t) - \omega t)}}
\right]
}\right\}  \ee
Here $V = \pi L w_0^2 / 4$ is the mode volume, with $L$ the cavity mirror
spacing.  

Writing out $z(t)$, we can see the splitting of the two cavity
frequencies $\omega \pm kv$:
\be \hat{E} = {1 \over 2} \sqrt{{2 \pi \hbar \omega} \over {V}}
\left[{\hat{a} e^{i(kz_0 - (\omega + kv)t)} + \hat{a} e^{i(-kz_0 - 
(\omega-kv) t)} + \mathrm{c.~c.} }\right]
\ee
Suppose $v=0$.  Then
\be \hat{E} = {1 \over 2} \sqrt{{2 \pi \hbar \omega} \over {V}}
\left[{\hat{a} e^{i\omega t}a( e^{ikz_0} + e^{-ikz_0}) +
\mathrm{c.~c.} }\right]
\ee 
\be = \sqrt{{2 \pi \hbar \omega} \over {V}}
\left[{\hat{a} e^{i\omega t} + \hat{a}^\dag e^{-i \omega t}} \right]\cos(kz_0)
\ee 
This is the familiar standing wave field operator.

For $v \neq 0 $ we can set $z_0 = 0$ by redefining the time origin.
(This works as long as $kvt_{\rm int} > 2\pi$) 
\be \hat{E} = {1 \over 2}\sqrt{{2 \pi \hbar \omega} \over {V}}
\left[{\hat{a} e^{i(\omega + kv) t} + \hat{a}^\dag e^{-i (\omega+kv) t}} \right]
+ {1 \over 2} \sqrt{{2 \pi \hbar \omega} \over {V}}
\left[{\hat{a} e^{i(\omega - kv) t} + \hat{a}^\dag e^{-i (\omega-kv) t}} \right]
\ee 
Suppose that only one these terms is resonant.  Since $\hbar g =
\hat{\mu} \cdot \hat{E}$, we have shown explicitly that the
traveling-wave case has a maximum coupling one-half that of the
standing-wave case.


\end{document}